# Exploration of new superconductors and functional materials and fabrication of superconducting tapes and wires of iron pnictides

Hideo Hosono[1, 2, 3], Keiichi Tanabe[4], Eiji Takayama-Muromachi[5], Hiroshi Kageyama[6], Shoji Yamanaka[7], Hiroshi Kumakura[5], Minoru Nahara[8], Hidenori Hiramatsu[2, 3], and Satoru Fujitsu[3]

1. Frontier Research Center, Tokyo Institute of Technology, Yokohama 226-8503, Japan
2. Materials and Structures Laboratory, Tokyo Institute of Technology, Yokohama 226-8503, Japan
3. Materials Research Center for Element Strategy, Tokyo Institute of Technology, Yokohama 226-8503, Japan
4. Superconductivity Research Laboratory, International Superconductivity Technology Center (ISTEC), 2-11-19 Minowa-cho, Kohoku-ku, Yokohama, Kanagawa, 223-0051, Japan
5. National Institute for Materials Science, 1-2-1 Sengen, Tsukuba, Ibaraki 305-0047, Japan
6. Department of Energy & Hydrocarbon Chemistry, Graduate School of Engineering, Kyoto University, Nishikyo-ku, Kyoto, 615-8510, Japan
7. Department of Applied Chemistry, Graduate School of Engineering, Hiroshima University, Higashi-Hiroshima, 739-8527, Japan
8. Department of Physics, Okayama University, Okayama 700-8530, Japan

E-mail: hosono@msl.titech.ac.jp

**Abstract.** This paper reviews the highlights of a 4-years-long research project supported by the Japanese Government to explore new superconducting materials and relevant functional materials. The project found several tens of new superconductors by examining ~1000 materials, each of which was chosen by Japanese team member experts with a background in solid state chemistry. This review summarizes the major achievements of the project in newly found superconducting materials, and the wire and tape fabrication of iron-based superconductors. It is a unique feature of this review to incorporate a list of ~700 unsuccessful materials examined for superconductivity in the project. In addition, described are new functional materials and functionalities discovered during the project.

Contents

## 1. Introduction

It has been a dream for human beings to realize a room temperature superconductor since the discovery of superconductivity by Heike Kammerling Onnes in 1911 [1]. Although the fundamental theoretical framework for superconductivity was established in 1957 by BCS theory, there exists no theory which can quantitatively predict the critical temperature ($T_c$) even now [2]. Thus, exploring for high $T_c$ superconductors is like a voyage in a big ocean without a precise compass, *i.e.*, researchers have to move ahead believing their materials sense and/or intuition referring to what the theorists say. In this sense, exploration for high $T_c$ superconductors is a truly challenging subject in condensed matter research. Not an insignificant number of people say that this is typical "*all or nothing*" research.

It is a historical fact that materials leading to breakthroughs have been discovered in most cases by chance amidst concentrated research efforts undertaken with a unique but flexible view. This is particularly true for the exploration of new superconductors. The group of one of the present authors (HH) group discovered LaFePO in 2006 [3] and $LaFeAsO_{1-x}F_x$ in 2008 [4] through LaNiPO ($T_c$ =3 K) [5] in 2007 in the course of exploring magnetic semiconductors, which started from his extensive research of transparent p-type semiconductors LaCuO*Ch* (where *Ch*=S and Se) with the same crystal structure as the so-called 1111-type layered compounds. P-type conduction in LaCuO*Ch* originates from the mobile holes at the top of the valence band which is composed of Ch p orbitals and Cu 3d orbitals [6]. It was his idea for approach to novel magnetic semiconductors to utilize strong *d-p* interactions in LaCuO*Ch* by replacing nonmagnetic $Cu^+$ ion with a magnetic 3d transition metal cation with a +2 charge state. In order to keep electro-neutrality upon this substitution, $Ch^{-2}$ is required to be replaced by $Pn^{3-}$ [7]. This is the reason why his group started to examine electronic and magnetic properties of La*TM*O*Pn* (where *TM*=3*d* transition metal, *Pn*=P and As). This effort resulted in the discovery of IBScs through the concentrated effort for finding high performance p-type transparent semiconductors which is a branch of his research home ground, transparent oxide semiconductors [8].

The discovery of iron-based superconductors (IBSc) was accepted with surprise by the condensed matter community because iron with a large magnetic moment was widely believed to be most harmful to the emergence of superconductivity. Extensive research in these materials started globally, especially in China [7]. As a result, the discovery of IBSc was chosen as a breakthrough of the year 2008 by *Science Magazine* and the paper (*J. Am. Chem. Soc.* 2007) reporting $T_c$ =26 K in $LaFeAsO_{1-x}F_x$ became the most cited report among all the papers published in 2008.

In early 2009, the Japanese Government announced the launch of a new large funding program -FIRST (<u>F</u>unding Program for World-Leading <u>I</u>nnovative <u>R</u>&D on <u>S</u>cience and <u>T</u>echnology). The aim of the FIRST Program is to advance the kind of leading-edge research and development that will strengthen Japan's international competitiveness while contributing to society and people's welfare through the application of its results. Hideo Hosono's proposal "exploration for novel superconductors and relevant functional materials, and development of superconducting wires for industrial applications" was fortunately selected as one of 30 projects ranging a very broad area of science and technology out of ~800 applications.

It was his expectation to find novel functionalities and materials with high potential through this tough and really challenging work just as IBScs were found though exploration of magnetic semiconductors. Hideo Hosono organized the research team mainly composed of solid state chemists who have much experience and many achievements not only in superconductors, but also in the relevant functional materials fields. Since research in finding new superconductors typically belongs to the domain of condensed matter physics, this team organization is a unique feature of this project. It is his belief that excellent solid state chemists will find new properties by serendipity, even if they fail to succeed in the hunt for new high $T_c$ materials. This phylosophy was set at the beginning of the project, i.e., *"All or something"!*

This paper reviews the major research achievements obtained in our FIRST project (FIRST PJ) performed over the 4 years from March 2010 through March 2014, along with some background for the research. We have examined more than 1,000 materials to seek new superconductors. The fraction of success was relatively small (~3%), but is just as we expected at the outset. So far, unsuccessful results in this field have not been presented in an open an academic journal. In this review, we have listed the records of materials that we examined in this project, including the unsuccessful materials, based on a consensus among the members of the research team that the consideration of unsuccessful trials will be good fertilizer, leading to fertile crops in near future.

This review consists of 5 parts excluding the introduction. The parts are as follows: an overview, new iron-based superconductors, new Ti-based superconductors, new intercalation & cluster systems, thin films and wires of IBSc, new functionalities and materials, and a final perspective.

## 2. Overview

On the biggining of this project, Hosono laid down five research targets. These were: 1) The discovery of a new superconductor with $T_c$> 77 K, 2) The development of new superconductors with high performance, 3) the development of the related materials with outstanding functions, 4) The development of a meter class superconducting wire with $J_c$ >$10^5$ A/cm$^2$ based on IBSc or other novel materials, and 5) The production of prototype Josephson junction (JJ) and SQUID devices using IBSc thin film. This project consists of six research groups listed in Table 1. Four groups (HH, EM, HKa and SY) have concentrated on the exploration of new superconductors, and the other two groups (KT and HKu) have concentrated on the development of superconducting wire and tape. HH group collaborated with five groups in the fields of the exploration for superconductor (MN) and the discovery of catalysis using electride (MH, TS, AS and SK). The HKu group collaborated with YK in the field of superconducting wire. Each group has made an effort to achieve its purpose using its special skills.

Though the FIRST PJ could not obtain the superconductor with $T_c$ > 77 K (56-58 K in the maximum achieved), over 100 new superconductors have been developed and characterized as new type of IBSc (112) has been found, new dopant into IBSc ($H^-$) has been employed to induce superconductivity, intercalation type compounds have been found, cobalt based and titanium based superconductors have been found, etc. The exploration for new superconductors is the most important target of the FIRST PJ, and over 40 researchers in 4 groups have worked on this mission. They have examined more than 1000 materials to seek new superconductors. The number of new superconductors found is a relatively small percentage, as envisioned at the onset, which is part of the motivation for the extensive search. We believe that listing all materials examined, including both success and failure, is meaningful for the people who work in this field or will join this field in the future, and thus show these in Table 2. The details of some representative results will be described in Section 3.

This project achieved the production of superconducting wire and tape with $J_c$ >$10^5$ A/$cm^2$ by the PIT method and has developed efficient magnetic pinning center for thin film type wires and tapes. Furthermore, the FIRST PJ has succeeded in preparing JJ and the SQUID devices by using epitaxial thin film of IBSc, clarifying the physical properties of IBSc including small anisotropy and high durability in magnetic fields. The induction of metallic state from an insulating parent material of IBSc by the electrostatic method was also the result of the research on the IBSc thin film device. We will describe these details in Section 4.

The new functional materials developed in the project are rather diverse. The discovery of a highly efficient catalysis for ammonia synthesis, the $12CaO7Al_2O_3$ (C12A7) electride, is the most remarkable result [108], which has impact not only in the academic community, but also in industry. The discovery of the spontaneous decomposition of carbon dioxide gas on the C12A7 electride surface [109], the preparation of stable Perovskite titanium oxy-hydride [110], the development of bipolar oxide semiconductor and its complementary circuit device [111], the development of new class transparent oxide conductor, $SrGeO_3$ [94], the discovery of the first Slater insulator, $LiOsO_3$ which is a ferroelectric metal [106], the development of the material showing new type of giant magnetoresistance, $NaCr_2O_4$ [94], and the discovery of the 2 dimensional electride, $Ca_2N$ [83], are also representative results. It is interesting that some of these harvests have resulted in the exploration of new superconductors. These results are to be introduced briefly in Section 5.

This project has reported these results as more than 330 of original papers and numerous oral and poster Presentations including over 170 invited and plenary talks at international meetings. The researchers in the PJ have also applied for over 30 patents on the work performed.

## 3. New superconductors

In FIRST Project, four research groups have concentrated to explore and evaluate the novel superconductors, and contributed to the progress of the research of superconductivity through the finding many novel superconductors and phenomena. We will review these results.

### *3.1. Iron-based superconductors*

The history of Iron-based superconductors (IBSc) started from 2006 when LaFePO with $T_c$=5K was found by Hosono's group [3]. Only few researchers took notice on this new type of superconductor based on iron with a large magnetic moment. In 2008, the discovery of $LaFeAsO_{1-x}F_x$ with $T_c$=26K by the same group [4] rekindled global interest in this area and opened a new frontier of superconductivity. At the early stage, this superconductor family was called "Pnictide Superconductors". However, researchers now call them "Iron-based Superconductors" because several measurements and evaluations have clarified that they all have similar electronic structure where the 3*d* electrons derived from Fe ion dominate the Fermi level to play a primary role in superconductivity.

Magnetism had long been believed to be antagonistic for emergence of superconductivity. Thus, the use of the elements with large magnetic moment, typically Fe, Ni and Co, was intentionally avoided in the field of superconductivity. Hence, the discovery of a high-$T_c$ superconductor based on iron gave an impact on research in this field. Condensed matter scientists encountered a new frontier of superconducting materials.

Immediately after Hosono's group reported an increase in $T_c$ (to 43 K) for La FeAs $O_{0.89}F_{0.11}$ under high pressure [112], two groups in China [113-115] reported a higher-$T_c$ (=55 K) for $SmFeAsO_{1-x}F_x$ under an ambient pressure. By now, it is reported that the highest $T_c$ (=55-58 K) in non-cuprate bulk superconductors are observed for some IBScs as $SmFeAsO_{0.74}F_{0.26}$ [116], $SmFeAsO_{0.85}$ [117], $SmFeAsO_{0.8}H_{0.2}$ [9], and $Gd_{0.8}Th_{0.2}SmFeAsO$ [118].

In July 2008, Johrendt's group in Germany reported $(Ba,K)Fe_2As_2$ ($T_c$=38 K) [119]. Thank to the ease of growing its single crystal which has a lateral size of several millimeters using Sn or FeAs as a flux, its physical properties have been well elucidated. In addition, several types of IBScs have been found, and researchers have identified their superconducting properties from various aspects [120-127].

In this section, we review the recent progress of IBScs focusing on the results of this project.

*3.1.1. Features of IBScs*

*3.1.1.1. Crystal structure*

Although approximately 100 materials of IBScs have been reported, their parent materials may be classified into seven types in terms of crystal structures (Fig. 1). These materials contains a common structural unit of the Fe*Pn* (or Fe*Ch*) layer formed by the square net of $Fe^{2+}$ (as the formal charge) which is tetrahedrally coordinated by four 7conicity (*Pn*) and/or chalcogen (*Ch*) atoms (see Fig. 1(a)). Unlike cuprate superconductors, where the parent materials are Mott insulators, this layer shows metallic conductivity without doping. An insulating blocking layer composed of *M*, *M*O or *M*F etc. where *M* indicates a metallic element such as an alkali, alkaline earth, or rare earth metal that lies between Fe*Pn* (or Fe*Ch*) layers. Similar to cuprates, this layered structure provides quasi-two-dimensional carrier transport properties although the magnitude of anisotropy rather differs depending on the blocking layer. The local structure of the Fe*Pn* layer is affected directly by the atomic (or ionic) size of *M* because *M* elements in the blocking layer bond to Fe elements. The crystal structures and the brief introductions of the seven different parent materials for IBScs are described below.

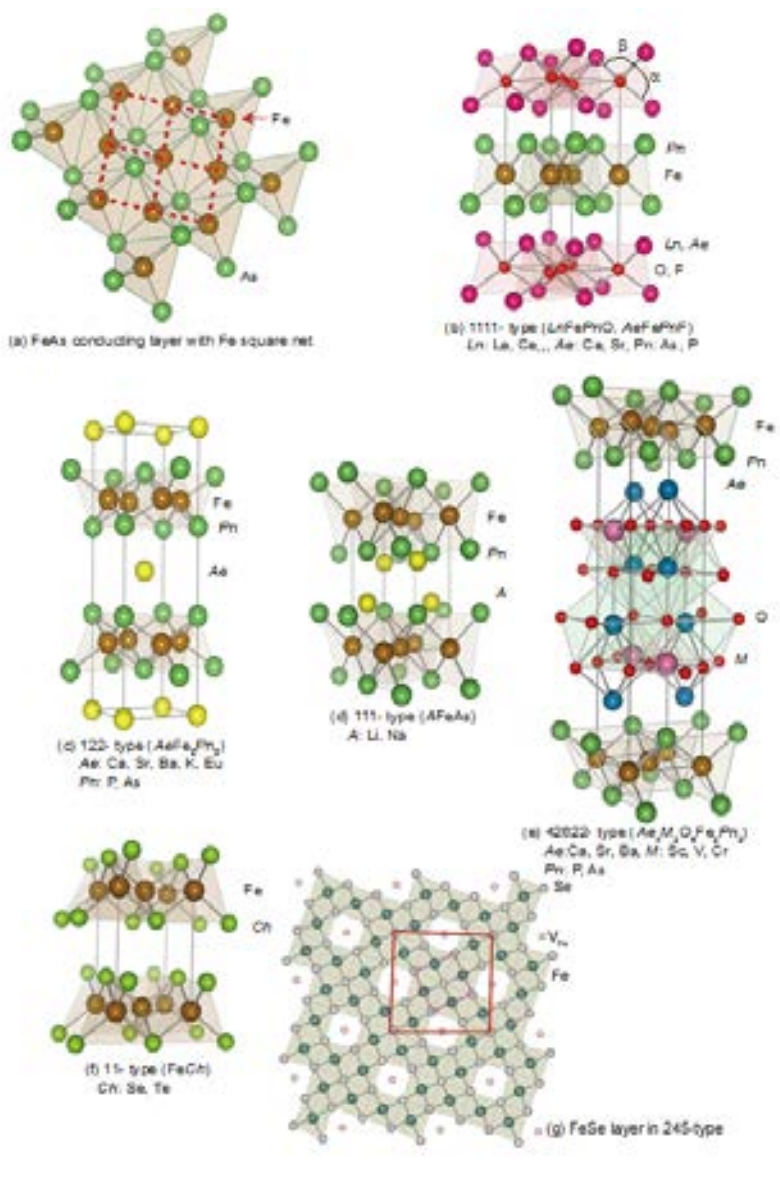

**Figure 1.** Crystal structures of iron based superconductors. a) structure of the FeAs conducting layer, which is common to all IBSc's. Dotted line indicates the Fe square net. b) 1111-type, c) 122-type, d) 111 type e) 42622-type and f) 11-type structures. g) structure of $K_{0.775}Fe_{1.613}Se_2$ from the [001] direction in the $5^{1/2}$x$5^{1/2}$x1) cell showing fully occupied Fe sites decorated with ordered vacancy sites. Dotted square indicates the basal plane of 122-type unit cell. [125]

*(i) 1111-type materials (LnFePnO, Ln: lanthanide, AeFeAsF, Ae: alkaline earth, Pn: P, As)*

1111-type compounds have the same structure as LaFeAsO, and is the prototype version of IBScs. Due to their atoms composition ratios, they are called the "1111- type". Fig. 1(b) show their crystal structure, which is a ZrCuSiAs-type structure [128, 129] with a tetragonal *P*4/*nmm* space group. Although LaFePO and LaFeAsO along with their crystal structures were identified by Zimmer *et al.* on 20 years ago [130], they were discovered to be superconductors in 2006 [3] and 2008 [4], respectively. Moreover, their two-dimensionality is relatively high among the seven types, and only this group has $T_c$ values above 50 K as a bulk form.

The 1111-type compound is composed of an alternating stack of positively charged *Ln*O layers and negatively charged FeAs (or FeP) layers along the c-axis. As mentioned above, the local structure of the FeAs layer is the same in all types of IBScs. The distance between the FeAs layers corresponds to the length of the *c*-axis (~0.8-0.9 nm). The formal valence state of each atom is $Ln^{3+}$, $Fe^{2+}$, $As^{3-}$, and $O^{2-}$. A $Fe^{2+}$ contains six electrons in its 3d orbital, and these electrons play an essential role in driving the superconductivity and magnetism. The electronic and magnetic properties of La$T_M$*Pn*O ($T_M$: 3*d* transition metal (Cr-Zn)) are summarized in Table 3 [120]. The 1111-type compounds composed of Fe and Ni reveal the superconductivity.

The lanthanide elements from La to Gd can occupy the *Ln* site for the 1111-type of material with *Pn*=P [130]. In the case of *Pn*=As, La to Ho and Y can also occupy the *Ln* site [136]. Additionally, $Ca(Fe_{1-x}Co_x)AsF$ ($T_c$=22 K) is a fluoride-containing superconductor of this type [137].

In this project, the effect of hydride ion as a new electron donor to this type was studied earnestly and its result is described in section 3.1.2.

*(ii) 122-type materials ($AEFe_2Pn_2$, AE: alkaline earth (alkali metal, Eu))*

122-type materials have a "$ThCr_2Si_2$" type crystal structure with a tetragonal *I*4/*mmm* space group [137]. This group contains the largest number of compounds among the 5 parent families.

In the case of $AEFe_2P_2$, not only alkaline earth elements but also lanthanides (La-Pr, Eu) can occupy the *AE* site. In $AeFe_2As_2$, the *Ae* site can be occupied by alkaline earth, alkali metal, or $Eu^{2+}$. Figure 1(c) shows the crystal structure of the 122-type. The layer composed by *An* ions, which is thinner than the *Ln*-O layer of the 1111-type, is sandwiched by the FeAs conducting layers. The distance between the FeAs layers of the 122-type (0.5-0.6 nm) is shorter than that of the 1111-type (0.8-0.9 nm). Because the nearest FeAs layers face each other with a mirror plane, the lattice parameter *c* is twice the FeAs-FeAs distance. The lattice parameter *a* (~0.4 nm) is almost the same as that of the 1111-type. Consequently, both 1111 and 122-type materials have similar Fe-Fe distance in the FeAs layer. Since single crystals of several millimeters can be obtained using Sn or FeAs as a flux, the physical properties of 122-type are well evaluated relative to other types of FeSCs. Johrendt's group of Germany was the first to report superconductivity for 122-type materials [119].

In this project, the lanthanide element doped 122 superconductors were prepared and evaluated for their bulk and thin film, and these results are described in section 3.1.3.

*(iii) 111-type materials (AFePn, A: alkali metal)*

While *AE* ion (alkaline earth ion with formal charge of 2+) is sandwiched by Fe*Pn* layers alternately in the 122-type, 111-type compounds contain two *A* ions (*A*: $Li^+$, $Na^+$) between Fe*Pn* layers in an unit cell. The crystal structure of this type is known as the "CeFeSi" type, with a tetragonal *P*4/*nmm* space group (Fig. 1(d)). This type is compatible with the structure of the 1111-type where all the oxygen atoms are removed, and the *Ln* site is occupied by $Li^+$ or $Na^+$. Wang et al. [139] ($T_c$=18 K: LiFeAs) and Parker et al. [140] ($T_c$=10 K: NaFeAs) first reported superconductivity for 111-type materials.

*(iv) Materials with thick blocking layer*

(32522-type ($AE_3M_2O_5Fe_2Pn_2$, *M*: Al, Sc))

(42622-type ($AE_4M_2O_6Fe_2Pn_2$, *M*: Sc, V, Cr))

(homologous type ($Ca_{n+1}Sc_nO_y$ $Fe_2As_2$: n=3, 4, 5))

The distance between the Fe*Pn*/*Ch* layers is in the order of the 1111, 122, 111 and 11-types. In contrast, these 3 types of iron oxy-pnictide have a thick blocking layer composed of a quasi-Perovskite structure assembled by $MO_5$ pyramids and *AE* (see Fig. 1(e) for the 42622-type ($Sr_4Sc_2O_6Fe_2As_2$). The FeAs-FeAs distance is 1.55 nm and 2.45 nm for $Sr_4Sc_2O_6Fe_2As_2$ and $Ca_6(Sc_{0.4}Ti_{0.6})_5O_yFe_2As_2$, respectively. Kishio-Shimoyama's group has studied this type of materials systematically [141-143]. The highest $T_c$ reported so far is 43 K [143]. Considering the thick blocking layer, this type should have the highest two-dimensionality, but the concrete value of anisotropic properties has not been reported yet because of difficulty of single crystal growth. The 32522-type has been proposed as a promised parent material [144, 145], and the emergence of superconductivity in the 32522-type was reported by Shirage et al. at 2011 for $(Ca_3Al_2O_{5-y})(Fe_2Pn_2)$ (*Pn*= As ($T_c$=30.2 K) and P ($T_c$=16.8 K)) [146].

*(v) Materials containing additional arsenic*

($Ca_{1-x}La_xFeAs_2$)

($Ca_{10}(M_4As_8)(Fe_2As_2)_5$, (*M*: Pr, Ir))

These new types of iron pnictide superconductors were found by Nohara's group of this project. The details for ($Ca_{1-x}La_xFeAs_2$) ($T_c$=43 K) and ($Ca_{10}(M_4As_8)(Fe_2As_2)_5$, (*M*: Pr, Ir)) ($T_c$=38K) are described on section 3.1.4 and 3.1.5, respectively.

*(vi) 11-type materials ($Fe_{1+x}Ch$, Ch: Se, Te)*

The 11-type crystal has the simplest structure among the parent compounds and is essentially the alkali metal-free 111-type. This crystal structure is known as "α-PbO" type with a tetragonal *P*4/*nmm* space group (Fig. 1(f)). A typical 11-type superconductor is β-FeSe ($T_c$=8 K) [147]. Medvedov et al. reported the 11-type may exhibit a high $T_c$ (=37 K) under 8.9 GPa [148].

Furthermore, FeSe attracts attention as one of the candidates showing higher $T_c$ than boiling temperature of Liq. $N_2$. Several groups in China reported that the mono layer of FeSe deposited on $SrTiO_3$ substrate showed high $T_c$ to be 65K in 2012 and they raised $T_c$ to 100K [149-153]. Though this superconductivity emerges for only mono layer of FeSe deposited on $SrTiO_3$ substrate so far, new route to high $T_c$ materials is expected to find.

*(vii) 245-type materials ($A_{1-x}Fe_{2-y}Se_2$: A=K, Cs, Rb, Tl)*

In 2010, Guo et al. reported a potassium-intercalated iron selenide superconductor with relatively high $T_c$ value (30 K) [154]. The crystal structure changed from 11 to quasi-122-type upon intercalation, of which the space group is assigned to *I*4/*m* due to vacancy ordering as shown in Fig.1(g). Though Guo et al. noted its chemical notation as $K_xFe_2Se_2$, the detailed chemical and structural analyses for its optimal material showed the composition to be $A_{0.8}Fe_{1.6}Se_2$ (=$A_2Fe_4Se_5$) and ordering of Fe vacancies with $\sqrt{5} \times \sqrt{5} \times 1$ supercell in the 122-type crystal structure [155]. This type of materials shows a wide range of non-stoichiometry and that with low Fe concentration is an antiferromagnetic insulator. The superconductivity in $A_xFe_ySe_2$ emerges in the proximity of an AFM Mott insulating state, similar to the cuprate high temperature superconductors [156]. Many unique properties let us to classify this as an independent type apart from the 122-type. Ivanovskii reviewed this material [157].

In this project, we intercalated Na to FeSe employing ammonothermal method, which cannot be prepared using a conventional thermal treatment at high temperatures. The result is described in section 3.1.6. Using this type of material with low Fe content is Mott insulator, we examined the effect of electrical field on its electrical transport properties. Its result is described in section 4.5.

*3.1.1.2. Electronic structure*

Figure 2 shows the photoemission spectra of LaFeAsO and $LaFeAsO_{0.94}F_{0.06}$, and calculated PDOS for Fe3*d* and As4*p* [158]. The Fermi level ($E_F$) controlling the transport property is primarily formed by a complex tangle of five Fe 3*d* orbitals, due to the small contribution of As, which is unlike cuprate superconductors where only Cu$d_{x2-y2}$ contributes to the $E_F$. With five bands comprising $E_F$, multi-pockets, i.e., disconnected Fermi surface (FS), appear on the FS. The energy levels of $d_{x2-y2}$, $d_{xy}$, and $d_{yz}$ are sensitive to both the changes in the symmetry of the $FeAs_4$ tetrahedron and the carrier density. Such an electronic structure dominates high $T_c$ with the unique pairing mechanism.

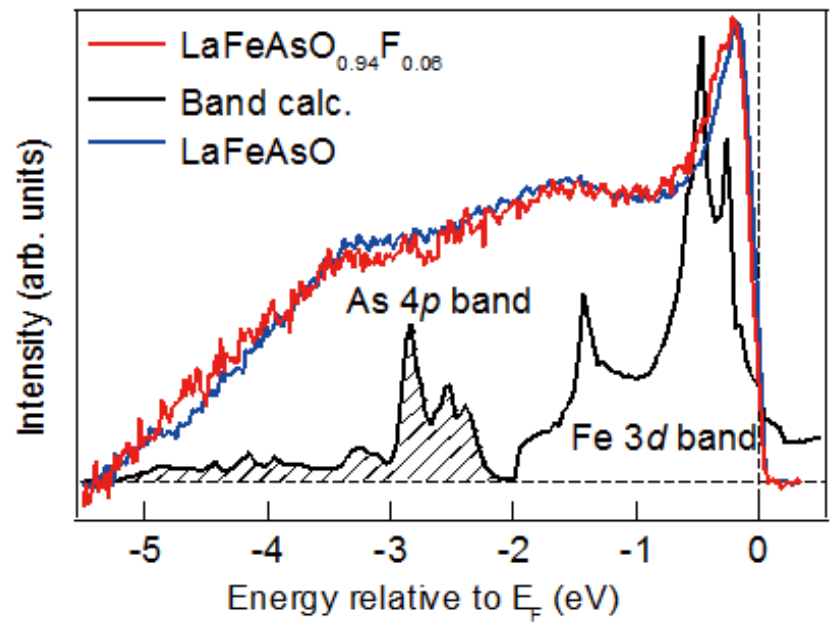

**Figure 2.** Photoemission spectra of $LaFeAsO_{1-x}F_x$ and calculated PDOS. The near-$E_F$ peak and the weak peak at about -1.5 eV corresponding to Fe 3*d* bands survive, and a broad peak corresponding to the As 3*p* band appears in the range -(3~4) eV [158].

At the early stage of the theoretical approach for the pairing mechanism, several physicists [159-166] suggested a possibility of spin fluctuation mediated pairing, where the spin fluctuation arises around the nesting vector (π, 0) (see Fig. 3 [160]). The spin fluctuation mediates $s_{\pm}$-wave pairing, where the gap function has s-wave symmetry, but its sign is reversed between the electron and hole Fermi surfaces.

In contrast, recent experimental results showed that high $T_c$ revealed when the nesting is degraded, or even in the absence of the nesting by heavily doping of impurities [11, 149, 151, 154, 167]. To explain the robust superconducting state against impurities, Kontani and Onari [168] proposed a mechanism of the $S_{++}$-wave superconducting state induced by orbital fluctuations, due to the phonon-mediated electron-electron interaction. On the other hand, Suzuki et al. [169] succeeded in reproducing the general trend of composition dependence of $T_c$ in *Ln*FeAs$O_{1-x}H_x$ (*Ln*: La, Ce, Sm and Gd) by diagonal (next nearest neighbor) electron hopping model where the next nearest neighbor (diagonal) hoppings between iron sites dominate over the nearest neighbor ones, plays an important role in the enhancement of the spin fluctuation and thus superconductivity. The theoretical and experimental evaluation for the superconducting mechanism will continue from now on.

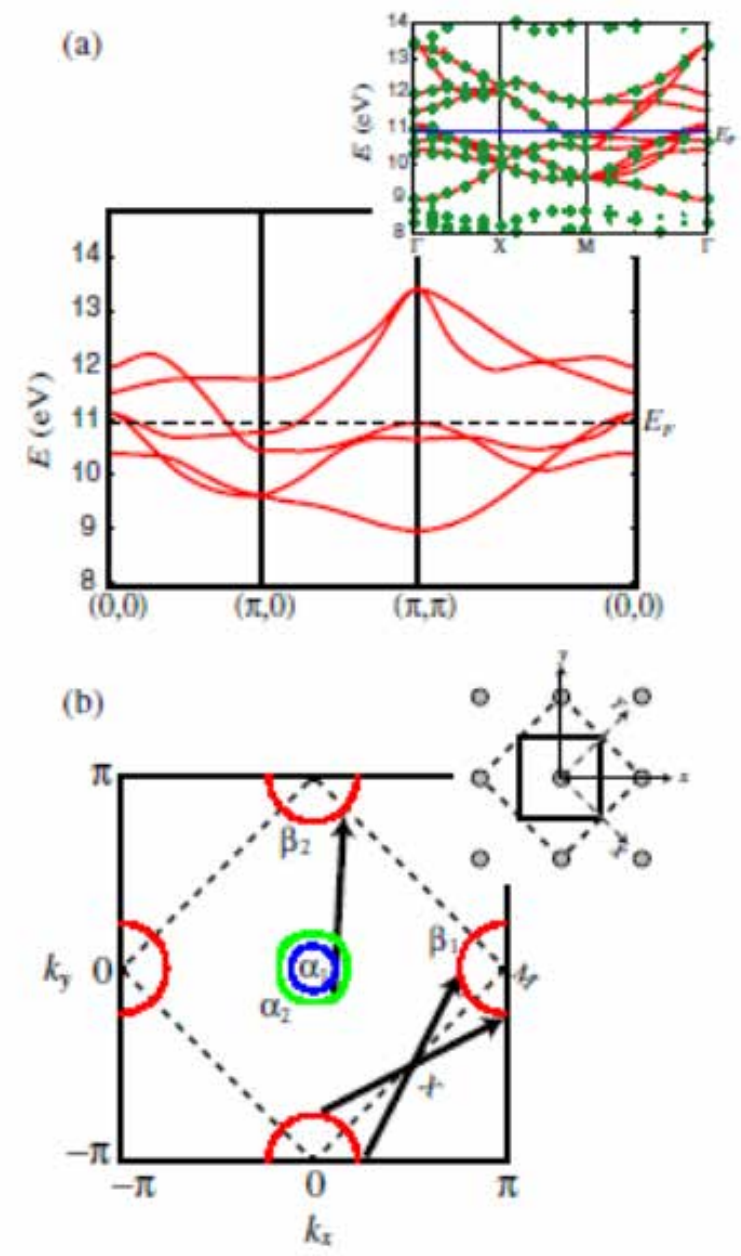

**Figure 3.** (a) The band structure of the five-band model in the unfolded BZ, where the interlayer hoppings are included. To compare with the ten-band model (thick red lines in the inset; the symbols are the present local-density approximation results), note the original (dashed lines) and the unfolded (solid lines) BZ shown in (b).
(b) Fermi surface for n = 6.1 (with the interlayer hoppings ignored), with the arrows indicating the nesting vectors. Inset depicts the original (dashed lines) and reduced (solid lines) unit cell in real space [160].

*3.1.1.3. Electronic phase diagram*

Unlike cuprate superconductors whose parent materials are Mott insulators, the parent materials of IBScs are antiferromagnetic metals with sufficient conduction carriers. Hence, it is considered that carrier doping into IBScs mainly alter the FS, which in turn leads to suppression of antiferromagnetism (AFM).

Here we mainly describe the electronic phase diagram for the 1111-type (*Ln*FeAsO, Ln: rare earth element). The parent materials for the 1111-type have tetragonal crystal structure at room temperature, but transform into orthorhombic structure at lower temperatures. In LaFeAsO, Pauli paramagnetism (PM) is shown around room temperature, and changes into AFM at a slightly lower temperatures ($T_N$~140 K) than that of the structural transitions ($T_s$~160 K) [170, 171].

Generally, the superconductivity occurs in the tetragonal phase and not in the orthorhombic phase due to antiferromagnetic ordering in the orthorhombic phase. With doping (e.g., substituting O for F in *Ln*FeAsO), the tetragonal–orthorhombic transition temperature decreases and is accompanied by the suppression of the AFM state and superconductivity emerges in succession. Electrons are doped into the bulk, when an element with more valence electrons is substituted. In contrast, holes are doped by substituting an element with fewer valence electrons. In many cases of both 1111 and 122-types, it is possible to substitute Fe or As in the conducting layer and *Ln*, *Ae*, O or F in the blocking layer for other elements. The former and the latter are called "direct doping" and "indirect doping", respectively.

The critical temperature ($T_c$) increases, reaches a maximum, and then decreases as the dopant levels increases. Since the decrease in $T_c$ in the over doping level is due to the precipitation of the secondary phase as SmOF in $SmFeAsO_{1-x}F_x$, the proposed phase diagram for 1111-type doped with F does not show the correct $T_c$ behavior in the over doping region [172-176]. In contrast, Hanna et al. [9] prepared $SmFeAsO_{1-x}H_x$ and showed its optimal $T_c$ (=55 K) at $x$=0.20 and decrease in $T_c$ by additional doping (over doping) without precipitation of secondary phase, indicating a wide superconducting dome in 1111-type. Figure 4(a) shows the schematic phase diagram for the 1111-type.

The first IBSc reported was formed by electron doping as $LaFeAsO_{1-x}F_x$ where F substituted the O site as ($F_O^{\bullet}$+e') [4]. In addition to the substitution of oxygen sites by F ($T_c$=55 K for $SmFeAsO_{0.9}F_{0.1}$) [115], various routes for the electron doping have been reported, including the formation of an oxygen vacancy ($V_O^{\bullet}$+e': $T_c$ =55 K for $SmFeAsO_{0.85}$ ) using a high-pressure synthesis [117, 177, 178] substitution of $H^-$ for an $O^{2-}$ site ($H_O^{\bullet}$+e': $T_c$=55 K for $SmFeAsO_{0.8}H_{0.2}$ ) [9], substitution of Th for *Ln* ($Th_{Ln}^{\bullet}$+e': $T_c$=56 K for $Gd_{0.8}Th_{0.2}FeAsO$) [118, 179], and substitution of Co, Ni or Ir for Fe ($Co_{Fe}^{\bullet}$+e': $T$c=14 K for $LaFe_{1-x}Co_xAsO$ [180, 181], $T_c$=22 K for $I_{1-x}Co_xAsF$ [137], $Ni_{Fe}^{\bullet\bullet}$+2e': $T_c$=6 K for $LaFe_{1-x}Ni_xAsO$ [182], $Ir_{Fe}^{\bullet}$+e': $T$c=18K for $SmFe_{1-x}Ir_xAsO$ [183]). The optimal amount of doping is 0.1-0.2/Fe atom for each route, and the indirect doping appears to be more effective than the direct doping in achieving a high $T_c$, which should be due to less structural perturbation to the conducting layer.

For the 122-type, the shape of the electronic phase diagram is similar to the 1111-type as general trend. The remarkable difference between the 1111- and the 122-types is whether AFM and superconducting phases are distinctly overlapped or not as shown in Fig. 4(b). In the 1111-type, the regions showing AFM and superconductivity are separated or barely overlap, whereas the 122-type materials have AFM regions with a high $T_c$. The optimal $T_c$ is apparently located around the temperature corresponding to the extrapolation of SDW curve to zero temperature, i.e., superconducting dome appears around the quantum critical temperature of SDW [184]. The emergence of SC by doping of isoelectronic dopant as P for As is also an unique property of the 122-type.

The comparison in doping between the 122- and the 1111-type is shown in Table 4.

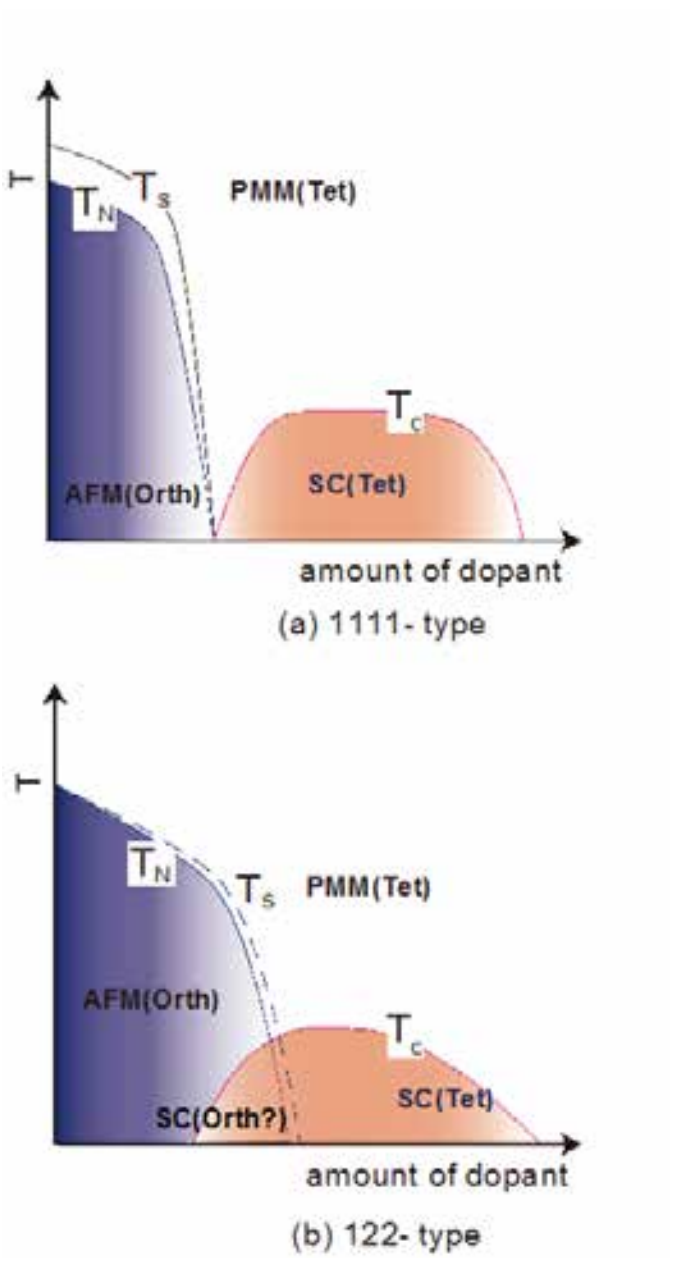

**Figure 4.** Schematic temperature versus composition phase diagram for (a) 1111-type and (b) 122-type ($T_s$: structural transition temperature; $T_N$: magnetic transition temperature; AFM: antiferromagnetic phase; PMM: paramagnetic metal phase; SC: superconducting phase) [125]

*3.1.1.4. Properties*

Compared to $MgB_2$ and cuprates, IBScs have several distinct characteristics. It is included in the unique characteristic of IBScs that the Fe 3d multi-orbital forms Fermi surface described in (b) and the parent material is the antiferromagnetic metal described in (c).

Generally, $T_c$ decreases upon doping with magnetic impurities such as Fe, Ni, and Co. In case of cuprate superconductors, $T_c$ of $YBa_2Cu_3O_{7-y}$ decreases from 90 K to 50 K by substituting Ni (17 %) for Cu, and that of $La_{1.85}Sr_{0.15}CuO_{4-y}$ also decreases from 40 K to 4.2 K by substituting Ni (5 %) for Cu [191]. The substitution of such elements for Fe on FeSCs with optimal state shows a similar trend. The superconductivity of $NdFeAsO_{0.89}F_{0.11}$ ($T_c$=48 K) disappears by substituting Co (>11 %) or Mn (>4 %) [192]. In contrast, the emergence of superconductivity by substitution of $Co^{2+}$ ($3d^7$), $Ni^{2+}$ ($3d^8$) or other transition metals for $Fe^{2+}$ ($3d^6$) in the non-superconducting parent materia1 described in (c) is also a unique nature for IBSc.

The high $T_c$, the large upper critical field ($H_{c2}$) and the small anisotropy are important merits to apply IBScs practically. Table 5 summarizes these values of IBScs along with those of $MgB_2$ and cuprates. Only cuptrates achieve the higher $T_c$ than the boiling point of liquid $N_2$ (77 K). It has been reported that the anisotropic ratio of the resistivity ($\gamma_\rho$) of the 122type of IBSc is compatible with that of $MgB_2$ and smaller than that of cuprates. The $H_{c2}(0)$ of IBSc is higher than that of $MgB_2$, but is smaller than that of a typical cuprate. The anisotropic ratio of the $H_{c2}$, $\gamma_H$, of IBSc is smaller than those of $MgB_2$ and cuprates. The $H_{c2}(0)$ is defined as the upper critical field at 0 K. The $\gamma_\rho$ means the ratio of the resistivity along the crystal axes directions, $a$ ($\rho(a)$) and $c$ ($\rho(c)$) measured just above $T_c$. The application of IBScs to superconducting wire and device will be described in Section 4.

*3.1.2. Hydrogen as electron donor*

The most effective route to achieve high $T_c$ is "indirect" and "electron" doping into the 1111-type (see Table 4). As mentioned in the previous section, the solubility limit of F which is used for "indirect" and "electron" dopant is 20% at most. Thus one could not know the effect on the emergence of SC by the impurity doping for its whole SC dome. So we have explored an appropriate dopant with high solubility limit.

Out of some candidates of monovalent anions ($X = X_O^{\bullet} + e$), we choose hydride ion ($H^-$) as an electron dopant which could substitute for $O^{2-}$ [9, 12]. Though it is considered generally that the ionic state of hydrogen is a proton ($H^+$) in condensed matters, it is not peculiar that hydrogen exists as hydride ion ($H^-$) which forms some stable materials as $LaH_2$, $CaH_2$ and NaH. Since its ionic radius varies to a large extent depending on the environment, *i.e.*, from 208 pm of Pauling's estimation, 129 pm of NaH to 106 pm of $CaH_2$. We considered that $H^-$ should be able to dissolve into the 1111-type superconductors with replacement of $O^{2-}$ because the blocking layer is composed of lanthanide cation which can form stable hydrides. To prevent the evaporation of hydrogen during the preparation process, we employed high pressure synthesis technique, *i.e.*, under 2 GPa at 1200°C. The experimental results showed clearly high solubility limit of $H^-$ doping. While the impurity phases as SmAs and/or SmOF precipitate in $SmFeAsO_{1-x}F_x$ ($x > 0.15$) [207, 208], such phases could not be observed in $SmFeAsO_{1-x}H_x$ ($x \leq 0.4$). Figure 5 compares the hydrogen content (x) and the deficient amount of oxygen ($y$) in the prepared samples per chemical formula ($SmFeAsO_{1-y}H_x$) as a function of nominal $x$ in the starting mixture [9]. The former value was determined by TG-MS, and the latter was measured using EPMA. For nominal $x \leq 0.4$, the hydrogen content agrees with $y$ and the nominal $x$, indicating the oxygen site ($O^{2-}$) was successfully substituted with hydrogen ($H^-$). Figure 6 shows the electronic phase diagram of $SmFeAsO_{1-x}H_x$ superimposed with that of $SmFeAsO_{1-x}F_x$ with the fluorine content x measured by EPMA as reported by Köhler and Behr [207] in which the $T_c$ vs x plots of $SmFeAsO_{1-x}H_x$ and $SmFeAsO_{1-x}F_x$ overlap at $x < 0.15$, indicating that hydrogen gives indirect electron doping to the FeAs layer just like fluorine. While the solubility limit of fluorine in the oxygen site is restricted to less than 20% ($x$ =0.2), [172, 207, 208] that of hydrogen can reach 40% for not only Sm-1111 but also all *Ln*-1111 compounds. The wider substitution range is useful for the optimization of the electron-doping level to induce superconductivity and to complete the electronic phase diagram, including the overdoped region.

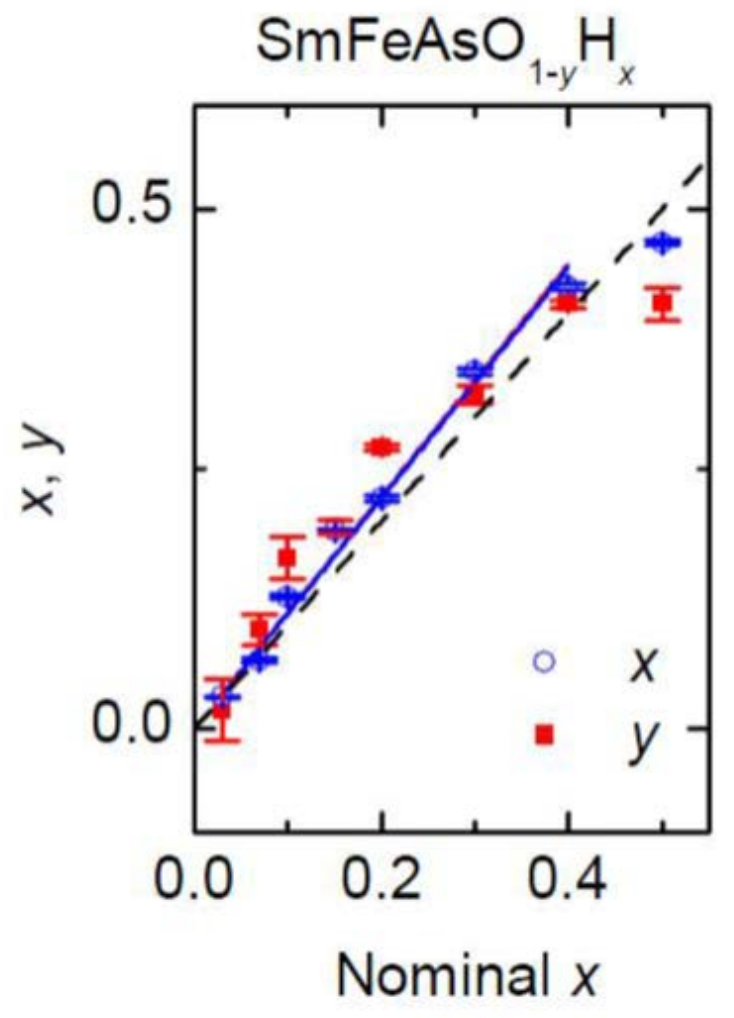

**Figure 5.** Oxygen deficiency content determined by EPMA measurement ($y$) and hydrogen content estimated by the TG-MS method ($x$) in $SmFeAsO_{1-y}H_x$ as a function of nominal $x$ in the starting mixture. The measured $x$ is almost equal to y and nominal $x$, indicating the deficiency of the oxygen site is wholly compensated by the occupation of hydrogen [9].

The most unique feature of revealed SC is observed in $LaFeAsO_{1-x}H_x$ [11]. While $T_c$-$x$ plots for $Ln$FeAs$O_{1-x}H_x$ ($Ln$=Ce, Sm, Gd) shows a single $T_c$ dome as shown in Fig. 6 (for Sm), $LaFeAsO_{1-x}H_x$ have two $T_c$ dome structure (see Fig. 7). When the amount of dopant ($x$) increased, the SC emerged from $x$ >0.04, $T_c$ increased and reached a maximum at $x$ =0.1 ($T_c$=26 K) and then decreased for a while. Surprisingly after reaching the minimum at $x$ =0.2, $T_c$ increased again and shows a broad peak (maximum $T_c$=36 K at $x$ =0.35), which cold not attained by F-doping. The two domes merged into a wider single dome with the optimal $T_c$ (=45 K at $x$=0.3) by applying high pressure of 3 GPa. The characteristics of these two $T_c$ domes in $LaFeAsO_{1-x}H_x$ are listed on Table 6.

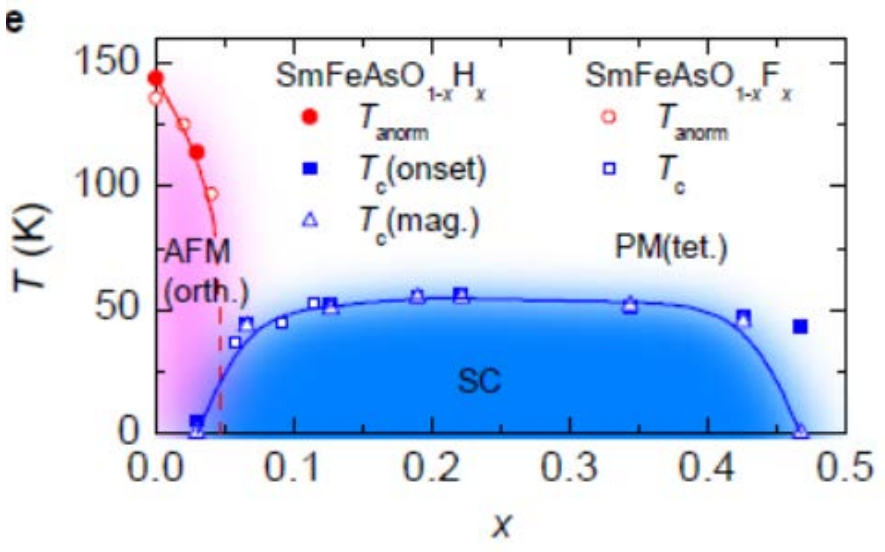

**Figure 6.** $x$-$T$ diagram of $SmFeAsO_{1-x}H_x$ superimposed by that of $SmFeAsO_{1-x}F_x$ [9].

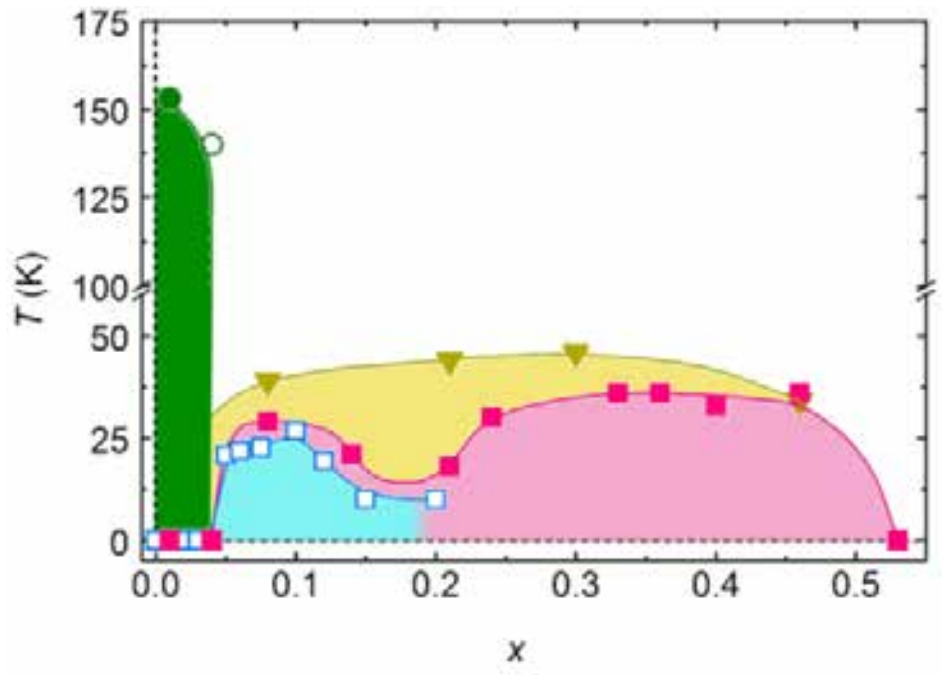

**Figure 7.** Electronic phase diagram for $LaFeAsO_{1-x}H_x$ (filled symbols) and $LaFeAsO_{1-x}F_x$ (open symbols). The $T_c$ under ambient (squares) and 3 GPa (inverted triangles) was determined from the intersection of the two extrapolated lines around superconducting transition and $T_s$ (circles) was taken as the anomaly kink in the ρ-$T$ curve [11].

Iimura *et al* who found this two dome structure considered initially that the superconductivity in low $x$ region (first dome) was due to the spin fluctuation mechanism and that in the high $x$ region (second dome) was due to the orbital fluctuation mechanism. Their density functional theory (DFT) calculations showed the strong Fermi surface (FS) nesting between the hole (at Γ point) and the electron (at M point) pockets at the low x region, which was the most important glue in the spin fluctuation model (see Fig. 8(a-h)). On the other hand, the FS nesting weakened with increasing of x, while the difference between the energy levels of Fe3$d$ bands ($3d_{xy}$, $3d_{yz}$, and $3d_{xx}$) decreased and became almost zero(degeneracy) at $x$ =0.30-0.35 where the maximum $T_c$ of the second dome occur (see Fig. 8(i)).

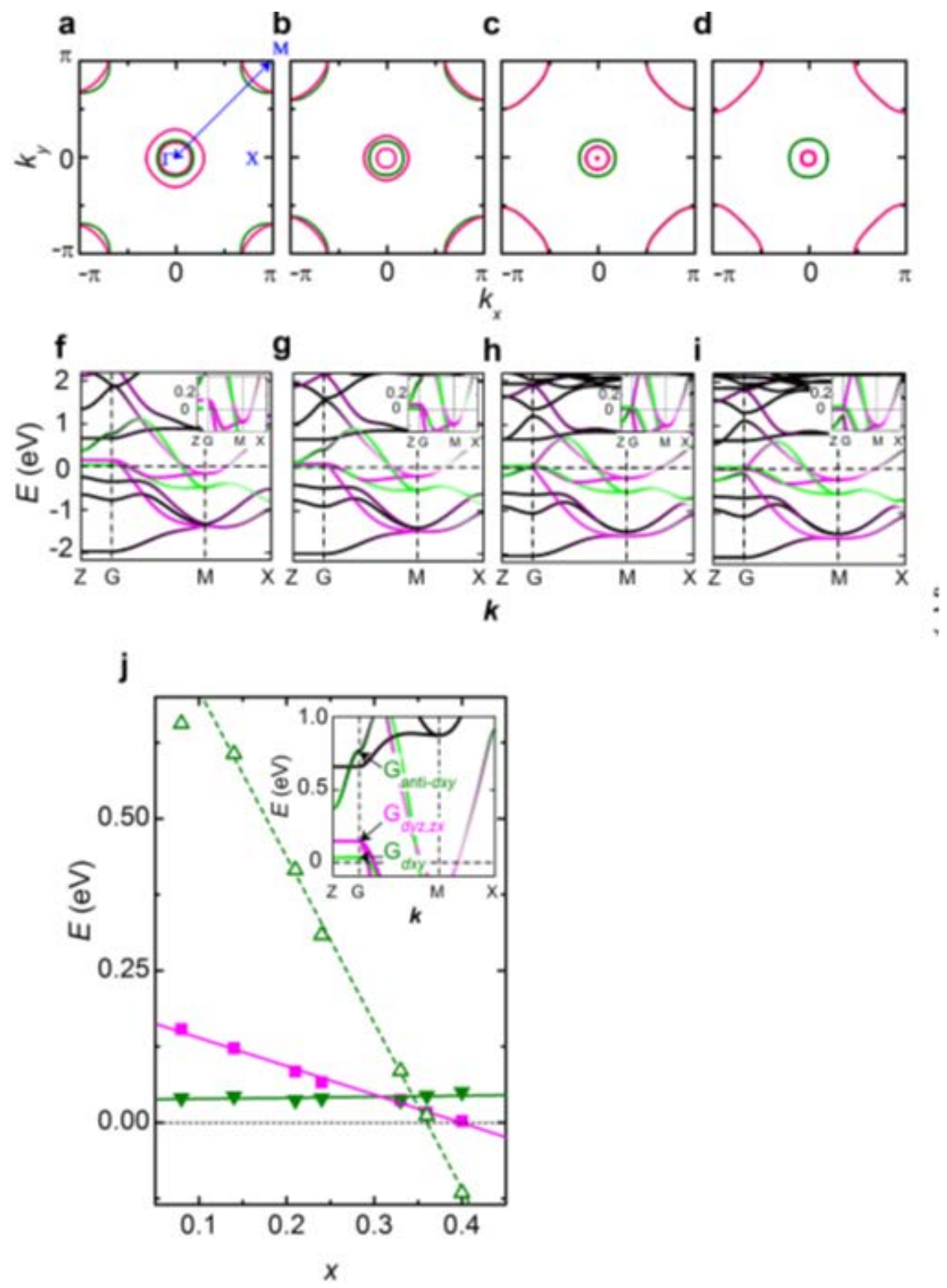

**Figure 8.** Electronic structure of $LaFeAsO_{1-x}H_x$. (a–d) Two-dimensional Fermi surface of $LaFeAsO_{1-x}H_x$ with x = 0.08 (a), 0.21 (b), 0.36 (c) and 0.40 (d). The blue arrow represents the nesting vector in the (π π) direction. The contribution of Fe-$d_{xy}$ and $d_{yz,zx}$ orbitals are colored green and pink, respectively. (f–i) Band structures of $LaFeAsO_{1-x}H_x$ with x = 0.08 (f), 0.21 (g), 0.36 (h) and 0.40 (i). Insets show close-up views of the low energy region. The contribution of Fe-$d_{xy}$ and $d_{yz,zx}$ orbitals are colored green and pink, respectively. (j) Variation in energy level of relevant Fe 3d bands at Γ point with x. The inset is the band structure of $LaFeAsO_{0.92}H_{0.08}$. The $\Gamma d_{xy}$ (filled green inverted triangles) and $\Gamma_{anti\text{-}dxy}$ (open green triangles) signify the bonding and anti-bonding states, respectively, for a bond primary composed of two Fe $d_{xy}$ orbitals in a unit cell. Also shown is the energy level of degenerate $d_{yx,zx}$ band ($\Gamma_{dyz,zx}$ indicated by filled pink squares). The solid and dashed lines are as a visual guide [11].

The importance of degeneracy in IBSc is deduced from the empirical plots reported by Lee et al., where the more the bonding angle of As-Fe-As (α: see Fig. 1(b)) is 109.5° (regular $FeAs_4$ tetrahedron) near, the higher $T_c$ reveals [209]. The regular $FeAs_4$ tetrahedron is achieved by the degeneracy of Fe3*d* bands and hence it should be reasonable to consider that the orbital fluctuation mechanism to mediate superconductivity emerges effectively in such a condition. The effect of the degeneracy of Fe3*d* bands for emergence of superconductivity is clear from the comparison of lanthanide cation substituted with La. The α's of the parent phases of $Ln$FeAs$O_{1-x}H_x$ are 114° (La), 112° (Ce), 111° (Sm) and 110° (Gd) and these values increases with doping of F or H. As shown in Fig. 9, the material of which α is close to 109.5° achieves the degeneracy of Fe3*d* bands and regular tetrahedral structure by small amount of doping and shows $T_c$ maximum simultaneously. From this viewpoint, Iimura et al. proposed that the high $T_c$ of 1111-type superconductors was due to the orbital fluctuation mechanism mainly.

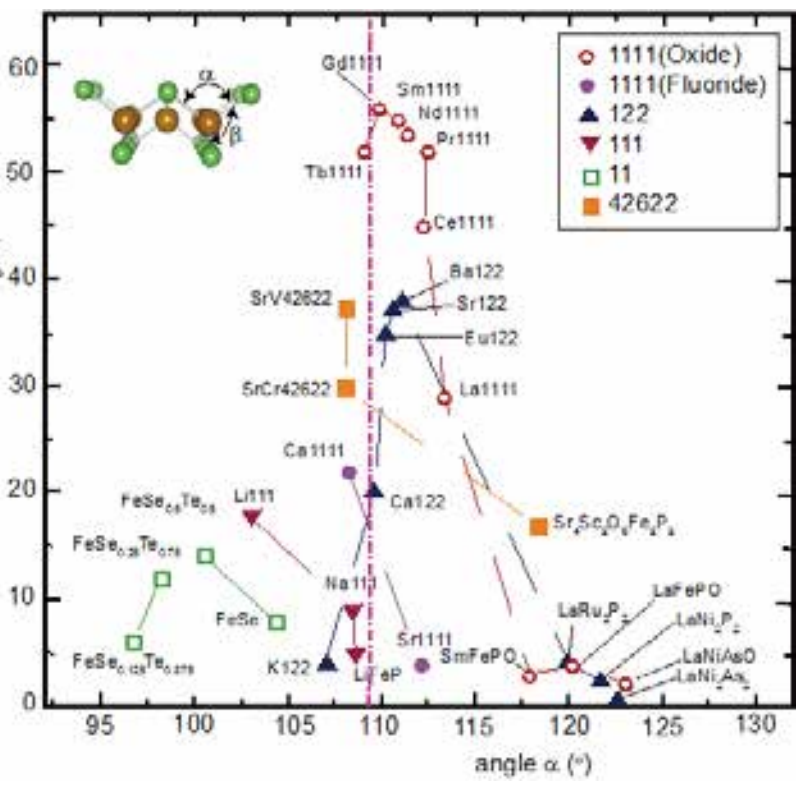

**Figure 9.** Correlation between $T_c$ and bonding angle a of *Pn(Ch)*-Fe-*Pn(Ch)*. α is adopted mainly from the data of the parent materials measured at room temperature. $T_c$ shows the highest reported value [125].

On the other hand, the experimental result using inelastic neutron scattering for $LaFeAsO_{1\text{-}x}D_x$ suggested the presence of the spin fluctuation in both low and high $x$ superconducting regions with different wave number [210]. Such spin fluctuation disappeared at the boundary of $x$=0.2. The relationships between $T_c$ and $E_R$ of several cuprate materials and IBScs containing these two regions are plotted in Fig. 10, where the $E_R$ is the measured value indicating the strength of spin fluctuation. For superconductors deriving from the spin fluctuation mechanism, it is known that the value of $E_R/k_BT_c$ is 4-6. As shown in figure, all plots lines up roughly on the straight line with the gradient of 5.7, which indicates that the spin fluctuation mechanism acts primarily to reveal the superconductivity in not only the low $x$ but also the high $x$ region. The theoretical calculations based on the random-phase approximation indicate that the spin fluctuations at x=0.1 are due to intra-orbital nesting within Fe3$d_{YZ,ZX}$, whereas, the spin fluctuations at $x$ =0.4 originate from intra-orbital nesting with in Fe3$d_{X2\text{-}Y2}$. These results suggest the orbital multiplicity plays an important role in the doping and/or material dependence of the $T_c$ of the IBScs.

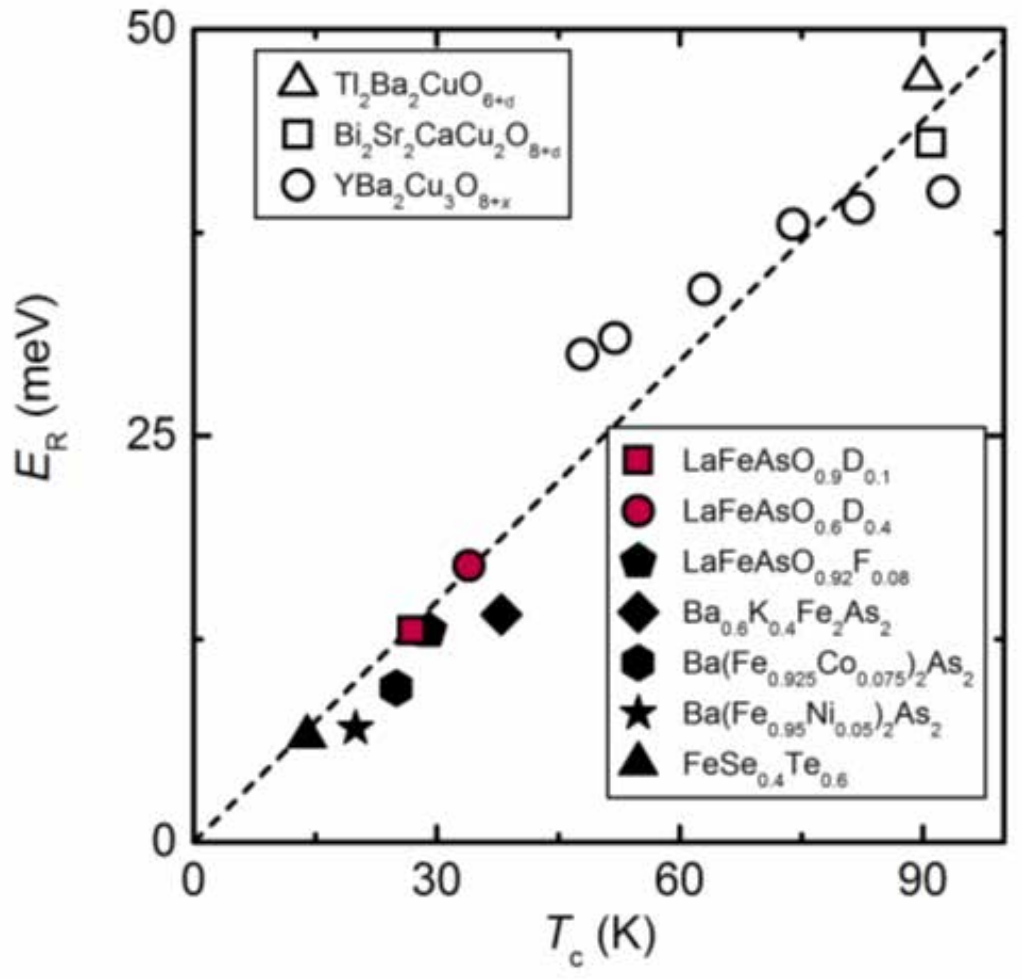

**Figure 10.** The $T_c$ dependence of ER in the iron pnictides (filled symbols) and cuprates (open symbols). The red square and circles are the present data for the samples with x = 0.1 and 0.4, respectively. The dashed line is the averaged slope of $5.7k_BT_c$ [210].

The discovery of two SC dome structure implies the existence of responsible two parent materials. Hiraishi *et al.* found a parent phase in the high x region with different magnetic structure from that on $x$=0 using multi-probe method composed by neutron, muon and synchrotron X-ray beams complementarily [211]. It was clarified by the muon spin relaxation (μSR) that new magnetic ordering phase developed with increasing dopant over $x$ ~0.4. From the experimental results using neutron and synchrotron X-ray diffraction, it was identified that this magnetic phase has a different antiferromagnetic ordering from that on $x$=0 and the structural transition occurs at $x$~0.5 where magnetic transition temperature shows maximum. Such features are shown in Fig. 11. This new magnetic phase is tentatively assigned to the parent phase in the high $x$ region.

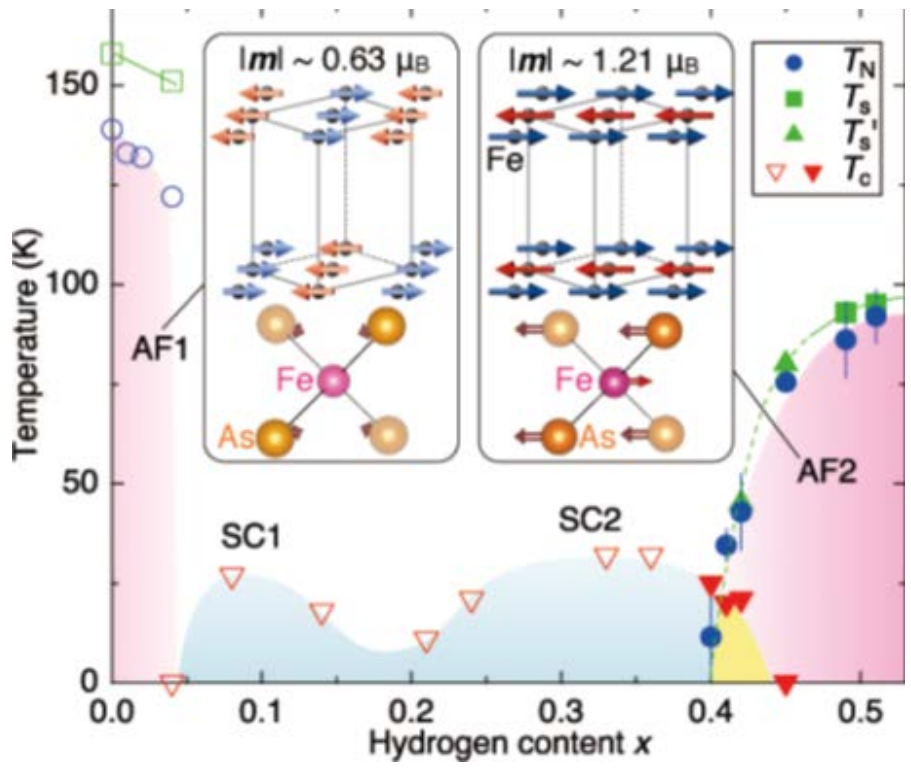

**Figure 11.** Magnetic, structural and superconducting phase diagram of $LaFeAsO_{1-x}H_x$. The original parent compound with $x$ = 0 exhibits a structural transition at $T_s$ =155 K, followed by an antiferromagnetic state (AF1) at $T_N$ = 137 K. With increasing $x$, two superconductivity domes appear: 0.05 ≤ $x$ ≤ 0.20 (SC1) with $T_{c,max}$ = 26 K, and 0.20 ≤ $x$ ≤ 0.42 (SC2) with $T_{c,max}$ = 36 K. Eventually, another antiferromagnetic phase (AF2) appears in the range 0.40 ≤ $x$ ≤ 0.51. In the advanced parent compound at $x$ = 0.51, structural and magnetic transitions occur at $T_s$ ~ 95 K and $T_N$ = 89 K, respectively. $T_s$' indicates the $c$ axis upturn temperature observed in X-ray measurements. The filled and open marks are obtained from the present and previous results, respectively. The magnetic structures of AF1 (left) and AF2 (right) are shown with their magnetic moments m, where the solid lines represent the tetragonal cell. The displacements of the Fe and As atoms across the structural transitions are schematically described by the arrows on the $FeAs_4$ tetrahedra from the view of the orthorhombic long axis, in which the Fe and As atoms move by 0.07 Å (0 Å) and 0.06 Å (0.01 Å) in $x$ = 0.51 ($x$ = 0), respectively. The error bars represent the uncertainty in the least-squares fitting routines [211].

In theoretical approach, Suzuki *et al.* successfully explained the composition dependence for $Ln$FeAsO$_{1-x}$H$_x$ ($Ln$: La, Ce, Sm, Gd) [169]. They showed that, besides the Fermi surface nesting, a peculiar motion of electrons, where the next nearest neighbor (diagonal) hopping between iron sites dominate over nearest neighbor ones, plays an important role in the enhancement of the spin fluctuation and thus superconductivity. According to their explanation, the crossover between the Fermi surface and this "prioritized diagonal motion" regime occurs smoothly with doping in the Ce, Sm and Gd case, while the two regimes are separated to give a double dome $T_c$ phase diagram in the La case. The feature of this change is schematically drawn in Fig.12.

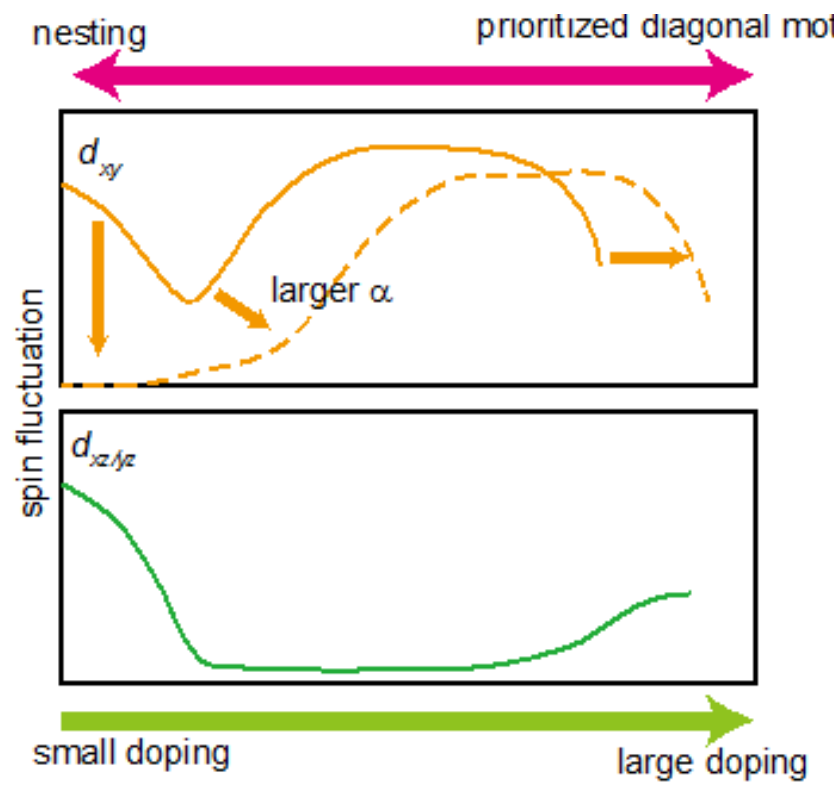

**Figure 12.** Schematic figure of the spin fluctuation contribution to superconductivity [169].

The hydrogen doping technique was applied to the alkaline earth metal 1111-type materials as CaFeAsF. It reveals superconductivity by replacing Fe to Co ($T_c$=26 K) [137]. Hanna et al. realized the electron doping by replacing F to H partially and heating them in He to selectively eliminate H as $CaFeAsF_{1-x}H_x$ → $CaFeAsF_{1-x}$+$x/2H_2$ [18]. The $T_c$ of 29K was attained by forming F vacancy (20% of F site). In the case of direct electron doping into CaFeAsH, the substitution of Fe with Co led this material to superconductor (max. $T_c$=29K) [16]. Indirect electron doping (substitution of Ca with La) showed higher $T_c$ than that of Co doping as shown in Fig. 13 [17]. The material obtained by substitution of 20 % of Ca with La showed the highest $T_c$ (47K) in IBScs without rare-earth element as a main component.

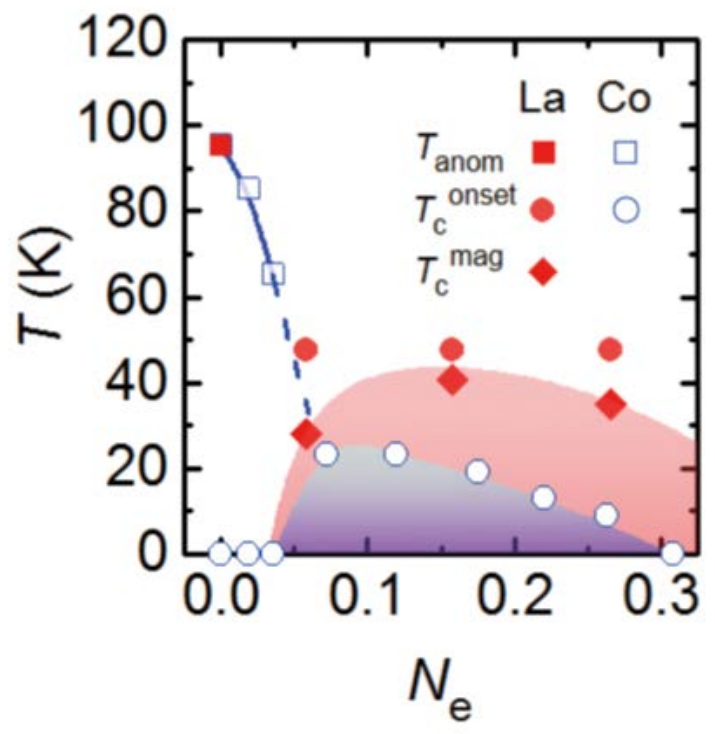

**Figure 13.** (a) Electronic phase diagram of $Ca_{1-x}La_xFeAsH_{1-y}O_y$ as a function of the total number of doped electron per iron ($N_e$=$x$-$y$), superimposed on that of $CaFe_{1-x}Co_xAsH$ [17].

*3.1.3. Rare-earth doped 122 iron arsenides*

The rare-earth (*RE*) doping of 122-type iron arsenides has been intensively studied as part of the FIRST PJ. The major achievements in this regard include the high-pressure synthesis of $(Sr_{1-x}La_x)Fe_2As_2$ with a maximum $T_c$ of 22 K [19] and the growth of thin films of $(Sr_{1-x}La_x)Fe_2As_2$ (maximum $T_c$ = 20.8 K) [22] and $(Ba_{1-x}La_x)Fe_2As_2$ (maximum $T_c$ = 22.4 K) by pulsed laser deposition (PLD) [20, 21]. Further, $(Ba_{1-x}RE_x)Fe_2As_2$ with *RE* = Ce, Pr and Nd has also been obtained by means of a nonequilibrium film growth process [21]. The success of *RE* doping enabled us to compare the phase diagrams of electron (La) and hole (K) doped $BaFe_2As_2$ as well as those of indirectly (La or K) and directly (Co) doped $BaFe_2As_2$. Another achievement is the simultaneous La and P doping of $CaFe_2As_2$ [23]: Melt-grown crystals of (Ca1–

$_x$La$_x$)Fe$_2$(As$_{1-y}$P$_y$)$_2$ exhibited bulk superconductivity with a maximum $T_c$ of 45 K for $0.12 \leq x \leq 0.18$ with $y$ = 0.06, while (Ca$_{1-x}$La$_x$)Fe$_2$As$_2$ without phosphorus exhibited filamentary superconductivity [23]. Characteristic two-dimensional Fermi surfaces were observed by means of angle-resolved photoemission spectroscopy (ARPES) in the 45 K phase of La- and P-doped CaFe$_2$As$_2$ [212].

To induce superconductivity in 122-type *AE*Fe$_2$As$_2$ (where *AE* = alkaline-earth elements), both types of doping carrier, i.e., holes and electrons, are typically used by substituting appropriate aliovalent elements, which include alkali metals (*A*) such as K that substitute for *AE* sites as in hole-doped (Ba$_{1-x}$K$_x$)Fe$_2$As$_2$ [119] and transition metals (*TM*) such as Co that substitute for Fe sites as in electron-doped Ba(Fe$_{1-x}$Co$_x$)$_2$As$_2$ [186, 193]. The doping can be classified into two types for the 122-type *AE*Fe$_2$As$_2$, namely, "indirect doping" for doping at sites other than the Fe sites and "direct doping" for doping at the Fe sites [21]. Because the superconducting FeAs and intermediary *AE* layers are spatially separated, direct doping has a major influence on carrier transport and thus superconductivity. Tables 7 and 8 summarize the maximum $T_c$ of directly doped *AE*(Fe$_{1-x}$*TM*$_x$)$_2$As$_2$ and indirectly doped (*AE*$_{1-x}$*A*$_x$)Fe$_2$As$_2$, respectively. The indirectly hole-doped (Ba$_{1-x}$K$_x$)Fe$_2$As$_2$ exhibits a maximum $T_c$ of 38 K, which is considerably higher than those for directly electron-doped AE(Fe$_{1-x}$TM$_x$)$_2$As$_2$. Therefore, we expected that a new indirect "electron" doping at the AE sites for AEFe$_2$As$_2$ would lead to high- $T_c$ superconductivity, as expected from the markedly higher $T_c$ observed for indirectly electron-doped SmFeAs(O$_{1-x}$F$_x$) (55 K) [115] than that for directly electron-doped Sm(Fe$_{1-x}$Co$_x$)AsO (17 K) [232].

However, indirect electron doping of Sr(Ba)Fe$_2$As$_2$ by substituting the divalent Sr(Ba) sites with trivalent *RE* ions was difficult to perform by means of conventional solid-state reactions. While Muraba *et al* [19] and Wu *et al* [233] examined the La substitution for SrFe$_2$As$_2$ and BaFe$_2$As$_2$, respectively, solid-state reactions of the ingredient mixture for (Sr$_{1-x}$La$_x$)Fe$_2$As$_2$ or (Ba$_{1-x}$La$_x$)Fe$_2$As$_2$ did not yield the La-substituted 122 phase upon using the conventional glass-tube technique.

In contrast, the indirect RE doping of CaFe$_2$As$_2$ was possible by a conventional melt-growth technique for *RE* = La, Ce, Pr and Nd [234, 235, 236, 237]. These materials exhibited superconductivity at $T_c$ = 40–49 K: Saha *et al* [234] reported $T_c$ = 47 K in (Ca$_{1-x}$Pr$_x$)Fe$_2$As$_2$; Gao *et al* [235] reported $T_c$ = 42.7 K in (Ca$_{1-x}$La$_x$)Fe$_2$As$_2$; and Lv *et al* [236] reported $T_c$ = 49 K in (Ca$_{1-x}$Pr$_x$)Fe$_2$As$_2$. These values of $T_c$ are considerably higher than those reported for directly electron-doped Ca(Fe$_{1-x}$*TM*$_x$)$_2$As$_2$ and indirectly hole-doped (Ca$_{1-x}$*A*$_x$)Fe$_2$As$_2$ listed in Tables 7 and 8, respectively. However, the shielding volume fractions of *RE*-doped CaFe$_2$As$_2$ were as low as <1% at 40 K [234, 235] or the shielding signal around 40 K was completely suppressed by the application of a weak magnetic field of 20 Oe [236].

Under such circumstances, Muraba *et al* have succeeded in the indirect La doping of SrFe$_2$As$_2$ by applying a high-pressure synthesis process to obtain (Sr$_{1-x}$La$_x$)Fe$_2$As$_2$ polycrystals [19]. The ionic radius of La$^{3+}$ (116 pm) is smaller than that of Sr$^{2+}$ (126 pm) and therefore, it is natural that a high-pressure synthesis is effective for obtaining La-substituted SrFe$_2$As$_2$. Muraba *et al* used a belt-type anvil cell to generate a pressure of 2 or 3 GPa at 1000 ºC for 2 h and obtained solid solutions of (Sr$_{1-x}$La$_x$)Fe$_2$As$_2$ for $0 \leq x \leq 0.5$. Further, (Sr$_{1-x}$La$_x$)Fe$_2$As$_2$ exhibited bulk superconductivity over a narrow range around $x$ = 0.4. The almost $x$-independent $T_c$ (shown in Fig. 14) suggested the inhomogeneous replacement of La dopants at the Sr sites. Figure 14 compares the electronic phase diagram of (Sr$_{1-x}$La$_x$)Fe$_2$As$_2$ [19] with those of Sr(Fe$_{1-x}$Co$_x$)$_2$As$_2$ [215] and (Sr$_{1-x}$K$_x$)Fe$_2$As$_2$ [229]. Here, the doped carrier number per Fe, i.e., $\Delta n$/Fe = $x$/2 for (Sr$_{1-x}$La$_x$)Fe$_2$As$_2$ and $(Sr_{1-x}La_x)Fe_2As_2$ and $\Delta n$/Fe = $x$ for $Sr(Fe_{1-x}Co_x)_2As_2$, is plotted in place of the doping composition $x$. The directly electron-doped $Sr(Fe_{1-x}Co_x)_2As_2$ exhibits superconductivity for $0.1 \leq \Delta n/\mathrm{Fe} \leq 0.2$ with a maximum $T_c$ of 19 K at $\Delta n$/Fe ~0.1, at which point the antiferromagnetic ordering vanishes [215]. The maximum $T_c$ of the indirectly electron-doped $(Sr_{1-x}La_x)Fe_2As_2$ is slightly higher than but close to that of directly electron-doped $Sr(Fe_{1-x}Co_x)_2As_2$. These results contrast markedly with the observed higher $T_c$ and wider superconducting range of the direct hole-doped system $(Sr_{1-x}K_x)Fe_2As_2$ (37 K) [228, 229], shown in Fig. 14, and $(Sr_{1-x}Cs_x)Fe_2As_2$ (37 K) [228]. The superconducting dome continues to the end member of this family, i.e., $KFe_2As_2$ [228, 229] and $CsFe_2As_2$ [228] at $\Delta n$/Fe = 0.5.

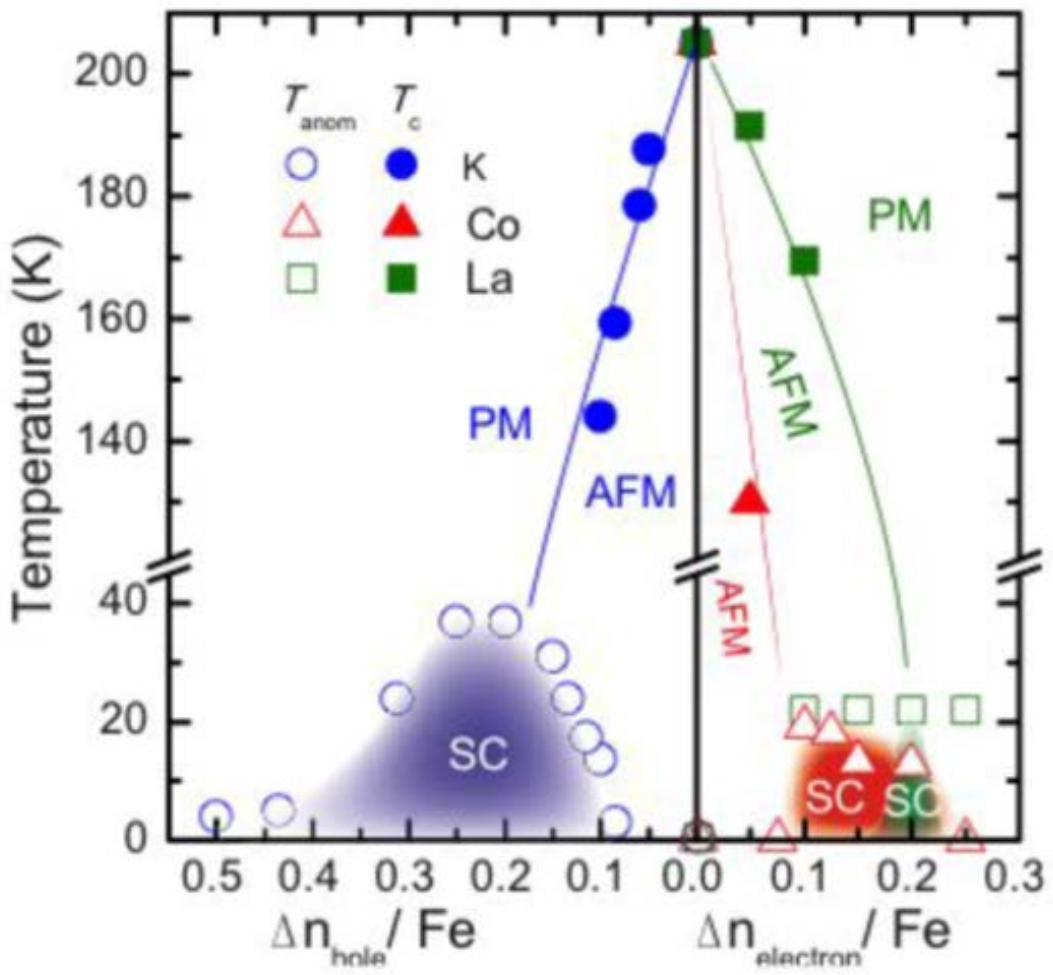

**Figure 14.** Electronic phase diagrams for electron-doped $(Sr_{1-x}La_x)Fe_2As_2$ [19] and $Sr(Fe_{1-x}Co_x)_2As_2$ [215] and for hole-doped $(Sr_{1-x}K_x)Fe_2As_2$ [229]. The ratio $\Delta n_{electron}$/Fe and $\Delta n_{hole}$/Fe denote the injected number of electrons and holes per Fe atom, respectively. Reprinted from [19].

Another promising technique that we have developed for the indirect *RE* doping of $SrFe_2As_2$ and $BaFe_2As_2$ is the non-equilibrium pulsed laser deposition (PLD) method [20-22]. Hiramatsu *et al* [22] and Katase *el al* [20, 21] have succeeded in the homogeneous doping of RE dopants in the films fabricated by PLD, and they have reported observing a superconducting dome from underdoped to overdoped regions irrespective of the largely different ionic radii of $Ba^{2+}$ (142 pm), $Sr^{2+}$ (126 pm) and $La^{3+}$ (116 pm). Here, we mention that $CaFe_2As_2$ epitaxial films could not be obtained by PLD [238]. Thin films of $(Sr_{1-x}La_x)Fe_2As_2$ (thickness = 200 nm) were grown on (La,Sr)(Al,Ta)$O_3$ (LSAT) (001) single crystals at a film-growth temperature of 750 ºC [22], while thin films of $(Ba_{1-x}RE_x)Fe_2As_2$ (thickness = 150–250 nm) were grown on MgO (001) single crystals at an optimized film-growth temperature of 850 ºC [20, 21]. Further, $(Sr_{1-x}La_x)Fe_2As_2$ thin films were successfully obtained for $0.0 \leq x \leq 0.48$ and $(Ba_{1-x}La_x)Fe_2As_2$ for $0.0 \leq x \leq 0.44$. The lattice parameters showed monotonic decrease with increasing $x$ (La content), thereby indicating that the substitution of the $La^{3+}$ ion for the $Sr^{2+}$ and $Ba^{2+}$ ions was achieved in the epitaxial films [20, 22]. Figure 15 shows the electronic phase diagrams of $(Sr_{1-x}La_x)Fe_2As_2$ [22] and $(Ba_{1-x}La_x)Fe_2As_2$ [20] epitaxial films. Here, the doped carriers per Fe (= $x$/2) for these two compounds are plotted for comparison with the phase diagram of the directly electron-doped $Sr(Fe_{1-x}Co_x)_2As_2$ [215]. The antiferromagnetic ordering at $T_{anom}$, which is determined by the anomaly in resistivity, is suppressed as the La content ($x$) increases. The maximum onset superconducting transition temperatures $T_c^{onset}$ are 20.8 and 22.4 K at $x/2$ = 0.16 and 0.07 for $(Sr_{1-x}La_x)Fe_2As_2$ [22] and $(Ba_{1-x}La_x)Fe_2As_2$ [20], respectively. The maximum values of $T_c^{onset}$ and the corresponding $x$ values of $(Sr_{1-x}La_x)Fe_2As_2$ and $(Ba_{1-x}La_x)Fe_2As_2$ are very close to those reported for directly doped $Sr(Fe_{1-x}Co_x)_2As_2$ [215] (shown in Fig. 15) and $Ba(Fe_{1-x}Co_x)_2As_2$ [193] (not shown), respectively. In addition, the suppression rate of $T_{anom}$ is similar for both compounds. These results sharply contrast with those of the 1111-type system, in which the maximum $T_c$ of indirectly electron doped $SmFeAs(O_{1-x}F_x)$ (55 K) [115] is considerably higher than that for direct electron-doped $Sm(Fe_{1-x}Co_x)AsO$ (17 K) [232]. These results can yield a clue to elucidate the mechanism of superconductivity in iron-based materials.

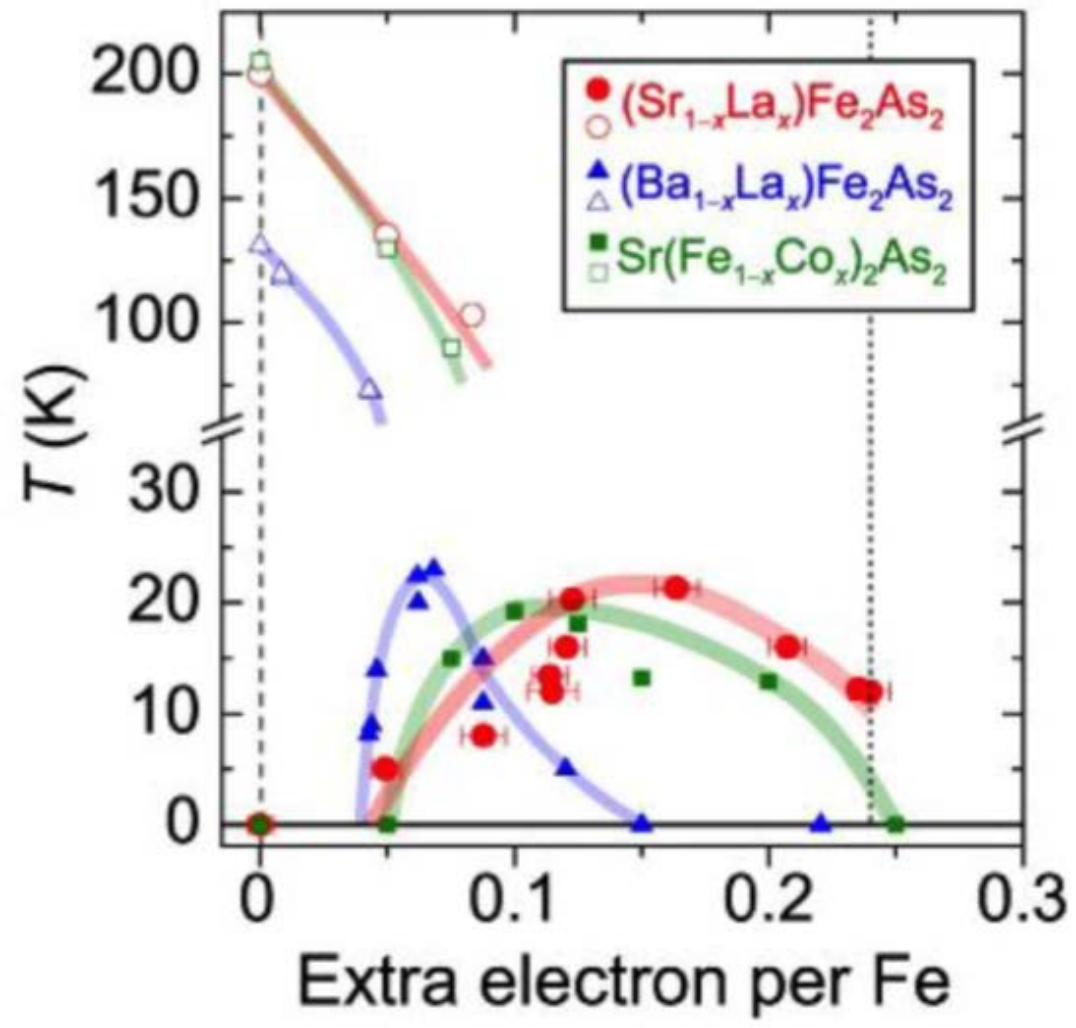

**Figure 15.** Electronic phase diagrams of $(Sr_{1-x}La_x)Fe_2As_2$ [22] and $(Ba_{1-x}La_x)Fe_2As_2$ [20] epitaxial films. The $T_{anom}$ and $T_c^{onset}$ values are indicated by open and closed symbols, respectively. Those of $Sr(Fe_{1-x}Co_x)_2As_2$ [215] are shown for comparison. Reprinted from [22].

Katase *et al* [21] have also succeeded in obtaining $(Ba_{1-x}RE_x)Fe_2As_2$ thin films with RE = Ce, Pr and Nd with reduced solubility limits of x = 0.29, 0.18 and 0.13, respectively. Thin films of $(Ba_{1-x}RE_x)Fe_2As_2$ exhibited superconductivity at 13.4, 6.2 and 5.8 K for Ce, Pr and Nd dopants, respectively [21]. The $T_c$ values of RE-doped $SrFe_2As_2$ and $BaFe_2As_2$ are listed in Table 9.

A conventional melt-growth method can be used to achieve indirect *RE* doping of $CaFe_2As_2$, although the obtained samples do not exhibit bulk superconductivity. In this project, two kinds of indirect doping, i.e., aliovalent La doping at the Ca sites and isovalent P doping at the As sites were combined to induce bulk superconductivity in $CaFe_2As_2$ [23]. The substitution of aliovalent La for Ca resulted in electron doping without leading to change in the lattice parameters because the ionic radius of $La^{3+}$ (116 pm) and $Ca^{2+}$ (112 pm) are similar, while the substitution of isovalent P for As resulted in a decrease in the lattice parameters without leading to change in the number of carriers. This enabled us to tune the number of charge carriers and lattice parameters independently, and thus to optimize superconductivity in $(Ca_{1-x}La_x)Fe_2(As_{1-y}P_y)_2$, which resulted in bulk superconductivity at $x = 0.17$ and $y = 0.06$. The resistivity shows a sharp drop at 48 K and becomes zero at 45 K [23], as shown in Fig. 16. A clear diamagnetic signal, together with robust diamagnetism against the increase in magnetic field, is the reason for the bulk superconductivity at 45 K in this material [23]. Figure 17 shows the $T$–$x$–$y$ phase diagram of $(Ca_{1-x}La_x)Fe_2(As_{1-y}P_y)_2$ [23]. Bulk superconductivity emerges in the range of $0.12 \leq x \leq 0.18$ and $y = 0.06$.

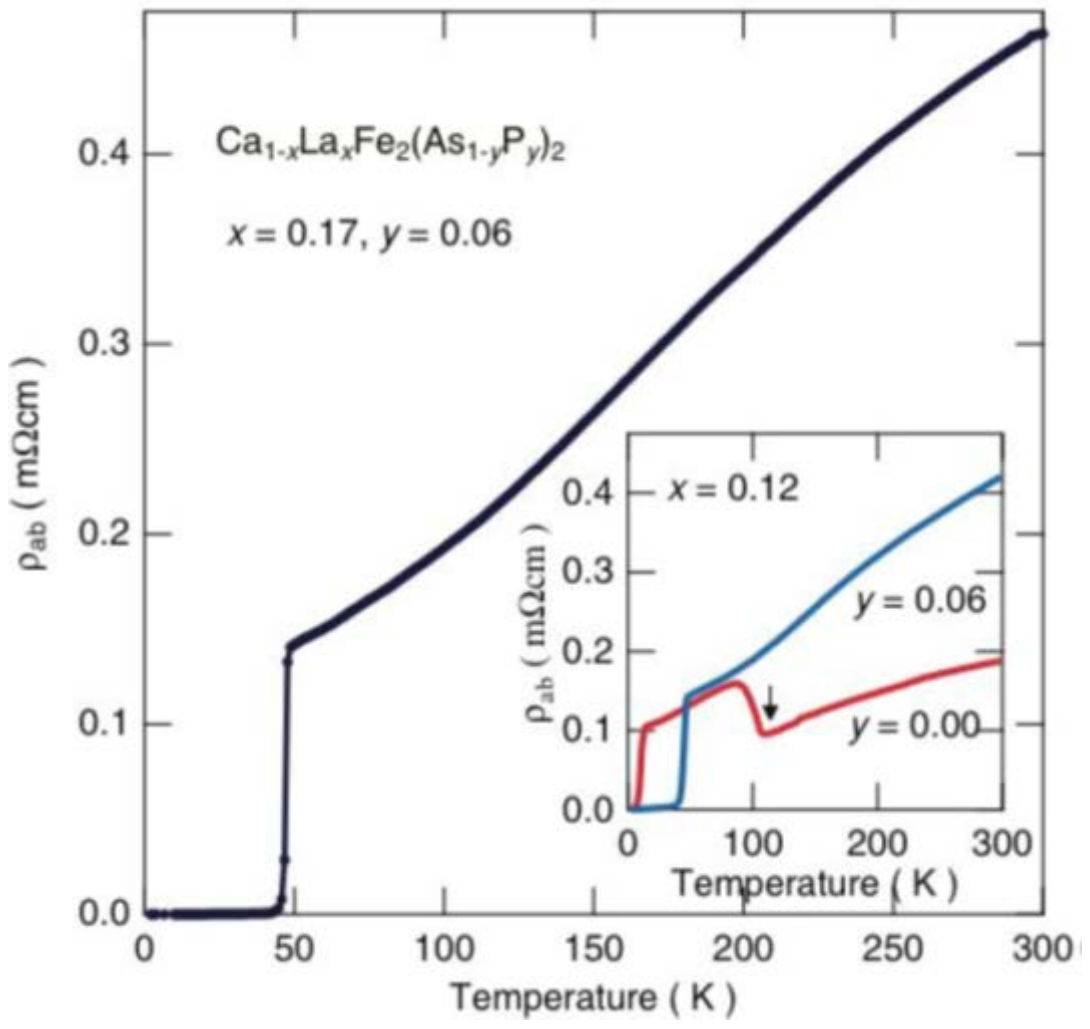

**Figure 16.** Temperature dependence of electrical resistivity $\rho_{ab}$ for $(Ca_{1-x}La_x)Fe_2(As_{1-y}P_y)_2$ ($x$ = 0.17 and $y$ = 0.06) [23]. The resistivity starts to decrease at a transition temperature $T_c^{onset}$ of 48 K and becomes zero below 45 K. The inset shows $\rho_{ab}$ for $x$ = 0.12 and $y$ = 0.00 and 0.06. The arrow indicates the antiferromagnetic/tetragonal-orthorhombic structural transition.

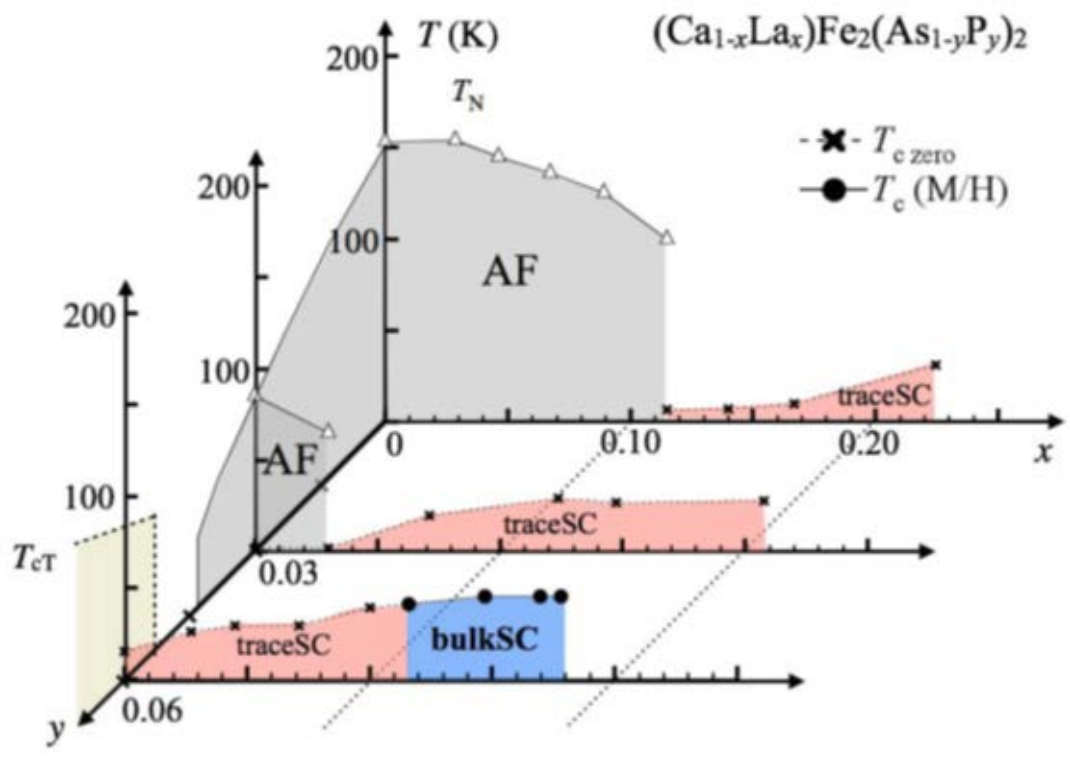

**Figure 17.** $T$–$x$–$y$ electronic phase diagrams for $(Ca_{1-x}La_x)Fe_2(As_{1-y}P_y)_2$ [23]. $T_c(M/H)$ denotes the bulk superconducting transition temperature determined from the magnetization. $T_{c\ zero}$ denotes the temperature below which the electrical resistivity becomes negligibly small. $T_N$ represents the antiferromagnetic and structural transition temperature. $T_{cT}$ denotes the transition temperature at which the high-temperature uncollapsed tetragonal (ucT) phase transforms into the low-temperature collapsed tetragonal (cT) phase. AF and SC indicate the antiferromagnetic and superconducting phases, respectively.

Angle-resolved photoemission spectroscopy (ARPES) [212] revealed that $(Ca_{0.82}La_{0.18})Fe_2(As_{0.94}P_{0.06})_2$ with $T_c$ = 45 K possesses only cylindrical hole- and electron-like Fermi surfaces (FSs). The size of the β hole-like FS is nearly the same as that of the ε electron-like FS, and both FSs have a weak $k_z$ dispersion, thus giving rise to a quasi-nesting. This feature is similar to that for directly electron-doped $Ba(Fe_{1-x}Co_x)_2As_2$ ($T_c$ = 23 K). Sunagawa *et al* [212] pointed out that a noticeable difference between $(Ca_{0.82}La_{0.18})Fe_2(As_{0.94}P_{0.06})_2$ and $Ba(Fe_{1-x}Co_x)_2As_2$ is the dimensionality of the inner hole-like FS; the inner hole-like FS ($\alpha_2$) of the former shows a cylindrical shape, while that of the latter shows a strong $k_z$ dispersion and is closed near the Γ point. It has been suggested that the tendency toward quasi-nesting between $\alpha_2$ and β, together with β and ε, can induce high $T_c$ in $(Ca_{0.82}La_{0.18})Fe_2(As_{0.94}P_{0.06})_2$ [212].

*3.1.4. New type of 112 IBSc*

The prominent achievements of the FIRST PJ include the discovery of the 112-type iron arsenide superconductor $(Ca_{1-x}La_x)FeAs_2$ [25], whose structure is shown in Fig. 18, and the enhancement of superconducting transition temperature $T_c$ up to 47 K by the simultaneous La and Sb doping of the 112 phase [26, 27]. In this subsection, we overview the crystal structure and superconducting properties of the newly discovered 112 phase.

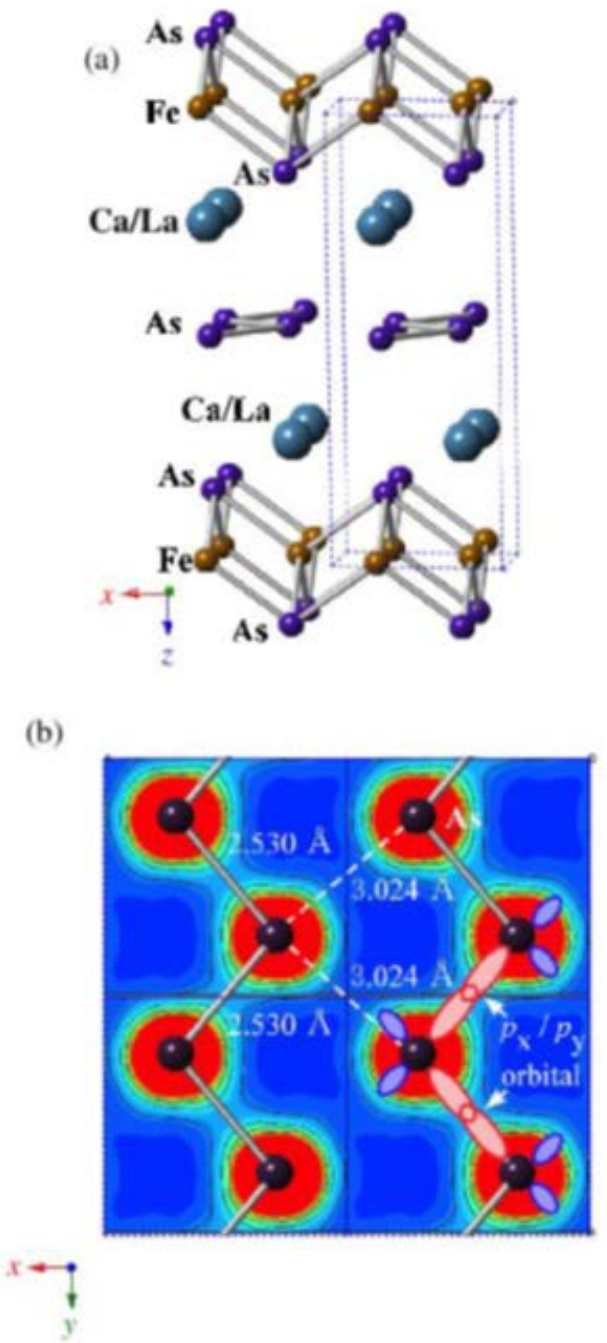

**Figure 18.** (a) The crystal structure of $(Ca_{1-x}La_x)FeAs_2$ (monoclinic, space group $P2_1$) [25]. (b) Top view of the arsenic zigzag chains. Arsenic $4p_x$ and $4p_y$ orbitals are schematically shown. The background color contour map shows the charge distributions obtained by synchrotron X-ray diffraction analysis [25]. The charge accumulation between the adjacent As atoms in the zigzag chains suggests a formation of covalent bonds.

The superconductivity of the 112 phase was first reported in $(Ca_{1-x}La_x)FeAs_2$ and $(Ca_{1-x}Pr_x)FeAs_2$ by Katayama *et al* [25] and Yakita *et al* [239], respectively. The substitution of a rare-earth element is essential to obtaining the 112 phase [25, 239]. A conventional melt-growth technique was used to obtain tiny single crystals [25, 239]. $(Ca_{1-x}La_x)FeAs_2$ exhibited superconductivity at $T_c$ = 34 K [25], while $(Ca_{1-x}Pr_x)FeAs_2$ exhibited $T_c$ of ~20 K with a broad resistive transition [239]. In a subsequent strudy, Sala *et al* [240] performed a high-pressure synthesis and obtained $(Ca_{1-x}RE_x)FeAs_2$ for *RE* = La–Gd. Moreover, $(Ca_{1-x}RE_x)FeAs_2$ for *RE* = Pr, Nd, Sm, Eu and Gd showed superconductivity at 10–15 K with a small shielding fraction of 5–20% [27, 240], while $(Ca_{1-x}Ce_x)FeAs_2$ did not exhibit superconductivity [27, 240]. The $T_c$ of $(Ca_{1-x}RE_x)FeAs_2$ is summarized in Table 10.

The 112-type $(Ca_{1-x}RE_x)FeAs_2$ compound crystallizes in a monoclinic structure with the space group of $P2_1$ (No. 4) [25] or $P2_1/m$ (No. 11) [239, 240]. The structure consists of alternately stacked FeAs and zigzag As bond layers with a Ca/La layer between them, as shown in Fig. 18(a). The most prominent feature of this structure is the presence of the one-dimensional zigzag As chains along the $b$-axis, as shown in Fig. 18(b). The short As-As bond length of approximately 2.53 Å indicates the formation of arsenic single bonds with a arsenic formal valence of $As^{-}$ ($4p^4$ configuration). The presence of two unpaired electrons in $As^{-}$ underlies the formation of two chemical bonds per As atom that yield a zigzag chain. In contrast, the arsenic at FeAs layers forms the $As^{3-}$ valence state with the filled $4p^6$ configuration. Thus, the chemical formula of the 112 phase can be written as $(Ca^{2+}_{1-x}RE^{3+}_x)(Fe^{2+}As^{3-})As^{-}\cdot xe^{-}$ with excess charge of $xe^{-}$/Fe, which is injected into the superconducting FeAs layers. This formula can be compared with that of the 1111-type fluoride, $(Ca^{2+}_{1-x}RE^{3+}_x)(Fe^{2+}As^{3-})F^{-}\cdot xe^{-}$ [137, 241], where $F^{-}$ with filled $2p^6$ orbitals forms an undistorted square network. In this manner, the 112-type structure can be related to the 1111-type structure of CaFeAsF. However, the chemical bonding of the intermediary layer is completely different between CaFeAsF and $CaFeAs_2$; the intermediary CaF layers consist of strong ionic bonds, while the CaAs layers consist of zigzag As chains with covalent bonds, which is weakly coupled to the adjacent Ca layers. Thus, the interlayer distance between the adjacent FeAs layers of $(Ca_{1-x}La_x)FeAs_2$ (~10.35 Å) [27] is considerably larger than that of CaFeAsF (~8.6 Å) [137], but is comparable to that of $Ca_{10}(Pt_4As_8)(Fe_{2-x}Pt_xAs_2)_5$ (~10 Å) with the $Pt_4As_8$ intermediary layers exhibiting a covalent nature [242, 243].

Kudo *et al* [26, 27] examined simultaneous doping, i.e., aliovalent La doping at the Ca sites and isovalent P or Sb doping at the As sites, and they observed that $T_c$ increased up to 47 K for La- and Sb-doped $(Ca_{1-x}La_x)Fe(As_{1-y}Sb_y)_2$ [27]. Figure 19(a) shows the temperature dependence of the magnetization of $(Ca_{1-x}La_x)Fe(As_{1-y}Sb_y)_2$. The La-doped sample without Sb ($y$ = 0) shows superconductivity at $T_c$ = 34 K. The Sb doping results in an increase in $T_c$ to 43 K for $y$ = 0.01 and 47 K for $y$ = 0.10. The enhancement of $T_c$ is also evident in the electrical resistivity $\rho_{ab}$ of the La-doped $y$ = 0.10 sample. The resistivity $\rho_{ab}$ exhibits a sharp drop at 49 K, and zero resistivity is observed at 47 K, as shown in Fig. 19(b). $T_c$ was also enhanced by Sb doping of $(Ca_{1-x}RE_x)Fe(As_{1-y}Sb_y)_2$ for RE = Ce, Pr and Nd [27], as summarized in Table 10. Figure 20 shows the $T$–$x$ phase diagram of $(Ca_{1-x}RE_x)Fe(As_{1-y}Sb_y)_2$ [27]. The 112 phase can be obtained at $x \geq 0.15$ for $y = 0.0$. The superconducting transition temperature $T_c$ is highest (35 K) at the lowest boundary of $x$ = 0.15, and it decreases monotonically with the La content $x$. Superconductivity disappears at $x \geq 0.25$. The lower limit of $x$ is extended down to $x$ = 0.12 for $y$ = 0.10, at which the highest $T_c$ of 47 K is observed [27].

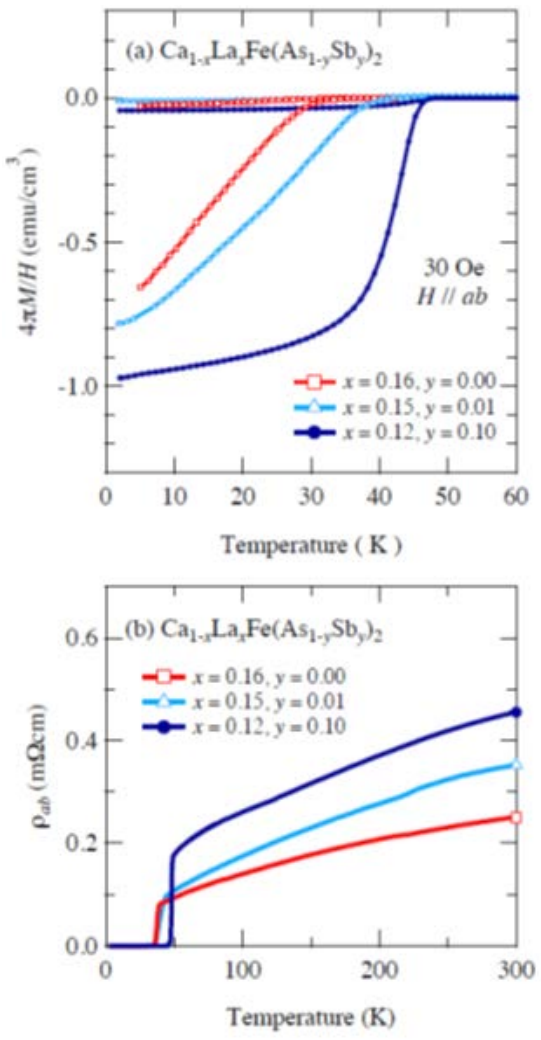

**Figure 19.** (a) Temperature dependence of the magnetization $M$ of $(Ca_{1-x}La_x)Fe(As_{1-y}Sb_y)_2$ measured at a magnetic field $H$ of 30 Oe parallel to the $ab$ plane under zero-field-cooling and field-cooling conditions. (b) Temperature dependence of the electrical resistivity $\rho_{ab}$ of $(Ca_{1-x}La_x)Fe(As_{1-y}Sb_y)_2$ parallel to the $ab$-plane [27].

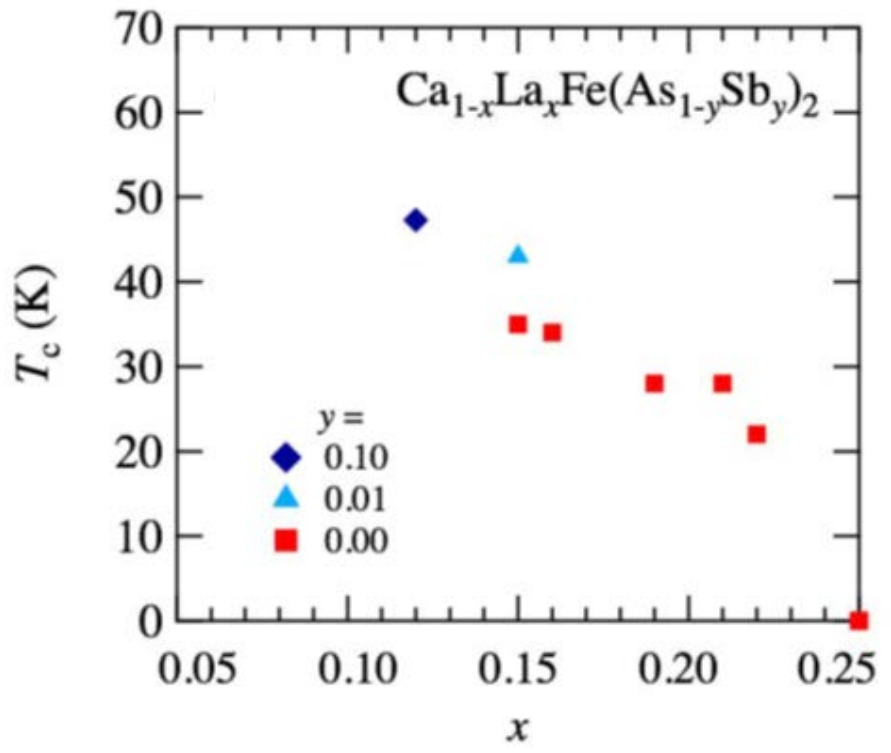

**Figure 20.** Dependence of the superconducting transition temperature $T_c$ of $(Ca_{1-x}La_x)Fe(As_{1-y}Sb_y)_2$ on the La content $x$ [27].

Recently, Zhou *et al* [244] have successfully grown large single crystals of $(Ca_{1-x}La_x)FeAs_2$ with $T_c$ = 42.6 K. Using these crystals, they estimated a critical current density $J_c$ of 3.5 × $10^5$ A/cm$^2$ from the magnetic hysteresis loops at 5 K, which indicates a strong bulk pinning. The anisotropic upper critical field $H_{c2}(0)$ was estimated to be 39 and 166 T for the out-of-plane ($H$ // $c$) and in-plane ($H$ // $ab$) directions, respectively. The $H_{c2}$ anisotropy parameter γ was ~2 near $T_c$. The moderate anisotropy and high $T_c$ indicates the potential of $(Ca_{1-x}La_x)FeAs_2$ for practical applications.

*3.1.5. 10-3-8 and 10-4-8 iron arsenide superconductors*

The rich chemistry of arsenic allowed us to develop various iron arsenide superconductors. Arsenic exhibits a wide variety of chemical networks, which is known as catenation, depending on the number of valence electrons. For instance, neutral arsenic has three unpaired electrons ($4p^3$), and thus, it forms three single bonds per As atom, thereby resulting in a buckled honeycomb network of α-As (trigonal, space group $R\bar{3}m$). Monovalent arsenic ($As^-$) has two unpaired electrons ($4p^4$), and thus forms two single bonds per As atom, which results in either one-dimensional zigzag chains or cis-trans chains, or molecular-like $As_4^{4-}$ tetramers. The 112-type iron arsenide superconductor $(Ca_{1-x}La_x)FeAs_2$ consists of As zigzag chains [25], as described in the previous subsection. Cis-trans chains can be observed in $LaAgAs_2$ [245] and $As_4$ tetramers in skutterudite $CoAs_3$, for instance. Divalent arsenic ($As^{2-}$) has one unpaired electron ($4p^5$), and thus forms a single bond per As atom, which results in molecular-like $As_2^{4-}$ dimers, as seen in pyrite-type $PtAs_2$. The 122-type iron arsenide superconductor $CaFe_2As_2$ exhibits the formation of molecular $As_2$ between adjacent FeAs layers along the $c$-axis at the collapsed tetragonal phase transition [218, 234, 246]. The $As_2$ molecular bonds are broken in the uncollapsed tetragonal phase. Here, the formal valence of arsenic is $As^{3-}$, and the $4p$ orbitals are completely occupied, and thus, no direct chemical bonds are formed between As. The 10-3-8- and 10-4-8-type iron arsenide superconductors $Ca_{10}Pt_3As_8(Fe_{2-x}Pt_xAs_2)_5$ and $Ca_{10}Pt_4As_8(Fe_{2-x}Pt_xAs_2)_5$ consist of $As_2$ dimers in the $Pt_3As_8$ and $Pt_4As_8$ intermediary layers, respectively [242, 247, 248].

Superconductivity in the 10-3-8 and 10-4-8 compounds was first reported by Kakiya *et al* [242], Löhnert

*et al* [247] and Ni *et al* [248] for Pt-based $Ca_{10}Pt_3As_8(Fe_{2-x}Pt_xAs_2)_5$ and $Ca_{10}Pt_4As_8(Fe_{2-x}Pt_xAs_2)_5$. The 10-4-8 compound exhibited a maximum $T_c$ of 38 K [242, 249]. The 10-3-8 compound showed a lower $T_c$ of 13 K [242] by Pt doping of $Ca_{10}Pt_3As_8(Fe_{2-x}Pt_xAs_2)_5$. The $T_c$ value was enhanced up to 30 K by La doping of $(Ca_{0.8}La_{0.2})_{10}(Pt_3As_8)(Fe_{2-x}Pt_xAs_2)_5$ [249]. Three polymorphs have been identified in the 10-4-8 phase, i.e. tetragonal ($P4/n$), triclinic ($P\bar{1}$) and monoclinic ($P2_1/n$) structures, while the 10-3-8 compound crystallizes in a triclinic ($P\bar{1}$) structure. Hieke *et al* [250] reported the observation of the 10-3-8 phase for Pd-based $Ca_{10}Pd_3As_8(Fe_2As_2)_5$, which exhibited superconductivity at 17 K by La substitution for Ca.

Under the aegis of the FIRST PJ, Kudo *et al* [28] discovered a new member of the 10-4-8 family with $Ir_4As_8$ intermediary layers, i.e. $Ca_{10}Ir_4As_8(Fe_{2-x}Ir_xAs_2)_5$. The compound crystallizes in the tetragonal structure with the space group $P4/n$, which is isotypic to one of the polymorphs of Pt-based 10-4-8. The crystal structure, shown in Fig. 21, consists of characteristic $IrAs_4$ squares, which are rotated alternately to form $As_2$ dimers. Figure 22 shows the temperature dependence of the in-plane electrical resistivity $\rho_{ab}$ of $Ca_{10}Ir_4As_8(Fe_{2-x}Ir_xAs_2)_5$ with $x$ = 0.07–0.08 [28, 251]. The resistivity $\rho_{ab}$ reached zero at 17 K. Magnetic measurements demonstrated large shielding signals below $T_c$ = 16 K. The small and almost $T$-independent Hall coefficient $R_H$, shown in the inset of Fig. 22, suggests overdoping, which most probably resulted in the low $T_c$. Another feature of interest is the kink in $\rho_{ab}$ at approximately 100 K. This kink is not due to antiferromagnetic ordering, since the singlet-peak structure of the $^{57}$Fe-Mössbauer spectra remained unchanged down to 50 K (as shown in the inset of Fig. 22), but due to a structural phase transition [251]. Katayama *et al* [251] performed single-crystal X-ray diffraction and identified that the transition is characterized by the displacement of Ir2 at the non-coplanar sites along the *c*-axis, thereby resulting in doubled periodicity along the *c*-axis without breaking the $P4/n$ symmetry. The structural phase transition suggests that either Ir charge or orbital degrees of freedom are active in the intermediary $Ir_4As_8$ layers. Sawada *et al* [252] performed ARPES measurements of $Ca_{10}Ir_4As_8(Fe_{2-x}Ir_xAs_2)_5$, thereby demonstrating that the Fe 3*d* electrons in the FeAs layers form hole-like and electron-like Fermi surfaces at the zone center and corners, respectively, as commonly observed in iron arsenide superconductors; Ir 5*d* electrons are metallic and glassy most probably due to atomic disorder related to the Ir 5*d* orbital instability.

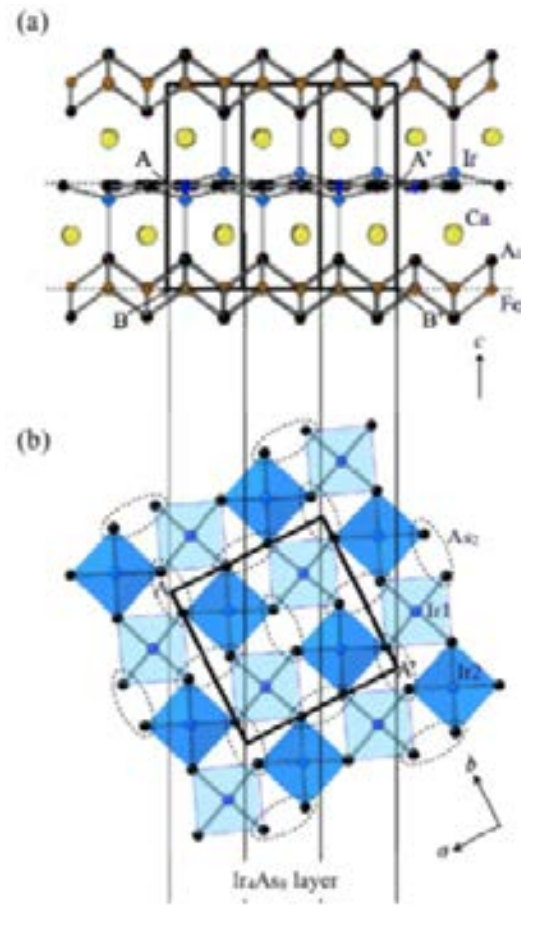

**Figure 21.** Crystal structure of 10-4-8 type $Ca_{10}Ir_4As_8(Fe_{2-x}Ir_xAs_2)_5$ with tetragonal structure (space group $P4/n$) [28]. (a) and (b) show the schematic overviews and $Ir_4As_4$ layer, respectively. The blue and dark-blue hatches in (b) indicate $IrAs_4$ squares with coplanar Ir1 and non-coplanar Ir2, respectively. The dashed ellipsoids in (b) represent $As_4$ dimers.

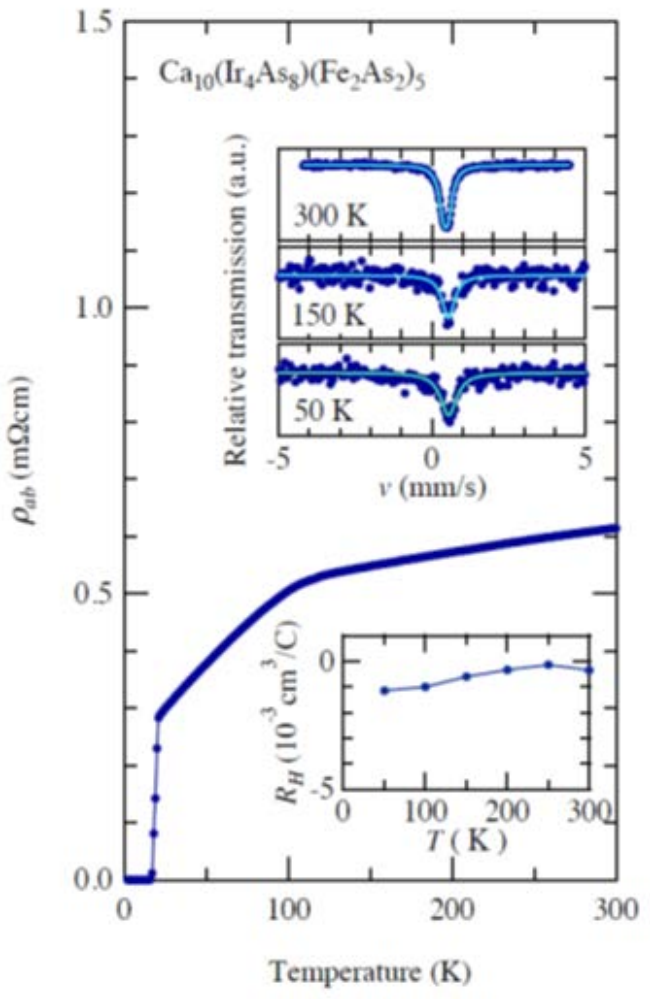

**Figure 22.** Temperature dependence of the electrical resistivity $\rho_{ab}$ for $Ca_{10}Ir_4As_8(Fe_{2-x}Ir_xAs_2)_5$ [28]. The upper inset shows $^{57}$Fe-Mössbauer spectra together with fitting curves. The lower inset shows the temperature dependence of the Hall coefficient $R_H$.

Another achievement of the FIRST project is the growth of superconducting nanowhiskers of $Ca_{10}Pt_4As_8(Fe_{1.8}Pt_{0.2}As_2)_5$ by Li *et al* [253]. The typical whiskers obtained had a length of 0.1–2.0 mm, width of 0.4–5.0 μm and thickness of 0.2–1.0 μm. High-resolution transmission electron microscopy (TEM) images showed that the whiskers exhibited excellent crystallinity and that whisker growth occurred along the *a*-axis of the tetragonal (*P*4/*n*) structure. The whiskers exhibited superconductivity with $T_c$ of 33 K, $H_{c2}$ ($H$ // $c$) of 52.8 T and $J_c$ of $6.0 \times 10^6$ A/cm$^2$ (at 26 K). Since cuprate high-$T_c$ whiskers are fragile ceramics, the present intermetallic superconducting whiskers with high $T_c$ are more suitable for device applications.

### *3.1.6. Intercalated FeSe superconductors*

The structurally simplest FeSe with space group *P*4/*nmm* has becomes very attractive material. Though pure FeSe shows low $T_c$ (8 K), it can be drastically enhanced by a factor of 5 under external high pressure [147, 148]. Moreover, the single-layer FeSe film epitaxialy grown on $SrTiO_3$ substrates with huge superconducting gap (~20 meV) implies that the potential $T_c$ could reach as high as 65 K even its bulk SC decreases to ~40K [149]. More recently, high $T_c$ over 100K has been reported in this type of the single-layer FeSe [153]. Another high-$T_c$ bulk SC derived from FeSe is intercalated $A_xFe_{2-y}Se_2$ with $ThCr_2Si_2$ structure (but classified to 245 type due to the exact composition) that is synthesized at a high temperature (~1300K) but is only available for the large-sized monovalent metals as *A* (K, Rb, Cs and Tl) [154-157, 254-256]. The bulk $T_c$ of $A_xFe_{2-y}Se_2$ is ~30K and the average crystal structure is a body-centered tetragonal phase (S.G. *I*4/*mmm*) [154]. However, the synthesis for FeSe intercalated by smaller alkali metals as Li and Na has not succeeded using conventional high temperature processes.

The low-temperature method is another way to approach the intercalated phase that is broadly applied and suitable for intercalating alkaline, alkali earth metals even those small ionic radius. Many superconductors such as $A_xC_{60}$ and $A_xMNX$ (*A*: Li-K, Ca-Ba, Yb and Eu; *M*: Ti, Zr and Hf; and *X*: Cl, Br and

I) were obtained through this method [55, 257]. Among them, the ammonothermal method which uses liquid ammonia as a solvent and make the starting materials react under high pressure in an autoclave is useful way to prepare the meta-stable and/or non-equilibrium materials. The relatively mild reaction keeps the host structure intact; therefore, the pure charge transfer without destroying the conductive layer is expected to be favor the higher $T_c$.

Application of this method to intercalate *A* into FeSe was first carried out by a group of the Institute of Physics, China [258]. They obtained several superconductors with a higher $T_c$ of 30-46K (FeSe intercalated by Li, Na, K, Ca, Sr, Ba, Eu, and Yb) compared with the samples prepared by the high-temperature method. In this report, they demonstrated that the ammonothermal method was useful to synthesize the intercalated FeSe superconductors with relatively high $T_c$, while they did not mention about remaining ammonia molecules or ions in their early stage of this research. Subsequently, they studied $K_xFe_2Se_2$+$NH_3$ systematically and found two superconducting phases, $K_{0.3}Fe_2Se_2(NH_3)_{0.47}$ ($T_c$ =44 K) and $K_{0.6}Fe_2Se_2(NH_3)_{0.37}$ ($T_c$ =30 K), where they noted that the most important factor to control $T_c$ was potassium content [259]. Then, they prepared ammonia-free $K_{0.3}Fe_2Se_2$ and $K_{0.6}Fe_2Se_2$ by removing $NH_3$ at 200°C completely and showed the same $T_c$. The theoretical and/or empirical reason for higher $T_c$ of $K_{0.3}Fe_2Se_2(NH_3)_{0.47}$ has not been reported yet.

In contrast, a group of the University of Oxford, UK concentrated on the FeSe co-intercalated by Li and ammonia, and analyzed its precise crystal structure using the powder neutron diffraction [260, 261]. They synthesized $Li_{0.6}(ND_3)_{0.8}(ND_2)_{0.2}Fe_2Se_2$ ($T_c$ =43 K) by ammonothermal method and showed its crystal structure as shown in Fig. 23. This crystal structure indicates that the resulting compound intercalates as not only ammonia molecule but also the amide anion. Furthermore they found reversible adsorption and desorption of ammonia by controlling pressure of ammonia at <-10 °C and obtained $Li_{0.6}(ND_{2.7})_{1.7}Fe_2Se_2$ (ammonia rich phase) with $T_c$ =39 K by exposing $Li_{0.6}(ND_3)_{0.8}(ND_2)_{0.2}Fe_2Se_2$ (ammonia poor phase) to 1 atm of ammonia($ND_3$) at -10°C (see Fig. 24). By intercalating more ammonia, the spacing between Fe layers (*d*) increased from 8.26 Å of the poor phase to 10.59 Å of the rich phase. Though the $T_c$ of the ammonia poor-phase is higher than that of the ammonia-rich phase, the lithium content is the same in both phases. This result looks to be inconsistent with Ying's result [259] who noted that the $T_c$ depends on the intercalated alkaline metal content. While the difference of $T_c$ between the poor and the rich phase is not so large (4 K) compared with the case of potassium (14 K), the effect of intercalated amide anion is also unclear.

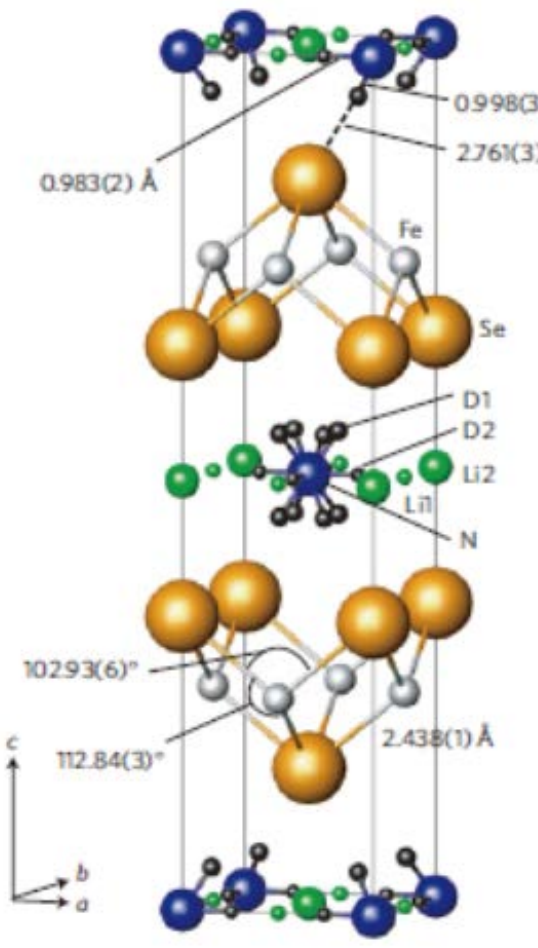

**Figure 23.** The 298 K crystal structure of $Li_{0.6(1)}(ND_2)_{0.2(1)}(ND_3)_{0.8(1)}Fe_2Se_2$. Refinement was against neutron powder diffraction data (GEM instrument). In the model each square prism of Se atoms contains either an $[ND_2]$ anion or an $ND_3$ molecule and these are both modelled as disordered over four orientations. The sizes of the spheres representing the Li atoms are in proportion to their site occupancies [260].

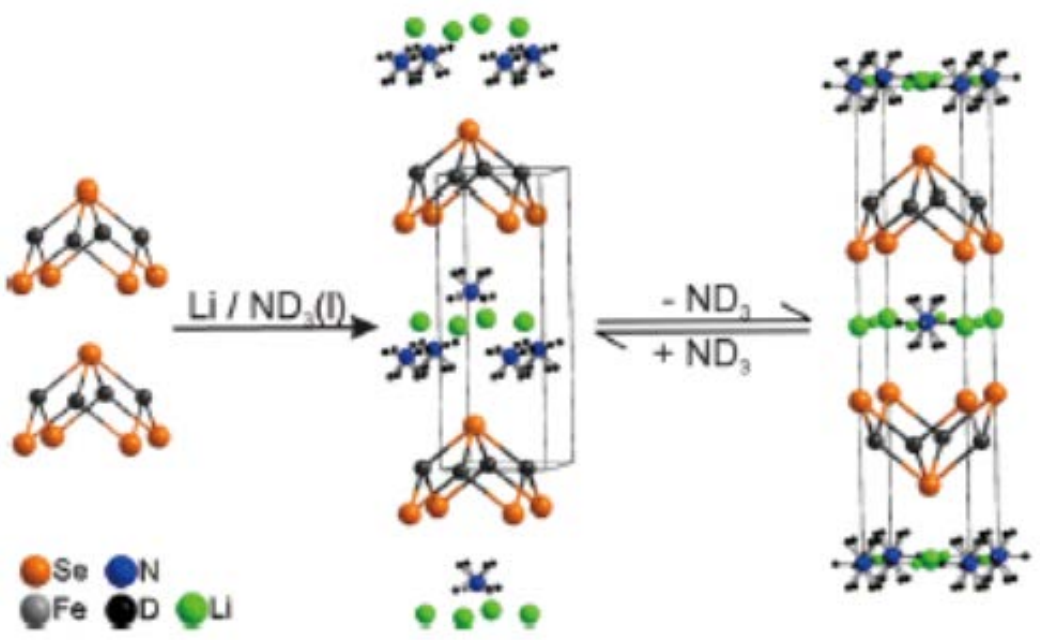

**Figure 24.** Schematic of the intercalation of lithium and ammonia into FeSe [261].

In our research project, the group of Tokyo Institute Technology, Japan synthesized three samples intercalated by sodium and ammonia as ammonia-free (phase I), ammonia-poor (phase II) and ammonia-rich phases (phase III) using the ammonothermal method and showed superconductivity with $T_c$ =37, 45 and 42 K, respectively (see Fig. 25) [29]. They prepared these phases by changing Na/$NH_3$ ratio (0.03 to 0.3 $mol^{-1}$). For phase I, the reaction vessel was evacuated to ~$10^{-2}$ Pa after immersion for 3 h at 223-243 K. The chemical compositions measured by EPMA was $Na_{0.65}Fe_{1.93}Se_2$ for phase I and $Na_{0.80}(NH_3)_{0.60}Fe_{1.86}Se_2$ for phase II. The phase III was so unstable which decomposed easily even at 250 K that the composition of phase III could not be defined. From powder X-ray diffraction measurement, the spacing between Fe layers ($d$) was 6.8, 8.7 and 11.1 Å for phase I, II and III, respectively. The lower $T_c$ and the wider $d$ of ammonia-rich phase (phase III) than those of ammonia-poor phase (phase II) are similar to the case of Li.

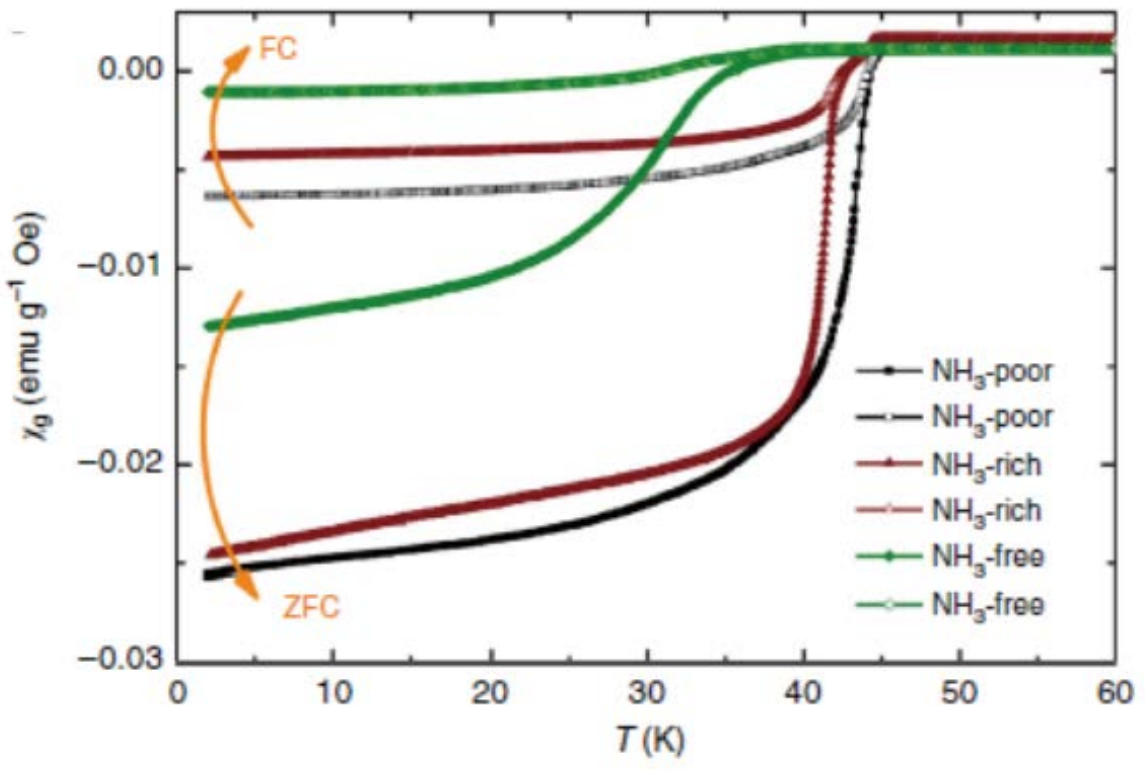

**Figure 25.** The magnetization curves of three Na/$NH_3$ intercalated FeSe measured with the zero-field-cooling (ZFC) and field-cooling (FC) modes at $H$=10 Oe [29].

These results are summarized in Table 11. It has been considered that the $d$ value could be a possible guideline to explore the high $T_c$ superconductor. In IBScs, the $d$'s of FeSe ($T_c$ =8 K), LiFeAs ($T_c$ =18 K), $BaFe_2As_2$ ($T_c$ =38 K), and SmFeAsO ($T_c$ =56 K) are 5.5, 6.4, 6.5 and 8.7 Å, respectively, which looks similar to the general rule for cuprate superconductors. Expecting that the material with wider *d* could reveal higher $T_c$, Ogino et al. [141-143] prepared IBScs with thick blocking layer with $d$ =15.5 to 24.5 Å. Unfortunately, the maximum $T_c$ in their attempt was 43 K for $Ca_6(Sc,Ti)_5O_yFe_2As_2$ ($d$ =24.5 Å) (see Section 3.1.1). In this case of this intercalates, the optimal *d* is 7.5-8.8 Å of ammonia-poor phase. It looks that the *d* is not a critical factor to control $T_c$ widely.

An important feature of these intercalates is none or very small deficiency of Fe compared to those synthesized by conventional high temperature method. In the case of conventionally prepared (Tl, K)$Fe_xSe_2$, superconductivity emerges over $x$=1.7 and the compound with smaller Fe content is an antiferromagnetic insulator [156]. The maximum Fe content of the compound prepared by high temperature method is reported to be $x$ ~1.9 and the maximum $T_c$ is 31 K. Zhang *et al.* reported that the compound with excess Fe ($K_{0.87}Fe_{2.19}Se_2$) which was prepared by the Bridgemen method using $Fe_{1+x}Se$ as a starting material revealed the sharp drop of resistivity at 44 K, but zero resistivity was observed at a lower temperature (25 K) and the shielding volume fraction at >25 K was very small (<1%) [262]. These results suggest that small deficiency of Fe is favorable to achieve high $T_c$, which is due to the large amount of the indirect electron doping efficiently. The ammonothermal method is profitable to prepare such a condition.

### *3.2. Superconductivity in layered titanium compounds*

Since the discovery of high-$T_c$ superconductivity in a layered perovskite oxide $(La,Ba)_2CuO_4$ with $CuO_2$ square lattice [263], numerous efforts have been dedicated to obtain new superconducting families. Although this led to a series of discoveries of other high-$T_c$ families such as $MgB_2$ [196], fullerenes [264] and iron pnictides [4], the transition temperatures never exceed liquid nitrogen temperature. In addition, the mechanism of high $T_c$ superconductivity in the cupper oxides remains unsolved and still under debate despite intensive investigations.

Searching parent structures having a $d^1$ square lattice may be a plausible strategy to access novel superconductivity, which is hole-electron symmetric with the $d^9$ square lattice. $Sr_2VO_4$ with $V^{4+}$ ($d^1$), which is isostructural with $La_2CuO_4$ (Fig. 26(a)), is a promising candidate material though all attempts to inject carriers were so far unsuccessful [265, 266]. When the ligand field splitting energy (LFSE) for an octahedral environment is taken into account, however, one notices that the electronic structures of $Sr_2VO_4$ and $La_2CuO_4$ are quite different from each other. In $d^9$ case, one unpaired electron occupies at one of the anti-bonding $e_g$ orbitals, $x^2 - y^2$, which is widely separated in energy from the $z^2$ orbital due to the Jahn-Teller effect and gives a half-filled electronic configuration. In $d^1$ case, one electron occupies at the non-bonding $t_{2g}$ ($xy$, $yz$, $zx$) orbitals. A certain octahedral distortion may be present, but is not enough to lift the orbital degeneracy significantly, so that the system should be better approximated by a 1/6-filled configuration.

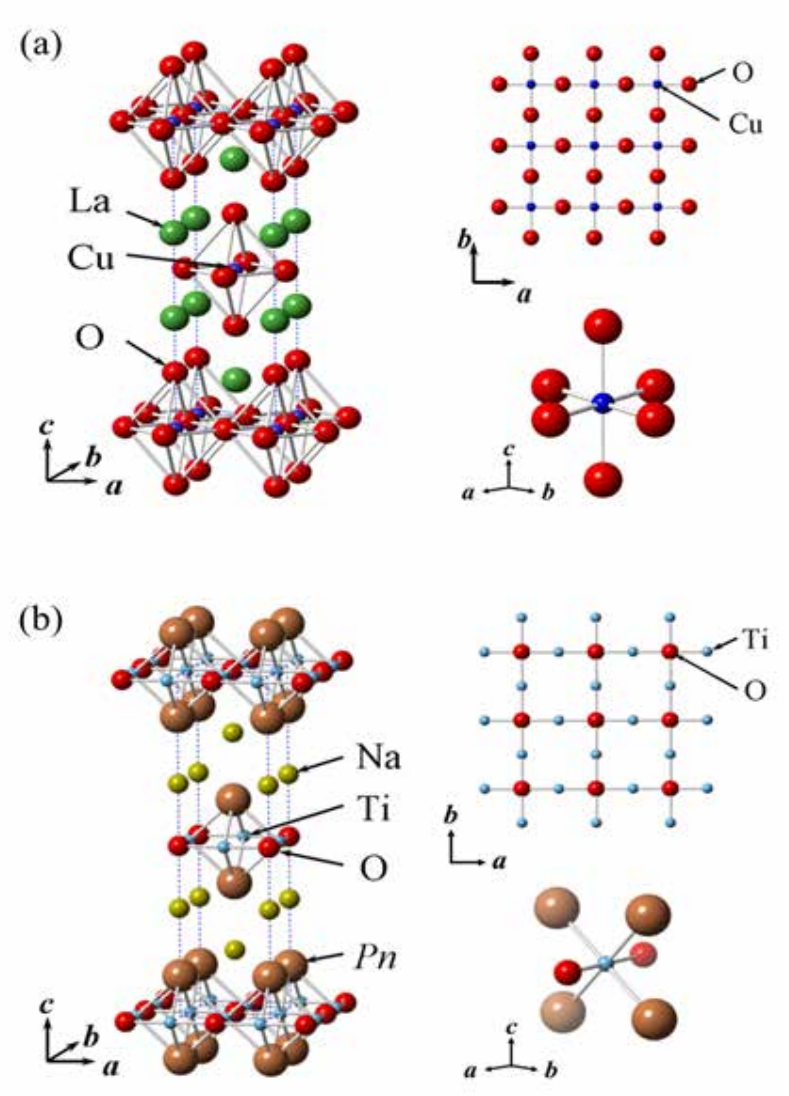

**Figure 26.** (a) Crystal structure of $La_2CuO_4$, $CuO_2$ sheets (upper) and $CuO_6$ octahedron (lower). (b) Crystal structure of $Na_2Ti_2Pn_2O$, $Ti_2O$ sheets (upper) and $TiO_2Pn_4$ octahedron (lower).

In order to tune the $t_{2g}$ orbital levels more drastically (than possible in oxides), we looked for mixed anionic compounds, where a $d^1$ metal is coordinated octahedrally by two kinds of anion species. Due to the large difference between anions in terms of valence, electronegativity and ionic radius, a mixed-anion coordination geometry might provide a unique opportunity to split $t_{2g}$ orbitals to a greater extent. Such a situation is realized in layered titanium oxypnictides $Na_2Ti_2Pn_2O$ ($Pn$ = As, Sb), the structure of which is illustrated in Fig. 26(b) [267]. The $Na_2Ti_2Pn_2O$ structure appears to be similar to that of $La_2CuO_4$, but contains an inversed $Ti_2O$ square lattice. The trivalent titanium ion ($d^1$) has an octahedral $TiO_2Sb_4$ coordination, and the $TiO_2Sb_4$ octahedra share edges to form a two-dimensional network. Like cuprate superconductors, $[TiO_2Sb_2]^{2-}$ layers can be sandwiched by various block layers. For example, $BaTi_2As_2O$ with a $Ba^{2+}$ layer, $(SrF)_2Ti_2As_2O$ with a $[(SrF)_2]^{2+}$ layer, and $(Ba_2Fe_2As_2)Ti_2As_2O$ with a $[Ba_2Fe_2As_2]^{2+}$ layer are reported (Fig. 27) [268-271].

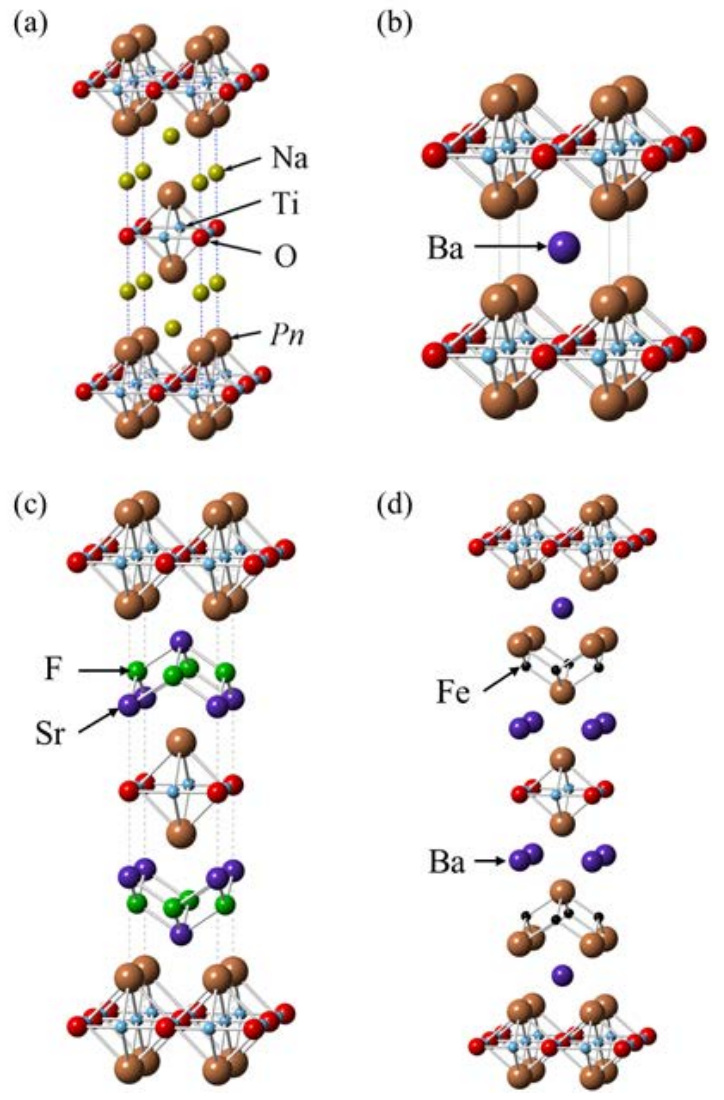

**Figure 27.** Crystal structures of (a) $Na_2Ti_2Pn_2O$, (b) $BaTi_2Pn_2O$, (c) $(SrF)_2Ti_2Pn_2O$, and (d) $(Ba_2Fe_2As_2)Ti_2As_2O$

Unlike cuprate superconductors, known "parent" compounds in the layered titanium oxypnictides are already metallic without carrier doping [272, 273]. Instead of superconductivity, they show anomalies in magnetic susceptibility and electric resistivity likely ascribed to charge density wave (CDW) and/or spin density wave (SDW) transitions. The density wave transition temperatures $T_{DW}$ are 330 K for $Na_2Ti_2As_2O$, 120 K for $Na_2Ti_2Sb_2O$ and 200 K for $BaTi_2As_2O$. It is hence expected that the suppression of density wave phase is a key to induce superconductivity. In 2012, we prepared a new compound $BaTi_2Sb_2O$ and observed a bulk superconductivity transition at 1 K [31]. Doan *et al.* independently showed the enhanced $T_c$ of 5.5 K in Na-doped $BaTi_2Sb_2O$ [274]. These reports have sparked a lot of investigations on the superconductivity in the layered titanium oxypnictides. In this chapter, we demonstrate the present status of our understanding of this new superconducting family from both experimental [31-33, 275-284] and theoretical [285-289] viewpoints.

### 3.2.1. Superconductivity in $BaTi_2Sb_2O$

A pure phase of $BaTi_2Sb_2O$ was synthesized by the conventional solid-state reaction method using BaO (99.99%), Ti (99.9%) and Sb (99.9%) in stoichiometric quantity [31]. A pellet specimen was wrapped by tantalum foil, sealed in a quartz tube, and was typically heated at 1000 ºC for 40 h, followed by controlled cooling at a rate of 50 °C/h to room temperature. The product is air and moisture sensitive. $BaTi_2Sb_2O$ is tetragonal with lattice constants of $a$ = 4.11039(2) Å, $c$ = 8.08640(4) Å at room temperature. $BaTi_2As_2O$ has larger cell parameters, $a$ = 4.046 Å, $c$ = 7.272 Å. Figure 28(a) shows the result of the synchrotron X-ray diffraction refinement for $BaTi_2Sb_2O$ with the space group of $P4/mmm$. Magnetic susceptibility and resistivity for $BaTi_2Sb_2O$ (Fig. 29(a) and 4(b)) shows a distinct anomaly at around $T_{DW}$ = 50 K, which should be related to a density wave transition. Upon further cooling, the $\rho$-$T$ curve showed zero resistivity, indicating a superconducting transition. A large diamagnetic signal associated with the shielding effect was observed at $T_c$ = 1 K. The shielding volume fraction is as large as 58%, providing firm evidence for bulk superconductivity in $BaTi_2Sb_2O$. Heat capacity measurements revealed a distinct peak at around $T_c$, further supporting the bulk superconductivity.

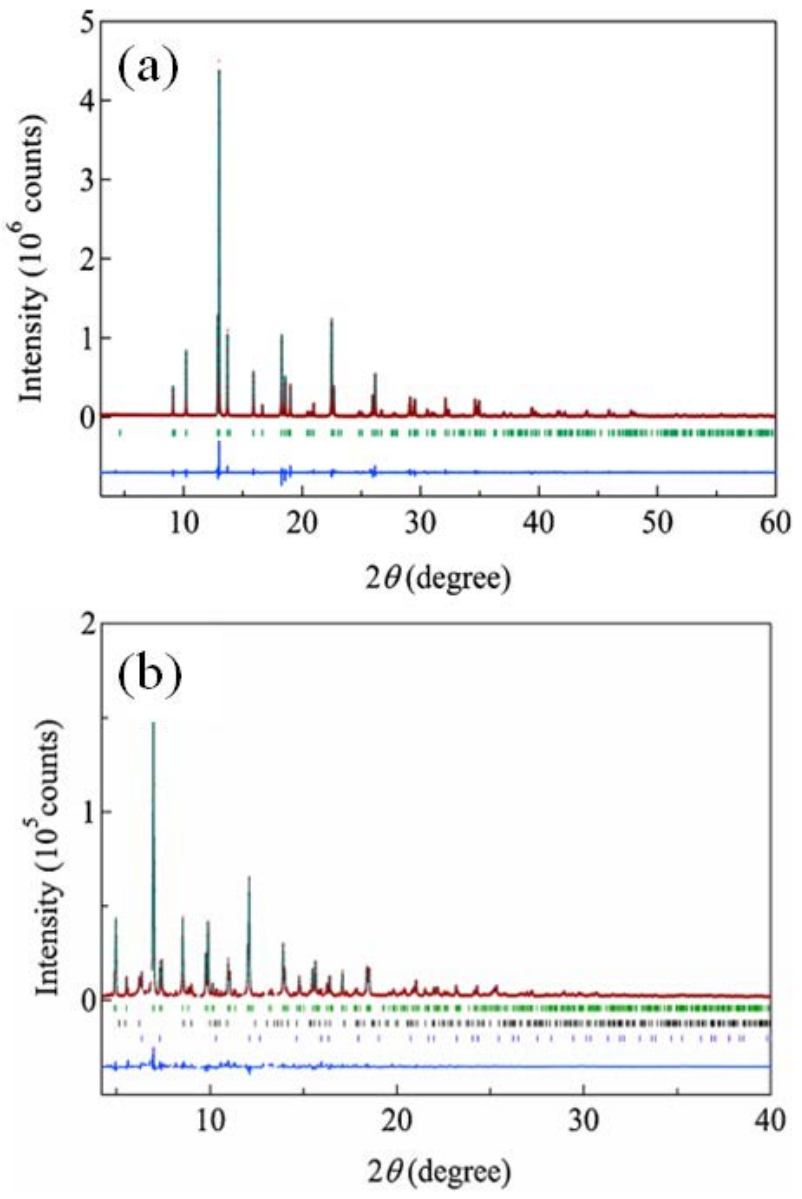

**Figure 28.** Refined synchrotron X-ray patterns of (a) $BaTi_2Sb_2O$, taken from [31] and (b) $BaTi_2Bi_2O$, taken from [33].

*3.2.2. Nature of superconductivity in $BaTi_2Sb_2O$*

It has been revealed from various experiments on $BaTi_2Sb_2O$ that the superconducting state is classified to a fully gapped *s*-wave state. The analysis of the specific heat at $T_c$ gives $\Delta C(T_c)/\gamma T_c \sim 1.36$, which is consistent with the BCS (Bardeen Cooper Schrieffer) weak coupling limit value of 1.43 [31]. Gooch *et al.* demonstrated that the temperature variation of the electronic heat capacity on $Ba_{1-x}Na_xTi_2Sb_2O$ ($x$ = 0 and 0.15) is well described by a weak coupling BCS function with $2\Delta/k_BT_c = 2.9$, as shown in Fig. 29(c) [278]. Kitagawa *et al.* showed that the temperature dependence on $1/T_1$ of $^{121}$Sb-NQR shows a coherence peak just below $T_c$ and decreases exponentially at low temperatures. From the slope of the plot, the magnitude of the superconducting gap is estimated to be $2\Delta/k_BT_c = 4.4$ (Fig. 29(d)) [275]. A superconductivity-induced muon relaxation rate $\sigma_{sc}$, which is proportional to the penetration depth as $\lambda^{-2}$, shows robust bulk superconductivity below $T_c \sim 1$ K (Fig. 29(e)) [277]. A fit to a BCS *s*-wave model in the weak coupling limit gave $T_c = 0.87 \pm 0.03$ K and $\sigma_{sc}$ ($T$ = 0) = 0.37 ± 0.01 μs$^{-1}$. Strong electron correlation in the present material is suggested from the Wilson ratio of $R_W = 2.21$, which is much larger than $R_w = 1$ for free-electron approximation [31].

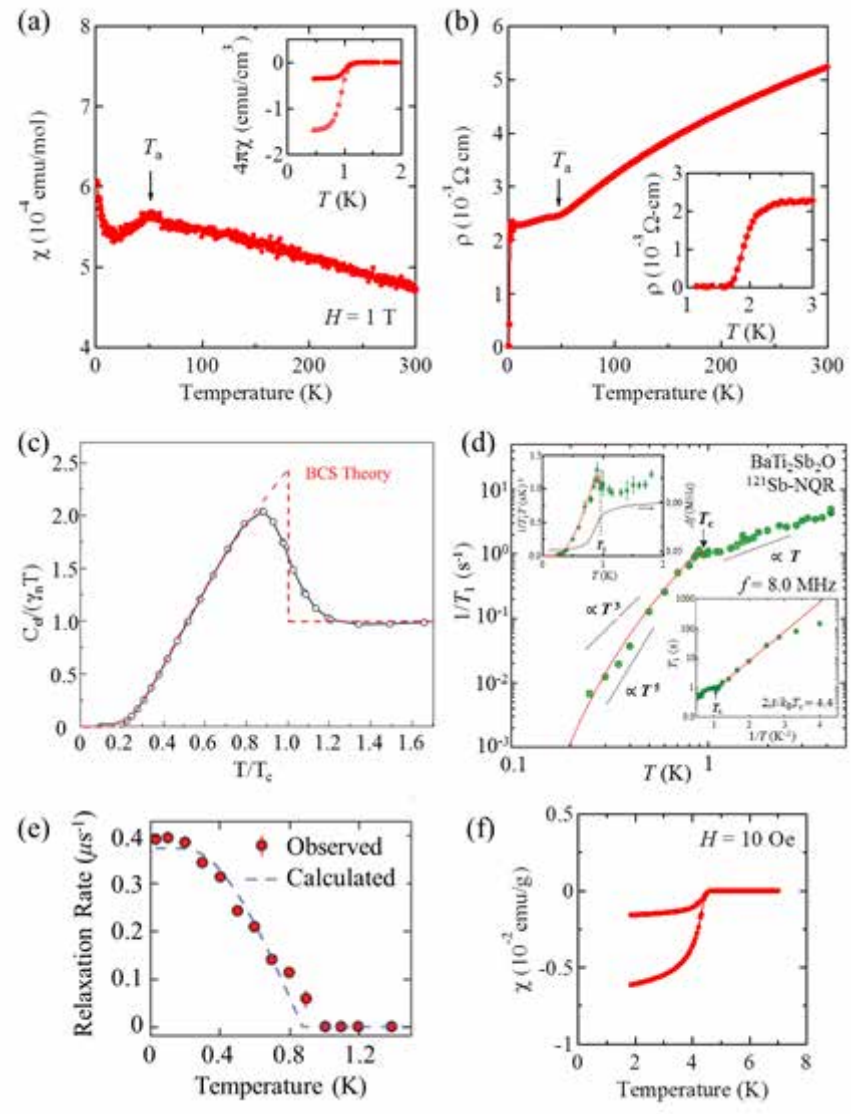

**Figure 29.** (a) Magnetic susceptibility and (b) electric resistivity of $BaTi_2Sb_2O$, taken from [31]. (c) Specific heat capacity of $Ba_{0.85}Na_{0.15}Ti_2Sb_2O$, taken from [278]. (d) $^{121}$NQR result: spin-lattice relaxation rate of $BaTi_2Sb_2O$, taken from [275]. (e) $\mu$SR result: Superconducting relaxation rate of $BaTi_2Sb_2O$. (f) Magnetic susceptibility of $BaTi_2Bi_2O$ taken from [33].

*3.2.3. Density wave state*

While the superconducting state is described well by the weak coupling BCS scheme, the nature of the pairing mechanism, namely, whether the superconductivity in $BaTi_2Sb_2O$ is driven by the electron-phonon coupling or spin-fluctuation, remained unclear until recently. Theoretically, for $BaTi_2Sb_2O$, a magnetic instability associated with Fermi surface nesting, leading to a SDW state with a propagation vector of ($\pi$, 0) (Fig. 30(a)), was theoretically proposed by D. Singh [290]. A sign-changing *s*-wave state within a scenario of spin-fluctuation mediated superconductivity was also suggested. The same type of a bi-collinear antiferromagnetic state was shown for $Na_2Ti_2Sb_2O$ by X. W. Yan *et al.*, while $Na_2Ti_2As_2O$ was suggested to have a blocked checkerboard antiferromagnetic state with a 2 × 2 magnetic unit cell (Fig. 30(b)) [291]. On the other hand, first-principles calculations of the phonon dispersions and electron-phonon coupling for $BaTi_2Sb_2O$ by Subedi [292] revealed lattice instability near the zone corners, which leads to a charge-density wave phase with a √2 × √2 × 1 superstructure arising from a coherent distortion corresponding to elongation or compression of the Ti squares without an enclosed O such that the Ti squares with O rotate either clockwise or counterclockwise as shown in Fig. 30(c).

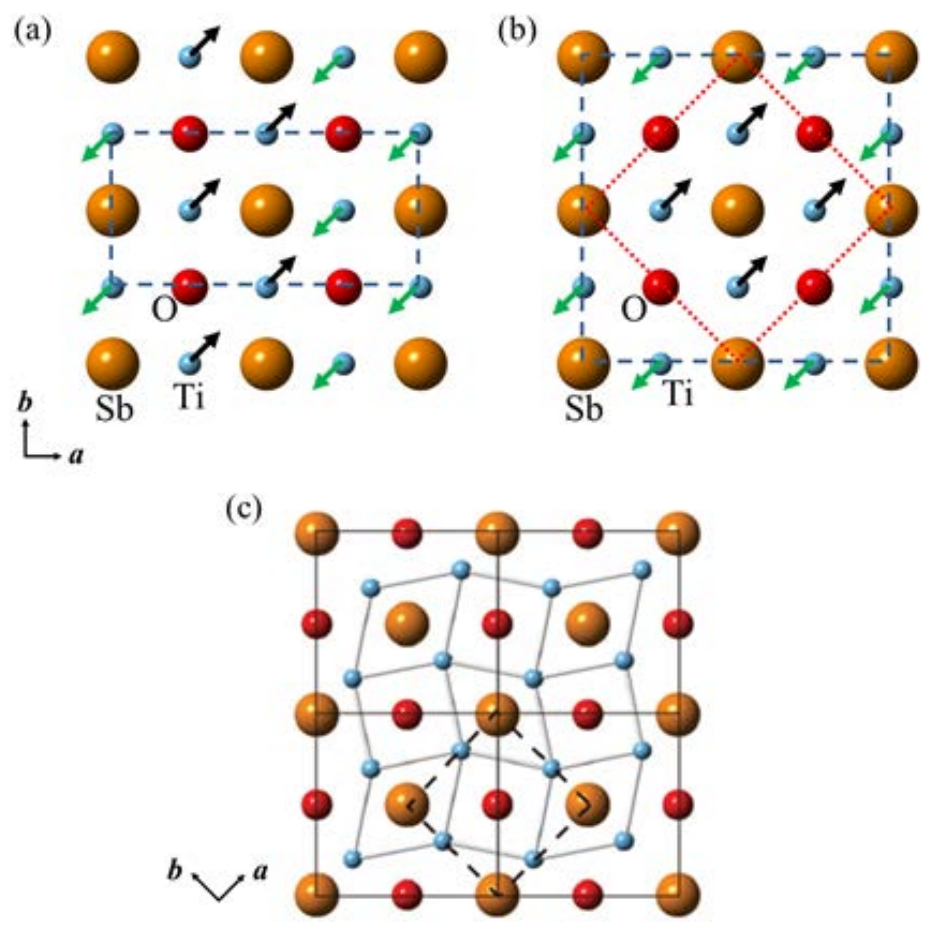

**Figure 30.** Theoretically proposed spin/charge density wave models for $BaTi_2Sb_2O$ or $BaTi_2As_2O$. (a) A bicolinear type and (b) blocked checkerboard type SDW model [290, 291]. (c)√2 × √2 × 1 superstructure as a result of CDW transition [292].

As shown in Fig. 31(b), zero-field muon spin relaxation measurements showed no significant increase in relaxation rate at the density wave ordering temperature, indicating that the density wave is of the charge rather than spin type [277]. $^{121/123}$Sb-NQR measurements revealed that the in-plane four-fold symmetry is broken at the Sb-site below $T_{DW}$ ~40K, without an internal field appearing at the Sb site, indicating a commensurate CDW ordering [275]. However, the absence of any superstructure peaks in high-resolution electron and neutron diffraction below $T_{DW}$ (Fig. 31(a)) signifies that the charge density wave does not involve modulation of atomic arrangement, implying a nontrivial nature of the CDW state. Recent in-depth structural studies have further suggested that $BaTi_2As_2O$ form a symmetry-breaking nematic ground state that can be naturally explained as an intra-unit-cell nematic charge order with d-wave symmetry, pointing to the ubiquity of the phenomenon [293]. These findings, together with the key structural features in these materials being intermediate between the cuprate and iron-pnictide high- temperature superconducting materials, render the titanium oxypnictides an important new material system to understand the nature of nematic order and its relationship to superconductivity.

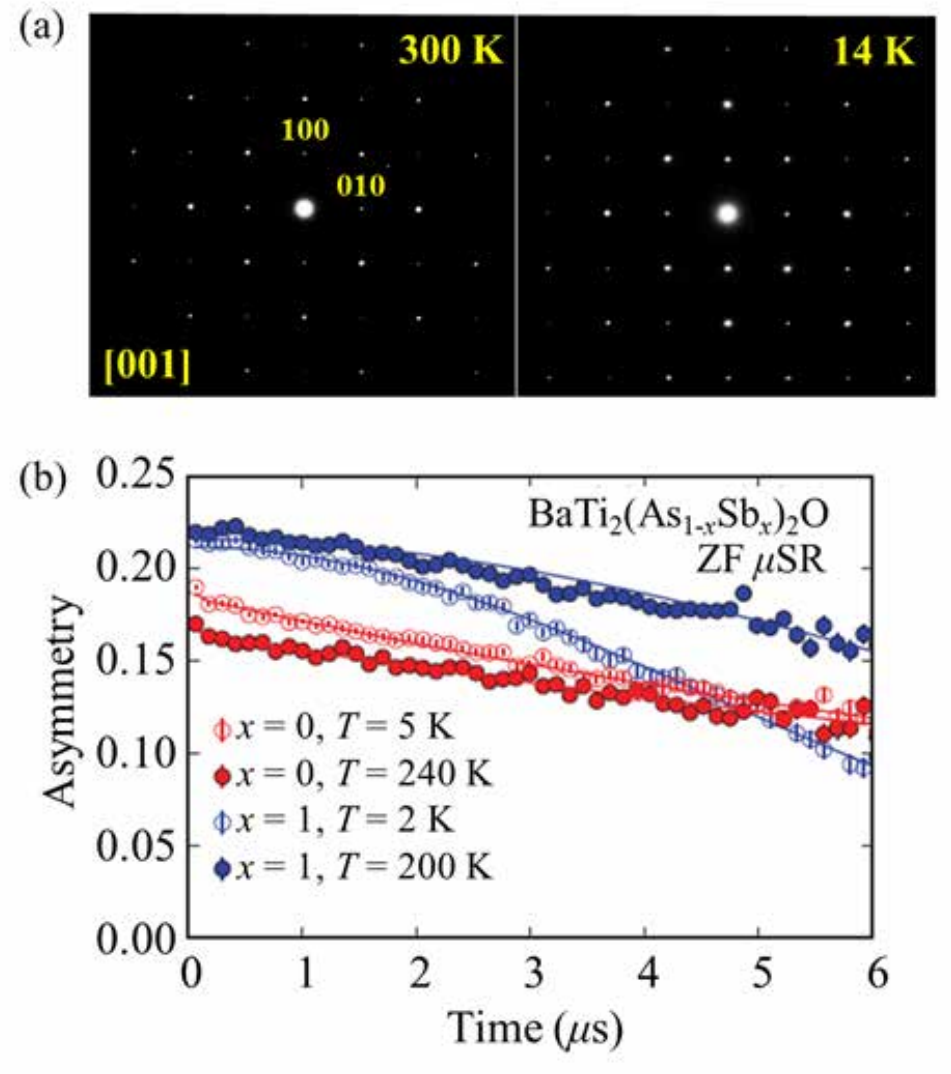

**Figure 31.** (a) Selected area electron diffraction patterns from [001] for $BaTi_2Sb_2O$ at 300 K and 14 K. (b) Asymmetry versus time spectra from zero-field muon spin relaxation measurements on $BaTi_2As_2O$ and $BaTi_2Sb_2O$. The figures are taken from [277].

*3.2.4. Aliovalent cation/anion substitution*

Since the "parent" phase $BaTi_2Sb_2O$ is a 1 K superconductor, coexisting with the CDW state with $T_{DW}$ ~ 50 K, it can be expected that the superconducting transition temperature can be enhanced by destabilizing the CDW phase. It was shown by Doan *et al.* that the divalent Ba site in $BaTi_2Sb_2O$ is substitutable by monovalent Na ions up to 33% [274]. The $T_c$ in $Ba_{1-x}Na_xTi_2Sb_2O$ gradually increases with increasing $x$ (hole concentration) and attained a maximum $T_c$ of 5.5 K at $x = 0.33$. Although the CDW state is gradually destabilized by the Na substitution, it still persists (e.g., $T_{DW}$ = 30 K at $x = 0.25$): $T_{DW}$ forms a downward concave curve in the region of a higher $x$, showing a saturation tendency. The alkali metal substitution by K [279] and Rb [280] is also effective in raising $T_c$. The K substitution ($Ba_{0.88}K_{0.12}Ti_2Sb_2O$) provides the highest $T_c$ of 6.1 K, which is due to smaller chemical disorder because of the similarity in ionic radius between $K^+$ and $Ba^{2+}$.

It is also possible to control physical properties by aliovalent anion Sb substitution. We have prepared $BaTi_2(Sb_{1-x}Sn_x)_2O$ for $x \leq 2.5$ and obtained the electronic phase diagram as a function of $x$ as shown in Fig. 32(b) [33]. A qualitative resemblance is seen between phase diagrams of hole doped systems, $BaTi_2(Sb_{1-x}Sn_x)_2O$ and $(Ba_{1-x}A_x)Ti_2Sb_2O$ (A = alkali metals). The saturated behavior of $T_c$ in the highly doped regime may be related to the robustness of the CDW phase. From quantitative point of view, however, $BaTi_2(Sb_{1-x}Sn_x)_2O$ differs from $(Ba_{1-x}Na_x)Ti_2Sb_2O$. The maximum $T_c$ in $BaTi_2(Sb_{1-x}Sn_x)_2O$ is 2.5 K at $x = 0.3$, which is remarkably smaller than $T_c$ = 5.5 K in $(Ba_{1-x}Na_x)Ti_2Sb_2O$ at $x$ ~ 0.33. The reduced $T_c$ in the former system is attributed to the greater chemical disorder induced by the *Pn*-site substitution and also to the less 36conicity in the Sb/Sn atoms in comparison with Ba/Na atoms.

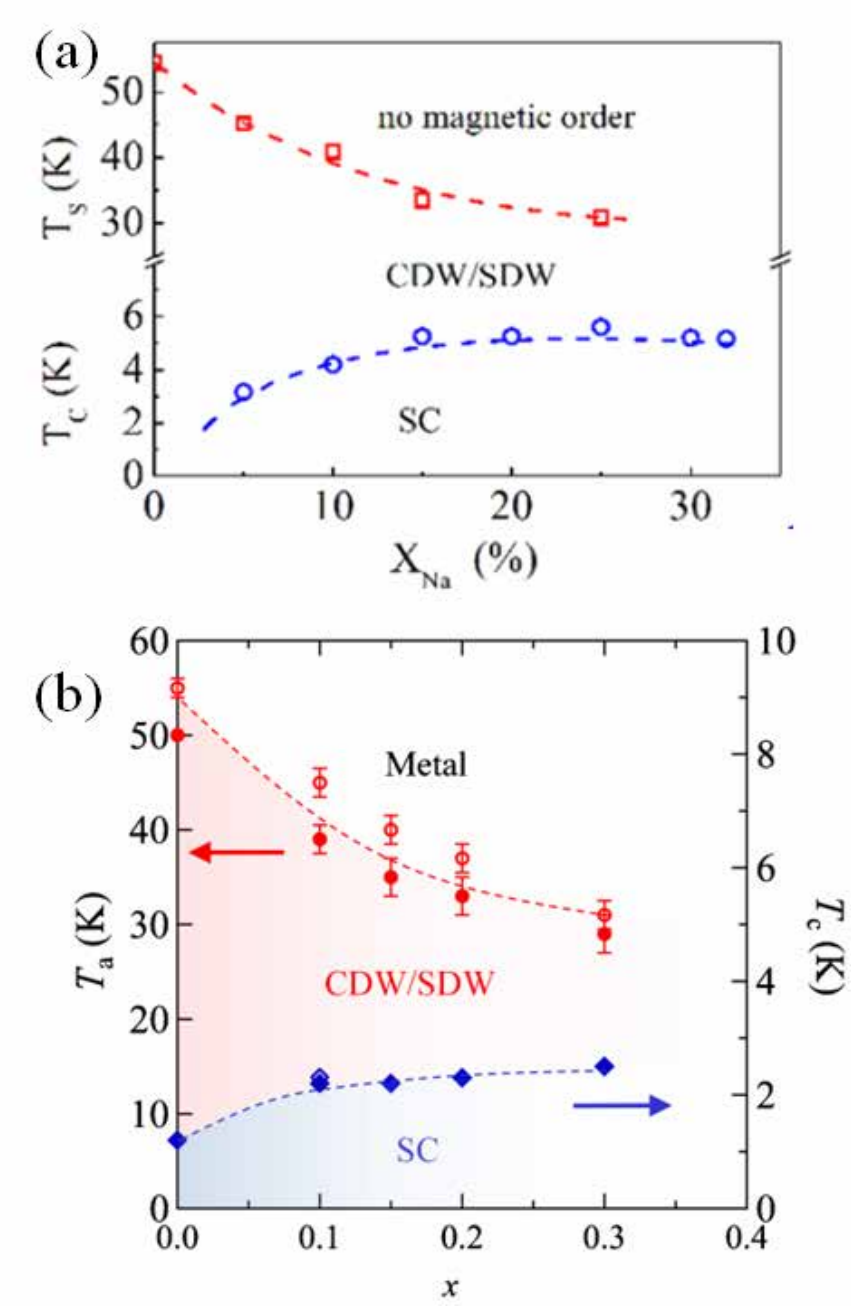

**Figure 32.** Phase diagram of (a) $Ba_{1-x}Na_xTi_2Sb_2O$, taken from [274] and (b) $BaTi_2(Sb_{1-x}Sn_x)_2O$, taken from [33].

*3.2.5. Isovalent anion substitution*

The comparison between $BaTi_2As_2O$ ($T_{DW}$ = 200 K, $T_c$ = 0 K) and $BaTi_2Sb_2O$ ($T_{DW}$ = 50 K, $T_c$ = 1 K) led us to prepare the isovalent anion substitution system $BaTi_2Bi_2O$ as so as to destabilize the CDW state and stabilize the superconducting state. It is shown from the structural refinement (Fig. 28(b)), $BaTi_2Bi_2O$, the first oxybismuthide in this family, is isostructural with $BaTi_2Sb_2O$ [31]. The magnetic susceptibility and electric resistivity indeed showed no anomaly indicative of CDW transition, meaning that hole doping is less effective in terms of destabilizing the CDW state than the isovalent anionic substitution. As a result, the enhanced $T_c$ of 4.6 K is successfully observed in $BaTi_2Bi_2O$. The lattice parameters of $BaTi_2Bi_2O$ ($a$ = 4.12316(4) Å, $c$ = 8.3447(1) Å) are not so different from those of $BaTi_2Sb_2O$, indicating that the lattice expansion is not the primary factor to destabilize the CDW state. Compared with Sb, Bi is less electronegative and provides more covalent Ti-*Pn* bonding. Furthermore, the Bi-6*p* orbital is more diffuse than the Sb-5*p* orbital. Therefore, in $BaTi_2Bi_2O$ the Bi-6*p* orbital possibly contributes more to the density of states at the Fermi surface than the Sb-5*p* orbital in $BaTi_2Sb_2O$. It results in weaker nesting of the Fermi surface for $BaTi_2Bi_2O$ and the suppressed CDW instability.

The possession of three compounds $BaTi_2Pn_2O$ (*Pn* = As, Sb, Bi) allows us to prepare isovalent anionic solid solutions, $BaTi_2(As_{1-x}Sb_x)_2O$ ($0 \le x \le 1$) and $BaTi_2(Sb_{1-y}Bi_y)_2O$ ($0 \le y \le 1$) [32]. Despite the Vegard's law behavior of the unit cell parameters in both solid solutions, it was unexpectedly observed a novel electronic phase diagram as summarized in Fig. 33(b), in marked contrast to that of the aliovalent substitution system (Fig. 32(a)). The gradual destabilization of the CDW state is seen going from $x$ = 0 to $x$ = 1 in $BaTi_2(As_{1-x}Sb_x)_2$, whereas superconductivity appears at $T_c$ = 0.5 K for $x$ = 0.9. The $T_c$ increases with increasing $y$ (Bi), takes a maximum value of 3.5 K for $y$ = 0.2 and further Bi substitution lead to a decrease in $T_c$. For $y$ = 0.4 and 0.5, no superconductivity is seen down to the lowest temperature measured (1.85 K).

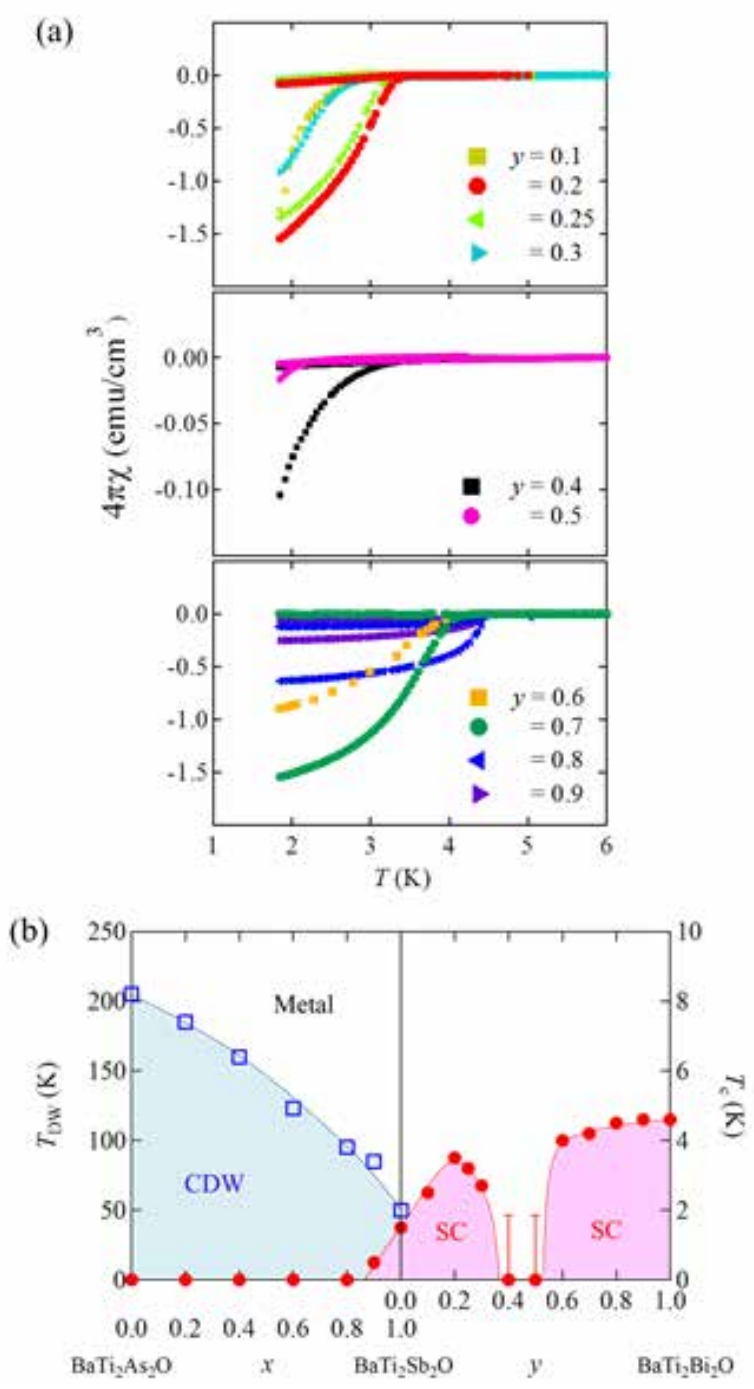

**Figure 33.** (a) Low-temperature magnetic susceptibility for $BaTi_2(Sb_{1-y}Bi_y)_2O$, taken from [32]. (b) The electronic phase diagram of $BaTi_2(As_{1-x}Sb_x)_2O$ and $BaTi_2(Sb_{1-y}Bi_y)_2O$, demonstrating the presence of two superconducting phases.

Most remarkably, superconductivity revives at $y = 0.6$, with an increased $T_c$ toward $y = 1$ (4.6 K). The presence of the second superconducting state for $0.6 \leq y \leq 1$ is quite unusual and implies novel mechanism behind the superconductivity. It is theoretically shown that a multiband structure involving three Ti-3*d* orbitals ($d_{xy}$, $d_{z2}$, and $d_{x2-y2}$) is present at the Fermi surface [290, 294]. It is thus likely that the second superconducting phase is different in nature from the first superconducting phase. Recently, a two-dome structure in $T_c$ has also been reported in the iron pnictide superconductor $LaFeAs(O_{1-x}H_x)$ [11], where the multiband structure comprising of Fe-3*d* orbitals is proposed. It should be noted that the two-dome structure in $LaFeAs(O_{1-x}H_x)$ is induced by the aliovalent substitution and the two superconducting regions are not separated. Hence, $BaTi_2Pn_2O$ might give a better opportunity to understand the nature of multiband Fermi surface.

### 3.2.6. Interlayer interactions

In $BaTi_2Pn_2O$, the CDW state becomes destabilized as $Pn$ = As ($T_{DW}$ = 200 K) → Sb ($T_{DW}$ = 50 K) → Bi ($T_{DW}$ = 0 K). The magnetic susceptibility and electric resistivity of $(SrF)_2Ti_2Bi_2O$ (Fig. 29(f)) has no signature of CDW transition [76]. However, zero-resistivity and diamagnetic signal are absent in $(SrF)_2Ti_2Bi_2O$ in spite of the fact that the in-plane lattice parameters are similar to each other. Since both $(SrF)_2Ti_2Bi_2O$ and $BaTi_2Bi_2O$ contain the $[Ti_2Bi_2O]^{2-}$ unit, it is possibly essential to consider the role of the $A^{2+}$ unit, or the interlayer coupling in order to explain the difference in their superconducting properties. The $(SrF)_2^{2+}$ unit is bulky with metal halide double layers, which provides an elongated interlayer distance of 10.685 Å, as compared with 8.345 Å in $BaTi_2Bi_2O$. Hence, the interlayer interaction, at least to a certain extent, is a key component for the appearance of superconductivity. Indeed, the application of external pressure on $Ba_{1-x}Na_xTi_2Sb_2O$ results in the increase in $T_c$ [286].

### *3.2.7. Mixed anion compounds*

Most of functional materials in our hands explored are single anion compounds such as oxides, sulfides, chlorides, and bromides. Hence, exploring materials with mixed anion configurations around a transition metal would be promising toward realizing new or improved functional properties including high $T_c$ superconductivity. For example, a oxyhydride cubic perovskite $BaTiO_{3-x}H_x$ ($x \sim 0.6$) with $TiO_{6-n}H_n$ ($n$ = 0, 1, 2) octahedral coordination (Fig. 34(a)), obtained by a topochemical reducing reaction using $CaH_2$ [268], shows a novel hydride exchangeability with hydrogen gas at moderate temperature [110]. Electron doping to Ti 3$d$ $t_{2g}$ band by hydride reduction makes $ATiO_{3-x}H_x$ (A = alkali earth metal) metallic, with high conductivity of $10^2$–$10^4$ S/cm, although superconductivity is absent [95]. $SrCrO_2H$ with $CrO_4H_2$ octahedra, directly prepared from high temperature and high pressure reaction, also adopts the cubic perovskite structure and exhibits the highest magnetic transition among chromium oxides [96]. An oxychloride layered perovskite $(CuCl)LaNb_2O_7$ with $Cu^{2+}$ ($S$ = 1/2) in $CuO_2Cl_4$ octahedral coordination (Fig. 34(b)), prepared by a topochemical ion-exchange reaction of $RbLaNb_2O_7$ [295], shows spin-liquid behavior with a finite gap in the excitation spectrum, due to quantum fluctuations enhanced by two-dimensional structure [295]. $Ba_2BiSb_2$, a hypervalent compounds with a unique "square-honeycomb" lattice (Fig. 34(c)) exhibits a CDW transition [54]. Isovalent anion substitution by Bi (i.e., $Ba_2Bi(Sb_{1-x}Bi_x)_2$ leads to the destabilization of the CDW phase and superconductivity wit the maximum $T_c$ of 4.4 K for $Ba_2Bi_3$ ($x$ = 1) as shown in Fig. 34(d).

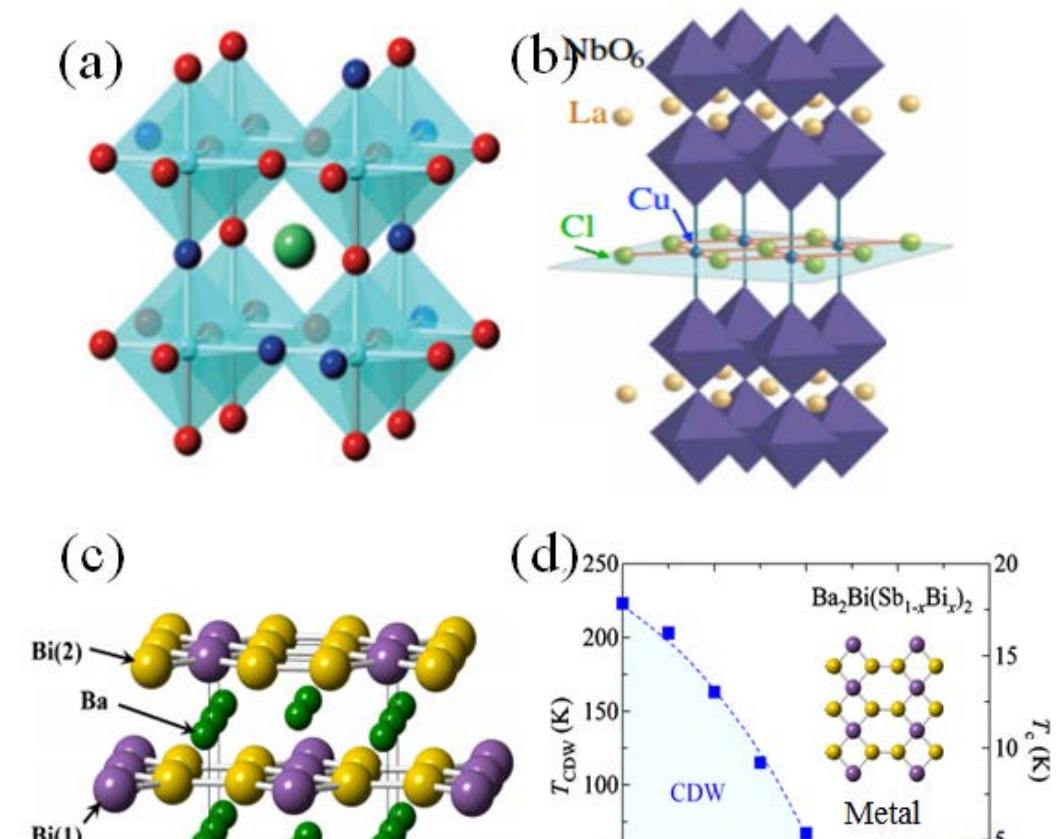

**Figure 34.** Crystal structure of (a) $BaTi(O,H)_3$, (b) $(CuCl)LaNbO_7$ and (c) $BaBiSb_2$. (d) The phase diagram of $Ba_2Bi(Sb_{1-x}Bi_x)_2$, taken from [54].

### *3.2.8. Conclusion*

We have demonstrated recent studies on layered titanium compounds $ATi_2Pn_2O$ ($Pn$ = As, Sb, Bi) with $Ti^{3+}$ in a square lattice, where Ti is coordinately octahedrally with four $Pn$ and two O atoms. A newly synthesized $BaTi_2Sb_2O$ shows superconductivity at $T_c$ = 1.2 K. The superconducting phase is of fully gapped BCS-type and is competing/coexisting with a CDW phase with $T_{DW}$ of 1.2 K. The aliovalent cantion/anion substitution leads to the stabilization of the superconducting state while the CDW state is destabilized only partially. The isovalent anion solid solution, $BaTi_2(As_{1-x}Sb_x)_2O$, shows a conventional phase diagram with the superconducting phase competing with the CDW phase. However, another isovalent anion solid solution, $BaTi_2(Sb_{1-y}Bi_y)_2O$, reveals the appearance of the second superconducting state for $0.6 < y$. The presence of the two superconducting phases strongly indicates the multi-orbital contribution to superconductivity, as also seen in iron arsenic superconductors. The nature of the second superconducting phase is not clear. This phase is possibly competing with other (presently unseen) phases. Further experiments as well as theories are necessary. It is important to note that superconductivity is only observed for $A$ = Ba. Further exploratory studies may find superconducting materials with different blocking layers. Finally, we believe that mixed anionic materials are fruitful playground for novel functional properties including high $T_c$ superconductivity.

### *3.3. Intercalation compounds with layered and cage-like structures*

High-$T_c$ superconductors recently developed have layered or cage-like structures such as cuprates [263], iron pnictides[4], $MgB_2$ [196], alkali metal doped fullerides [297], and so on. Those have intercalated structures composed of charged (doped) layers or frameworks coupled with charge reservoirs in the interlayer or cage-like space. In this study, we will also explore new superconductors, focusing on layered and cage-like structures with covalent networks, and dope electrons via intercalation. Layer structured metal nitride halides, and alloys with clathrate related structures have been developed; electrons are doped by means of intercalation using the interstices between the layers, and the cages.

#### *3.3.1. Intercalation compounds of metal nitride halides*

There are two kinds of layered polymorphs in metal nitride halides $MNX$ ($M$ = Ti, Zr, Hf; $X$ = Cl, Br, I), α- and β-forms with the FOCl and the SmSI structures, respectively [298, 299]. The α-form layered polymorph has an orthogonal $M$N layer network separated by halogen layers as shown in Fig. 35(a). The β-form consists of double honeycomb-like $M$N layers sandwiched by close-packed halogen layers as shown in Fig. 35(b). Both polymorphs are band semiconductors with gaps larger than 2.4-4 eV. We have already reported that high-$T_c$ superconductivity is obtained on β-HfNCl upon electron doping by intercalation of alkali metals. The highest $T_c$ was observed in the lithium and tetrahydrofuran (THF) cointercalated compound $Li_{0.48}(THF)_y$HfNCl at 25.5 K [300]. The Zr homolog $Li_x$ZrNCl also exhibits superconductivity at $T_c$ ~14 K [301]. The electron doped $M$NCl ($M$ = Zr, Hf) show unconventional superconductivity [302,303]; the electron carrier concentration is very low, but the electron-phonon coupling constant observed ($\lambda_{e\text{-}ph} << 1$) is too small to explain the relatively high $T_c$'s [304]. Unexpectedly large gap ratios have been observed by specific heat ($2\Delta/k_BT_c = 6.5$) [305] and tunneling spectroscopy ($2\Delta/k_BT_c$ = 7-10) [306] measurements. The small isotope effect is also unconventional [307]. Since these superconductors do not contain any magnetic ions, a magnetic pairing mechanism is excluded. Charge and spin fluctuations have been proposed as the possible candidates for the pairing mechanism [308-310].

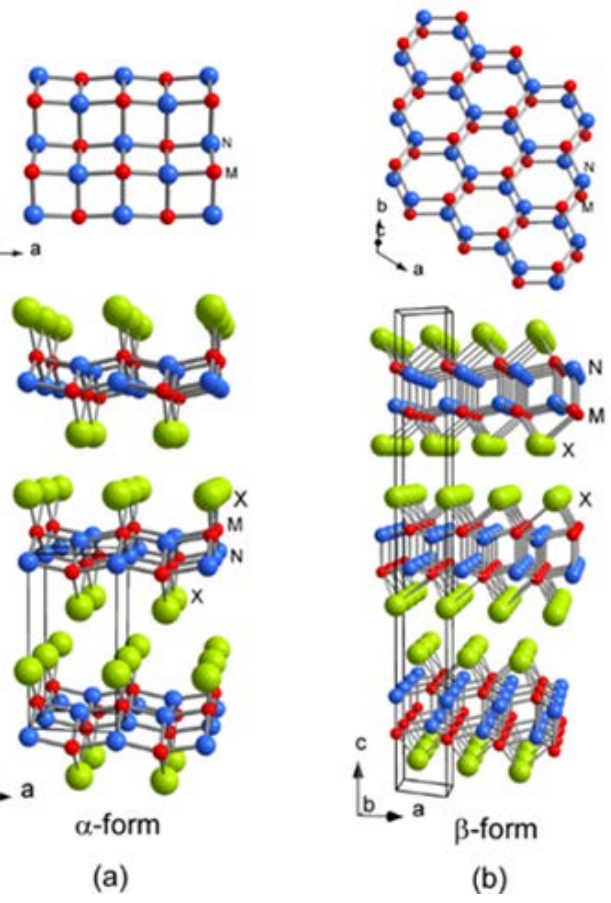

**Figure 35.** Schematic illustration of the crystal structures of (a) the α- and (b) the β-forms of MNX (M = Ti, Zr, Hf; X = Cl, Br, I): small red balls, M; blue balls, N; and large green balls, X. The lower part of the illustration shows the views along the *b*-axes, and the upper part shows the two dimensional nitride layers of each form [299].

In this study the β-form layered compounds have been highly electron-doped by using liquid ammonia solutions of alkaline earth as well as rare-earth metals. TiNCl with the α-form layered structure has been changed into superconductors by electron doping with alkali metals and organic bases such as pyridine and alkylene diamines.

*3.3.1.1. Intercalation of alkaline and rare-earth metals in β-HfNCl and β-ZrNCl*

In alkali metal (*A*) intercalated compounds, *A* atoms occupy the octahedral or trigonal prism sites between chlorine layers of β-*M*NCl [58, 311, 312]. The largest metal concentration attained is expected to be $x = 0.5$ for $A_xM$NCl ($M$ = Zr, Hf). We are interested in the electron doping to a much higher concentration using multivalent metals. Alkaline earth (*AE* = Ca, Sr, Ba) [59] and rare earth metals (*RE* = Eu, Yb) [60] have been successfully introduced into the interlayer space of parent materials by using liquid ammonia solutions, which make it possible to study the effect of high doping concentration and additional magnetic spin, respectively. The dependence of the metal doping concentration $x$ of $AE_x(Solv)_y$HfNCl (*AE* = Ca, Sr, Ba; *Solv* = $NH_3$, THF) onto the $T_c$ is shown in Fig. 36 [59]. Note that the $T_c$ is hardly influenced by the doping concentration $x$ up to $x$ ~0.4, corresponding to $x$ ~0.8 for monovalent alkali metal intercalated compounds $A_x(Solv)_y$HfNCl. The superconductors are not yet over-doped. The $T_c$ decreases in the following order of the increasing basal spacing ($d$) with intercalation of Sr, Ba ($d$ = ~10 Å) > $Ca_x(NH_3)_y$, $Ba_x(NH_3)_y$, $Sr_x(NH_3)_y$ ($d$ = 12 Å) > $Ca_x(THF)_y$ ($d$ = 15 Å), although a minimum level of doping is certainly necessary for the superconductivity. The as-prepared compounds are cointercalated with ammonia used as solvent, which can be replaced with THF. With varying electron-doping concentrations and interlayer spacings, the highest $T_c$ of 26.0 K was obtained for the Ca and THF cointercalated compound $Ca_{0.11}(THF)_y$HfNCl, a new record of high $T_c$ in the electron-doped metal nitride chloride system [59].

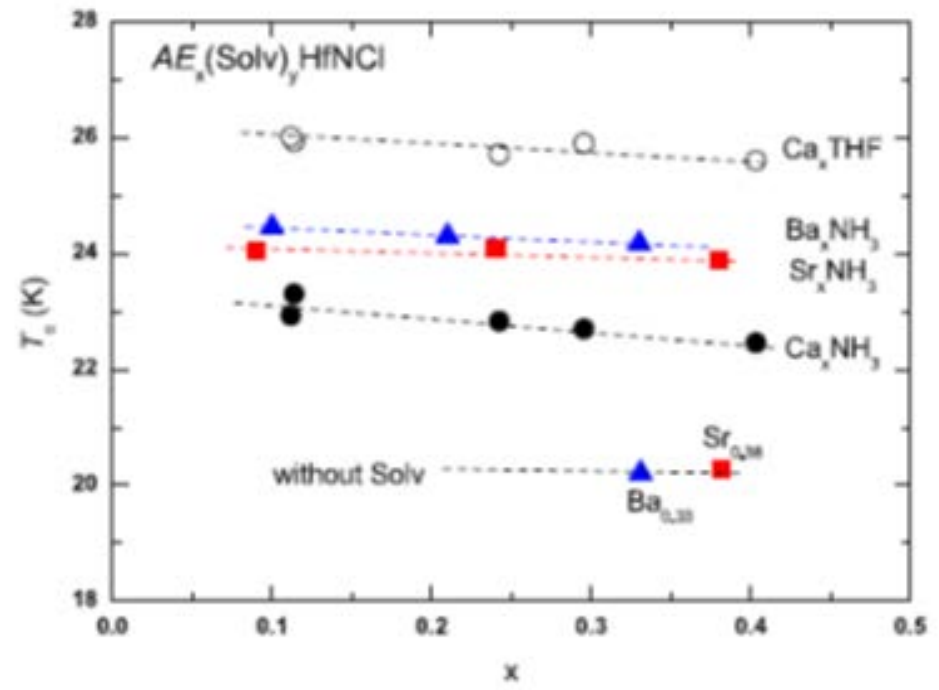

**Figure 36.** $T_c$ of the as-prepared $AE_x(NH_3)_y$HfNCl ($AE$ = Ca (●), Sr (■), Ba (▲)) as a function of concentration $x$. $T_c$'s of the evacuated $AE_x$HfNCl ($AE$ = Sr, Ba) and the THF-cointercalated $Ca_x(THF)_y$HfNCl are compared in the same figure [59].

Multivalent rare-earth metals (Eu and Yb) can be solved in liquid ammonia, and intercalated into β-$M$NCl ($M$ = Zr, Hf) from the solutions [60]. The $T_c$'s of the ammonia cointercalated compounds $RE_x(NH_3)_y$HfNCl ($RE$ = Eu, Yb) are comparable with those of $AE_x(NH_3)_y$HfNCl (Fig. 36); 24.1 and 23.0 K for $Eu_{0.13}(NH_3)_y$ and $Yb_{0.11}(NH_3)_y$, respectively. The temperature dependence of the magnetic susceptibility of $RE_x(NH_3)_yM$NCl measured in a temperature range of 100-300 K suggests that the $RE$ metals exist as paramagnetic ions $Eu^{+2}$ and $Yb^{+3}$. The anisotropic magnetoresistance measurement has evidenced that the paramagnetism of $Eu^{2+}$ and $Yb^{3+}$ can coexist with the superconductivity even under high magnetic fields up to 14 T as shown in Fig. 37 for $Eu_{0.08}(NH_3)_y$HfNCl. The anisotropic upper critical fields ($H_{c2}^{\parallel ab}$ and $H_{c2}^{\parallel c}$) were determined with the magnetic field parallel to $ab$ plane and $c$ axis, respectively on the uniaxially preferred oriented pellet sample. The anisotropic parameter $\gamma = (dH_{c2}^{\parallel ab}/dT)/(dH_{c2}^{\parallel c}/dT) = 4.1$, comparable to those of other electron-doped β-$M$NCl superconductors, 3.7 for $Li_{0.48}(THF)_y$HfNCl [313] and 4.5 for $ZrNCl_{0.7}$ [314].

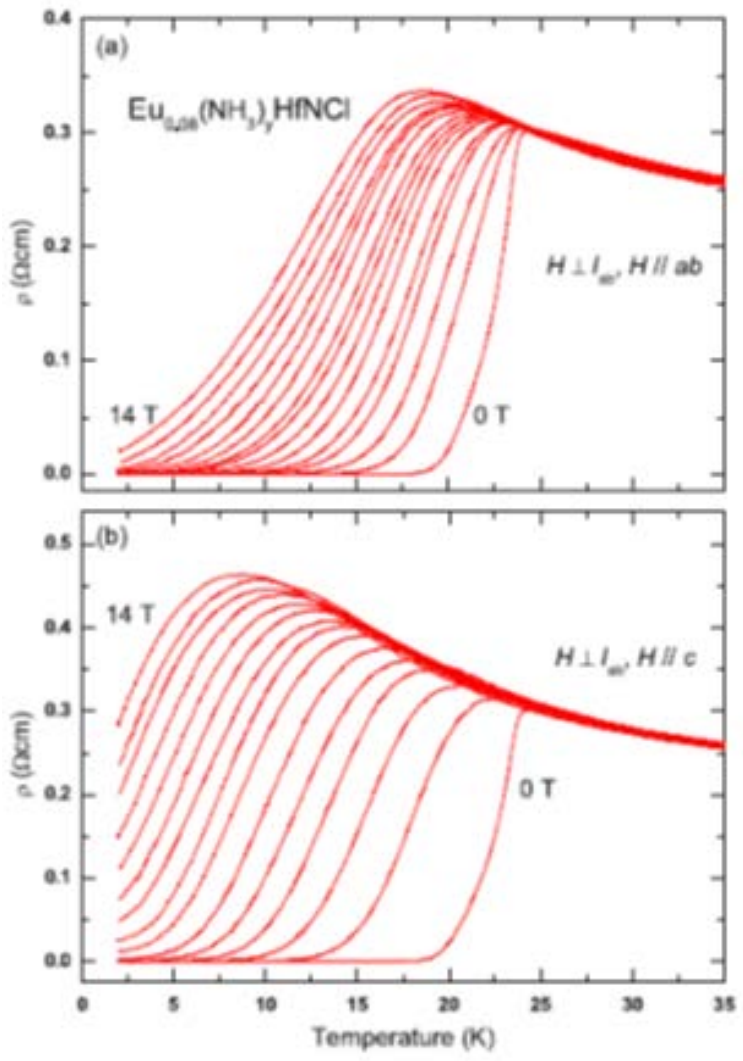

**Figure 37.** Temperature dependence of the resistivity of $Eu_{0.08}(NH3)_y$HfNCl under varying magnetic field for (a) $H \parallel ab$ ($H \perp I_{ab}$) and (b) $H \parallel c$ ($H \perp I_{ab}$). The applied magnetic field varies from 0 to 14 T with an interval of 1 T [60].

All the $T_c$'s of the electron doped β-HfNCl so far determined can fit on a single curved line as a function of the basal spacing ($d$) as shown in Fig. 38 [59], suggesting that the alkali, alkaline-earth and rare-earth metals act as similar electron dopants, and the $T_c$ is hardly dependent on the doping concentration. A very similar trend was also observed on the electron doped β-ZrNCl [315]. It should be noted that the $T_c$ increases with increasing $d$ upon cointercalation of $NH_3$ and THF, and then decreases gradually with the further increase of $d$ upon cointercalation of PC.

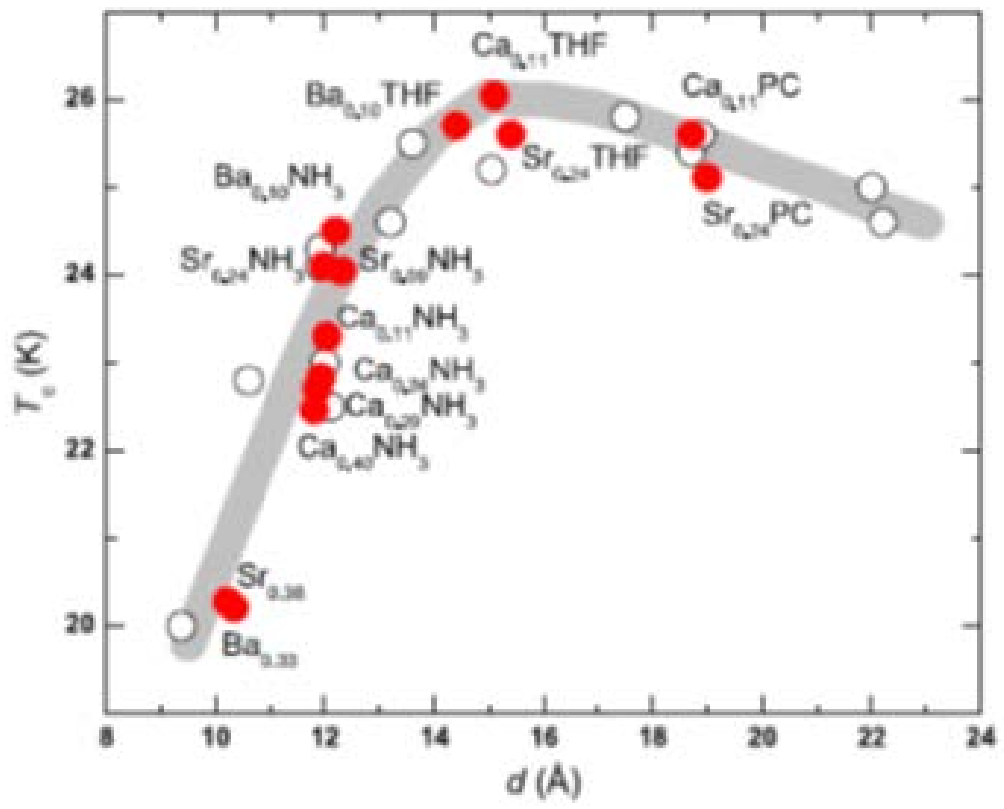

**Figure 38.** $T_c$ dependence on the basal spacing $d$ of $AE_x(Solv)_y$HfNCl (●), and from alkali metal and rare earth metal intercalated compounds (○) [59].

*3.3.1.2. Intercalation compounds of TiNCl with alkali metals*

In a previous study we prepared the alkali metal intercalation compounds of the α-form structured HfNBr, $A_x(THF)_y$HfNBr ($A$ = Li, Na) [316]. The intercalation was successful, and the color of the layered crystals changed from pale yellow to black. However, the resulting compounds were found to be not superconductors, but insulators with the resistivity > $10^7$ Ωcm. This is a quite contrast to the fact that the electron doped β-HfNBr shows high-$T_c$ superconductivity like β-HfNCl. In this study, we have used the α-form layered crystals TiNCl, which are prepared by the ammonolysis of $TiCl_4$ at elevated temperatures, followed by purification via chemical transport [317]. TiNCl can be intercalated with alkali metals as well as neutral organic molecules such as pyridine and diamines. The intercalation compounds become superconductors. The reason of the quite different behaviors of the two kinds of the α-form crystals TiNCl and HfNBr is not clear.

*(i) Alkali metal intercalation in TiNCl*

In the first attempt of the intercalation, TiNCl was subjected to reaction with various kinds of metal azides $AN_3$ ($A$ = Li, Na, K, Rb) at elevated temperatures under vacuum [317]. The azides are thermally decomposed to metal and nitrogen, and the resulting metal is intercalated into the interlayer space between chloride layers. A part of alkali metal is also used to extract or deintercalate chloride ions from the interlayer space, forming metal chloride;

$$\mathrm{TiNCl} + (x+y)\,M\mathrm{N_3} \rightarrow M_x\mathrm{TiNCl}_{1\text{-}y} + y\,M\mathrm{Cl} + 3/2\,(x+y)\,\mathrm{N_2}.$$

The metal intercalated compounds show superconductivity with $T_c$ ~16 K, although the superconducting volume fractions determined from the diamagnetic expulsion were found to be as low as 5-30%. The Rietveld analysis of the X-ray powder diffraction data revealed that the TiNCl crystalline layers are mutually shifted to accommodate the metal atoms between the chloride layers as shown in Fig. 39. The space groups of the resulting new polytypes are *Bmmb* for Na, and *Immm* for K or Rb intercalated compounds. The space group of $Li_xTiNCl$ is unchanged from *Pmmn* of the pristine crystal.

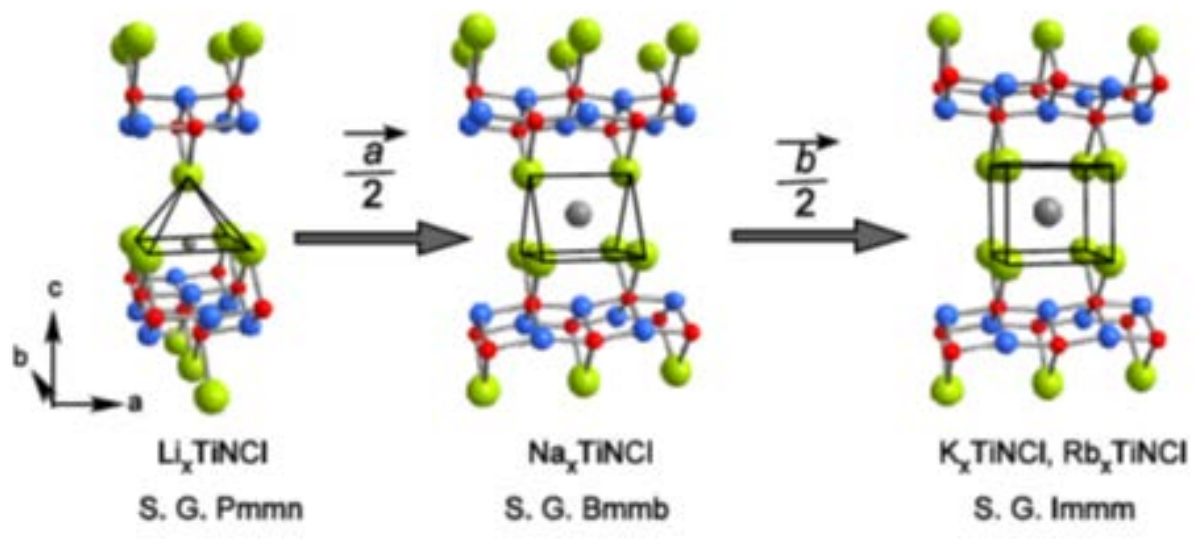

**Figure 39.** Schematic structural illustration of the TiNCl polytypes with intercalated cations between the layers [317].

A mild intercalation reaction using metal naphthalene solutions in THF can produce superconductors with a much higher volume fraction > 60% [56]. The as-prepared compound is co-intercalated with THF. As shown in Fig. 40, the basal spacing (*d*) of the compounds varies depending on the cointercalation conditions. $Na_{0.16}(THF)_yTiNCl$ has a basal spacing of 13.10 Å ($T_c$ = 10.2 K), which decreases to 8.44 Å ($T_c$ = 18.0 K) upon removal of THF by prolonged evacuation at 90 °C. The THF molecules can be replaced with larger size solvent molecules PC (propylene carbonate); the basal spacing increases to 20.53 Å with decreasing $T_c$ to 7.4 K. Figure 41 shows the $T_c$'s as a function of $1/d$ for alkali metals and solvent cointercalated compounds. The $T_c$ decreases with the increase of the basal spacing (*d*). The data fit on a linear line passing through the origin, suggesting the importance of the Coulomb interlayer coupling in the pairing mechanism in this system. It should also be noted that the $T_c$ of uncointercalated compounds $A_x$TiNCl (*A* = Na, K, Rb) also fit on this line except $Li_{0.13}TiNCl$. The as-prepared sample $Li_{0.13}(THF)_yTiNCl$ has a basal spacing of 13.1 Å similar to that of $Na_{0.16}(THF)_yTiNCl$, and a $T_c$ = 10.2 K. The uncointercalated compound $Li_{0.13}TiNCl$ was obtained by evacuation at 150 °C, which has the smallest basal spacing of 7.8 Å, the same as that of the pristine TiNCl. Unexpectedly, $T_c$ was found to be ~6.0 K, much lower than the value expected for the small basal spacing of Fig. 41. Li ions are small enough in size to penetrate into chlorine layers, forming double LiCl layers between $[TiN]_2$ layers, $[TiN]_2(Li_{0.13}Cl)(ClLi_{0.13})[TiN]_2$, in which Li ions are located close to TiN superconducting layers in parallel with Cl atoms. On the other hand, in the THF cointercalated compound, Li ions can be coordinated with THF molecules between chlorine layers $[ClTi_2N_2Cl]Li_{0.26}(THF)_y[ClTi_2N_2Cl]$. The low $T_c$ of $Li_{0.13}TiNCl$ against the small *d* suggests that the location of positive centers may also influence the Coulomb interlayer coupling for superconductivity. The linear relation shown in Fig. 41 appears to be applied to the structure where the positive centers are located between the chlorine layers as shown in Fig. 40 for the $Na_{0.16}(THF)_y$ cointercalated compound. Another systematic study on the relation between the basal spacing and $T_c$ has been performed on TiNBr, and again a similar linear relation was found [57].

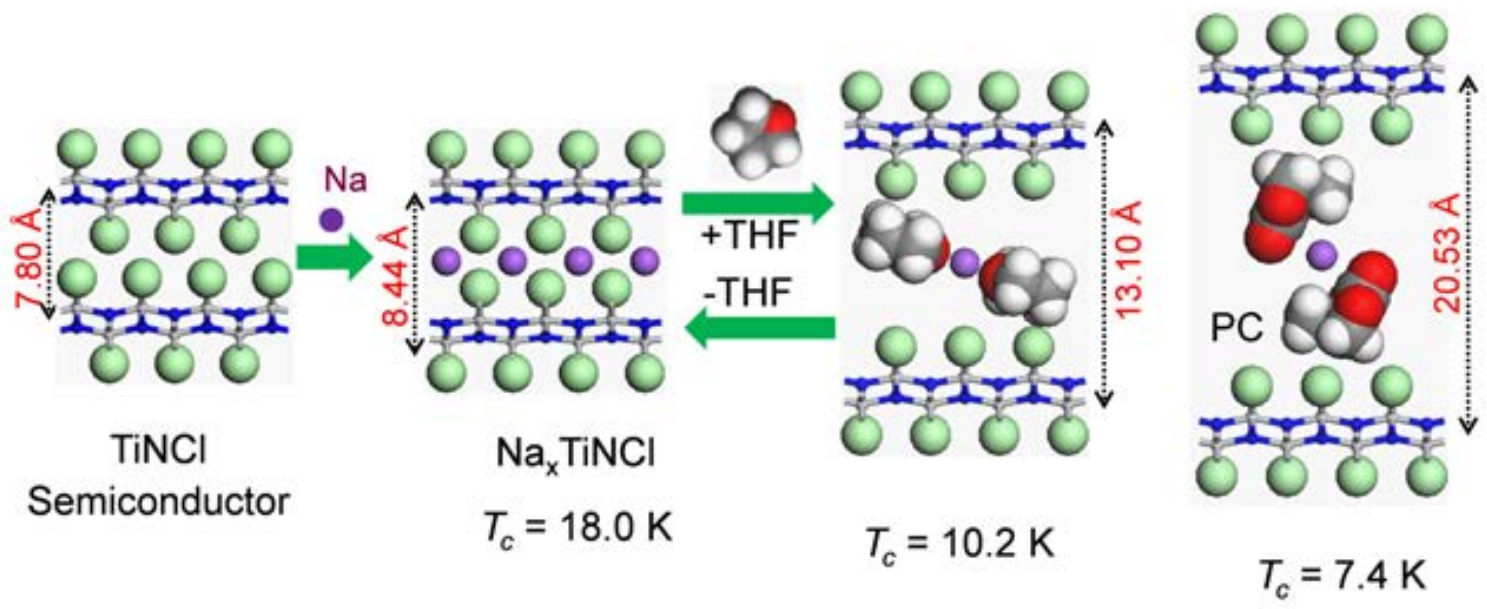

**Figure 40.** Schematic illustration showing the expansion of the basal spacing of TiNCl upon intercalation of Na and cointercalation of solvent molecules [303].

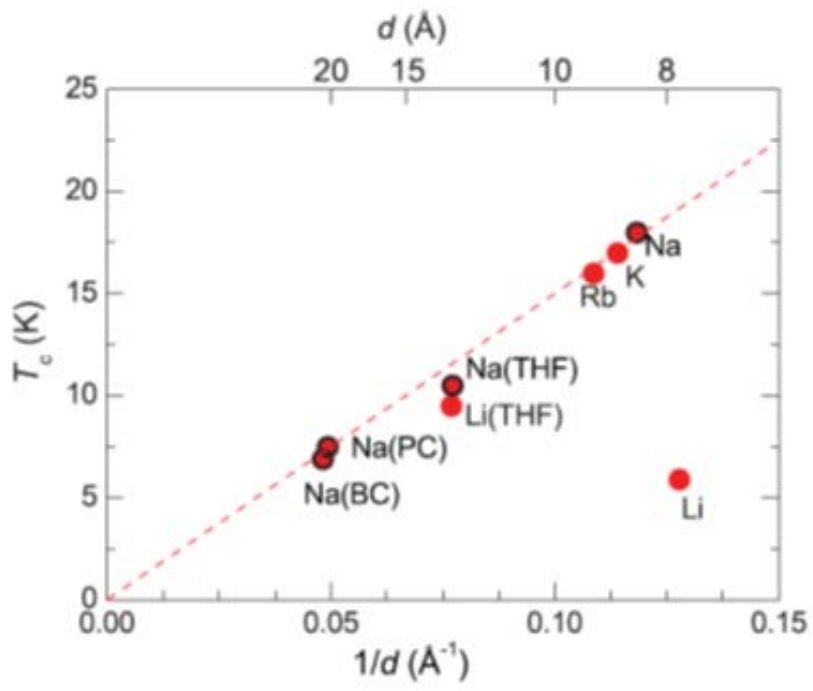

**Figure 41.** $T_c$ vs 1/*d* for *Ax*TiNCl (*A* = Li, Na, K, Rb) superconductors with and without cointercalation [56].

*(ii) Anisotropic superconducting properties*

The anisotropic magnetic susceptibility of the layered superconductors $Na_{0.16}(THF)_yTiNCl$ was measured on the highly oriented pellet sample with the magnetic field parallel and perpendicular to the *c* and *ab* plane [56]. Figure 42 shows the anisotropic $H_{c2}$ thus determined as a function of temperature. The anisotropic parameter was calculated to be $\gamma$ = 1.5 and 1.2 for $Na_x(THF)_yTiNCl$ and $Na_xTiNCl$, respectively. Note that the TiNCl superconductors exhibit rather isotropic or 3D character. The expansion of the basal spacing from 8.44 to 13.10 Å by cointercalation of THF has little effect on the anisotropy. The superconductor derived from TiNBr also shows a similar $\gamma$ value [57]. The characteristic superconducting parameters of the α- and the β-structured layered nitride superconductors are compared in Table 12. The anisotropic parameter $\gamma = \xi_{ab}/\xi_c$ (ratio of the coherence lengths in the *ab* plane and along the *c* axis) for $K_{0.21}TiNBr$ is calculated to be ~1.3, close to that found in electron-doped TiNCl; $Na_{0.16}TiNCl$ ($\gamma$ = 1.2, $T_c$ = 18.1 K) and $Na_{0.16}(THF)_yTiNCl$ ($\gamma$ = 1.5, $T_c$ = 10.2 K) [56]. On the contrary the β-structured nitrides show the anisotropic parameter $\gamma$ as large as 3.7–4.5 [59, 60, 313, 314]. The small anisotropic parameter $\gamma$ appears to be one of the characteristic features of the α-structured superconductors. The coherence length along the *c* axis ($\xi_c$) of β-$Li_{0.48}(THF)_yHfNCl$ is about 16 Å, comparable with the basal spacing 17.8 Å, i.e., the separation of the superconducting layers. This suggests that the superconducting β-form layers may be weakly Josephson coupled. On the other hand, in the α-form TiNBr and TiNCl, the $\xi_c$'s are more than three times larger than the basal spacing, implying that the nitride layers are more strongly coupled.

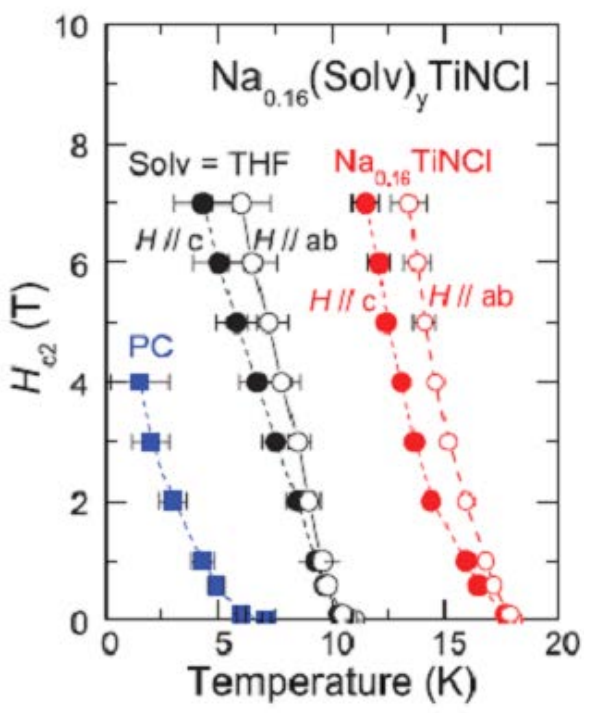

**Figure 42.** $H_{c2}$-$T$ phase diagram of Na-intercalated compounds with and without cointercalation [56].

High-$T_c$ layered cuprate superconductors have large anisotropic parameters γ on $H_{c2}$, varying in a range of 3–30 [318]. Iron pnictide superconductors *Ln*FeAs(O, F) (*Ln* = La, Sm, Nd) recently discovered also show a large anisotropy of γ = 4–9 [319-323]. The large anisotropic parameters of the layered compounds have been considered to be important to realize high-$T_c$ superconductivity. The electron-doped β-ZrNCl and β-HfNCl are also classified into this category. However, the Ba-122 superconductor such as (Ba, K)$Fe_2As_2$ with $T_c$ = 38 K has been developed, which has a small isotropic parameter γ = 1.5–1.9 [231, 319, 324-327]. The layer coupling through intervening (Ba, K) atoms seems to be stronger than those of the 1111 pnictides coupled through metal oxide layers. It is interesting to note that β-$ZrNCl_{0.7}$ (Table 12), which is electron-doped by a partial deintercalation of chlorine atoms from the interlayer space, has an anisotropic parameter as large as 4.5. In β-$ZrNCl_{0.7}$ with $d$ = 9.8 Å, the nitride layers should be directly coupled without intervening alkali atoms. Nevertheless, the anisotropic parameter is comparable to, or even larger than, that of the cointercalated compound β-$Li_{0.48}(THF)_y$HfNCl. It is evident that the interlayer separation is not a decisive parameter for the anisotropy on $H_{c2}$. The small anisotropy on $H_{c2}$ and the Coulomb coupling between the superconducting layers should be the relevant nature of the superconductivity of the α-form layered nitrides [57]. For more discussion on the superconducting mechanisms and the anisotropy of the two different kinds of layered nitrides, a theoretical study including the electric band structure is required [328].

*3.3.1.3. Intercalation compounds of TiNCl with neutral amines*

In the formation of intercalation compound of β-form layered compounds, organic solvent molecules are cointercalated with metal atoms; organic molecules alone cannot be intercalated. On the contrary α-form TiNCl can intercalate neutral organic molecules without metal atoms. TiNCl can form intercalation compound with pyridine from liquid and gas phases, $Py_{0.25}$TiNCl. The basal spacing increases to 13.5 Å with the molecular plane oriented perpendicular to the layers as schematically shown in Fig. 43 [317]. The compound becomes a superconductor with $T_c$ = 8.6 K. The $T_c$ is different from those of the alkali metal intercalated compounds. It is interesting to develop different kinds of organic compounds which can be intercalated to obtain high-$T_c$ superconductors. Although the doping mechanism is not yet clear in the intercalation compound with pyridine, it would be reasonable to estimate that the lone pair electrons of nitrogen atoms in pyridine may act as electron donors to the TiNCl layers. It was reported that FeOCl isotypic with TiNCl forms an intercalation compound with pyridine, and the electrical conductivity increases by about seven orders due to the charge transfer from the organic Lewis base to the FeOCl layers [329]. In this study, various kinds of aliphatic amines have been intercalated into TiNCl to develop new superconductivity.

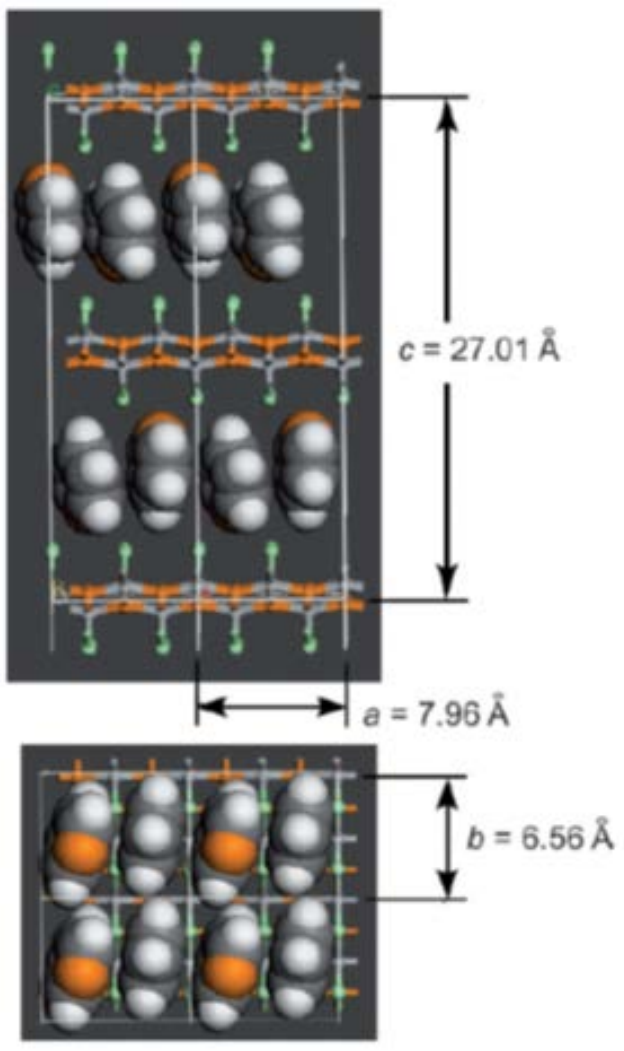

**Figure 43.** The arrangements of pyridine (Py) molecules in $Py_{0.25}TiNCl$ obtained by geometrical optimization. A 2 x 2 x 1 supercell is used; the views along the *b*-axis (top) and along the *c*-axis (bottom). Nitrogen atoms are orange-colored [317].

*n*-Alkyl monoamines ($C_nH_{2n+1}NH_2$, $3 \leq n \leq 12$) can form intercalation compounds with TiNCl, expanding the basal spacing to a value in the range of 12.0 to 37 Å, with the alkyl chains oriented in various ways. All of the compounds with *n*-alkyl monoamines are not superconductors down to 2 K [55]. It is interesting that ethylene diamine ($NH_2CH_2CH_2NH_2$) can form a similar intercalation compound with TiNCl with a basal spacing of 11.12 Å, which shows superconductivity with $T_c$ = 10.5 K [55]. Systematic studies have been done using alkylene diamines with different numbers of carbon atoms, $NH_2C_nH_{2n}NH_2$ ($2 \leq n \leq 12$). The results are shown in Fig. 44 [55]. Most of the diamine intercalation compounds are superconductors. The basal spacings are near 12.5 Å irrespective of the chain length of diamines, suggesting that the alkylene chains are aligned with the molecular axis parallel to the layers, and oriented along the *b*-axis as shown in Fig. 45. Diamines with even number of carbon atoms appear to have larger superconducting volume fractions, and the diamines with longer alkylene chains are suitable for higher $T_c$. The compound with *n* = 10 (decamethylene diamine, DMDA) shows a large volume fraction > 50%, and $T_c$ = 17.1 K, which is comparable with $T_c$ = 18.1 K of $Na_{0.16}TiNCl$. Mechanisms for the superconductivity are not clear, and remain open problems for physicists as well as chemists.

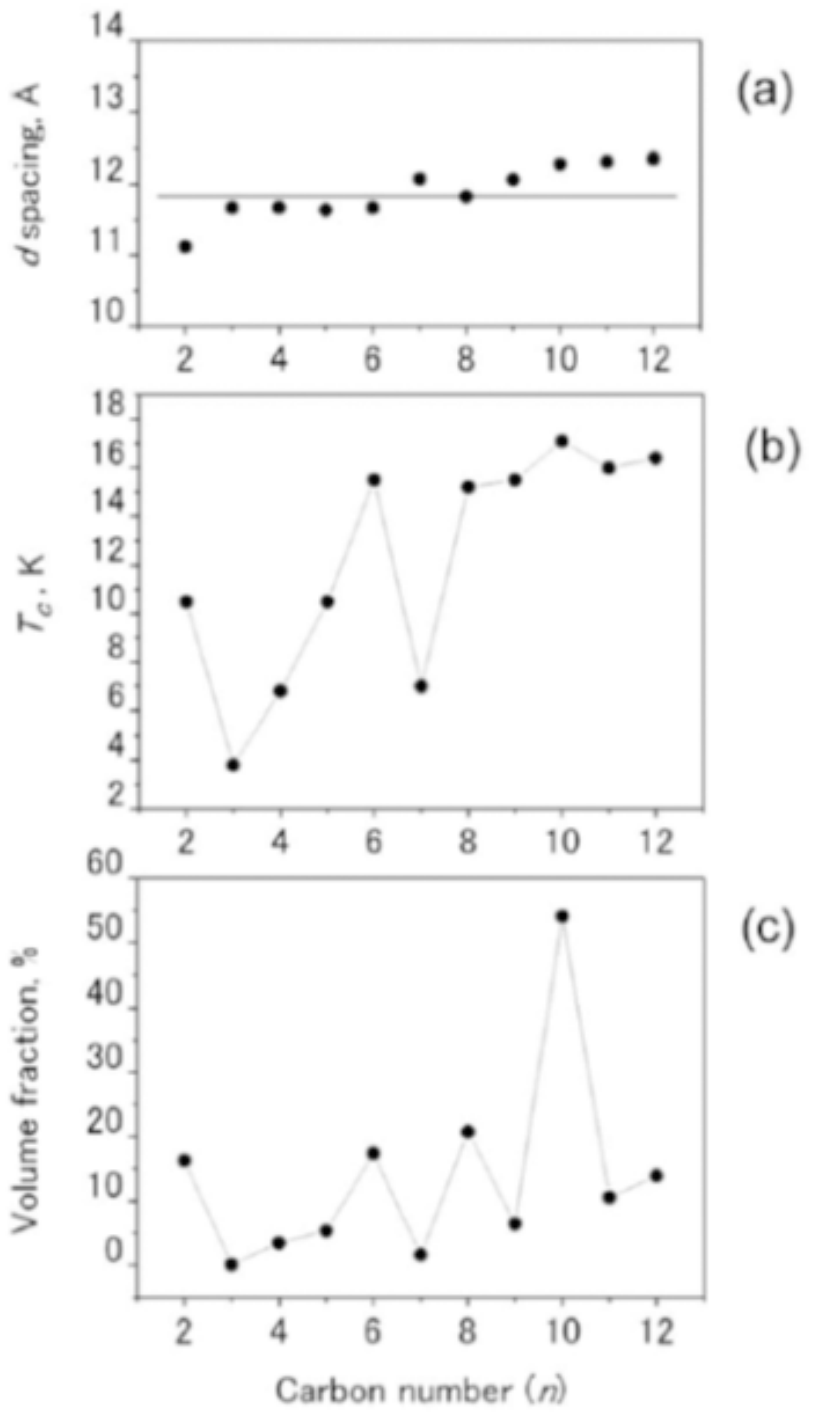

**Figure 44.** (a) Basal spacing ($d$), (b) superconducting transition temperature ($T_c$) and (c) superconducting volume fraction as a function of the number of carbon atoms ($n$) in the alkylene chains of diamine in the $(NH_2C_nH_{2n}NH_2\ (2 \leq n \leq 12))_x$-TiNCl compounds [55].

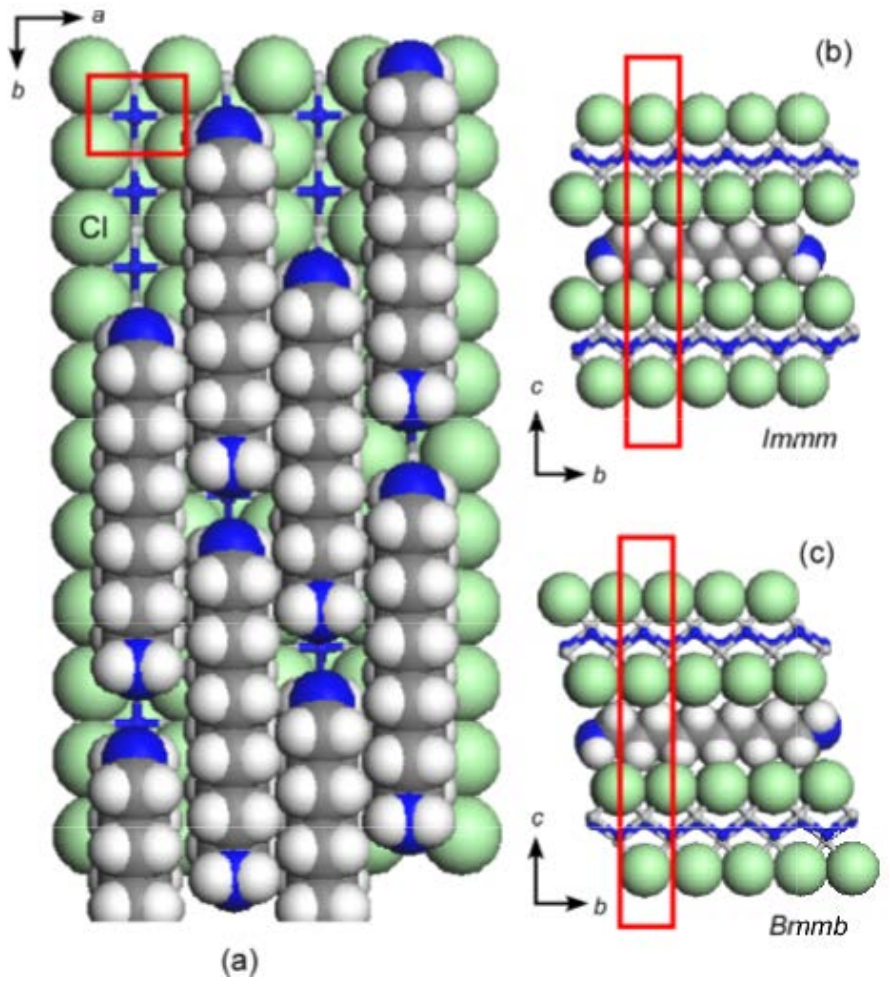

**Figure 45.** Schematic illustration showing the arrangements of alkylene diamine molecules between TiNCl layers; (a) *ab* projection: the linear alkylene diamine molecules are oriented parallel to the layers and aligned along the *b*-axis, (b) the *bc* projection for the arrangement of nonamethylene diamine (NMDA) molecules in the interlayer space of the TiNCl host structure with the space group *Immm*, and (c) the *bc* projection for the arrangement of decamethylene diamine (DMDA) molecules in the interlayer space of the host structure with the space group *Bmmb*. The unit cells are shown in red color in the projections [55].

### *3.3.2. Silicon clathrates and related compounds with cage-like structures*

We have prepared the barium containing silicon clathrate compound $Ba_8Si_{46}$ by using high pressure and high temperature (HPHT) conditions, which shows superconductivity with $T_c = 8.0$ K [330]. This is the first superconductor with a clathrate structure. The application of HPHT conditions is favorable to synthesize silicon rich binary phases such as $LaSi_5$, $LaSi_{10}$, $BaSi_6$, $Ba_{24}Si_{100}$, $Ba_8Si_{46}$, $NaSi_6$, $Na_8Si_{46}$, and $Na_xSi_{136}$ [98, 331-334]. The silicon rich compounds are generated under high pressure by obeying Le Chatelier's principle; the molar volume of the reactant of the system decreases in the product by forming covalent networks and high coordination environments. Electrons are doped from metals into the covalent networks. Most compounds are found to become superconductors. In this study ternary systems of ubiquitous (commonly found) elements have been developed using HPHT conditions.

*3.3.2.1. Ternary system Ca-Al-Si under HPHT conditions*

Ca-Al-Si ternary metal mixtures were pre-melted using an arc furnace, followed by remelting at 1000-1200 °C using radio frequency induction heating under an Ar atmosphere in an *h*-BN crucible. The cooled ternary mixtures were supplied for the HPHT treatment up to 5-13 GPa and 600-1000 °C by using a Kawai-type multianvil apparatus. A new ternary compound $Ca_2Al_3Si_4$ was obtained above 650 °C under a pressure of 5 GPa. It crystalizes with the space group $Cmc2_1$ and the lattice parameters $a = 5.8846(8)$, $b = 14.973(1)$, and $c = 7.7966(8)$ Å [45]. The structure is composed of aluminum silicide framework [$Al_3Si_4$] and layer structured [$Ca_2$] network interpenetrating with each other as shown in Fig. 46. The [$Ca_2$] subnetwork has the isomorphous structure with black phosphorus. The compound shows superconductivity with $T_c$ of 6.4 K.

Under a higher pressure of 13 GPa at 1000 °C, solid solutions $Ca(Al_{1-x}Si_x)_2$ ($0.35 \leq x \leq 0.75$) isomorphous with the cubic Laves phase were obtained [71]. As shown in Fig. 47 the structure can be regarded as a kind of clathrate compound composed of face-sharing truncated tetrahedral cages with Ca atoms at the center, $Ca@(Al,Si)_{12}$. The compound with a stoichiometric composition CaAlSi shows superconductivity with $T_c$ of 2.6 K [71]. This is the first superconducting Laves phase compound composed solely of commonly found elements.

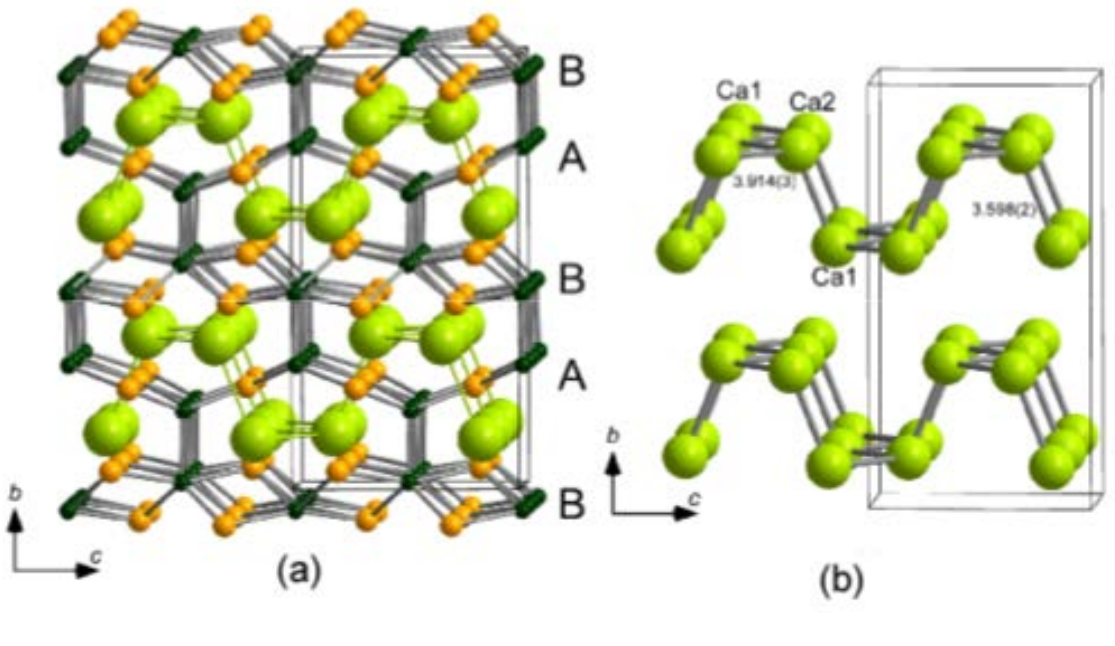

**Figure 46.** Schematic representation of the crystal structure of (a) $Ca_2Al_3Si_4$ composed of **A** [$AlSi_2$] and **B** [$Al_2Si_2$] layers with Ca atoms: small black balls, Si; orange balls, Al; green balls, Ca. (b) [$Ca_2$] layer emphasizing the subnetwork isomorphous with black phosphorus [45].

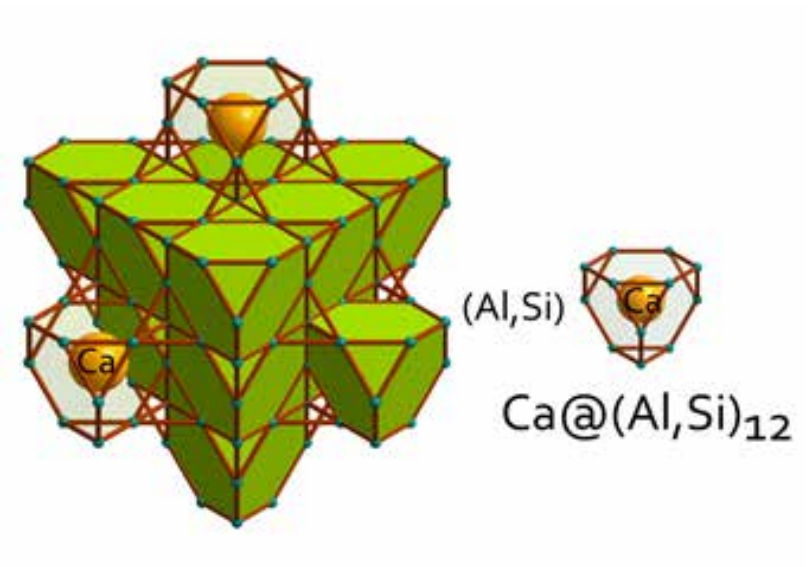

**Figure 47.** Crystal structure of the Laves phase $Ca(Al,Si)_2$ composed of face-sharing truncated tetrahedral $Ca@(Al,Si)_{12}$ [71].

*3.3.2.2. Ternary system Al-Mg-Si under HPHT conditions*

Although the ternary system Al-Mg-Si is an important subject in the development of commercial Al alloys [335], $Mg_2Si$ with the antifluorite structure is the only one compound known as bulk phase in the ternary system under ambient pressure. However, various kinds of fine coherent precipitates are formed in the Al matrix during low temperature aging, which are considered to play an important role for the hardening of commercial Al alloy. It is reasonable to assume that the ternary and binary precipitates found in Al-based alloys are formed under high pressure generated by Al matrix of the alloy. In this study, new binary and ternary compounds in the ternary system Al-Mg-Si have been prepared using HPHT conditions, and the structures are determined using single crystals. Some of them become superconductors.

Ternary compounds $Mg(Mg_{1-x}Al_x)Si$ ($0.3 < x < 0.8$) have been prepared under HPHT conditions of 5 GPa at 800−1100 °C. The single crystal study revealed that the compound ($x = 0.45$) is isomorphous with the anticotunnite, or the TiNiSi structure, and crystallizes with space group *Pnma*, with lattice parameters $a$ = 6.9242(2), $b$ = 4.1380(1), $c$ = 7.9618(2) Å, and $Z$ = 4. The compound with $x > 0.5$ shows superconductivity with a $T_c$ ~ 6 K [70]. The compound is a peritectic solid solution associated with other phases such as $Mg_9Si_5$, Al, and Si, depending on cooling protocols in the preparation. The band structure calculation on the composition MgAlSi suggests that the Al and Mg orbitals mainly contribute to the density of states near the Fermi level, and the substitution of Mg with Al favors the superconductivity.

Two kinds of magnesium-based compounds $Mg_9Si_5$ and $Mg_4AlSi_3$ have been prepared under a similar HPHT condition. Single crystal study revealed that $Mg_9Si_5$ crystallizes in space group $P6_3$ (No. 173) with the lattice parameters $a$ = 12.411(1) Å, $c$ = 12.345(1) Å, and $Z$ = 6 [46]. The structure can be derived from the high pressure form $Mg_2Si$ with the anticotunnite structure; excess Si atoms of $Mg_9Si_5$ form Si−Si pairs in the prismatic cotunnite columns running along the $c$ axis. $Mg_4AlSi_3$ is obtained by a rapid cooling of a ternary mixture Mg:Al:Si = 1:1:1 from ~800 °C to room temperature under a pressure of 5 GPa. The compound crystallizes in space group *P4/ncc* (No. 130) with the lattice parameters $a$ = 6.7225(5) Å, $c$ = 13.5150(9) Å, and $Z$ = 4 [46]. As shown in Fig. 48, the structure consists of an alternate stacking of [$AlSi_2$] layers having a Cairo pattern and [$Mg_4Si$] antitetragonal prismatic layers. It can be viewed as composed of hexa-Si-capped tetragonal prismatic cages $Mg_8Si_6$ with an Al atom at the center of each cage, $Al@Mg_8Si_6$. The compound shows superconductivity with a transition temperature $T_c$ = 5.2 K. The formation regions of the two kinds of new magnesium-based compounds have been proposed.

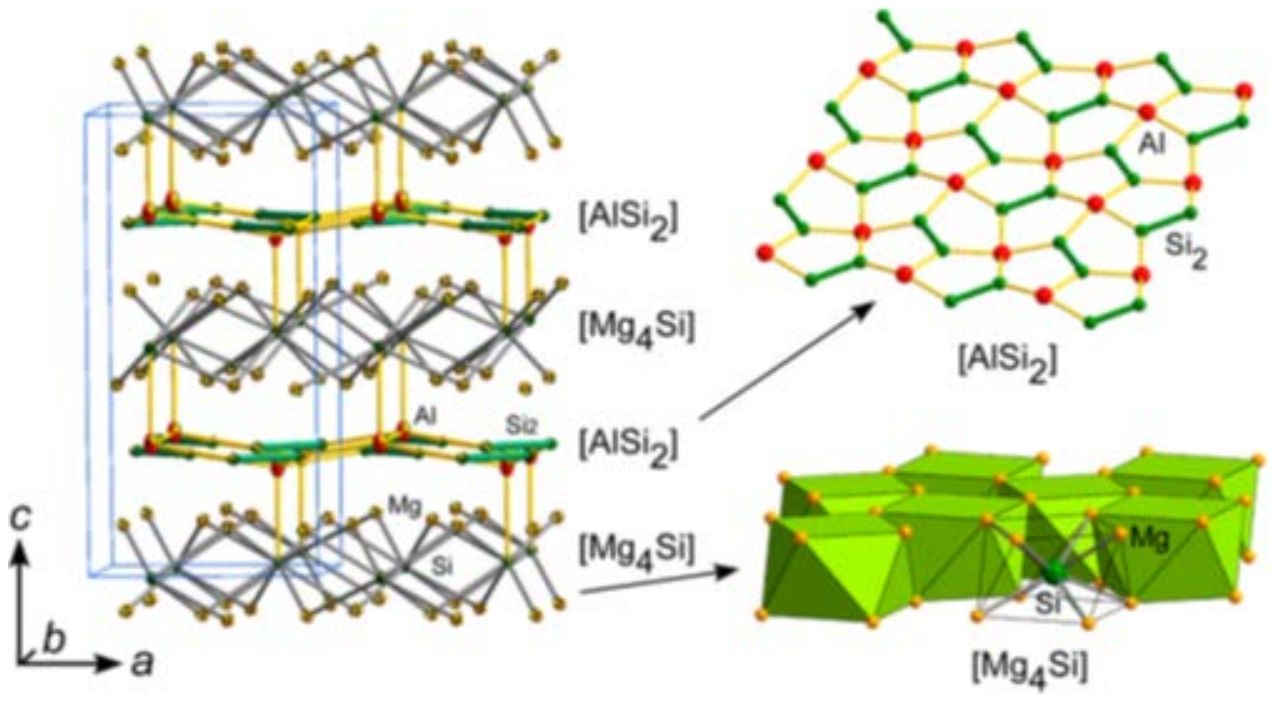

**Figure 48.** Structure of $Mg_4AlSi_3$ composed of alternate stacking of $[Mg_4Si]$ and $[AlSi_2]$ layers. The $[AlSi_2]$ layers form an almost coplanar Cairo pattern with a short Si−Si bond distance (2.43 Å). The $[Mg_4Si]$ layers consist of Mg antitetragonal prism polyhedra surrounding Si atoms [46].

*3.3.2.3. An attempt to prepare carbon analogs for silicon clathrate compounds*

Silicon and carbon chemistries are often discussed comparatively from the viewpoint of belonging to the same group in the periodic table [336, 337]. It is interesting to prepare metal doped carbon clathrate compounds, in which carbon forms $sp^3$ clathrate network like open diamond framework. A hole doped diamond was found to become a superconductor [338]. If the synthesis of an electron doped carbon clathrate is realized, it is expected that the carbon $sp^3$ framework with a high Debye temperature should exhibit high-$T_c$ superconductivity [339].

We have already obtained 3D carbon frameworks with cages by polymerization of $C_{60}$ crystals under HPHT conditions [340, 341]. An attempt has been made of preparing Ba doped clathrate-like structure from Ba doped fulleride, $Ba_3C_{60}$ using HPHT conditions.

The powder $Ba_3C_{60}$ sample was compressed using a Kawai-type multianvil press at 5-15 GPa and 500-1150 °C [97]. The X-ray powder diffraction (XRD) pattern showed that the $Ba_3C_{60}$ compressed under 15 GPa at 900 °C is changed into amorphous solids, which was found to be chemically stable in air even in water. The high micro Vickers hardness (1700 kg/mm$^2$) and the Raman spectra of the solids suggest that the $C_{60}$ molecules are collapsed to form amorphous 3D polymer encapsulating Ba atoms. As shown in Fig. 49, the solid obtained by compression under 15 GPa at 900 °C shows a semi-metallic conductivity. It is interesting to note that the Hall coefficient of this sample is positive in the whole temperature range < 300 °C, indicating that the dominant carriers are holes. As shown in Fig. 50, the covalent diameter of a Ba atom is too large to substitute one C atom in the carbon matrix. The diameter is rather comparable with the covalent diameter of a six-membered carbon ring. If a six-membered carbon ring with 12 electrons in the $sp^3$ hybridized orbitals is substituted by a Ba atom with a similar size having only two valence electrons, the carbon matrix should be efficiently hole doped. The electrical characteristics such as conductivity, Hall coefficient, carrier density, and mobility of the carbon matrix encapsulating Ba atoms were found to be comparable with those of the B-doped diamond [97]. The Ba encapsulated carbon matrix obtained in this study is amorphous. It is interesting to prepare a crystalline carbon analogs for silicon clathrate compounds.

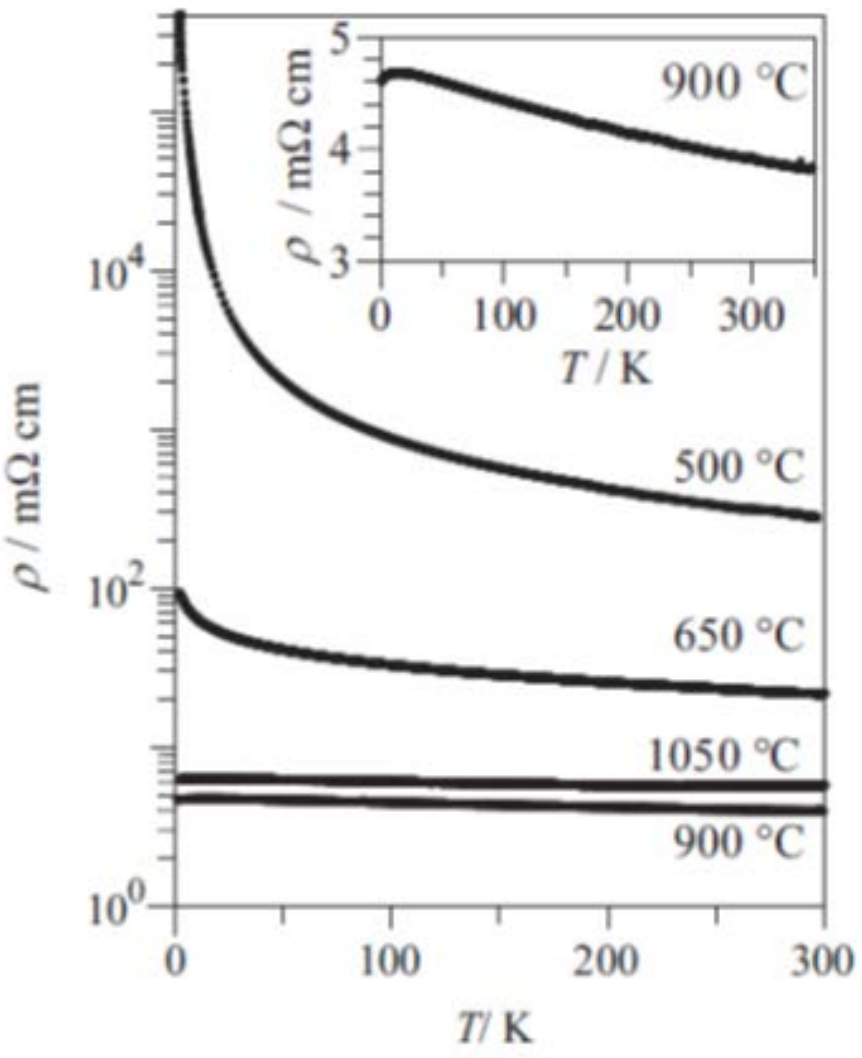

**Figure 49.** Temperature dependences of the electrical resistivity of $Ba_3C_{60}$ samples obtained by treatment at various temperatures under 15 GPa. The inset shows the resistivity in an enlarged linear scale for the sample obtained at 900 °C under 15 GPa [97].

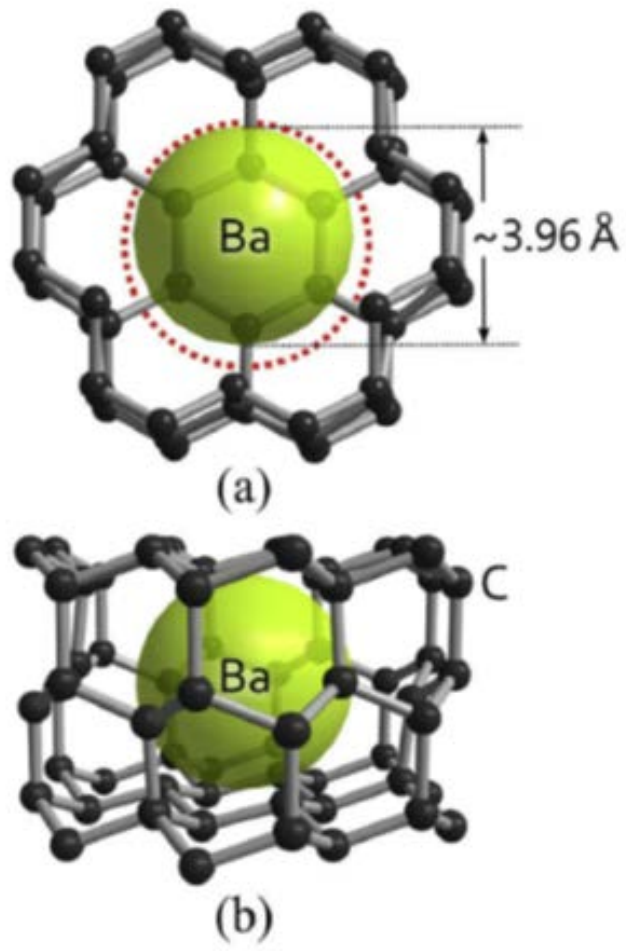

**Figure 50.** Schematic illustrations of a local arrangement of carbon network encapsulating a Ba atom in $Ba_3C_{60}$ collapsed by HPHT treatment, (a) top and (b) side views; the Ba atom (covalent diameter 3.96 Å ) replaces carbon atoms in a six membered ring with a comparable size. The covalent diameter (4.2 Å) of a six-membered carbon ring (red circle in (a)) includes the diameter of the six-membered ring (2.9 Å) and a fringe with a width of a half length of the C–C bonds formed with carbon atoms surrounding the ring [97].

### *3.4. Other new superconductors*

Besides superconductors described in the preceding sections, many new superconductors have been discovered in this project. In this section, some of them are overviewed.

#### *3.4.1. $AM_2X_2$-type (122-type, M≠Fe) superconductors*

The 122-type iron arsenide superconductors $AEFe_2As_2$ ($AE$= Ba, Ca, Sr and Eu) have been investigated intensively by many groups including the FIRST PJ team (see the preceding section). These compounds have the $ThCr_2Si_2$-type structure and non-iron 122-superconductors with the same structure have also been studied widely; this class of compounds include $SrNi_2P_2$ [342], $BaNi_2P_2$ [343], $BaRh_2P_2$, $BaIr_2P_2$ [344], $LaRu_2P_2$ [345], $LiCu_2P_2$ [346], $SrNi_2As_2$ [347], $BaNi_2As_2$ [348], etc. The $AEM_2X_2$ compounds sometimes crystallize in a different polymorph with $CaBe_2Ge_2$-type structure [349]. The two structures of $ThCr_2Si_2$-type and $CaBe_2Ge_2$-type are deeply concerned with each other as compared in Fig.51 [350]. In $ThCr_2Si_2$-type $AEM_2X_2$, $A$, $M$ and $X$ atom planes are stacked in a sequence of $A$-($X$-$M_2$-$X$)-… along the c-axis of the tetragonal cell, forming $MX_4$ coordination tetrahedron. In the $CaBe_2Ge_2$-type structure, on the other hand, the sequence of the atom planes is $AE$-($X$-$M_2$-X)-$AE$-($M$-$X_2$-$M$)… where $XM_4$ tetrahedron is formed as well as the $MX_4$ tetrahedron. Less number of compounds had been known for the $CaBe_2Ge_2$-type and $SrPt_2As_2$ with $T_c$ =5.2 K is a rare example of the superconductor having this structure [350] (strictly speaking, its structure is an incommensurate orthorhombic variant of the $CaBe_2Ge_2$-type structure [351]). Band structure calculation for $CaBe_2Ge_2$-type $SrPt_2As_2$ revealed that it has two two-dimensional (2D) like Fermi surfaces as well as two three-dimensional (3D) like ones in contrast to the strong 2D character in the electronic structure of $ThCr_2Si_2$-type $AFe_2As_2$ [352]. Several 122-type superconductors with $ThCr_2Si_2$-type, $CaBe_2Ge_2$-type or related structures were discovered in the FIRST PJ as overviewed below.

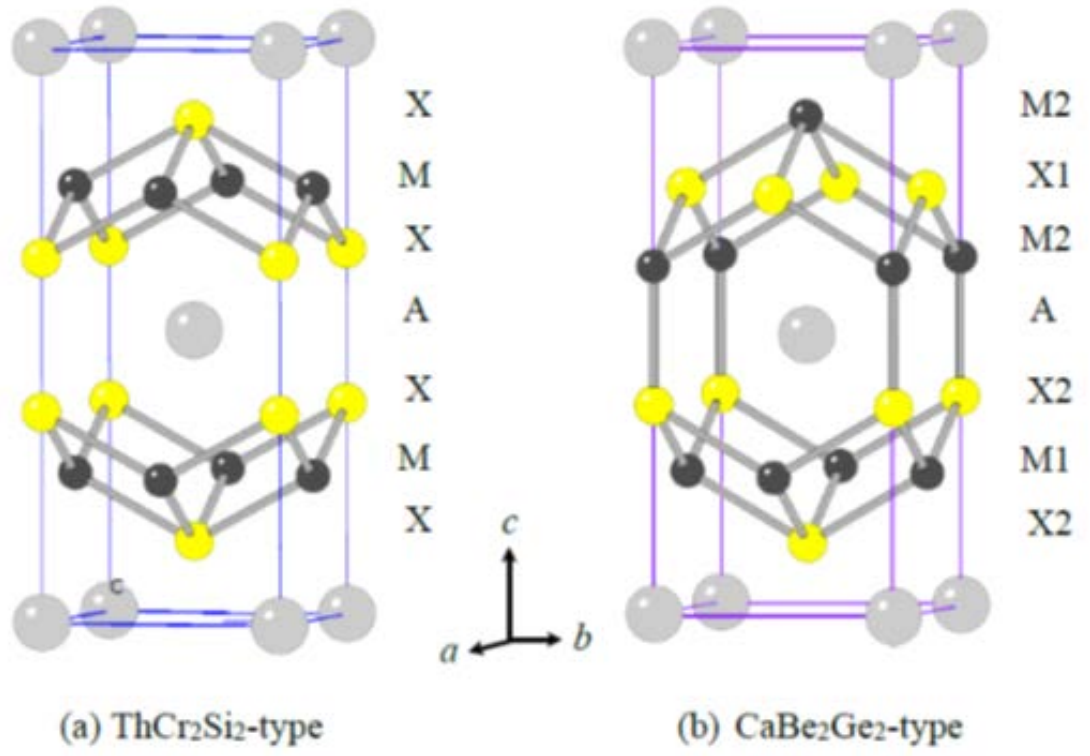

**Figure 51.** Comparison of the crystal structures of (a) $ThCr_2Si_2$-type and (b) $CaBe_2Ge_2$-type $AM_2X_2$ compounds. In the latter structure, two nonequivalent sites are present for both the $M$ and $X$ atoms as denoted by the suffixes of 1 and 2 [350].

*3.4.1.1. 122-type antimonides and a related compound*

The $SrPt_2Sb_2$ had been known to have the $CaBe_2Ge_2$-type structure with tetragonal cell of a=4.603 and c=10.565 Å [351] but any physical properties had not been reported. In the FIRST PJ, this material was revisited to elucidate its physical properties [41]. Sample was synthesized starting from Sr, Pt and Sb in two-step procedures: arc melting and re-melting of the arc-melted specimen. Though electron probe microanalysis confirmed the composition of $SrPt_2Sb_2$ (122) for the major part of the ingot obtained, powder X-ray pattern was not consistent with the $CaBe_2Ge_2$-type tetragonal lattice. Thus, $SrPt_2Sb_2$ has a different structure which may be some derivative of the $CaBe_2Ge_2$-type. As shown in Fig. 52, electrical resistivity, magnetization and specific heat measurements confirmed bulk superconducting transition at $T_c$ =2.1 K for $SrPt_2Sb_2$. It is the type-II superconductor with a lower critical field ($H_{c1}$) of 6 Oe and upper critical field ($H_{c2}$) of 1 kOe at 1.8 K. Debye temperature ($\Theta_D$) and electronic specific heat coefficient ($\gamma$) were derived from specific heat data to be $\Theta_D$ =183 K and $\gamma$ = 9.2 mJ/(mol K$^2$). The normalized specific heat jump at $T_c$ was calculated as $\Delta C(T_c)/\gamma T_c$ =1.29 in consistent with the BCS weak coupling limit of 1.43. Normal state electrical resistivity of $SrPt_2Sb_2$ exhibited anomalies around 250 K with thermal hysteresis which corresponded to a certain structural transition, though its detail has not been elucidated yet. The $SrPt_2Sb_2$ is the 122-type superconducting antimonide discovered for the first time and detailed studies for the structure and the phase transition are much desirable.

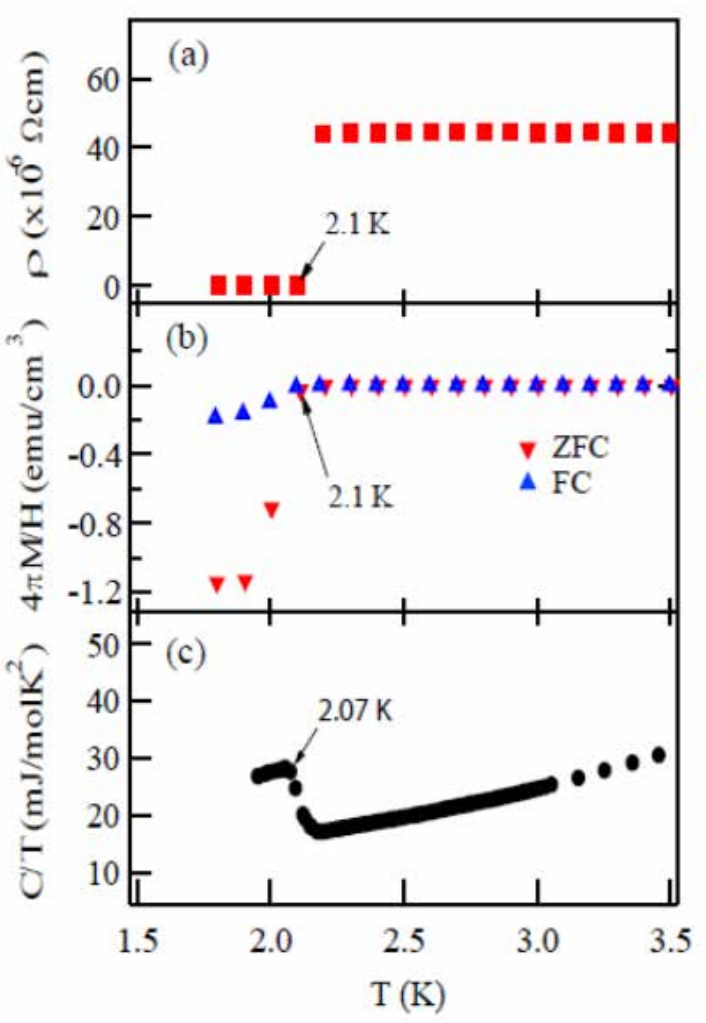

**Figure 52.** (a) Electrical resistivity (b) magnetization in zero field cooling (ZFC) and field cooling (FC) and (c) specific heat as functions of temperature for $SrPt_2Sb_2$ [41].

The Ba-derivative of the aforementioned compound, $BaPt_2Sb_2$ had not been reported, and was prepared first in the FIRST PJ by the arc melting method for a Ba, Pt and Sb mixture [42]. The X-ray pattern of the $BaPt_2Sb_2$ sample was consistent with a monoclinic lattice having parameters of a=6.702 Å, b=6.752 Å, c=10.47 Å and β=91.23°. The structure of $BaPt_2Sb_2$ is shown in Fig. 53 which can be interpreted as a monoclinic variant of the $CaBe_2Ge_2$-type structure (note that a-axis and b-axis of $BaPt_2Sb_2$ correspond to diagonal of the $a_0$-axis and $b_0$-axis of the original $CaBe_2Ge_2$-type lattice with the relationship of $a(b) \approx \sqrt{2}a_0$). Figure 54 gives temperature dependency of resistivity of $BaPt_2Sb_2$ which confirms superconducting transition at 1.8K. Specific heat data gave $\Theta_D$=146 K and $\gamma$ = 8.6 mJ/(mol $K^2$), deriving $\Delta C(T_c)/\gamma T_c$ =1.37 which is comparable with the BCS weak coupling limit. Magnetization measurements revealed type-II superconductivity with $\mu_0 H_{C2}(0)$=0.27 T and Ginzburg-Landau (GL) coherent length $\xi_{GL}(0)$=350 Å. Band structure calculation revealed that Fermi surfaces of $BaPt_2Sb_2$ resemble those of $SrPt_2As_2$ with two 2D like Fermi surfaces and two 3D like ones. Thus, it has more 3D like character compared with the $AFe_2As_2$ system which may account for the relatively lower $T_c$ of $BaPt_2Sb_2$.

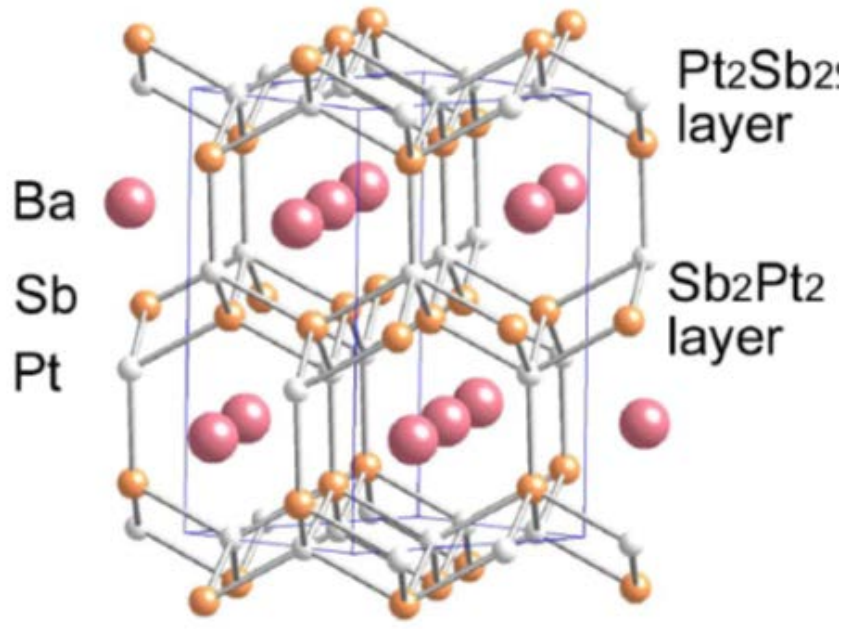

**Figure 53.** Crystal structure of $BaPt_2Sb_2$ [42].

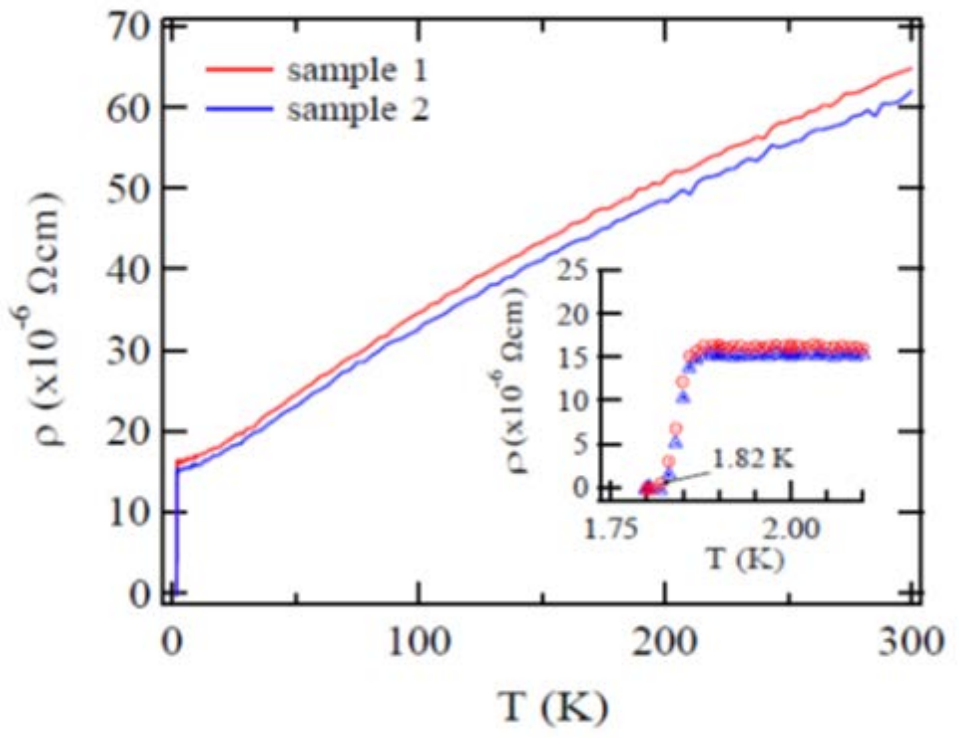

**Figure 54.** Temperature dependences of electrical resistivity of two $BaPt_2Sb_2$ samples. The inset is an expanded view of the resistivity at low temperatures ranging from 1.8 to 2.2 K [42].

Pd-based $CaBe_2Ge_2$-type antimonides such as $\mathit{Ln}Pd_2Sb_2$ (*Ln*=La, Ce, Pr, Nd, Eu) [353] were known to exist but superconductivity was not reported for this class of compounds. In the FIRST PJ, $LaPd_2Sb_2$ was selected among them and reinvestigated in details to find superconducting transition [39]. The $LaPd_2Sb_2$ sample was prepared at 900 °C starting from La, Sb and Pd in an evacuated silica tube. The resultant specimen was single phase of the $CaBe_2Ge_2$-type compound though its tetragonal lattice parameters (a=4.568 Å, c=10.266 Å) were slightly different from the previous report [353]. Figure 55 shows temperature dependence of resistivity for $LaPd_2Sb_2$ under zero and various magnetic fields. Superconducting transition occurs at onset temperature of 1.4 K and zero resistivity temperature of 1.2 K at $H$=0. Magnetization measurements revealed type-II nature of superconductivity with $\mu_0 H_{c2}(0)$=0.86 T and $\xi_{GL}(0)$=233 Å. Bulk superconductivity was confirmed by specific heat measurements which gave $\Theta_D$=210 K, $\gamma$ = 6.89 mJ/(mol K$^2$) and $\Delta C(T_c)/\gamma T_c$ =1.325 in consistent with the BCS weak coupling limit. Figure 56 gives total density of states for La, Pd1, Pd2, Sb1 and Sb2 (see Fig. 51 on the atom sites with the suffixes of 1 and 2) calculated by density functional theory (DFT). It is seen that Pd contributes the most to the total DOS at the Fermi level consistent with a general tendency that the DOS at the Fermi level are dominated by d-band of the M atom in $CaBe_2Ge_2$-type $AM_2X_2$. Figure 56 reveals hybridization between Pd 4d of Pd1(Pd2) and Sb 5p of Sb2(Sb1) as demonstrated by synchronized modulation in partial DOS of Pd1(Pd2) and Sb2(Sb1). The total DOS at the Fermi level is nearly 40 % of that of $SrPt_2As_2$ which may account for relatively lower $T_C$ of $LaPd_2Sb_2$ compared with $SrPt_2As_2$.

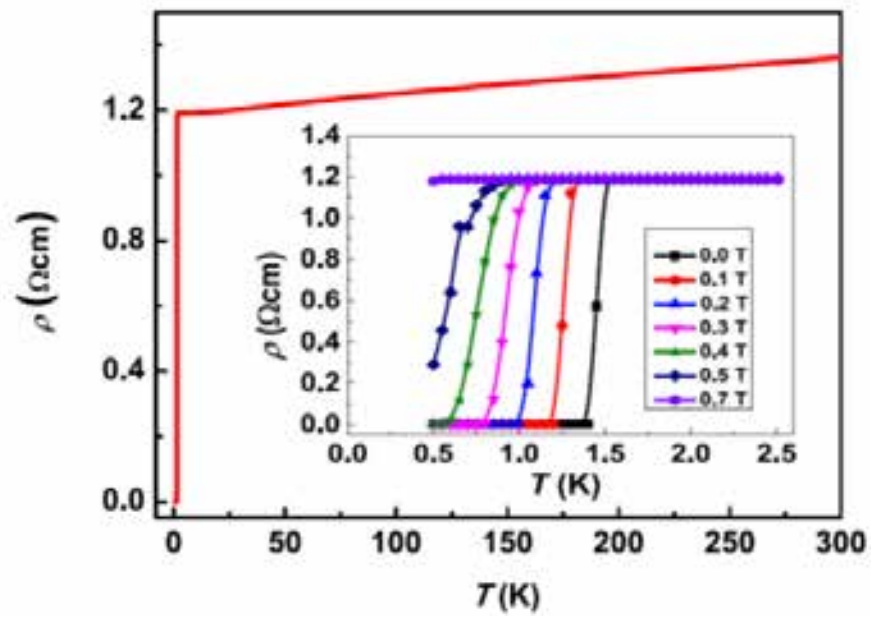

**Figure 55.** Temperature dependence of electrical resistivity for $LaPd_2Sb_2$. The inset shows the variation of resistivity with respect to magnetic field [39].

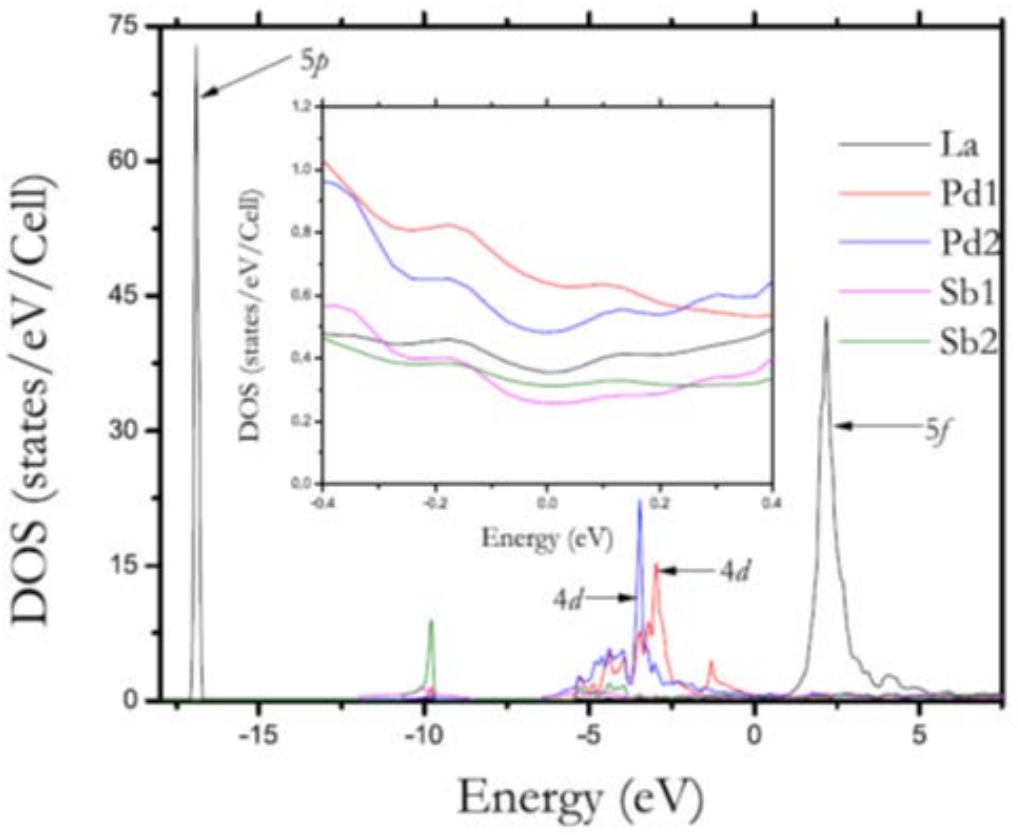

**Figure 56.** Calculated total density of states for La, Pd1, Pd2, Sb1 and Sb2 (see Fig. 51 on the atom sites with the suffixes of 1 and 2). The inset shows the contribution to Fermi level from each atom [39].

The As-derivative of the aforementioned compound, $LaPd_2As_2$, which was known to have a $ThCr_2Si_2$-type structure [353], was also studied in the FIRST PJ [40]. In the $ThCr_2Si_2$-type $AEM_2X_2$, interlayer *X-X* bonding is sometimes formed between the adjacent $M_2X_2$ layers to give the collapsed tetragonal (*cT*) structure with a 3D network. Superconductivity in the collapsed $Fe_2As_2$ planes is often attained under high pressure taking well known examples of $AEFe_2As_2$ (*AE*=Ca, Sr, Ba, Eu) which undergo both superconducting and collapsed transitions under high pressure [354-357]. Superconductivity under ambient pressure in the *cT* structure is rather rare and has been reported for $AEPd_2As_2$ (*AE*=Ca, $T_c$ =1.27 K; *AE*=Sr, $T_c$ =0.92 K) [358]. In the FIRST PJ, it was elucidated that $LaPd_2As_2$ has the *cT* structure under ambient pressure with the interlayer As-As distance of 2.318 Å, slightly shorter than the covalent single bond of 2.38 Å for As. Moreover, it was confirmed for the first time that this *cT* phase shows type-II superconductivity below 1 K. Superconducting and physical parameters obtained for $LaPd_2As_2$ are $\mu_0 H_{c2}(0)$ =0.402 T (by the WHH theory using $\rho_{90\%}$ point as $T_c$), $\xi_{GL}(0)$=137 Å, $\Theta_D$=261 K, $\gamma$ = 5.56 mJ/(mol K$^2$) and $\Delta C(T_c)/\gamma T_c$ =1.17. The density of states at the Fermi level calculated from the specific heat data is as small as ~0.84 states/(eV f.u.) which may explain the relatively low $T_c$ of this *cT* phase.

*3.4.1.2. Co-based superconductor $LaCo_2B_2$*

Only few reports are available for Co-based superconducting compounds [359, 360]. In the FIRST project, $LaCo_2B_2$, which was reported first in 1973 [361], was re-visited and was found to be superconducting after isovalent or aliovalent substitution for the constituent cations [34]. Polycrystalline samples of $(La_{1-x}Y_x)Co_2B_2$, $La(Co_{1-x}Fe_x)_2B_2$ and $LaCo_2(B_{1-x}Si_x)_2$ as well as the mother compound were prepared by the arc-melting method for La, Co, B, Y, Fe mixtures. Powder X-ray diffraction of the arc-melted $LaCo_2B_2$ sample was consistent with the tetragonal $ThCr_2Si_2$-type structure as reported previously [361], having lattice parameters of a=3.610 Å and c=10.20 Å. As shown in Fig. 57, 10% Y-doped sample showed superconductivity below 4.4 K. Magnetic susceptibility data revealed Pauli paramagnetism for the normal state of this compound and the *M-H* curve (Fig. 57 (b) inset) at 2K reveals type-II nature of superconductivity. Superconductivity with $T_c$ ~4K was also seen in the Fe-doped system of $La(Co_{1-x}Fe_x)_2B_2$ for x ≥0.1 while the Si-doped samples did not show superconductivity.

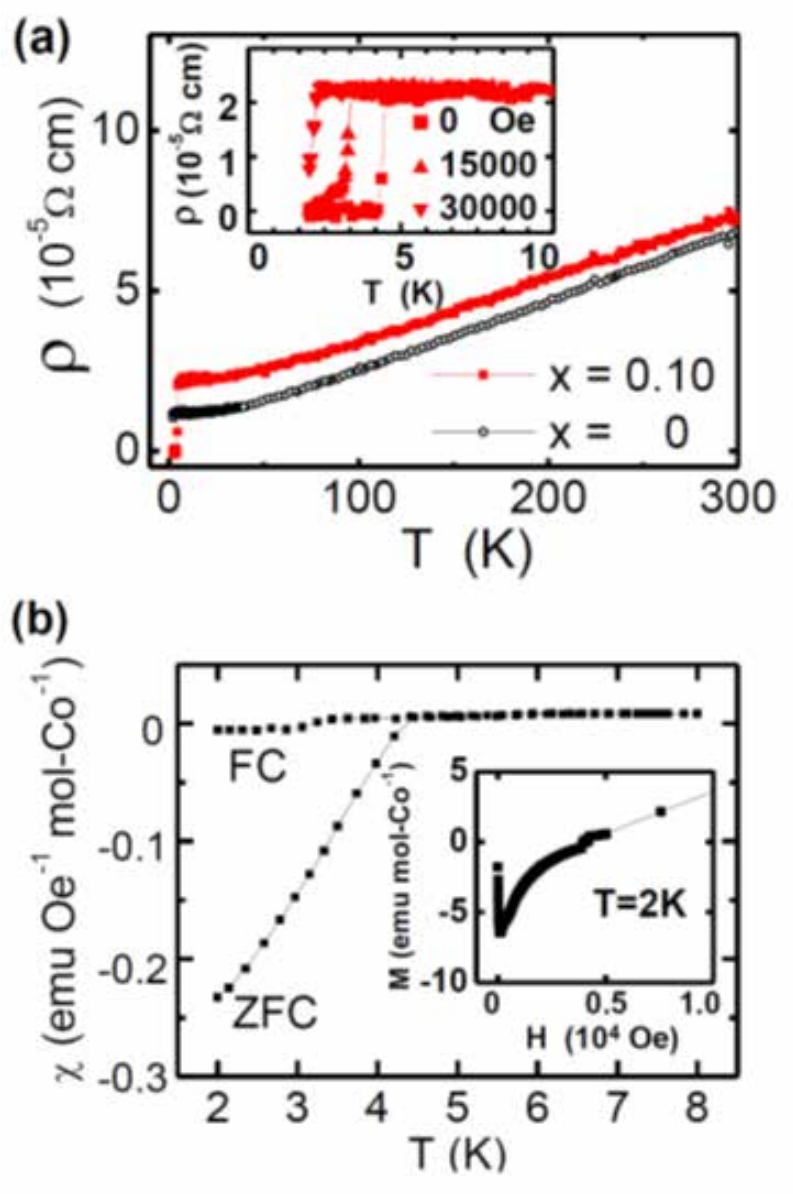

**Figure 57.** (a) Temperature dependence of electrical resistivity for $(La_{1-x}Y_x)Co_2B_2$. The inset shows temperature dependence of resistivity for the $x$=0.1 under various magnetic fields magnetic field. (b) Temperature dependence of the magnetic susceptibility for the $x$=0.1 sample under ZFC and FC conditions at 10 Oe. The inset shows the field dependence of magnetization at 2 K [34].

Electronic structure of $LaCo_2B_2$ was investigated theoretically by the DFT method; calculated DOS is shown in Fig. 58. From the DFT calculation, it was confirmed that La ions take +3 state and metallic conduction occurs in the CoB layer composed of highly covalent Co and B. This strong covalency suppresses the spin moment of the Co ion resulting in the Pauli paramagnetic state. Such a situation is caused by the relatively shallow B 2$p$ level compared with the As 4$p$ level in LaFeAsO where antiferromagnetc state is realized rather than the paramagnetic one.

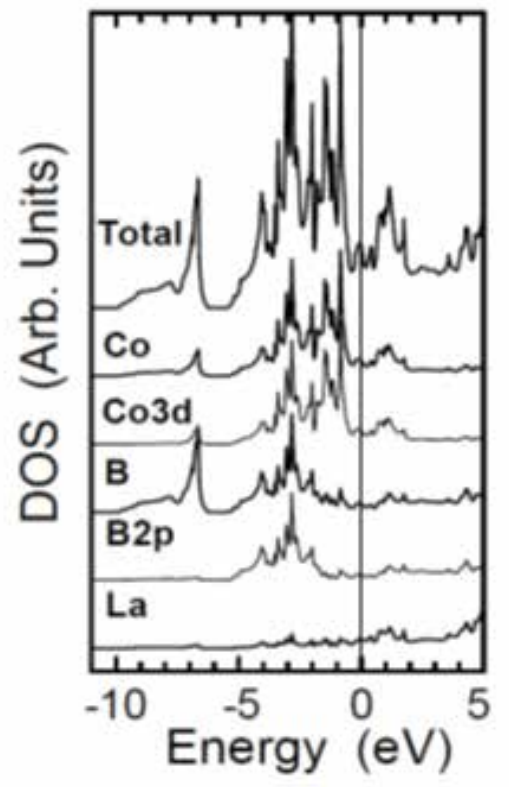

**Figure 58.** Calculated DOS of $LaCo_2B_2$, with the partial DOS (PDOS) for La, Co, Co 3$d$, B, and B 2$p$ [34].

*3.4.1.3. Enhancement of superconductivity by phosphorous doping in $BaNi_2As_2$*

$BaNi_2As_2$ has the tetragonal $ThCr_2Si_2$-type structure at room temperature [348, 362, 363] but it undergoes a structural transition at ~130 K to be a triclinic form where alternate Ni-Ni bonds are formed in the Ni plane (see Fig. 60 (a)) [362]. Below 0.7 K, the triclinic phase shows superconductivity which is believed to be of conventional BCS-type [348, 363-365]. In the FIRST project, $BaNi_2As_2$ was reinvestigated because of this unique structural transition, i.e., this material was expected to offer a stage for studying chemical tuning of soft phonons by elemental substitution and its effects on superconductivity [35].

Single crystals of $BaNi_2(As_{1-x}P_x)_2$ were grown by a self-flux method starting from a mixture of Ba, NiAs, Ni and P [35]. Solubility limit was determined to be x=0.13 and powder X-ray diffraction confirmed single-phase nature for all specimens with $x \le 0.13$. Figure 59 shows temperature dependence of the electrical resistivity parallel to the ab plane of the $BaNi_2(As_{1-x}P_x)_2$ crystal. The tetragonal-to-triclinic transition is clearly seen in this figure as the resistivity anomaly with thermal hysteresis. The transition temperature decreases with increasing x and finally the triclinic phase disappears for $x \geqq 0.07$. Enhancement of superconductivity by the phosphorous doping is striking; $T_C$ which is below 0.7 K for the triclinic phase with x<0.07, is suddenly risen up to 3.33 K in the tetragonal phase with x=0.077. Figure 60 shows phase diagram of the $BaNi_2(As_{1-x}P_x)_2$ system. The triclinic phase formation is suppressed by the phosphorous doping and instead superconductivity is enhanced drastically following the disappearance of the triclinic phase. Debye frequency $\omega_D$ and logarithmic averaged phonon frequency $\omega_{ln}$ calculated from the specific heat of the tetragonal phase exhibit significant softening near the tetragonal-to-triclinic phase boundary. The low-lying soft phonons seem to play, being strongly coupled with acoustic modes, an important role in the enhancement of superconductivity in the tetragonal phase. Indeed, the normalized specific heat jump $\Delta C(T_c)/\gamma T_c$ is increased from 1.3 in the triclinic phase to 1.9 in the tetragonal phase with x=0.077, i.e., the system is transformed from the weak coupling regime to the strong coupling one by the phosphorous doping. Such a mechanism of $T_c$ enhancement is common to other systems of $CaC_6$ [366-368] and Te [369] under high pressure.

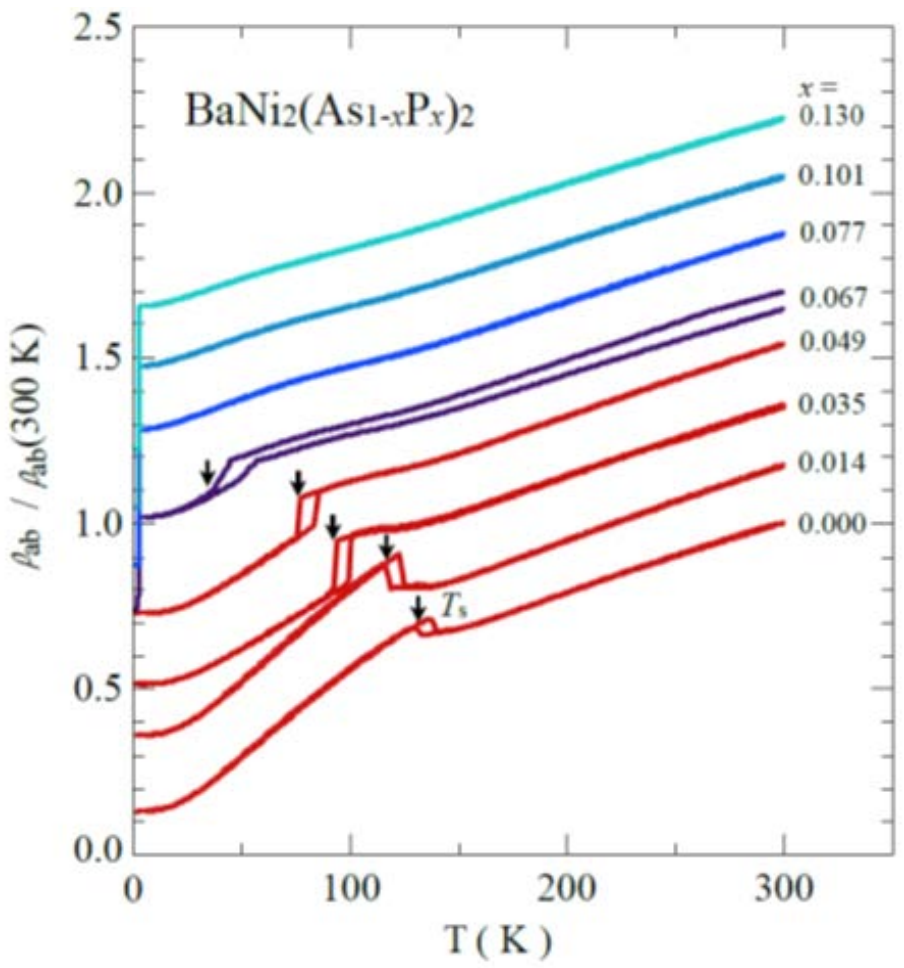

**Figure 59.** Temperature dependence of electrical resistivity parallel to the ab plane, $\rho_{ab}$, normalized by the value at 300 K for $BaNi_2(As_{1-x}P_x)_2$. The data measured upon heating and cooling are plotted. For the sake of clarity, $\rho_{ab}/\rho_{ab}(300\ K)$ is shifted by 0.175 with respect to all data. $T_s$ is the phase transition temperature at which the tetragonal-to-triclinic phase transition occurs [35].

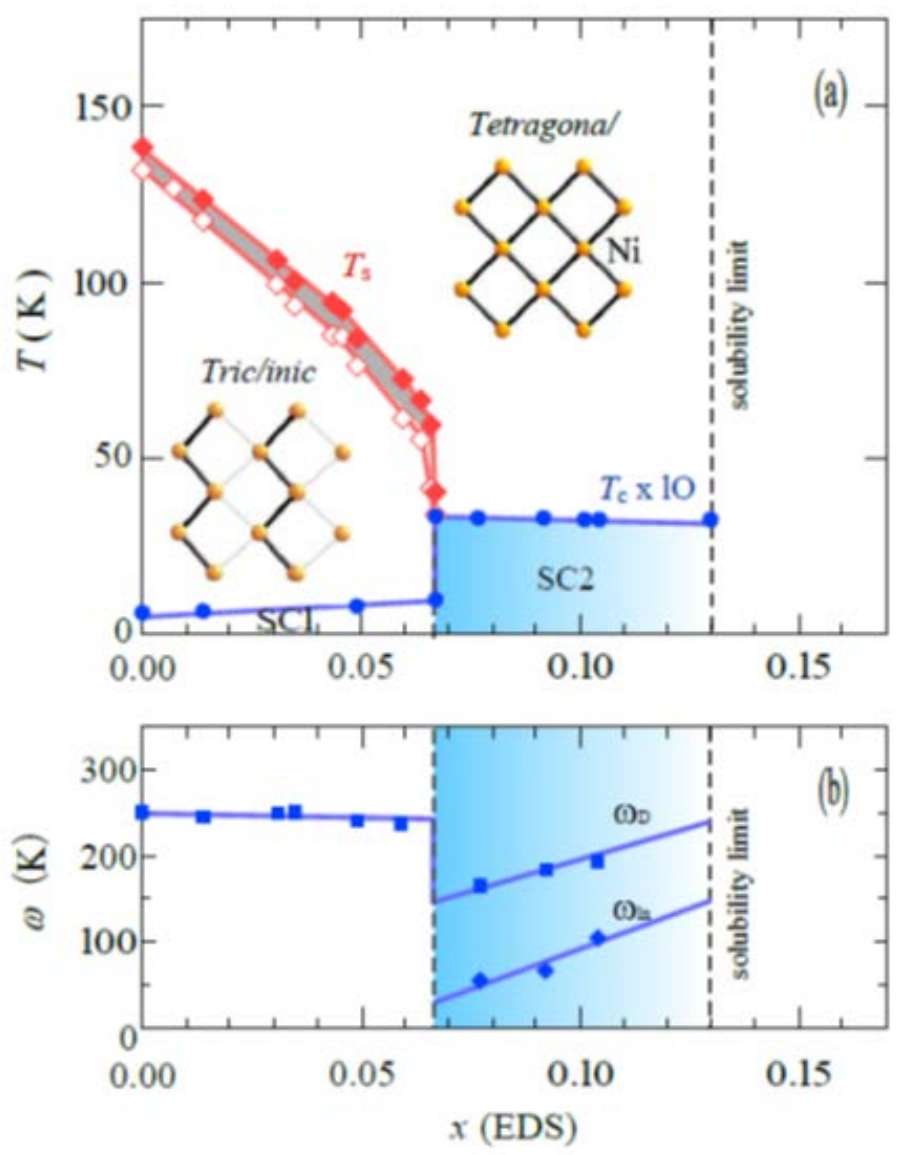

**Figure 60.** (a) Electronic phase diagram of $BaNi_2(As_{1-x}P_x)_2$. The closed circles represent the superconducting transition temperatures $T_c$. For clarity, the values of $T_c$ have been scaled by a factor of 10. SC1 and SC2 denote the superconducting phases. The open and closed diamonds represent the tetragonal-to-triclinic structural transition temperatures $T_s$ upon cooling and heating, respectively. The insets show schematic views of Ni planes in the triclinic and tetragonal phase. (b) Debye frequency $\omega_D$ and logarithmic averaged phonon frequency $\omega_{ln}$ as a function of phosphorous content $x$ [35].

### *3.4.2 Transition metal dichalcogenides*

The transition metal dichalcogenides, $MX_2$ have renewed interest recently and were studied intensively in this project. In this section, two topics are presented; one is the $CdI_2$-type telluride family and the other is pyrite ($FeS_2$)-type chalcogenide family.

#### *3.4.2.1 $CdI_2$-type tellurides*

$IrTe_2$ has a trigonal $CdI_2$-type structure (see Fig. 61) with the space group of P-3m1. The Ir atom is octahedrally coordinated by the Te atoms and the edge-sharing of the $IrTe_6$ octahedron forms the Te-Ir-Te composite layers which are stacked along the *c* axis of the trigonal lattice. Both the Ir and Te atoms form 2D regular triangular lattices within the ab-plane with three equivalent Ir-Ir (Te-Te) bonds. $IrTe_2$ undergoes a first-order transition at ~250 K transforming to a low temperature phase for which a monoclinic structure was proposed first [370]. Recently, a CDW-like superlattice modulation with wave vector of $\boldsymbol{q}$ =(1/5,0,-1/5) was observed in electron diffraction patterns for the low temperature phase [371]. Considering the partly filled *d*-orbitals of the Ir atoms, the orbital degree of freedom is believed to play an important role for this transition [372, 373]. In this project, the low temperature structure was analyzed in details using X-ray data for single crystals [374]. The structure at 20 K was found to be triclinic (space group *P*-1) as shown in Fig. 62. In this triclinic structure, one out of 5 Ir-Ir bonds along the trigonal *a* axis shrink considerably forming Ir-Ir dimers as illustrated by the yellow hatch in Fig. 62. The plane of the dimers propagates with the vector of $\boldsymbol{q}$ =(1/5,0,-1/5) consistent with the aforementioned electron diffraction data. This dimerization seems to affect the physical properties of the system seriously as the first-principal band calculations indicated that tilted two-dimensional Fermi surfaces emerge in the triclinic phase with a possible switching of the conduction plane from the vassal (*ab*) plane in the trigonal phase to tilted plane normal to $q$ in the triclinic phase [374].

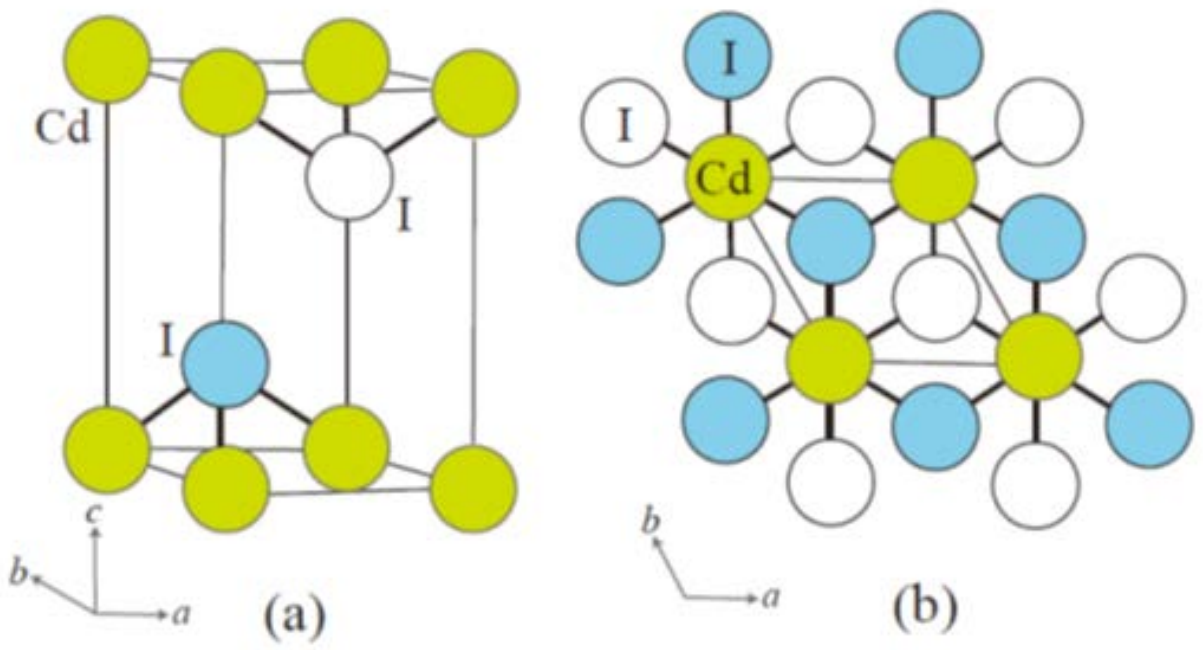

**Figure 61.** Crystal structure of $CdI_2$: (a) unit lattice and (b) composite plane consisting of the $CdI_6$ octahedra projected on the ab plane where the I atom represented by the white (blue) circle is located below (above) the Cd plane.

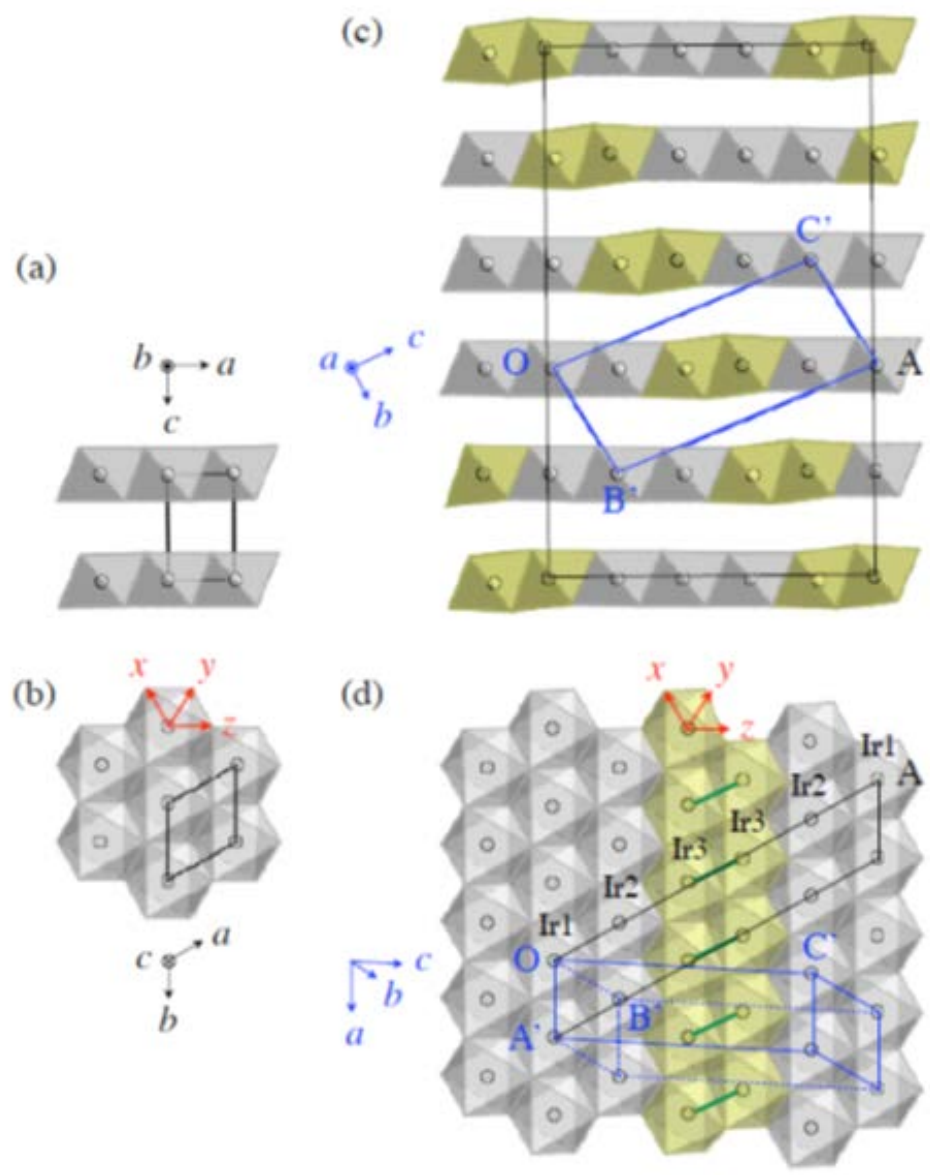

**Figure 62.** (a) and (b) The crystal structure of $IrTe_2$ at 300 K. Black lines represent the unit cell of the trigonal lattice. (c) and (d) The crystal structure of $IrTe_2$ at 20 K. Blue lines represent the unit cell of the triclinic lattice. Black lines represent the 5*a* x *b* x 5*c* supercell (*a*, *b*, and *c* are the high-temperature trigonal-cell parameters) [374].

It was found in 2011 that partial substitution of Pt for Ir in $IrTe_2$ suppresses the formation of the low temperature phase resulting in the appearance of superconductivity [375, 376]. A phase diagram of the $Ir_{1-x}Pt_xTe_2$ system was first determined for the polycrystalline samples [63] and then by single crystal data in the FIRST PJ [377]. Figure 63 is the phase diagram based on the single crystal data; superconductivity appears for ~0.04<$x$<~0.14 in $Ir_{1-x}Pt_xTe_2$ with the highest $T_c$ of ~3.2 K for $x$=0.04 at the phase boundary of the trigonal and triclinic phases (the monoclinic phase in this figure corresponds to the triclinic phase in Fig. 62, representing the simplified symmetry). For the polycrystalline sample with $x$=0.04, type-II superconductivity with $\mu_0H_{c2}(0)$ =0.17 T and $\Delta C(T_c)/\gamma T_c$ =1.5 has been reported [63]. The Pt substitution works as the electron doping shifting the Fermi level upward and affecting the DOS near the Fermi level (it causes increase of DOS near the Fermi level in the triclinic phase while decrease of DOS in the trigonal phase). Another effect of the substitution, which seems to be more essential, is the suppression of the triclinic phase by breaking the Ir-Ir dimers.

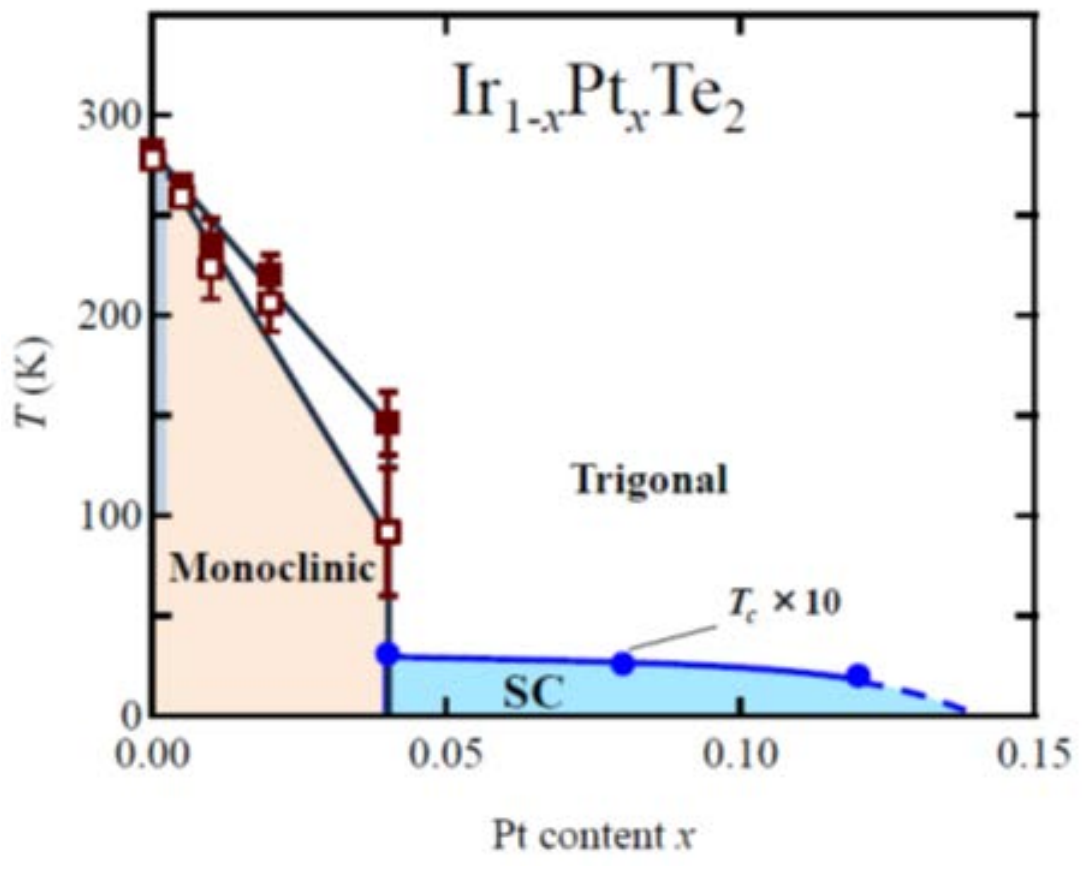

**Figure 63.** Phase diagram of $Ir_{1-x}Pt_xTe_2$ determined by single-crystal samples. Closed and open squares respectively represent trigonal-monoclinic transition temperature $T_s$ upon heating and cooling (the "monoclinic" phase corresponds to the triclinic phase in Fig. 62, representing the simplified symmetry). Closed circles represent the $T_c$ and SC represents the bulk superconducting region. Hatched area represents a temperature range of broad hysteresis in resistivity [63].

Similar results on suppression of the phase transition and appearance of superconductivity have been reported for several systems such as $Ir_{1-x}Pd_xTe_2$ [371], $Pd_xIrTe_2$ [371] and $Cu_xIrTe_2$ [378] where substitution or intercalation of Pd or Cu brings about essentially the same effect as the Pt substitution does. In the FIRST PJ, two more progresses were made on the $IrTe_2$ and related systems. The first one is the isovalent Rh doping for $IrTe_2$ [64]. The $Ir_{1-x}Rh_xTe_2$ system shows similar feature to $Ir_{1-x}Pt_xTe_2$, i.e., the triclinic phase formation is suppressed by the Rh doping resulting in the appearance of superconductivity with $T_c$ ~2.6 K, in spite that band filling is unchanged by the isovalent Rh doping as long as a rigid-band picture is concerned. A distinct difference was, however, seen between $Ir_{1-x}Rh_xTe_2$ and $Ir_{1-x}Pt_xTe_2$; in the former phase, a doping of $x$~0.1 is needed for the complete suppression of the triclinic phase which is three times larger than x=0.03 in the latter. This difference seems to be caused by the less volume expansion and the aforementioned unchanged band filling in the case of the Rh doping

Another progress was attained from the study of a related system of $Au_{1-x}Pt_xTe_2$ [62]. $AuTe_2$ has a monoclinically distorted $Cd_2I_2$-type average structure with a space group of $C2/m$ where Te-Te zigzag chains run along the a-axis [379]. In actuality, the structure is associated with incommensulate modulation with a wave vector $\boldsymbol{q}$=-0.4076 a* +0.4479c* and due to this modulation, Te-Te dimers having a short distance of 2.88 Å exist in the real structure instead of the zigzag chains [380, 381] . It was found that the Pt doping brings about structural change from the distorted $CdI_2$-type for x=0 to distortion-free $CdI_2$-type without the Te-Te dimers for x=0.35, via two-phase mixed region for x=0.1 and 0.15 [62]. The type-II superconductivity was observed for the distortion-free x=0.35 sample with $T_c$ =4.0 K, $H_{c2}(0)$ =12.9 kOe and $\xi_{GL}(0)$ =160 Å. The $\Delta C(T_c)/\gamma T_c$ is 1.57 exceeding the BCS weak coupling limit. It should be noted that superconductivity appears by the breaking of the Te-Te dimers which shows striking similarity with the $Ir_{1-x}Pt_xTe_2$ system where superconductivity appears by the breaking of the Ir-Ir dimers. It was also found that superconductivity is induced in $AuTe_2$ by application of pressure instead of the Pt doping [61]. As shown in the phase diagram of Fig. 64, application of mechanical pressure causes structural change from the distorted $CdI_2$-type to the distortion-free $CdI_2$-type via two-phase mixed region. Superconductivity appears above 2.12 GPa with the highest $T_C$ of 2.3 K at 2.34 GPa.

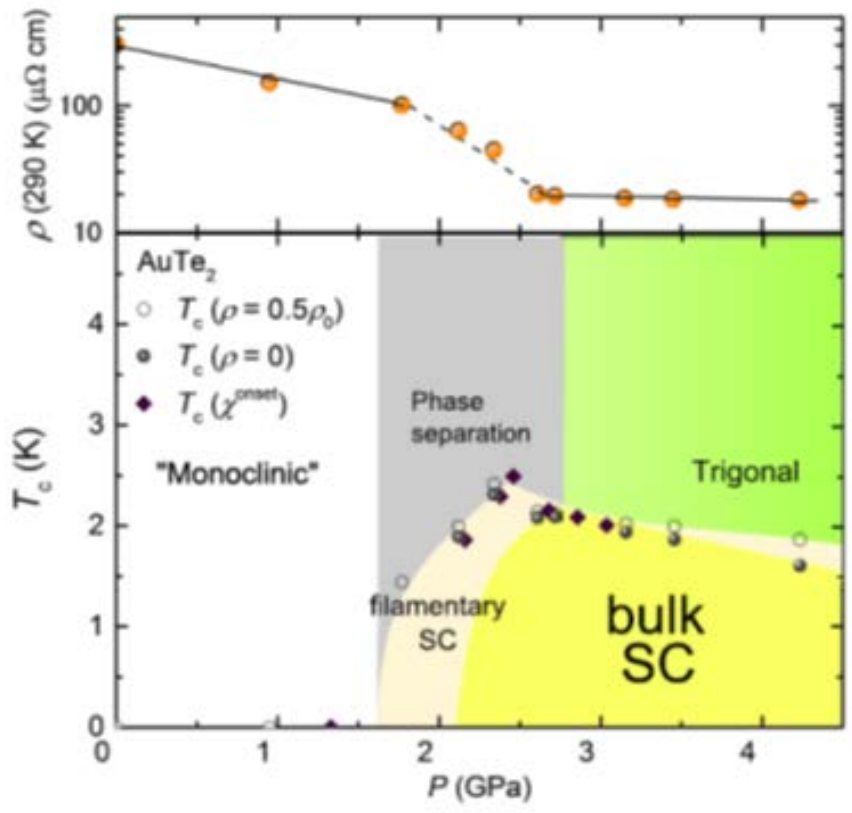

**Figure 64.** (Upper panel) Pressure dependence of resistivity at 290 K, ρ(290 K) (since the applied pressure decreases by approximately 0.2 GPa on cooling, the actual pressure at 290K is larger than the displayed pressure). (Lower panel) *T*–*P* phase diagram for $AuTe_2$. Open (filled) circles represent $T_c$ determined using $\rho_{50\%}$ point and filled diamonds indicate $T_c$ determined by the onset of the diamagnetic shielding signal [61]. (The "monoclinic" phase corresponds to the triclinic phase in Fig. 62, representing the simplified symmetry.)

*3.4.2.2. Pyrite-type chalcogenide family*

Iridium dichacogenides $IrX_2$ ($X$=Se and Te) sometimes take pyrite-type ($FeS_2$-type) structures after introduction of vacancies for the Ir sites [382]. It is also known that application of high pressure is effective to stabilize the pyrite-type form against the $Cd_2I_2$-type one [383]. In the Ir-deficient phase $Ir_xX_2$, Ir vacancies are distributed randomly forming the cubic pyrite-type structure with a space group of Pa-3 (Fig. 65). This structure can be interpreted as the NaCl-type constructed by the face centered cubic sublattice of Ir and the *X*-*X* dimers located at the center of each edge as well as at the body center of the cubic lattice. In the $Ir_xX_2$ phase with the particular value of $x$=0.25, Ir vacancies tend to be distributed in an ordered way in which one of four Ir atoms are regularly removed [382]. The vacancy ordering results in a stoichiometric phase of $Ir_3X_8$ having a rhombohedral structure with the space group of $R\bar{3}$ (in $Ir_3X_8$, the rhomboheral distortion is far less pronounced and the vacancy ordering may not be perfect compared with the corresponding phase of $Rh_3X_8$ [382]).

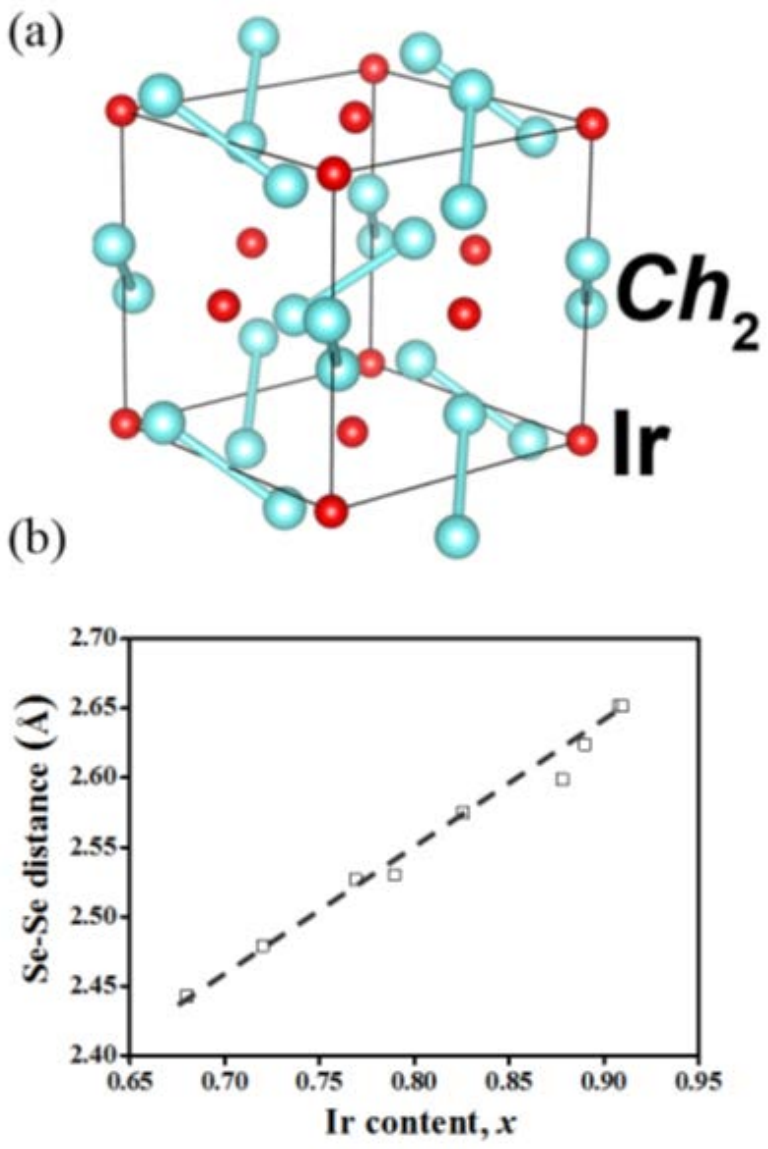

**Figure 65.** (a) Crystal structure of pyrite $Ir_xSe_2$. (b) Variation of the Se-Se distance in the dimer anions with the Ir content $x$ [65].

In the FIRST PJ, the pyrite-type $Ir_xX_2$ samples were synthesized under high pressure and their superconductivity was studied for the first time [65]. Figure 66 gives the electronic phase diagram obtained for the $Ir_xSe_2$ system. By increasing $x$, the system changes from insulating to metallic nature and superconductivity starts to emerge at the Ir content of $x$ ~0.75 with the highest $T_c$ of ~6.4 K at $x$ =0.91 (0.91 is the highest Ir content attained experimentally). The Se-Se distance of the Se dimer increases linearly with the increasing $x$ as shown in Fig. 65. Similar phase diagram was obtained for the $Ir_xTe_2$ system with the highest $T_c$ of ~4.7 K for $x$ =0.93. The DFT calculation was carried out for the vacancy-ordered structural model of $Ir_3Se_8$ where exist one long $Se_1$-$Se_1$ dimer (the edge site dimer with $r_{Se1\text{-}Se1}$=2.61 Å) and three short $Se_2$-$Se_2$ dimers (the body center site dimer with $r_{Se2\text{-}Se2}$=2.50 Å). The DFT calculations revealed that the band crossing the Fermi level consists mainly of σ* (anti bonding) orbital of the $Se_1$-$Se_1$dimer and $d_{z2}$ orbitals of the nearest Ir atoms with far less contribution from the σ* orbitals of the $Se_2$-$Se_2$ dimers. Such a nature of the electronic structure results in the half-filled narrow conduction band which easily becomes insulating by electron-electron correlation, electron-lattice interactions and/or the disordered Ir vacancies. The DFT calculations were also performed for the vacancy-free pyrite-type structure which is composed of equivalent Se-Se dimers. In this situation, the σ* orbitals of the Se-Se dimers contribute equally to form a wider conduction band. It should be noted that the Ir vacancy introduction causes the linear increase of the Se-Se distance of the Se dimer in correlation with the monotonous increase of $T_c$. In the $CdI_2$-type $IrTe_2$, breaking of the Te-Te dimer is essential for the appearance of superconductivity while in pyrite-type $Ir_xX_2$, control of the bonding state of the $X$-$X$ dimers by elongation and equalization is indispensable for inducing superconductivity. Superconducting parameters obtained for $Ir_{0.91}Se_2$ are $\mu_0H_{c2}(0)$ =14.3 T (type-II superconductor), $\xi_{GL}(0)$ =~48Å and $\Delta C(T_c)/\gamma T_c$ =3.1 (strong coupling superconductor).

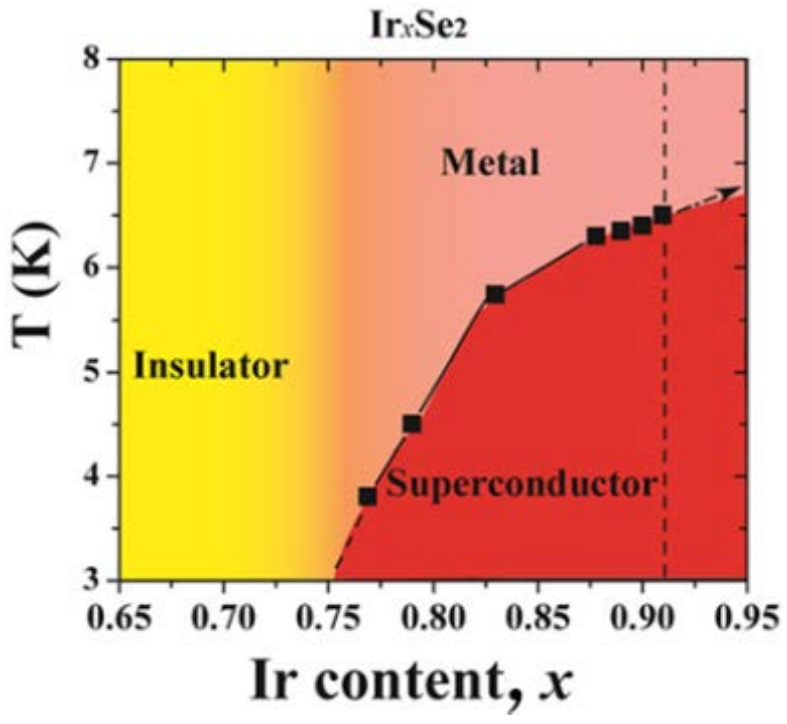

**Figure 66.** Electronic phase diagram of $Ir_xSe_2$ [65].

The strong correlation between the Se-Se distance and $T_c$ was also confirmed for the $Ir_{0.94-x}Rh_xSe_2$ system in the FIRST PJ [66]. Figure 67 indicates the phase diagram of the Rh-doped system in question. With increasing the Rh content, the system undergoes changes from non-metal state to the normal-metal state with $T$-square resistivity via strange-metal state with $T$-linear resistivity. Accompanied by this alteration, the Se-Se distance increases first then decreases taking the maximum at $x$~0.4. The striking correlation of the Se-Se distance and $T_c$, $\Theta_D$ and $\Delta C(T_c)/\gamma T_c$ is worth noting; both $T_c$ and $\Delta C(T_c)/\gamma T_c$ have the maximum values while $\Theta_D$ has the minimum value when the Se-Se distance is longest. This suggests strengthening of the electron-phonon coupling and softening of phonon due to the structural instability at the edge of weak dimer states when the Rh content is $x$ ~0.4.

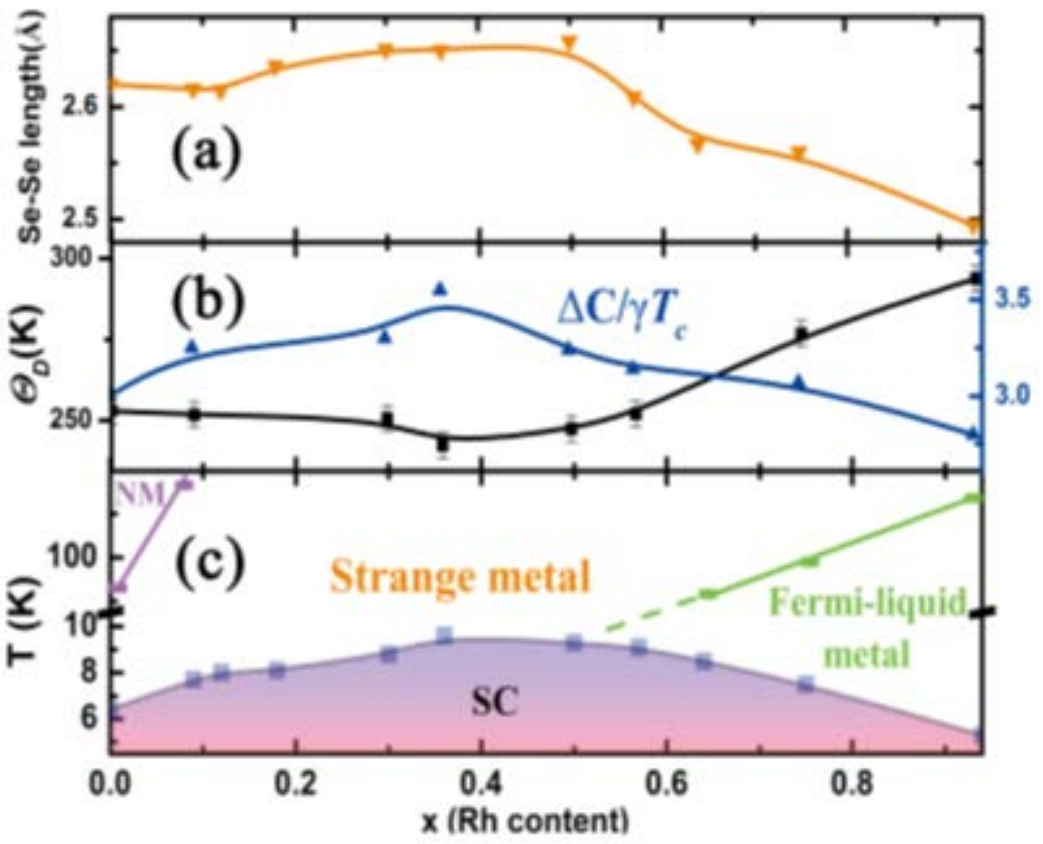

**Figure 67.** Summary and phase diagram of $Ir_{0.94-x}Rh_xSe_2$. (a) Variation of the Se−Se dimer bond length. (b) Debye temperature $\Theta_D$ and $\Delta C(T_c)/\gamma T_c$ as a function of the rhodium content $x$. (c) Electronic phase diagram of $Ir_{0.94-x}Rh_xSe_2$ [66].

Similar study has been carried out for the corresponding telluride system of $Ir_{0.95-x}Rh_xTe_2$ [67]. In this system, the Te-Te distance of the Te-dimer shows a parabolic change taking the maximum between $x$=0.2 and 0.3. Nevertheless, aforementioned correlation was not observed between the Te-Te distance and $T_C$ nor $\Delta C(T_c)/\gamma T_c$; both parameters decrease almost linearly with $x$. This result may be related to the reduced electron-electron repulsion in the telluride system compared with the selenide system, which is caused by a wider width of the conduction band in the telluride system consisting of the more spatially spread Te5$p$ orbital.

*3.4.3 Noncentrosymmetric superconductors*

Inversion symmetry of a crystal structure is deeply linked to the superconducting state induced in it. Usual superconductors are centrosymmetric with the inversion symmetry and in that case, state of the Cooper pair can be described by the multiplication of a spin part and an orbital part. In order to satisfy the odd symmetry of the wave function for the exchange of pairing electrons, the spin-singlet state (asymmetric spin part) of the Cooper pair should be combined with the symmetric orbital of *s*-wave or *d*-wave. On the contrary, the spin-triplet (symmetric spin part) state should be combined with the asymmetric *p*-wave. Such a simple picture is not applicable for noncentrosymmetric superconductors; the paring state is not classified any more as singlet or triplet, caused by asymmetric spin-orbit coupling. Theories predicted that mixing of singlet and triplet states (mixed parity state) occurs for a noncentrosymmetric superconducting system [384-386]. In addition to this striking feature, anisotropic gap structure with line nodes is predicted for a noncentrosymmetric superconductor [384-386]. The first example of this class of superconductors, $CePt_3Si$ was found in 2004 [387] followed by discoveries of a variety of compounds. Among them, Ce-containing heavy-fermion materials with $BaNiSn_3$-type structure such as $CeCoGe_3$ [388], $CeIrSi_3$ [389] and $CeRhSi_3$ [390] are worthy of special mention because of their unusually large upper critical fields far beyond the Pauli limit which seem to be concerned with the mixed parity state.

*3.4.3.1. Noncentrosymmetric silicides*

A wide variety of silicide superconductors have been known and most of them crystallize ceontrosymmetric structures showing conventional s-wave superconductivity. Some exceptions are Ce-containing heavy-fermion superconductors described above. In the FIRST PJ, we found two new noncentrosymmetric silicide superconductors, $SrAuSi_3$ [49] and $Li_2IrSi_3$ [50]. Here, their structural and physical properties are overviewed.

The $SrAuSi_3$ is the first noncentrosymmetric superconductor containing Au which is a heavy element and may cause strong spin-orbit coupling. The $SrAuSi_3$ is stable only under high pressure and its polycrystalline sample was prepared under high temperature-high pressure condition [49]. In Fig. 68, crystal structure of $SrAuSi_3$ is shown; it is a $BaNiSn_3$-type tetragonal structure in space group of *I*4*mm*. The structure can be interpreted as sequence of the atom planes along the c-axis as Sr-(Au-$Si_2$-Si)-Sr-(Au-$Si_2$-Si)-Sr… which has a close relationship with the $ThCr_2Si_2$-type and $CaBe_2Ge_2$-type structures (see Fig. 51). Figure 69 shows temperature dependence of electrical resistivity measured varying the magnetic field (*H*). $T_C$ of $SrAuSi_3$ is 1.6 K at *H*=0 and decreases with increasing magnetic field, giving $H_{c2}(0)$~2.2 kOe which is much lower than the Pauli limit ($H_p(0)$~30 kOe). This result suggests that $H_{C2}$ is governed by the orbital pair breaking mechanism. Indeed, the orbital limit estimated from the WHH theory is ~1.5 kOe comparable with the experimental value. However, $H_{c2}$ of $SrAuSi_3$ increases almost linearly with temperature deviated substantially from the WHH convex upward curve. This deviation may suggest a nonspherical Fermi surface or gap anisotropy in $SrAuSi_3$. The GL coherent length and penetration depth were estimated to be $\xi_{GL}(0)$=390 Å and $\lambda(0)$=4400 Å giving the GL parameter $\kappa_{GL}$=11 consistent with the type-II nature of superconductivity.

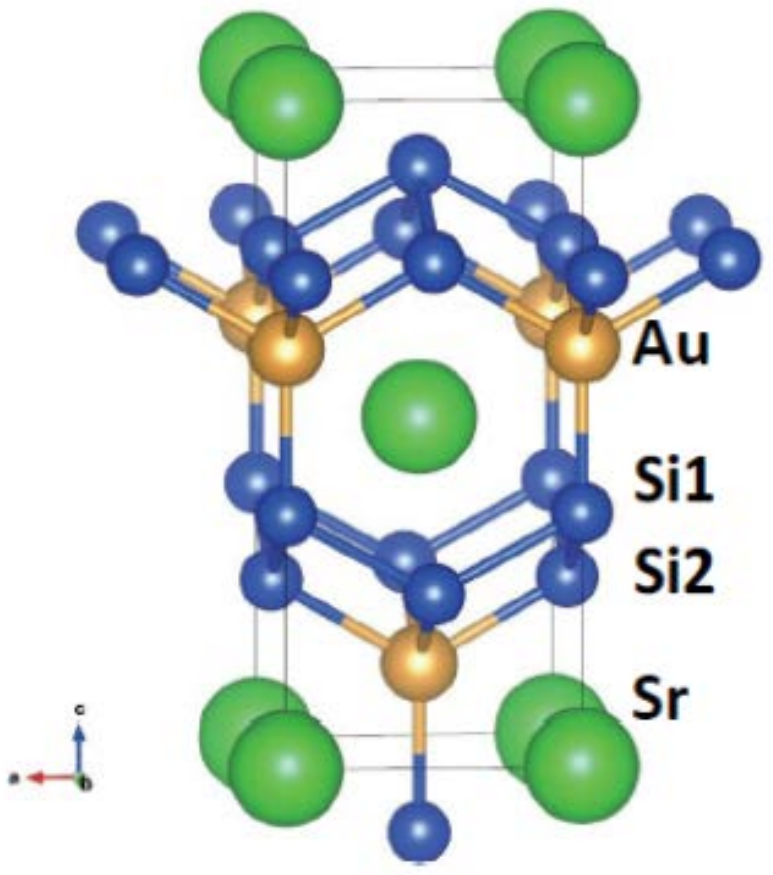

**Figure 68.** Crystal structure of $SrAuSi_3$ [49].

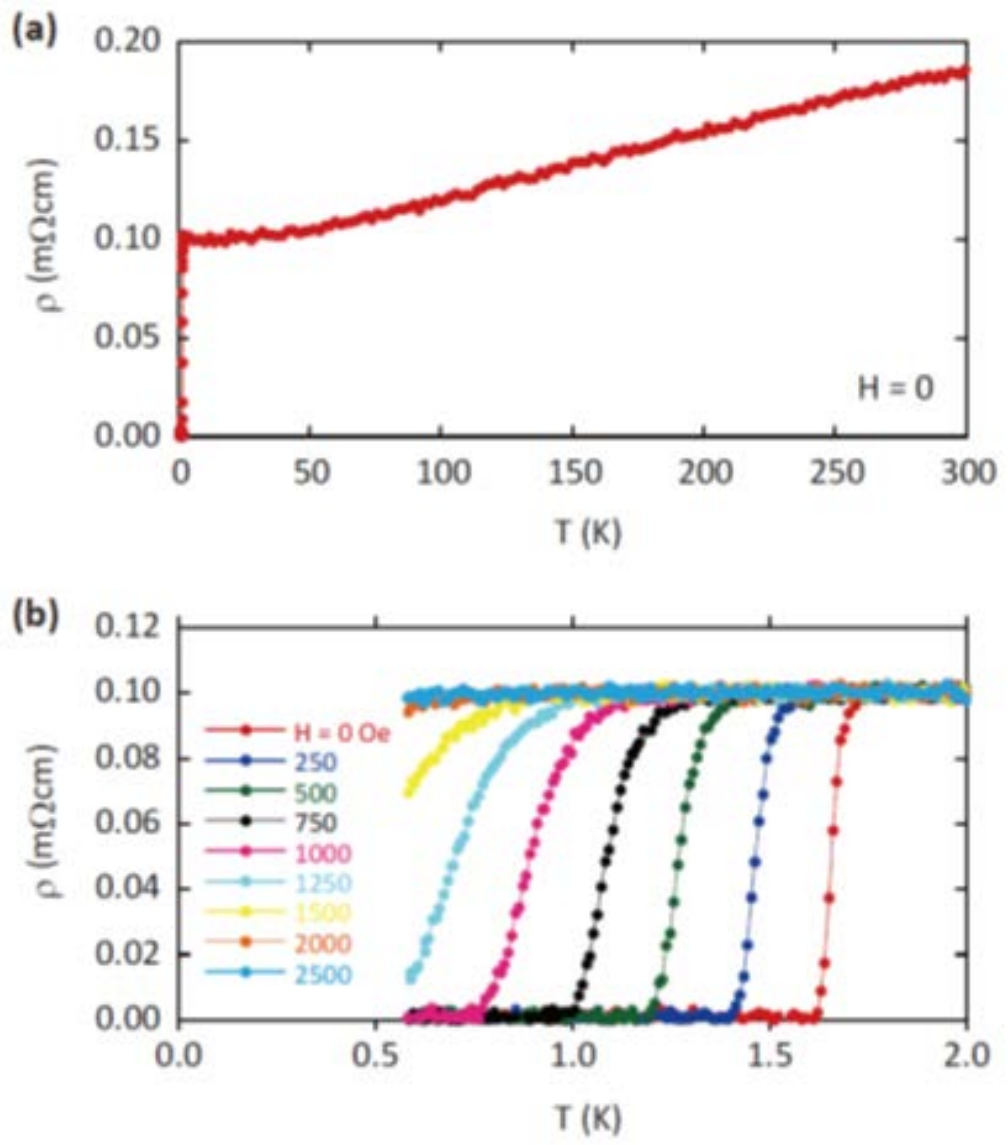

**Figure 69.** Temperature dependence of the electrical resistivity for a $SrAuSi_3$ polycrystalline sample: (a) data taken in the wide temperature range of 0.6−300 K under zero magnetic field ($H$ = 0) and (b) data below 2 K under the various magnetic fields ($H$ = 0−2500 Oe) [49].

Specific heat data gave $\gamma$=6.0 mJ/(mol $K^2$) and $\Theta_D$=410K. The normalized specific heat jump is calculated to be $\Delta C(T_c)/\gamma T_c$ =1.92 and the electron-phonon coupling to be $\lambda_{ep}\approx$0.97 using density of states at the Fermi level, 1.3 states/(eV f.u.) from the band calculation. These parameters indicate that $SrAuSi_3$ is a moderately strong coupling superconductor. $T_c$ was estimated by the McMillan's formula [391], $T_c = (\Theta_D/1.45) \times exp\{-1.04(1 + \lambda_{ep})/[\lambda_{ep} - \mu^*(1 + 0.62\lambda_{ep})]\}$, using the standard value of 0.13 for the Coulomb repulsion parameter $\mu^*$, resulting in ~19 K. The large difference between this value and experimental $T_c$ of 1.6K is striking; it may be caused by the parity mixing or the gap anisotropy in the nonocentrosymmetric superconductivity. The DFT band calculation for $SrAuSi_3$ revealed that two-types of carriers exist on the multiple Fermi surfaces. The major carriers conduct in the Si layers while other types of carriers conduct through the 3D network in the structure. Some of the latter carriers looked to have asymmetric spin-orbit coupling suggesting that they may cause the unusual behaviors in $SrAuSi_3$.

$IrSi_3$ crystallize in a hexagonal structure with the space group $P63mc$ [392] (Fig. 70) where planar layers

of four-fold Si (a distorted kagome network) are stacked along the c-axis sandwiching the Ir atoms. The Ir atoms are placed with unequal distances from the upper and lower neighboring Si planes leading to strong polar noncentrosymmmetry. In the FIRST PJ, Li atoms were intercalated to this structure for the first time to obtain $Li_2IrSi_3$ [50]. $Li_2IrSi_3$ has a trigonal structure with the space group of $P31c$ as given in Fig. 70 where essential nature of the Si plane stacking is preserved. In $Li_2IrSi_3$, positions of Ir and Li are not symmetric concerning the distances from the upper and lower neighboring Si planes, leading to the nonpolar asymmetry. However, displacement from the symmetrical equivalent position is $\Delta z/c$~0.007, which is one order of magnitude smaller than that of the mother compound of $IrSi_3$.

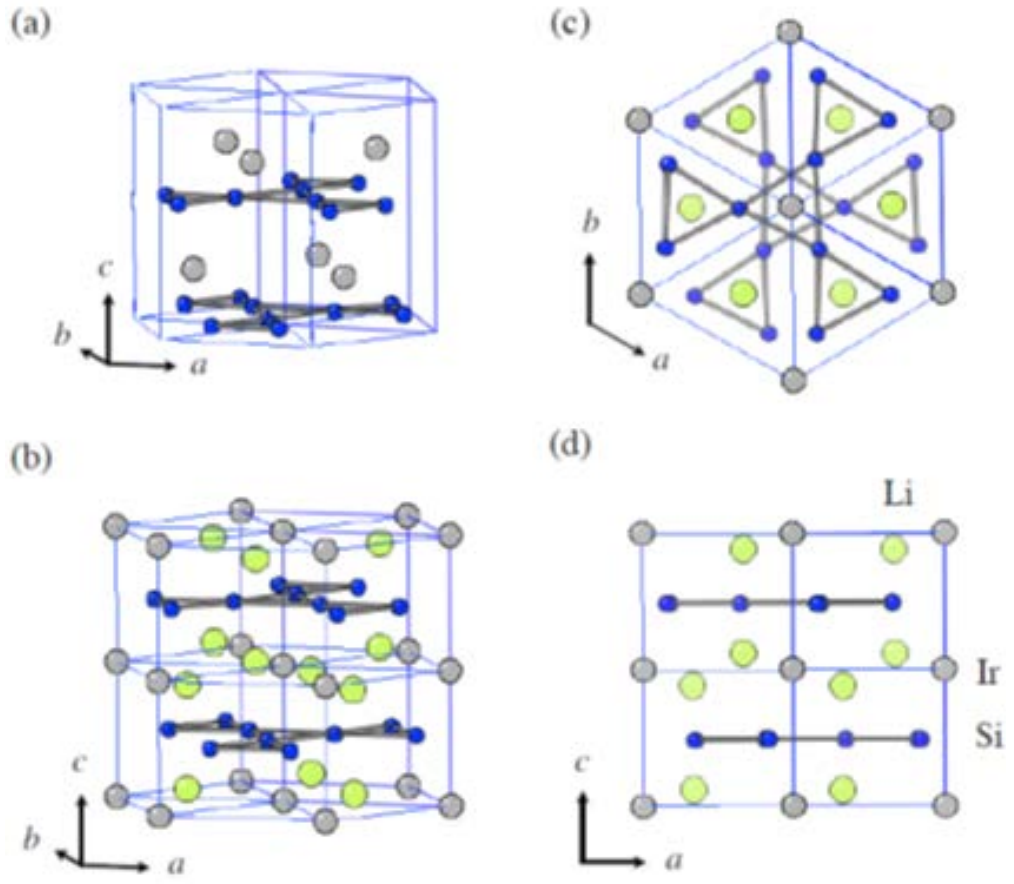

**Figure 70.** Crystal structures of (a) $IrSi_3$ (hexagonal, space group P63mc) and (b) $Li_2IrSi_3$ (trigonal, space group P31c). Blue circles, yellow circles, and gray circles denote Si, Li, and Ir, respectively. (c) and (d) show the top and side views of the crystal structure of $Li_2IrSi_3$ [50].

Figure 71 shows resistivity under various magnetic fields as functions of temperature. $Li_2IrSi_3$ exhibits type II-superconductivity with $T_c$ of 3.8 K. Temperature dependence of $\mu_0 H_{c2}$ gave $\mu_0 H_{c2}(0)$ =0.3 T and $\xi_{GL}(0)$ =330 Å (see the inset in Fig. 71) while the specific heat data gave $\gamma$ =5.3 mJ/(mol K$^2$) and $\Theta_D$=484 K which are worth to be compared with $\gamma$ =0.73 mJ/(mol K$^2$) and $\Theta_D$=516K for $IrSi_3$. The normalized specific heat jump at $T_c$ is $\Delta C(T_c)/\gamma T_c$=1.41 consistent with the BCS weak coupling limit. All these experimental data suggest conventional nature of superconductivity in $Li_2IrSi_3$; in particular, $\mu_0 H_{c2}(0)$ (0.3 T) is much lower than the Pauli limit of $\mu_0 H_P(0)$ =6.9T. The weakened inversion symmetry breaking seems to account for the less noticeable unique nature of superconductivity. The appearance of superconductivity in $Li_2IrSi_3$ may be concerned with the enhancement of $\gamma$, i.e., seven-fold increase of the electronic DOS at the Fermi level after the intercalation of Li for $IrSi_3$.

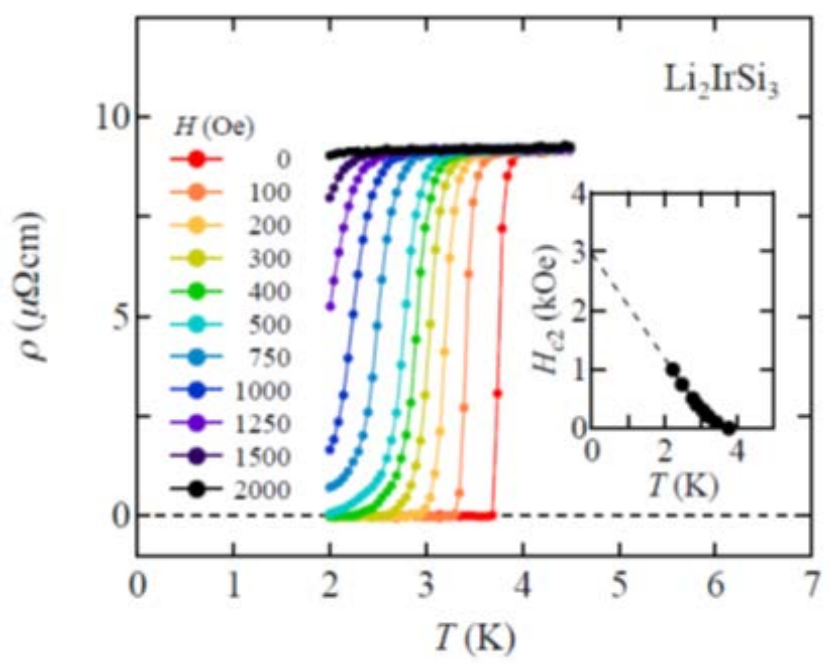

**Figure 71.** Temperature dependence of electrical resistivity for $Li_2IrSi_3$ at magnetic field $H$ up to 2000 Oe. The inset shows the temperature dependence of the upper critical field $H_{c2}$. The broken line represents the linear extrapolation of $H_{c2}$(T) [50].

*3.4.3.2. Ternary equiatomic pnictides, LaMPn (M=Ir, Rh; Pn=P, As)*

The ternary equiatomic pnictides with the general formula, *M'MPn* (*M'* is a large sized electropositive transition metal, *M* is a smaller transition metal, *Pn* is a pnictogen) form a large family of compounds. In the FIRST PJ, the equiatomic pnictides with 4*d* and 5*d* transition metals for the *M* site were studied because previous studies have been rather confines to the systems with the 3*d* transition metals for *M*. Consequently, type-II superconducting transitions were observed for the first time in LaIrP, LaIrAs and LaRhP which were prepared using the high-pressure synthesis technique [51]. Lattice parameters for LaRhAs and LaIrP were in good agreement with the previous reposts [393, 394]. In Fig. 72, crystal structure is shown for LaRhP which has tetragonal lattice with the space group of $I4_1md$ and is an ordered ternary derivative of the α-$ThSi_2$-type structure. In the structure, the Rh and P atoms are linked forming a 3D network with a trigonal planar coordination and the La atoms are placed in the cavities of the network. There are two sets of Rh-P zigzag chains running toward the perpendicular directions

In Table 13, superconducting parameters and the lattice parameters are shown for LaIrP, LaIrAs and LaRhP. Among them, LaIrP has the highest $T_c$ of 5.3 K and the highest $H_{c2}(0)$ of 13.8 kOe (WHH value) or 16.4 kOe (GL value) corresponding to the highest value of electron phonon coupling, $\lambda_{ep}$ =0.67. The $\Delta C(T_c)/\gamma T_c$ values are less than unity for all three compounds indicating the weak electron phonon coupling regime. For every compound, the experimental $H_{c2}(0)$ is much lower than the Pauli limit of field, suggesting that the asymmetric spin-orbital coupling induced by the noncentrosymmetry is not significant in the La*MPn* system.

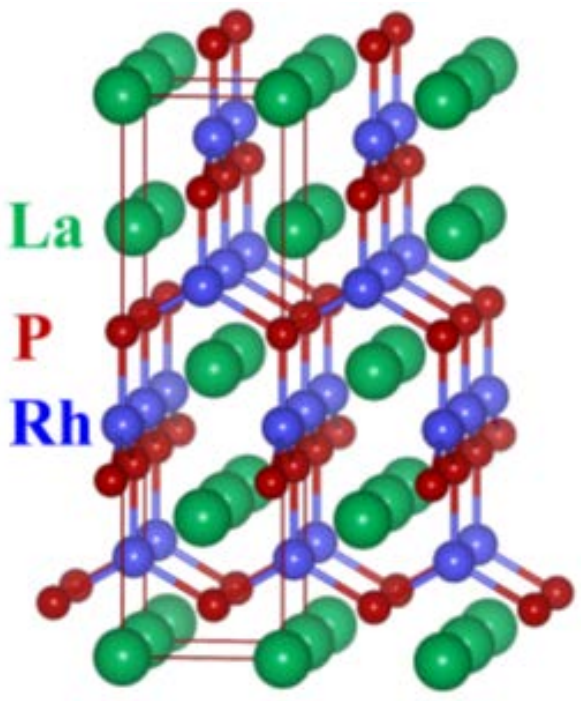

**Figure 72.** Crystal structure of LaRhP [51].

*3.4.4. Miscellaneous non-iron superconductors*

Besides the superconductors described in the preceding sections, various new non-iron superconductors were discovered through the FIRST PJ: they include NbSiAs ($T_c$ =8.2 K) [30], $CeNi_{0.8}Bi_2$ ($T_c$ =4 K) [37], LaNiBN ($T_c$ =4.1 K), LaPtBN ($T_c$ =6.7K), $La_3Ni_2B_2N_3$ ($T_c$ =15 K) [38], $La_2Sb$ ($T_c$ =5.3 K) [43], $Ba_{n+2}Ir_{4n}Ge_{12n+4}$ ($T_c$ =6.1 K for $n$=1; $T_c$ =3.2 K for $n$=2) [73], $Nb_4NiSi$ ($T_c$ =7.7 K) [72], and $Ca_2InN$ ($T_c$ =0.6 K) [74].

## 4. Thin films and wires of iron-based superconductors

### *4.1 Epitaxial thin film deposition*

#### *4.1.1. Deposition of Ba(Fe,Co)$_2$As$_2$ epitaxial films*

Fabrication of high-quality epitaxial films is essential for the study of new superconductors such as IBScs, from the viewpoints of not only the investigation of their anisotropic physical properties but also their application to superconducting wires or tapes as well as electronic devices. Among a variety of IBScs, the 122 type compounds have been expected as the most promising candidates for wire application, because of their anisotropy substantially smaller than that of the 1111 compounds and comparable to that of $MgB_2$ [395]. For thin-film growth, carrier doping by the substitution of Co or P seems easier than F doping in the 1111 compounds.

Actually, the first superconducting epitaxial films of the IBScs were realized in the 122 type compound Sr(Fe,Co)$_2$As$_2$ (Sr122:Co) films with a $T_c$ onset and $T_c$ zero of approximately 20 and 15 K on (La,Sr)(Al,Ta)$O_3$ (LSAT) (001) single-crystal substrates by Hiramatsu *et al.* of Tokyo Institute of Technology (Tokyo Tech) [396]. They employed pulsed laser deposition (PLD) using a second-harmonic neodymium-doped yttrium aluminum garnet (Nd:YAG) laser ($\lambda$ = 532 nm). Later Katase *et al.* succeeded in fabricating Co-doped Ba122 (Ba122:Co) epitaxial films on LSAT substrates by using the same PLD technique and demonstrated that they had much higher stability against water vapor than Sr122 films [397]. However, their rather large resistive transition width ($\Delta T_c$) of approximately 3 K and low critical current density ($J_c$) at 5 K of $10^5$ A/cm$^2$ indicated that their film quality was not good enough to apply them to electronic devices and superconducting tapes.

In order to obtain higher-quality Ba122:Co epitaxial films, Katase *et al.* of Tokyo Tech Hosono group tried improving the purity of the Ba122:Co target and homogeneity of substrate temperature [398]. A key point to obtain a high-purity Ba122:Co target was use of fine Ba metal pieces to synthesize the BaAs precursor. Figure 73 shows X-ray diffraction patterns for Ba122:Co films on LSAT obtained before and after such improvement. The diffraction peak from Fe impurities observed in the previous film almost disappears in the improved film. The in-plane alignment of the 122 grains was also found to be improved. The improved Ba122:Co film exhibited a substantially higher $T_c$ zero above 20 K, narrower transition width of about 1 K, and a higher self-field $J_c$ up to 4 MA/cm$^2$ at 4 K, as shown in Fig. 74.

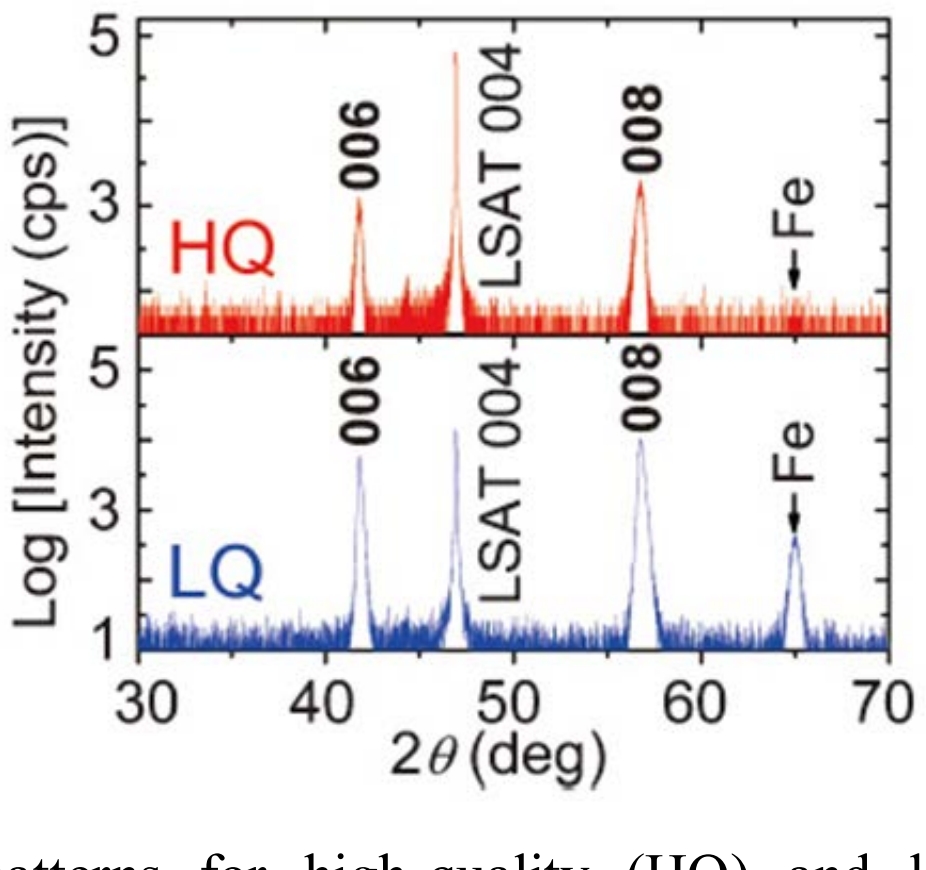

**Figure 73.** X-ray diffraction patterns for high-quality (HQ) and low-quality (LQ) Ba(Fe,Co)$_2$As$_2$ (Ba122:Co) epitaxial films on LSAT substrates fabricated by PLD. High-quality films were obtained by improving purity of targets and homogeneity of substrate temperature [398].

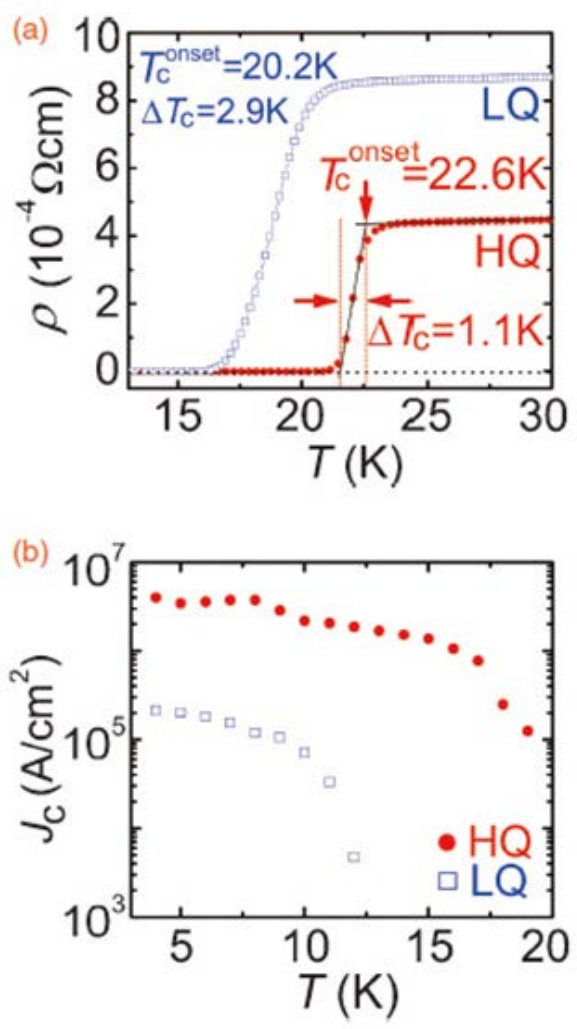

**Figure 74.** Temperature dependences of (a) resistivity and (b) $J_c$ for high-quality (HQ) and lower-quality (LQ) Ba122:Co epitaxial thin films on LSAT substrates [398].

They also comprehensively studied the deposition conditions for Ba(Sr)122 thin films [238] and found that Ba122 epitaxial films could be grown in a wider temperature range of 700 – 900 °C than that for Sr122. In particular, Ba122 epitaxial films with a two-dimensional growth mode, high crystallinity, and a high self-field $J_c$ at 4 K above 1 MA/cm$^2$ could be obtained in a temperature range of 800 – 850 °C at a growth rate between 0.28 and 0.33 nm/s. Investigation of the Co content dependence of the transport properties for $Ba(Fe_{1-x}Co_x)_2As_2$ (Ba122:Co) epitaxial films revealed that the highest $T_c$ of approximately 25.5 K was obtained for $x$ = 0.075 [238]. They also demonstrated that Ba122:Co epitaxial films with a self-field $J_c$ above 1 MA/cm$^2$ could be directly prepared on MgO (100) single-crystal substrates without using a conducting buffer layer such as a Fe buffer reported by Iida *et al.* [399].

Figure 75 shows the field angular dependence of $J_c$ for a high-quality Ba122:Co epitaxial film on LSAT [400]. A broad $J_c$ peak around the *c*-axis direction is observed at 6 T. This *c*-axis peak becomes less prominent at higher fields but still exists even at a high field of 15 T. This result suggested that naturally formed defects along the *c*-axis such as dislocations work as rather strong pinning centers in the Ba122:Co epitaxial film on LSAT. This is quite different from the result reported for the Ba122:Co epitaxial films on a Fe buffer layer where no *c*-axis peak was observed [401].

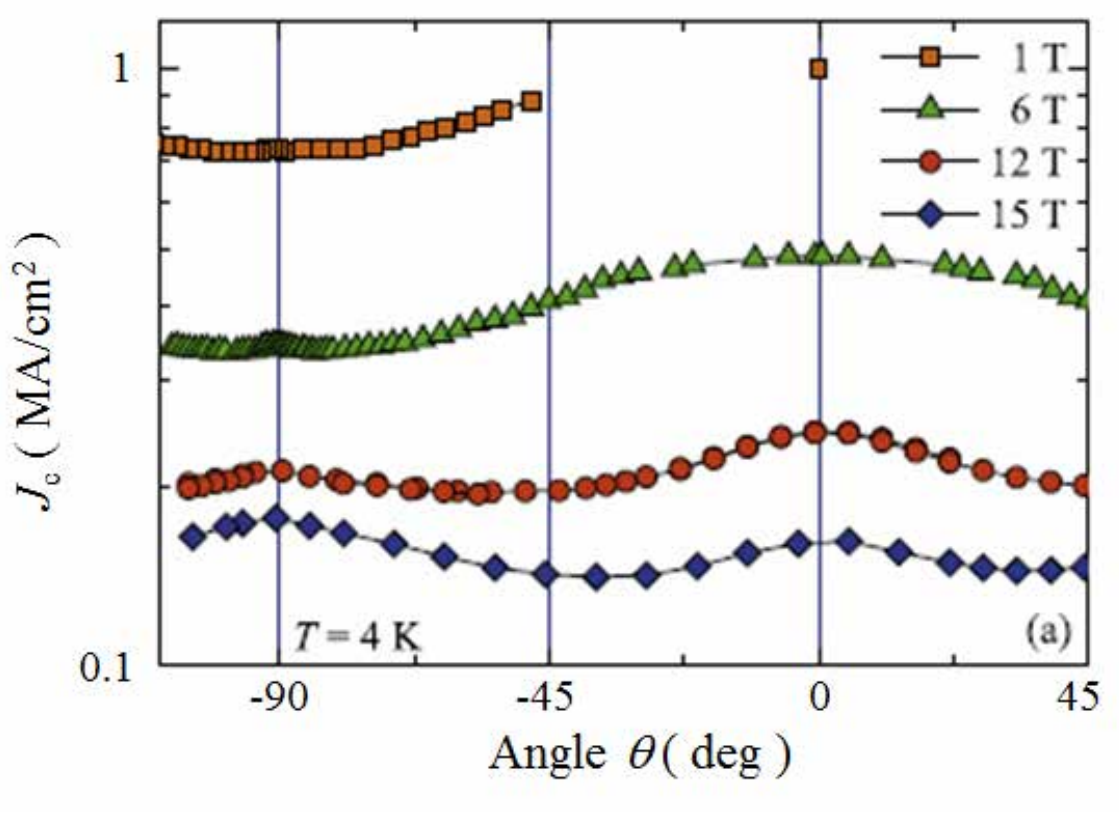

**Figure 75.** Angular dependence of $J_c$ at 4 K in fields of 1-15 T for a high-quality Ba122:Co epitaxial film on LSAT [400].

*4.1.2. Deposition of $BaFe_2(As,P)_2$ epitaxial films*

*4.1.2.1. Superconducting properties of $BaFe_2(As_{0.6}P_{0.4})_2$ epitaxial films*

As described in the former section, the 122 compounds seem the most suitable for application to superconducting wires or tapes among IBScs. The 122 compounds exhibit superconductivity by substituting Co for the Fe site, K for the Ba(Sr) site, or P for the As site. The K-doped Ba(Sr)122 has the highest $T_c$ and upper critical filed ($H_{c2}$) as summarized in Table 14. Although synthesis of K-doped Ba122 (Ba122:K) epitaxial films by a molecular beam epitaxy (MBE) method was reported [402], existence of volatile K makes it difficult to fabricate its films by a PLD method. It was also reported that the films were not stable in ambient atmosphere [402]. The P-doped Ba122 (Ba122:P) has a higher $T_c$ than Ba122:Co of approximately 30 K [403] and is expected to be stable in ambient atmosphere. Adachi *et al.* of ISTEC group chose this compound as a material candidate for production of superconducting tapes by a PLD method using a second-harmonic Nd:YAG laser, and examined the film preparation conditions on MgO single crystal substrates [404].

High-purity targets with a nominal composition of $BaFe_2(As_{0.6}P_{0.4})_2$ were carefully synthesized by a conventional solid-state reaction method. Epitaxial films were obtained at a substrate temperature of approximately 800 °C. The energy density on the target was relatively high, approximately 10 J/cm$^2$, leading to a deposition rate of 5 nm/s at the repetition rate of 10 Hz and the substrate-target distance of 7 cm. The average FWHM value of the peaks in the $\phi$-scan was about 1.5°. The obtained film exhibited a $T_c$ onset and $T_c$ zero of 26.5 and 24.0 K, as seen in the resistive transition curve of Fig. 76. An even higher $T_c$ of 27.0 K was also observed for a film without patterning. The first synthesis of Ba122:P epitaxial films was previously achieved by an MBE method [405]. The observed $T_c$ values of the PLD films are comparable to those reported for the MBE films [405] and the single crystals [403]. Figure 77 shows the dependence of $J_c$ on the applied field along the *c*-axis at different temperatures for the Ba122:P epitaxial film with the $T_c$ onset of 26.5 K and a self-field $J_c$ value at 4.2 K of 3.5 MA/cm$^2$. The film exhibited rather high in-field $J_c$ values, for example, approximately 1 MA/cm$^2$ at 4.2 K, 3 T and 10 K, 1 T, which are higher than those for Ba122:Co films [406].

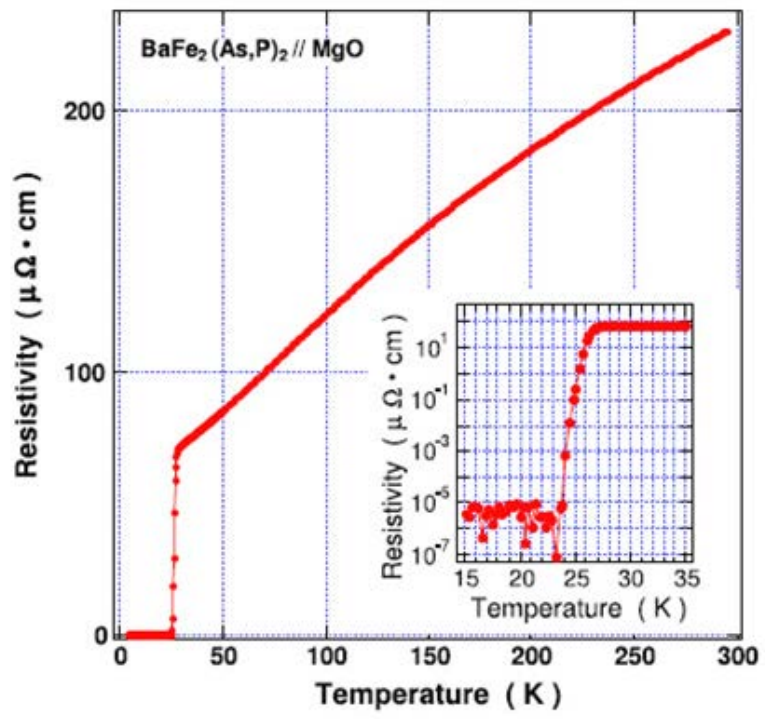

**Figure 76.** Temperature dependence of resistivity for a $BaFe_2(As_{1-x}P_x)_2$ (Ba122:P) epitaxial film on MgO substrate fabricated by a PLD method. The inset shows the magnified curve near $T_c$ [404].

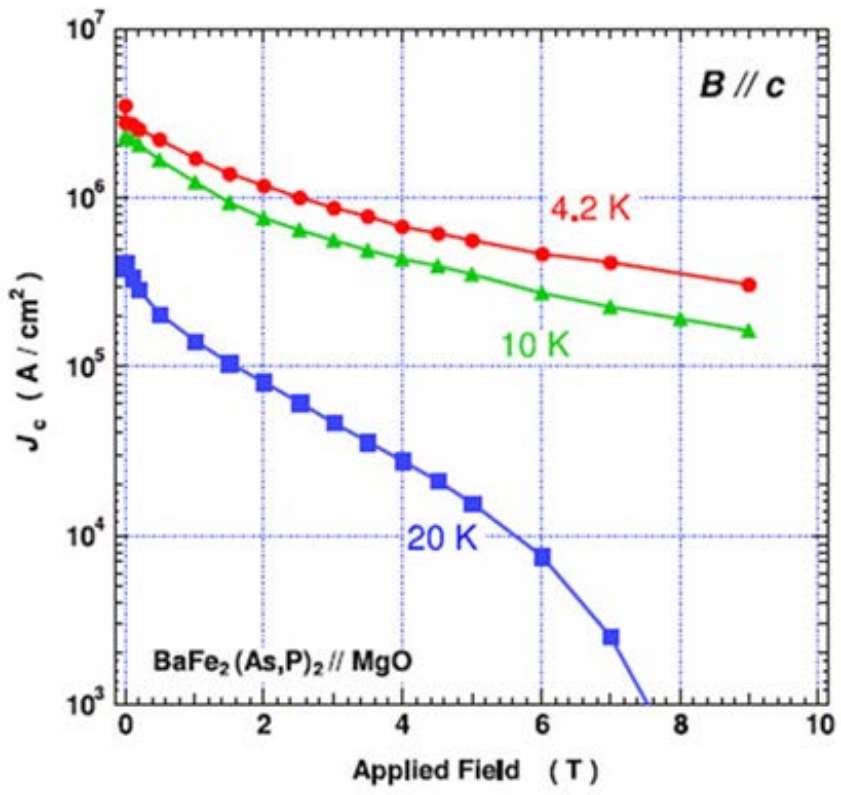

**Figure 77.** Magnetic field dependence of $J_c$ at 4.2, 10, and 20 K for a Ba122:P epitaxial film on MgO substrate fabricated by a PLD method. Magnetic field was applied parallel to the *c*-axis [404].

*4.1.2.2. Superconducting properties of $BaFe_2(As_{1-x}P_x)_2$ epitaxial films*

The dependence of superconducting properties of $BaFe_2(As_{1-x}P_x)_2$ (Ba122:P) epitaxial films on the P content was systematically investigated by Miura *et al.* by using Ba122:P targets with nominal P content $x$ of 0.25, 0.33, 0.40 and 0.50 [407]. The $x$ values in the films analyzed using an electron probe micro analyzer (EPMA) were found to be slightly smaller by approximately 0.05 than the nominal $x$ values in bulk targets. Figure 78 shows the temperature dependence of $H_{c2}$ for Ba122:P films with various analyzed $x$ values in magnetic fields up to 12 T. The $x = 0.28$ film exhibits a maximum $T_c$ zero of 26.5 K, and a further increase in $x$ leads to a reduction in $T_c$ zero, while the $x = 0.19$ film showed a broad transition and $T_c$ zero of about 12 K. This $x$ dependence of $T_c$ is similar to that observed in single crystals [403]. The $x = 0.28$ film also exhibits the highest $H_{c2}$ in both field directions. The inset shows the anisotropy of $H_{c2}$, $\gamma_H = H_{c2}{}^{ab}/H_{c2}{}^{c}$ for the films. The $x = 0.28$ film with the optimal $T_c$ shows the smallest $\gamma_H$ value of 1.54. This is different from the case of cuprate superconductors [408], and preferable from the viewpoint of wire or tape application. Figure 79 shows the angular dependence of $J_c$ for the $x = 0.28$ film at 10 K under different magnetic fields. This film also exhibits high in-field $J_c$ values. It is also found that $J_c$ shows minimum values in the $c$-axis direction at all fields from 0.5 to 7 T, indicating that no $c$-axis-correlated pinning centers are included in this film.

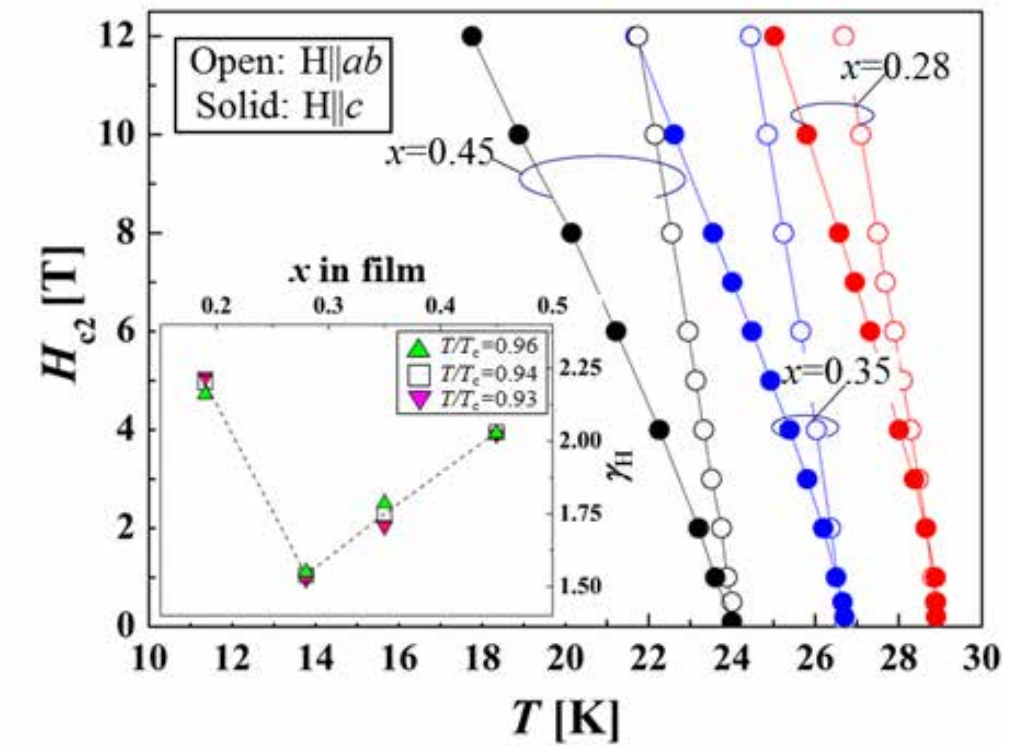

**Figure 78.** Temperature dependence of upper critical fields ($H_{c2}$) for Ba122:P epitaxial films on MgO with analyzed P content $x$ of 0.28, 0.35, and 0.45. The inset shows the $x$ dependence of anisotropy parameter $\gamma_H$ at $T/T_c = 0.93$, 0.94, and 0.96 [407].

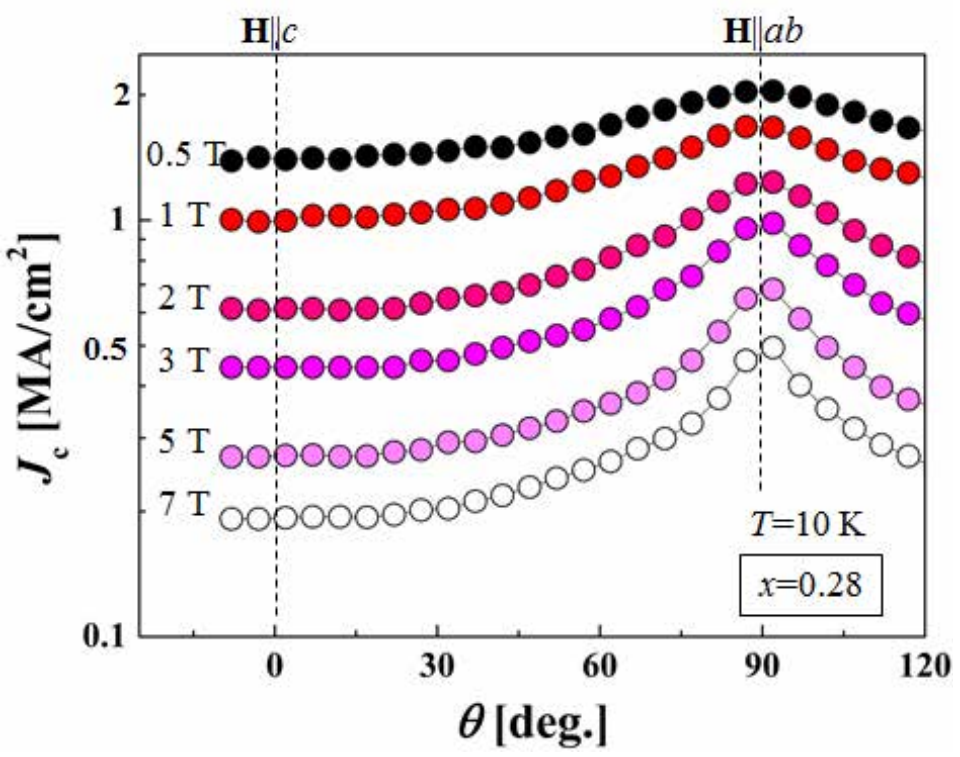

**Figure 79.** Angular dependences of in-field $J_c$ at 10 K in fields of 0.5 – 7 T for the Ba122:P film with $x$ = 0.28 [407].

*4.1.3. $BaFe_2(As,P)_2$ films with strong pinning centers*

*4.1.3.1 Addition of artificial pinning centers consisting of $BaZrO_3$ nanoparticles*

The Ba122:P epitaxial films on MgO substrates fabricated by a PLD method showed relatively high in-field $J_c$ values. However, these values seem to be not high enough for high-field application. Since the Ba122 superconductors have very high $H_{c2}(0)$ values over 60 T, remarkable improvement in in-field $J_c$ would be expected, if effective vortex pinning centers could be introduced into the materials. Actually amorphous tracks induced in Ba122:K single crystals by heavy-ion irradiation significantly enhanced their in-field $J_c$ and resulted in a matching field as high as 21 T without $T_c$ degradation [409], although heavy-ion irradiation is not practical for fabrication of long-length wires or tapes. For the case of Ba122 epitaxial films, naturally formed defects in the Ba122:Co films enhanced the $J_c$ in the fields parallel to the *c*-axis to some extent [400], as described in the previous section. It was also reported that naturally formed nanopillars along the *c*-axis in the Ba122:Co thin films prepared by PLD using a $SrTiO_3$ buffer layer and oxygen-rich targets significantly enhanced the in-field $J_c$ around the *c*-axis direction [410, 411]. However, the $J_c$ around the *ab* plane direction was not much improved.

For application to superconducting tapes, it is desirable to find a controllable and practical way of enhancing vortex pinning in Ba122 films in an isotropic way. In cuprate superconductor films such as $REBa_2Cu_2O_y$ (REBCO; *RE*=Y or rare-earth elements) films, where *RE* is Y or rare-earth elements, the addition of nanoparticles consisting of second oxide phases has been found to be effective to enhance their in-field $J_c$ in an isotropic way [412-414]. However, in order to apply a similar technique to Ba122 epitaxial films, the second phases need to be chemically stable and crystallographically compatible with the Ba122 phase, and its size should be small enough to avoid blocking of the current path.

Miura *et al.* tried introducing $BaZrO_3$ (BZO) nanoparticles, which are known to be effective pinning centers in REBCO films, into Ba122:P epitaxial films on MgO by a PLD method [415]. Approximately 80 nm thick BZO-doped Ba122:P epitaxial films were grown from 1 mol.% and 3 mol.% BZO-doped Ba122:P targets with the nominal P content of 0.33. The deposition conditions were similar to those employed by Adachi *et al.* for the synthesis of undoped Ba122:P films [404]. The cross-sectional elemental maps and the X-ray diffraction pattern shown in Fig. 80(a) of the film grown from the 3 mol.% BZO-doped target indicated the presence of homogenously dispersed BZO nanoparticles. The average nanoparticle size and the average spacing were found to be 8 nm and 24 nm, respectively, leading to a density $n$ of approximately 6.8 x $10^{22}$ $m^{-3}$. The cross-sectional TEM image of a typical BZO nanoparticle is shown in Fig. 80(b). The size of the nanoparticle is about 5 and 10 nm parallel and perpendicular to the *c*-axis, respectively. The periodicity of the Ba122 planes around the nanoparticles is only perturbed by creation of stacking faults. However, nano-beam diffraction patterns for the nanoparticle and the Ba122:P matrix indicated that the BZO nanoparticles are not epitaxially oriented along the Ba122:P matrix. The $T_c$ zero values for the undoped and doped films are 26.3 and 25.0 K, respectively, indicating that introduction of the BZO nanoparticles does not induce significant $T_c$ degradation.

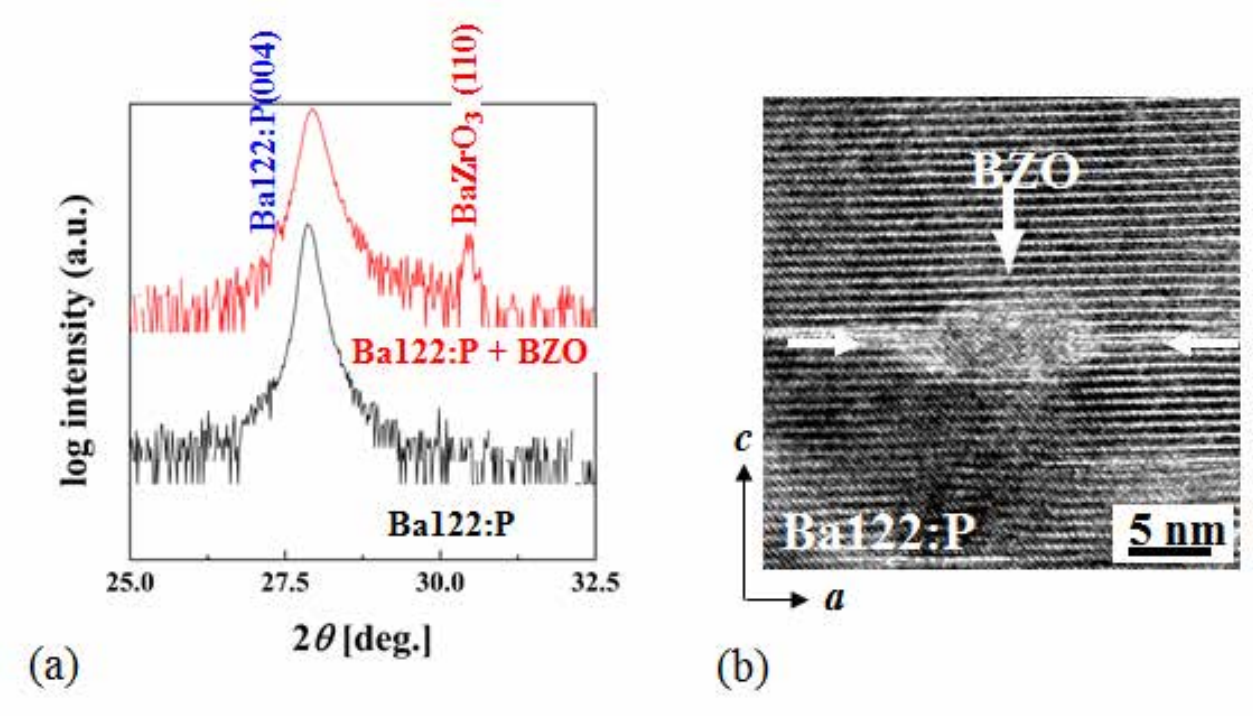

**Figure 80.** (a) X-ray diffraction patterns for a Ba122:P and a Ba122:P+3 mol% $BaZrO_3$(BZO) film prepared on MgO substrates. (b) High-resolution cross-sectional TEM image for a BZO nanoparticle in the Ba122:P+3 mol% BZO film [415].

Figure 81(a) shows the field dependence of $J_c$ ($H//c$) at 5 K for the undoped Ba122:P and BZO added Ba122:P films. The self-field $J_c$ ($J_c^{s.f.}$) increases monotonically with the amount of BZO additive and that the field decay of $J_c$ is greatly reduced within the measured field range. The Ba122:P film shows a characteristic crossover field $H^*$ (90% of $J_c(H)/J_c^{s.f.}$), followed by a power-law regime ($J_c \propto H^{-\alpha}$) with $\alpha \sim 0.40$ at intermediate fields. A more rapid decay of $J_c$ is observed as $H$ approaches the irreversibility field, $H_{irr}$. For the Ba122:P+BZO films, $H^*$ also increases with the BZO content. At intermediate fields, we find that both the films with BZO show a slower decay of $J_c(H)$, indicating the importance of BZO nanoparticles to enhance $J_c(H)$ in magnetic fields. The non-power-law dependence observed for the Ba122:P+BZO films is similar to that observed in RE123 films with strong pinning coming from uniformly dispersed nanoparticles [413, 414, 416].

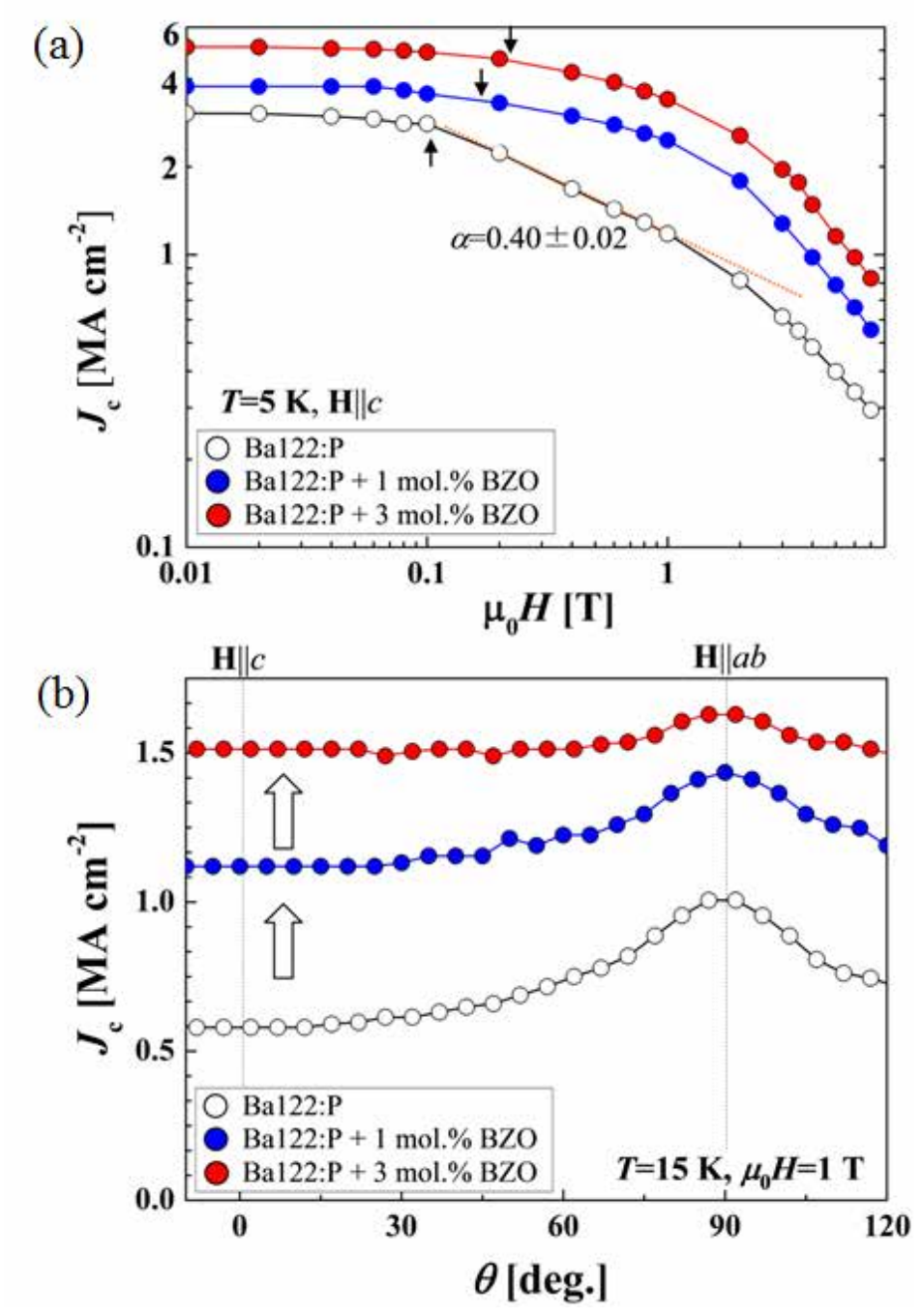

**Figure 81.** (a) Magnetic field dependence of $J_c$ at 5 K for Ba122:P, Ba122:P+1 mol% BZO and Ba122:P+3 mol% BZO films on MgO. Magnetic field was applied parallel to the *c*-axis. (b) Angular dependence of in-field $J_c$ at 15 K, 1 T for the Ba122:P, Ba122:P+1 mol% BZO and Ba122:P+3 mol% BZO films [415].

Figure 81(b) shows the angular dependence of $J_c$, $J_c(\theta)$, curves measured for the three films at 1 T, 15 K. By adding BZO the $J_c$ increases for all orientations with respect to the $J_c$ of the Ba122:P film. In particular, the Ba122:P+3mol%BZO film exhibits an almost isotropic $J_c$ with the value for *H*//*c* 2.6 times higher than that of the Ba122:P film. The minimum value of $J_c(\theta)$, $J_{c,min}$, of 1.5 MA/cm at 1 T, 15 K is over 28 times and 7 times higher than that of Ba122:Co films with *c*-axis columnar defects [410] and Ba122:Co films with super-lattice structures [417], respectively, in very similar field and temperature conditions, indicating strong isotropic pinning by the BZO nanoparticles.

In Fig. 82, the pinning force, $F_p=J_c(H)\times\mu_0H$ is compared with that of several superconductor materials. At 15 K, $F_p$ for the Ba122:P+3mol%BZO film is over 3 times higher than that for the Ba122:P film and higher than NbTi [418] at 4.2 K at all magnetic fields. Comparing with $MgB_2$ data at 15 K [419], $F_p$ is clearly higher for $\mu_0H > 0.5$ T. At $T$=5 K, the $F_p$ of the Ba122:P+3 mol.%BZO film, which is the minimum value in all field directions, reaches ~59 GN/m$^3$ for $\mu_0H > 3$ T up to the highest field we measured (9 T), a 50% increase over $Nb_3Sn$ at 4.2 K [420].

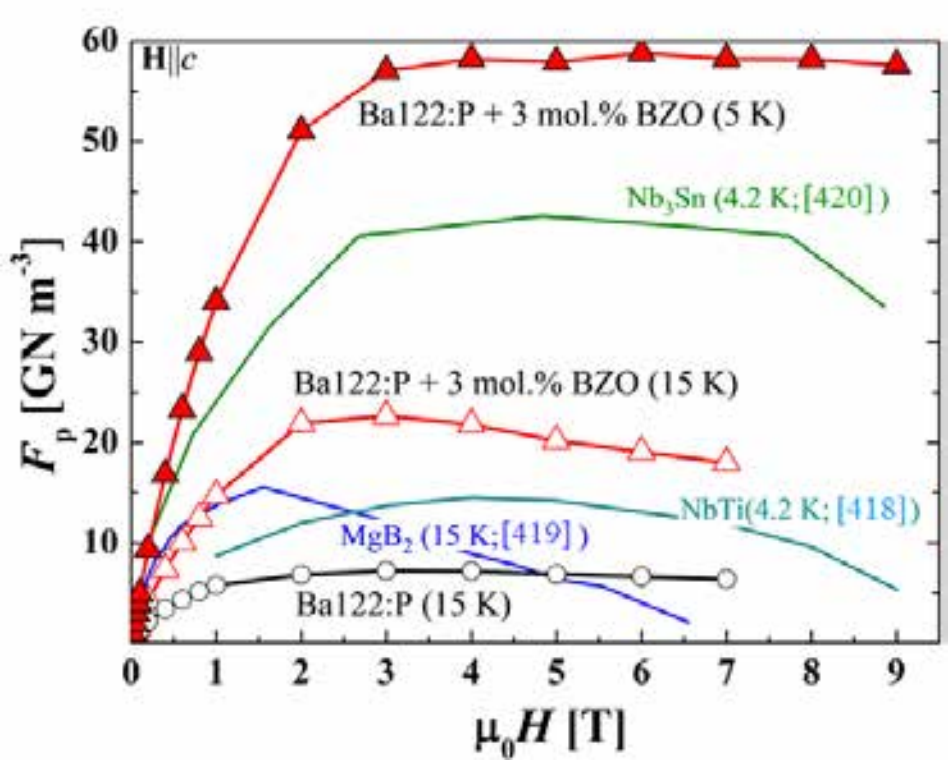

**Figure 82.** Pinning force $F_p$ at magnetic fields applied parallel to the *c*-axis for the Ba122:P+3mol%BZO film at 5 K and 15 K and the Ba122:P film at 15 K. For comparison, the data for NbTi at 4.2 K [418], $MgB_2$ (H||*ab*) at 15 K [419], $Nb_3Sn$ at 4.2 K [420] are included. Reprinted from [415].

Miura *et al.* also found that the characteristic magnetic field, where the maximum relative $J_c$ enhancement by BZO addition was observed, increased monotonically by increasing the nanoparticle density. From the comparison with the previous results of REBCO films with BZO nanoparticles [421], it was deduced that the effective way of optimizing the $J_c$ performance of IBScs as well as cuprates is dispersing nanoparticles with the average size smaller than ~3*(2$\xi(T)$) and with densities such that the nanoparticle spacing matches the intervortex distance. These results suggested a possibility of further enhancing in-field $J_c$ properties, in particular at higher fields, by optimizing the landscape of nanoparticles or nano defects in Ba122:P epitaxial films.

*4.1.3.2. Intentionally grown c-axis pinning centers*

Sato *et al.* carefully investigated the properties of Ba122:P epitaxial films on MgO (001) single-crystal substrates prepared by PLD using the second harmonic of an Nd:YAG laser and found that films with very high and less anisotropic in-field $J_c$ could be obtained at certain deposition conditions [422]. They employed a semiconductor infrared diode ($\lambda$ = 975 nm, and maximum power = 300 W) for substrate heating and achieved high substrate temperature ($T_s$) up to 1400 °C. High-purity Ba122:P targets with the nominal P content $x$ of 0.30 were used. It was found that 150-200 nm thick epitaxial films with high crystallinity ($\Delta\phi$ and $\Delta\omega$ well below 0.8°) were grown at rather high $T_s$ of 1000-1100 °C (optimum at 1050 °C) and low growth rate of 0.2-0.4 nm/s for laser fluence of 3.0 – 3.5 J/cm$^2$. Their lattice parameters slightly different from the single crystal data [403] indicated existence of tensile strain in the films. The optimum film exhibited a $T_c$ of 26.5 K, a narrow transition width $\Delta T_c$ of 1.5 K, and a very high self-field $J_c$ up to 7 MA/cm$^2$. This $J_c$ value is comparable to the recently reported high value for a film grown by an MBE method [423].

Figure 83 shows the angular dependence of $J_c$ at 12 K, 3 T for Ba122:P epitaxial films grown at the optimum $T_s$. The $J_c(\theta_H)$ curves exhibit a broad peak around the *c*-axis direction ($\theta_H$ =0°) in addition to the intrinsic $J_c$ peak at $\theta_H$ = 90°, indicating existence of pinning centers along the *c*-axis. With decreasing the growth rate from 0.39 nm/s to 0.22 nm/s, the $\theta_H$ =0° peak becomes more prominent and the $J_c$ values in all directions become remarkably higher, resulting in less anisotropic angular dependence. These results indicate that the vortex pinning properties and $J_c$ anisotropy of Ba122:P epitaxial films can be controlled by tuning the growth rate.

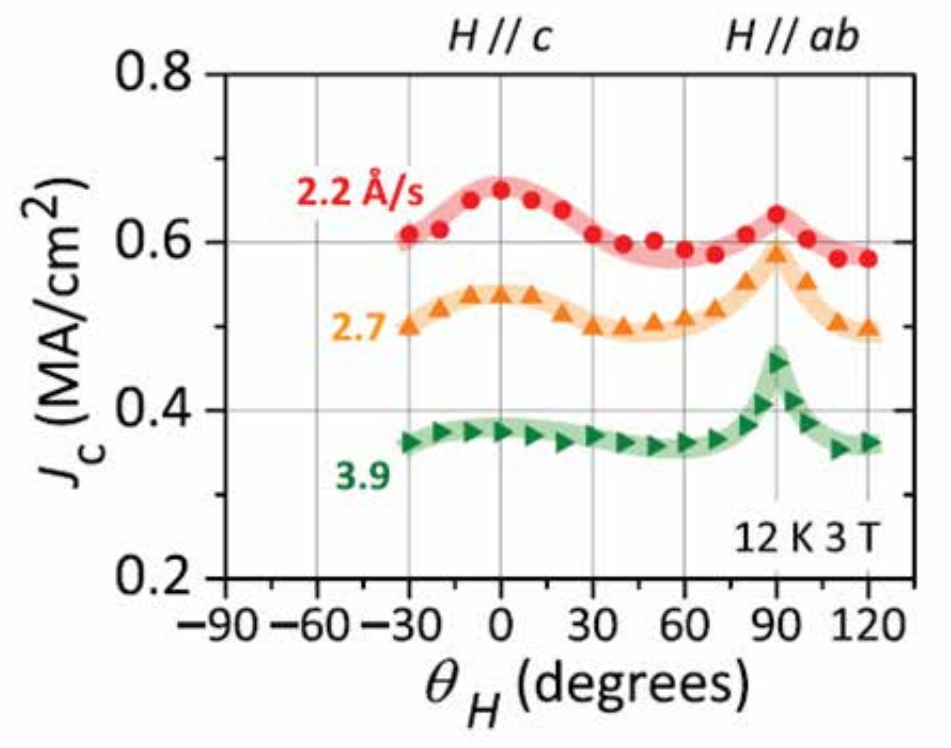

**Figure 83.** Angular dependence of in-field $J_c$ at 12 K, 3 T for high-quality Ba122:P epitaxial films grown on MgO by PLD at relatively high substrate temperatures and low deposition rates of 0.22, 0.27, and 0.39 nm/s [422].

In Fig. 84, the magnetic field dependence of $J_c^{H//ab}$ (closed symbols) and $J_c^{H//c}$ (open symbols) of the optimum Ba122:P epitaxial films is compared with those reported for other Ba122 epitaxial films with high $J_c$ [415, 417, 424-426] as well as $SmFeAsO_{1-x}F_x$ and Fe(Se,Te) films [427, 428]. The optimum Ba122:P film exhibits an even higher $J_c^{H//c}$ than the Ba122:P film with BZO nanoparticles at $H > 4$ T and a $J_c^{H//c}$ as high as 0.8 MA/cm$^2$ at 4 K, 9 T, resulting in the highest pinning force of 72 GN/m$^3$. It would also be the noteworthy that the $J_c^{H//ab}$ value of 1.1 MA/cm$^2$ at 9 T giving a pinning force of 99 GN/m$^3$ is the highest obtained for IBScs films.

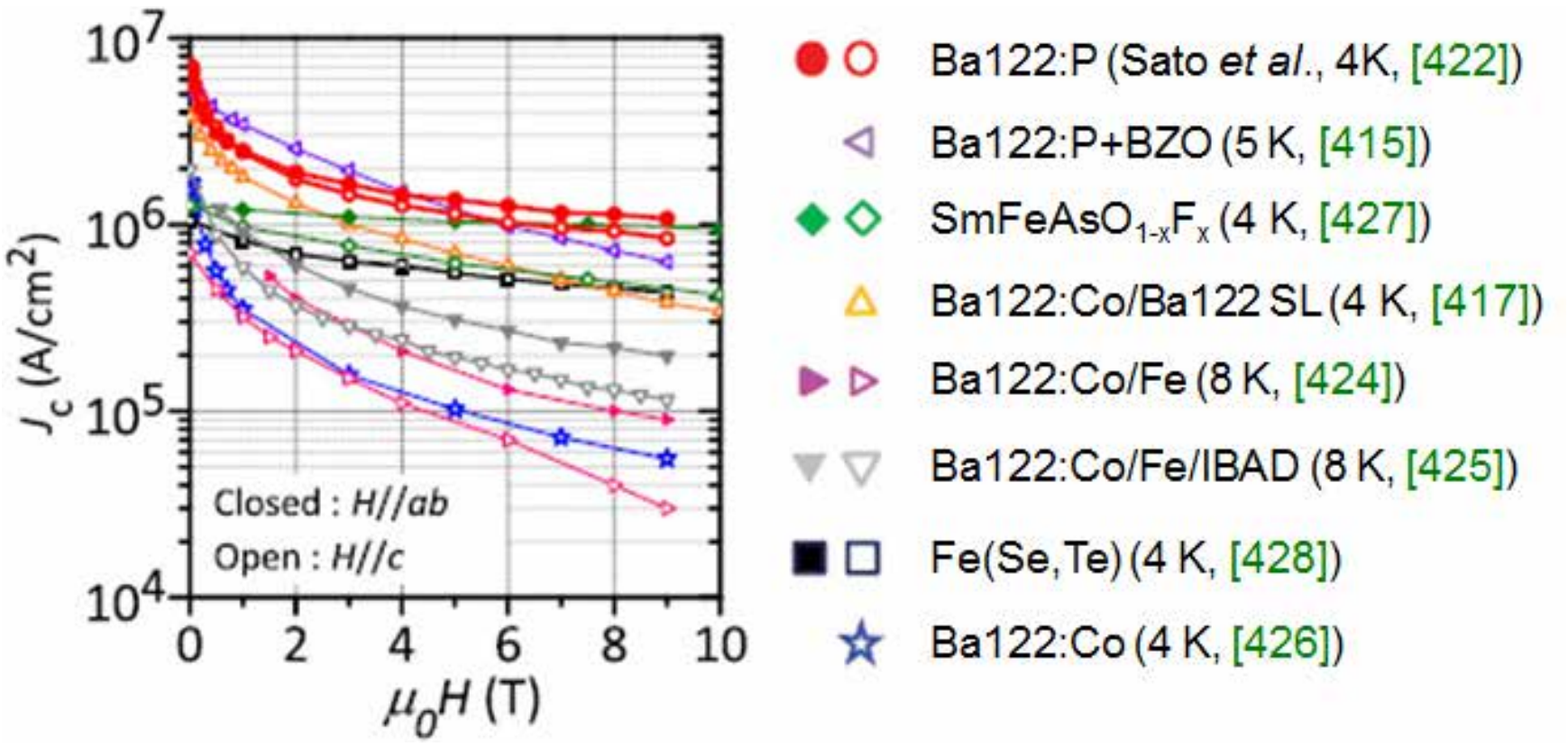

**Figure 84.** Comparison of magnetic field dependence of $J_c$ for the optimum Ba122:P epitaxial film with those reported for other Ba122 films as well as $SmFeAsO_{1-x}F_x$ and Fe(Se,Te) films with high $J_c$ values. The closed and open symbols represent $J_c$ for fields parallel to the *a*,*b*-axes and *c*-axis, respectively [422].

Figure 85 shows the cross-sectional bright-field STEM images for Ba122:P epitaxial films grown at the growth rate of 0.22 nm/s and 0.39 nm/s. As indicated by the vertical white arrows in Fig. 85, there are many vertical defects with a substantially higher density than that observed in the Ba122:Co epitaxial films by Katase *et al.* It is also found that most of the defects in the film grown at 0.22 nm/s start appearing at mid-thickness and are oriented parallel to the *c*-axis, while the defects in the latter film originate just at the substrate surface and are tiled with respect to the *c*-axis. Other planar or line defects in the *ab* plane, such as stacking faults, were not observed. STEM-EDX analyses revealed that the chemical composition of the defects is the same as that of the matrix region and that the impurity oxygen concentration in the films is less than the detection limit. These results indicated that the defects are not an impurity phase such as $BaFeO_2$ [429] but may be edge or threading dislocations and/or domain boundaries. The stronger vortex pinning along the *c*-axis for the film grown at the lower rate can be explained by a high density of straight defects with a lateral size (4 nm) close to the double of the $\xi_{ab}$ of Ba122:P at 4 K [407]. The very high in-field $J_c$ observed in these epitaxial films indicates a high potential of Ba122:P epitaxial films for application to superconducting tapes or wires, though the very high $T_s$ would not be favorable to film fabrication on metal substrates.

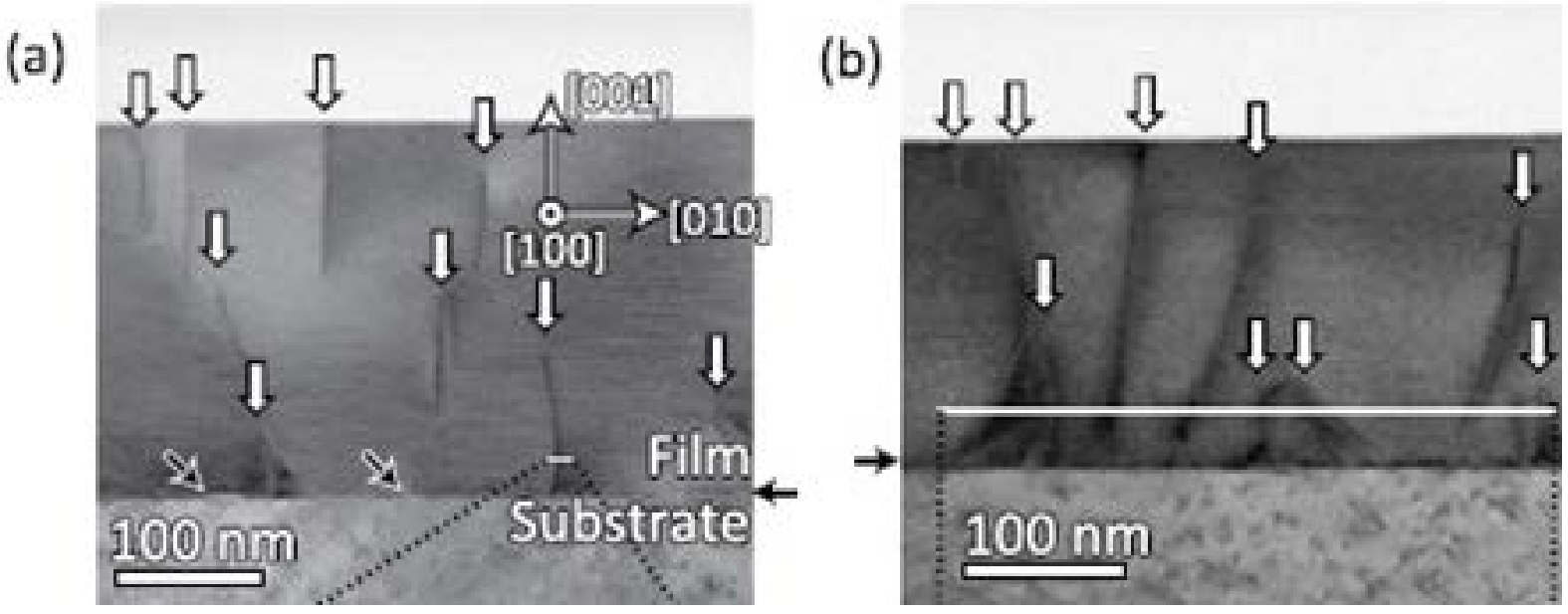

**Figure 85.** Cross-sectional bright-field STEM images for Ba122:P epitaxial films grown at the growth rate of (a) 0.22 nm/s and (b) 0.39 nm/s. The arrows indicate the positions of vertical defects which are considered as strong pinning centers [422].

*4.2. Thin film superconducting device fabrication*

*4.2.1. Bi-crystal Josephson junctions using $Ba(Fe,Co)_2As_2$ epitaxial films*

Josephson junction is the most important basic element in electronic application of superconductors. For low temperature application using liquid He or a cryocooler, Josephson junctions using Nb thin films and a thin Al oxide layer as a tunnel barrier [430] have been in practical use. This sandwich-type junction exhibits ideal superconductor-insulator- superconductor (SIS) type current-voltage (*I-V*) characteristics and is very reliable. For cuprate superconductors such as YBCO, it is very difficult to fabricate SIS type Josephson junctions because of their peculiar physical properties such as very short coherence length and *d*-wave symmetry of the superconducting gap. The very high substrate temperature of 700-800 °C for fabrication of their epitaxial films also makes it very difficult to realize a sharp film-barrier interface required for SIS junctions. However, Josephson junctions exhibiting weak-link-type or superconductor-normal metal-superconductor (SNS) type *I-V* characteristics can be readily obtained by utilizing weak links naturally formed at high-angle grain boundaries (GBs) of their thin films [431]. This weak-link behavior at a high-angle GB comes from carrier depletion due to local structural disorder near the GB. Actually GB Josephson junctions using an epitaxial film on a bicrystal substrate (bicrystal junction) or a substrate with an artificially formed step (step-edge junction) have been applied to electronic devices such as superconducting quantum interference devices (SQUIDs) operating at the liquid-nitrogen temperature [432].

IBScs also exhibit superconductivity only when a proper amount of charge carrier is doped, though their parent materials are not antiferromagnetic insulators but antiferromagnetic metals. Their short coherence length and rather high film growth temperature would make fabrication of SIS Josephson junctions difficult.

Katase *et al.* tried fabricating Josephson junctions using high-quality Ba122:Co epitaxial films and bicrystal substrates [433]. Ba122:Co thin films were deposited on LSAT bicrystal substrates with a symmetrical [001]-tilt boundary having a misorientation angle $\theta_{GB}$ of 30°. Figure 86 shows the *I-V* curve and magnetic field dependence of $I_c$ ($I_c$-$B$ curve) at 10 K for a 10 μm wide bridge patterned across the bicrystal GB (BGB). The bridge clearly shows a resistively-shunted-junction (RSJ) type curve. The magnitude of $I_c$ modulation, which is here defined as $[I_c(0) - I_c(0.9\text{ mT})]/I_c(0)$, is approximately 95%, indicating that most of the supercurrent originates from the Josephson current. This is actually the first demonstration of a thin film Josephson junction in IBScs. However, the hysteresis observed in the $I_c$-$B$ curves suggests that flux trapping occurs in the BGB regions, probably due to inhomogeneity in their microstructures. A 10 μm wide bridge patterned in a single grain region of the same substrate showed an $I_c$ of 40 mA at 10 K, implying that the $I_c$ across the GB is suppressed to less than 1/20 of the film $I_c$.

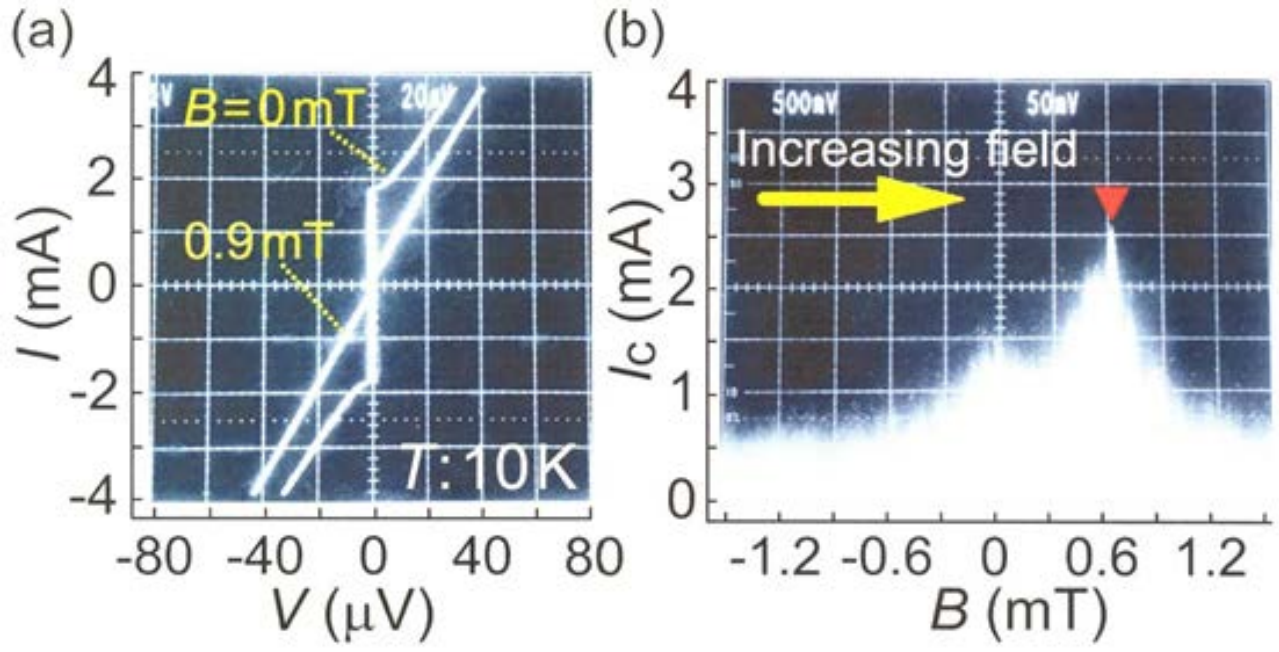

**Figure 86.** (a) Current-voltage (*I-V*) curve under $B$ = 0 and 0.9 mT and (b) magnetic field dependence of $I_c$ ($I_c$-$B$) curve at 10 K for a 10 μm wide bridge patterned across the bicrystal GB (BGB) in a Ba122:Co epitaxial film on an LSAT [001]-tilt bicrystal substrate with the misorientation angle $\theta_{GB}$ = 30° [433].

Katase *et al.* also fabricated Ba122:Co epitaxial films on [001]-tilt MgO bicrystal substrates with various misorientation angles and examined the GB transport properties [434]. Though the details will be described in section 4.2.4, *I-V* curves for BGB junctions with $\theta_{GB}$ = 30° and 45° were found to be well fitted by the Ambegaokar-Halperin (AH) model [435], or the RSJ model taking account of thermal fluctuation, confirming that the excess current ratio is minimal for these high-angle BGB junctions. In contrast, BGB junctions with lower $\theta_{GB}$ showed *I-V* curves containing flux-flow behavior. Figure 87 shows the *I-V* curves for BGB junctions with $\theta_{GB}$ = 16°, 24°, 30° and 45°. The dotted lines are the fits by the AH model, while the solid lines are fits by a phenomenological model previously proposed to explain the fractions of flux flow and RSJ behaviors [436]. The latter fits indicate that the fraction of the RSJ current are approximately 70%, 90%, 100% and 100% for the BGB junctions with $\theta_{GB}$ = 16°, 24°, 30° and 45°, respectively.

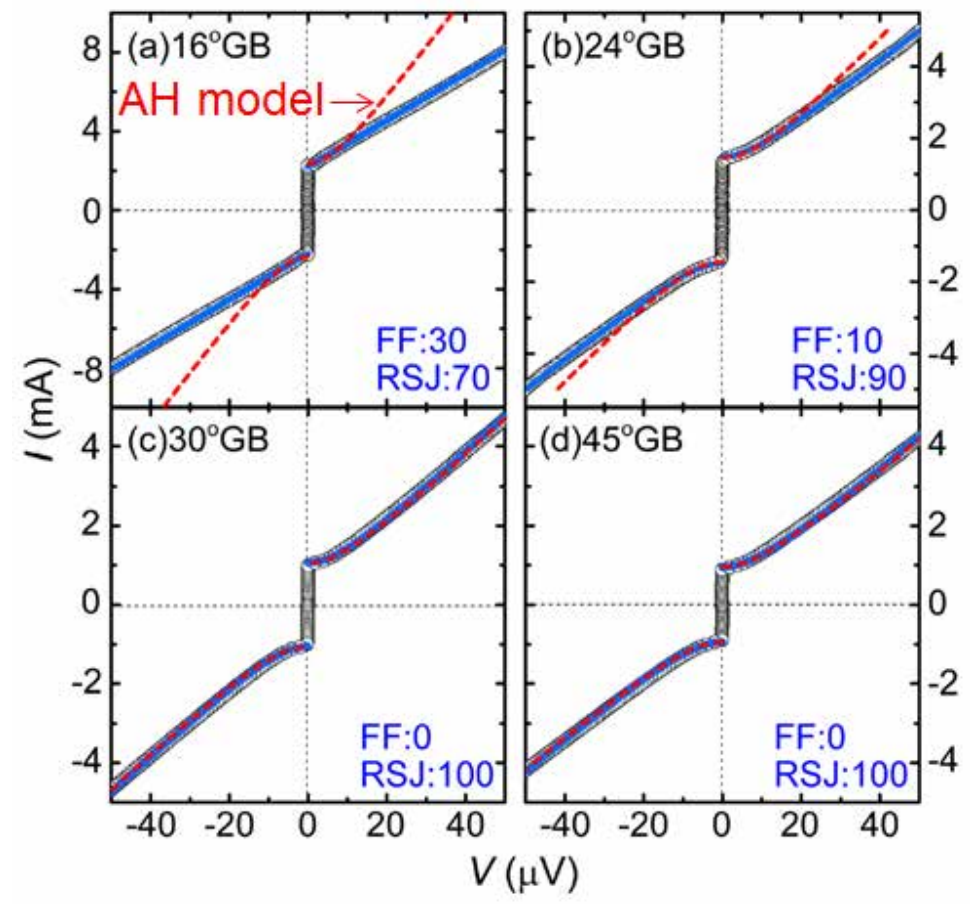

**Figure 87.** *I-V* curves at 12 K for Ba122:Co BGB junctions with $\theta_{GB}$ = 16°, 24°, 30°, and 45° grown on MgO bicrystal substrates. The red dotted lines indicate the fits by the A-H model, and the blue lines show the fits to the *I-V* curves with the phenomenological model combining the RSJ behavior and the FF behavior.

The BGB junctions using Ba122:Co epitaxial films were found to exhibit the specific resistance $AR_N$ ($A$ is the junction area and $R_N$ is the junctions normal resistance) of $10^{-10}$ - $10^{-9}$ $\Omega cm^2$, which is more than one order of magnitude lower than that for YBCO [001]-tilt bicrystal junctions [431], and metallic temperature dependence. Figure 88 shows the temperature dependence of $J_c$ for the BGB junctions with $\theta_{GB}$ = 16°, 24°, 30°, and 45° [434]. Clear quadratic temperature dependence is observed for these high-angle BGB junctions. The solid lines in the figure are the fits to de Gennes' theory based on a conventional proximity effect in the dirty limit [437],

$$I_c = I_0\left(1-\frac{T}{T_c}\right)^2 \frac{\kappa d}{\sinh(\kappa d)} = \frac{V_0}{R_N}\left(1-\frac{T}{T_c}\right)^2 \frac{\kappa d}{\sinh(\kappa d)} \quad ,$$

where $d$ is the barrier thickness, $\kappa^{-1}$ is the decay length for a normal metal, and $V_0$ is the characteristic voltage, which is approximately proportional to the energy gap near the barrier. By using the junction resistance $R_N$ estimated from the *I-V* characteristics, all the curves can be well fitted, confirming that the Ba122:Co BGB junctions are SNS type junctions. This is in quite contrast to the case of the YBCO BGB junctions which are basically SIS junctions as indicated by their quasi-linear temperature dependence of $I_c$ and small hysteresis at low temperatures [431, 438].

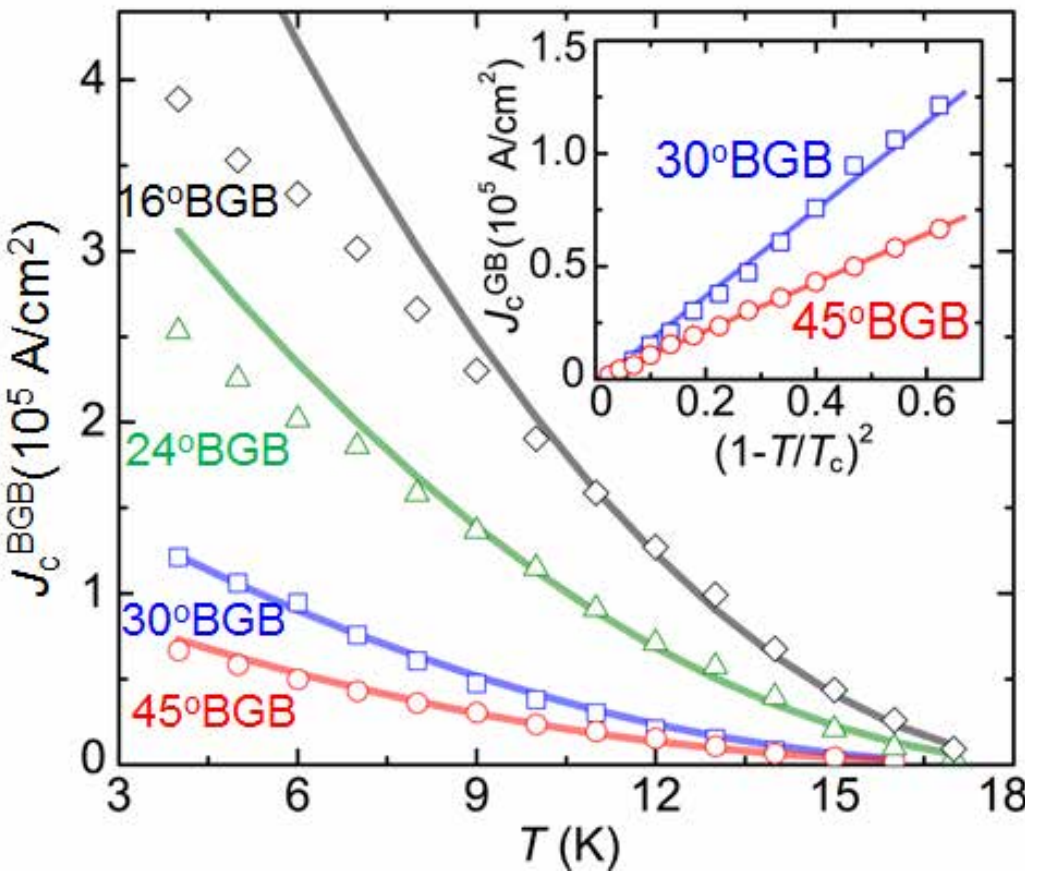

**Figure 88.** Temperature dependence of $J_c$ for the BGB junctions with $\theta_{GB}$ = 16°, 24°, 30°, and 45° grown on MgO bicrystal substrates. The solid lines show the temperature dependence of $J_c$ predicted from de Gennes' theory. The inset shows a linearized plot of the quadratic temperature dependence for the $\theta_{GB}$ = 30° and 45° junctions.

*4.2.2. Fabrication of DC SQUIDs using bi-crystal junctions shown in 4.2.1*

Katase *et al.* fabricated DC SQUIDs using Ba122:Co BGB junctions on LSAT [001]-tilt bicrystal substrates with $\theta_{GB}$ = 30° and demonstrated their operation for the first time [439]. As schematically shown in Fig. 89(a), the SQUIDs have a loop with 18 x 8 μm$^2$ size containing two 3 μm-wide BGB junctions. Figure 89(b) shows the voltage-flux ($V$–$\Phi$) characteristics at 14 K. A clear voltage modulation with $\Delta V$ = 1.4 μV is seen, though its magnitude is one order of magnitude smaller than that typically observed in practical SQUIDs [432].

Figure 90(a) shows the flux noise spectrum for the SQUID at 14 K measured using a commercial flux-locked-loop (FLL) circuit. The white noise level is 1.2 x 10$^{-4}$ $\Phi_0$/Hz$^{1/2}$ and onset of 1/$f$ noise is observed at about 20 Hz. The intrinsic white noise level was estimated to be 9.1 x 10$^{-5}$ $\Phi_0$/Hz$^{1/2}$ by subtracting the contribution of the amplifier noise. This is about one order of magnitude larger than the white noise typically observed for YBCO SQUIDs at 77 K [432, 440, 441]. As shown in Fig. 90(b), the SQUID or the BGB junctions exhibits a rather steep quadratic temperature dependence of $I_c$ due to the metallic nature of the barrier. The SQUID shows $I_c$ less than 100 μA, which is required for operation using the FLL circuit, only at temperatures very close to the junction $T_c$. However, $I_cR_N$ at 14 K for instance is as small as 10 μV. The observed small voltage modulation and the large white noise level for the Co-Ba122 SQUIDs can be explained by this small $I_cR_N$ as well as the small $R_N$. Thus improvement of BGB junction properties seems necessary to make a SQUID based on doped Ba122 practical one.

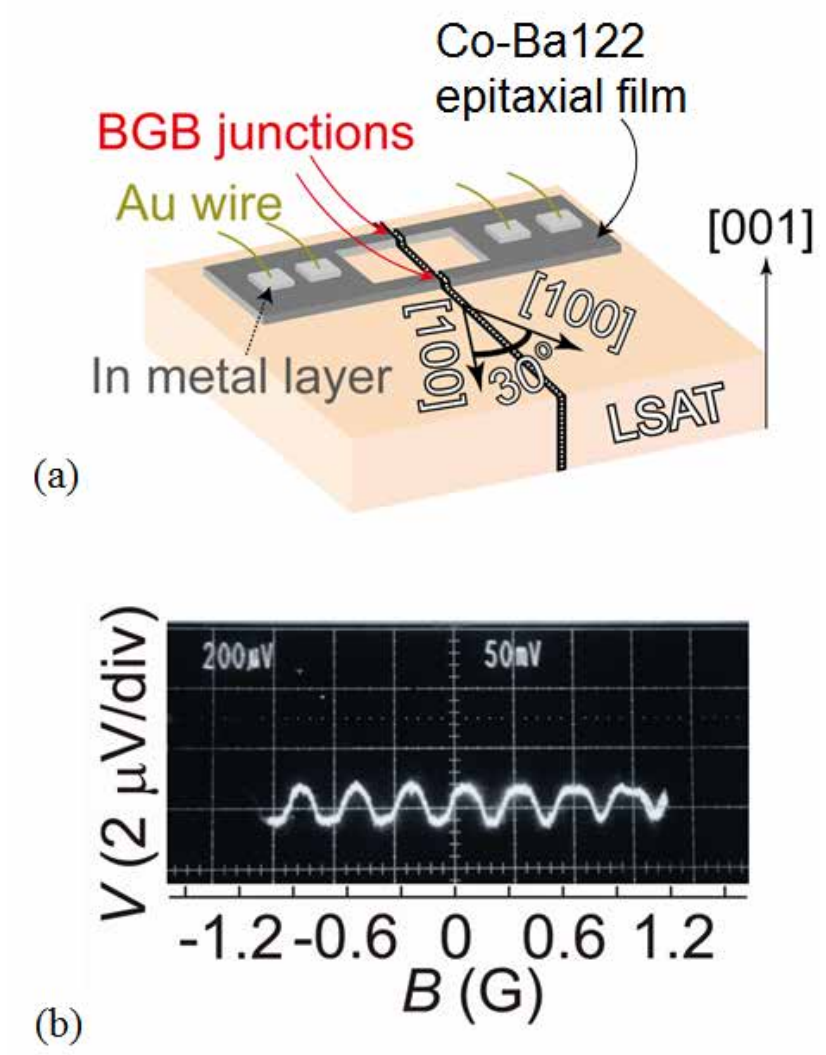

**Figure 89.** (a) Schematic dc SQUID structure fabricated using a Ba122:Co film on LSAT bicrystal substrate with $\theta_{GB}$ = 30°. (b) Voltage-flux ($V$-$\Phi$) characteristics of the dc SQUID measured at 14 K [439].

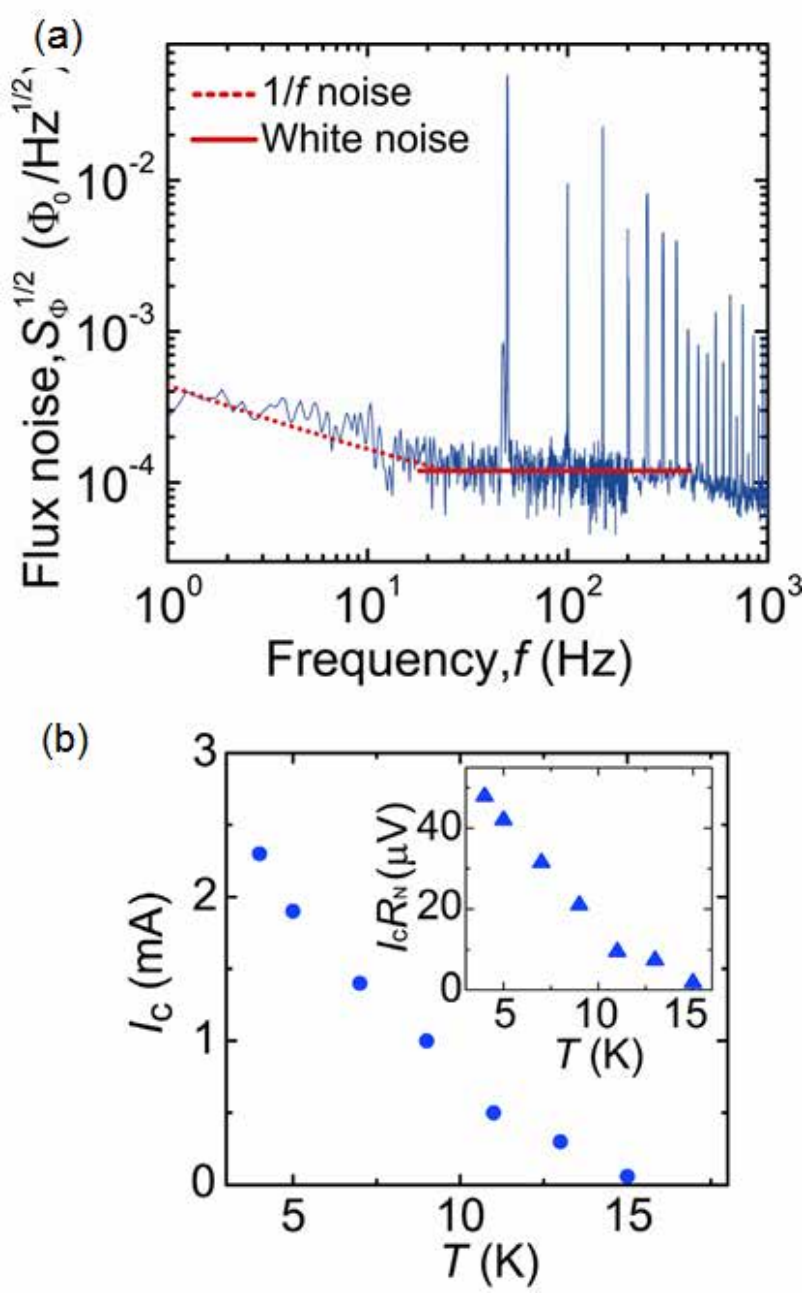

**Figure 90.** (a) Flux noise ($S_\Phi^{1/2}$) spectrum, including the equivalent input noise in the FLL circuit, measured at 14 K for a Ba122:Co dc SQUID. (b) Temperature ($T$) dependence of the critical current ($I_c$) for the SQUID. The inset shows the temperature dependence of the $I_cR_N$ product [439].

*4.2.3. Fabrication of BaFe$_2$(As, P)$_2$ step-edge junctions*

Ishimaru *et al.* tried fabricating step-edge junctions using Ba122:P epitaxial films [442]. One motivation is the fact that larger $I_cR_N$ products than BGB junctions were reported in REBCO step-edge junctions [443, 444], although this was more or less attributed to the $d_{x^2-y^2}$ symmetry of the order parameter in REBCO. Another motivation is that the GB properties in Ba122:P could be different from those for the Ba122:Co BGB junctions. Actually substantially higher BGB $J_c$ has been reported in Ba122:P epitaxial films fabricated by an MBE method [423].

A step structure with height of about 70 nm and an angle of 20° was fabricated on MgO (100) substrates by an Ar ion milling technique. An approximately 90 nm thick Ba122:P epitaxial film deposited on the substrate exhibited a self-field $J_c$ as high as 6.7 MA/cm$^2$ at 4.2 K. Figure 91 shows the *I-V* curves for a 10 μm wide bridge across the substrate step. The *I-V* curve without microwave irradiation is RSJ-type. The junction $I_c$ and $R_N$ are 0.41 mA and 1.35 Ω, respectively, resulting in an $I_cR_N$ product of 0.55 mV. This $I_cR_N$ product is one order of magnitude larger than that for the Ba122:Co BGB junctions [433, 439]. Upon irradiation of 10.02 GHz microwave, clear Shapiro steps are observed. By changing the irradiation frequency and power, the junction $I_c$ could be suppressed to zero, while Shapiro steps were observed up to the voltage of 0.7 mV. This confirms that the observed supercurrent has Josephson current origin. As shown in Fig. 92, the junction $I_c$ could be observed up to the temperature of approximately 30 K, which is very close to the film $T_c$. However, magnetic field modulation of $I_c$ was not observed for this step-edge junction. Cross-sectional TEM observation revealed that the Ba122:P film on the step-slope did not have single orientation and complicated grain structures, suggesting that Josephson junction is formed not along a single GB but at a GB with a very small area between certain grains. These results indicate a possibility of fabricating GB junctions with better performance for IBScs, though control of microstructure of the junction region is required.

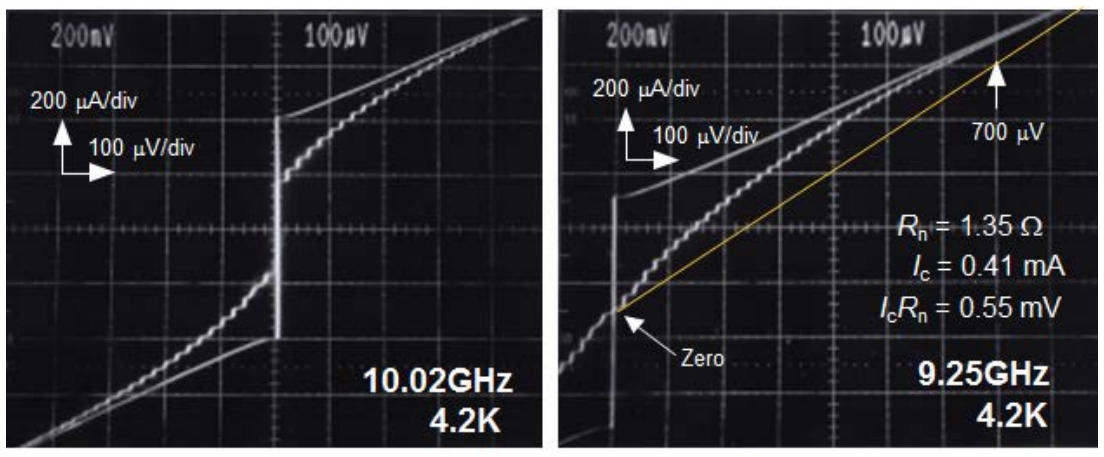

**Figure 91.** *I-V* characteristics at 4.2 K with and without microwave irradiation for a Ba122:P step-edge junction fabricated on an MgO substrate.

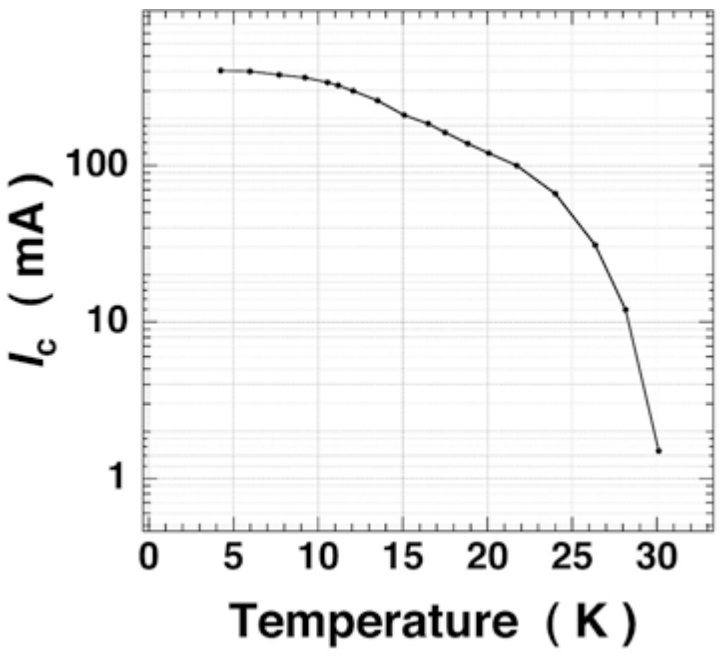

**Figure 92.** Temperature dependence of $I_c$ for a Ba122:P step-edge junction fabricated on an MgO substrate.

*4.2.4. Effect of grain boundary angle on inter-grain transport properties*

The properties of GBs are of great importance for the application of superconducting materials, in particular, to wires and tapes [445]. It is well known that GBs block the supercurrent in cuprate superconductors such as REBCO. For REBCO, $J_c$ across the grain boundary starts to decrease at a critical angle ($\theta_c$) of approximately 3–5° and shows nearly exponential rapid decay with further increasing $\theta_{GB}$ [431, 446]. Because of this weak-link behavior, strict control of the in-plane orientation by employing a biaxially textured buffer layer which is realized, for example, by an ion-beam-assisted deposition (IBAD) technique [447] is required to fabricate REBCO superconducting tapes on flexible metal substrates or coated conductors with a high $J_c$.

The first study on the GB properties of IBScs using Ba122:Co epitaxial thin films on STO [001]-tilt bicrystal substrates was reported by Lee *et al.* [448].They found that $J_c$ across the BGB, even with a low misorientation angle of 6°, was strongly suppressed in their low-temperature laser scanning microscope imaging and transport $J_c$ measurements in a magnetic field of 0.2–0.5 T. Katase *et al.* performed a more systematic study on the transport properties of BGBs with $\theta_{GB}$ = 3–45° using high-quality Ba122:Co epitaxial thin films with a self-field $J_c$ of well above 1 MA/cm$^2$ at 4 K prepared on both MgO and LSAT bicrystal substrates [434]. Figure 93 (a) shows the self-field $J_c$ of BGB ($J_{c\,BGB}$) at 4 and 12 K as a function of $\theta_{GB}$. Nearly exponential decay of $J_{c\,BGB}$ for $\theta_{GB}$ above approximately 10° at both temperatures is observed, indicating that high-angle BGBs act as weak links. However, the exponential decay is more gradual than that for YBCO BGBs. Figure 93 (b) shows the $\theta_{GB}$ dependence of $J_{c\,BGB}$ normalized by the film $J_c$ at 4 K for lower-angle BGBs. The ratio starts to decrease at a critical angle $\theta_c$ of approximately 9 – 10°, while there is no reduction for $\theta_{GB} < \theta_c$, indicating that BGBs in this region are strong links. This critical angle is substantially larger than the value of 3–5° reported for YBCO BGBs [431].

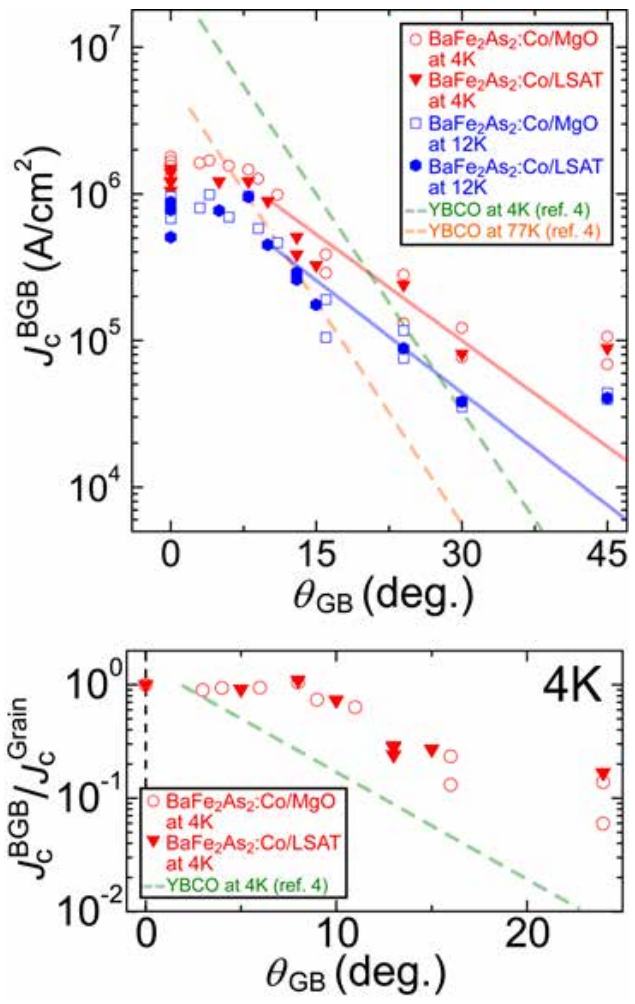

**Figure 93.** (a) Intergrain transport critical current density $J_c^{BGB}$ at 4 and 12 K in a self-field as a function of misorientation angle $\theta_{GB}$ for Ba122:Co BGB junctions grown on [001]-tilt bicrystal substrates of MgO and LSAT. The red and blue solid lines are fits to the empirical equation $J_c^{BGB} = J_{c0}\exp(-\theta_{GB}/\theta_0)$. The average data for the YBCO BGB junctions taken at 4 and 77 K [431] are also indicated by the green and orange dashed lines, respectively, for comparison. (b) Ratio of $J_c^{BGB}$ to the intragrain $J_c$ ($J_c^{Grain}$) in the range $\theta_{GB}$ = 0–25$^o$ at 4 K. The dashed green line shows the result for the YBCO BGB junctions [434].

Figure 94 shows [001] plan-view HR-TEM images of the Ba122:Co BGB junctions on MgO bicrystal substrates. The BGBs with $\theta_{GB}$ = 4° and 24° clearly indicated an array of misfit dislocations with the periodic distance of approximately 5.0 nm for $\theta_{GB}$ = 4$^o$ and 1.2 nm for $\theta_{GB}$ = 24$^o$. Using a geometric tilted boundary model, the grain boundary dislocation spacing $D$ is given by $D = (|b|/2)/\sin(\theta_{GB}/2)$, where $|b|$ is the norm of the corresponding Burgers vector. With the lattice constant $a$ = 0.396 nm of Ba122, $D$ is estimated to be 5.7 nm and 1.0 nm for $\theta_{GB}$ = 4$^o$ and 24$^o$, respectively. The estimated $D$ values are very similar to the $D$ values observed above. Energy dispersive spectroscopy (EDS) line spectra across the BGBs and parallel to the BGBs confirmed that the chemical compositions of the BGBs and the film region are homogeneous, and no secondary phase was observed in the BGB regions.

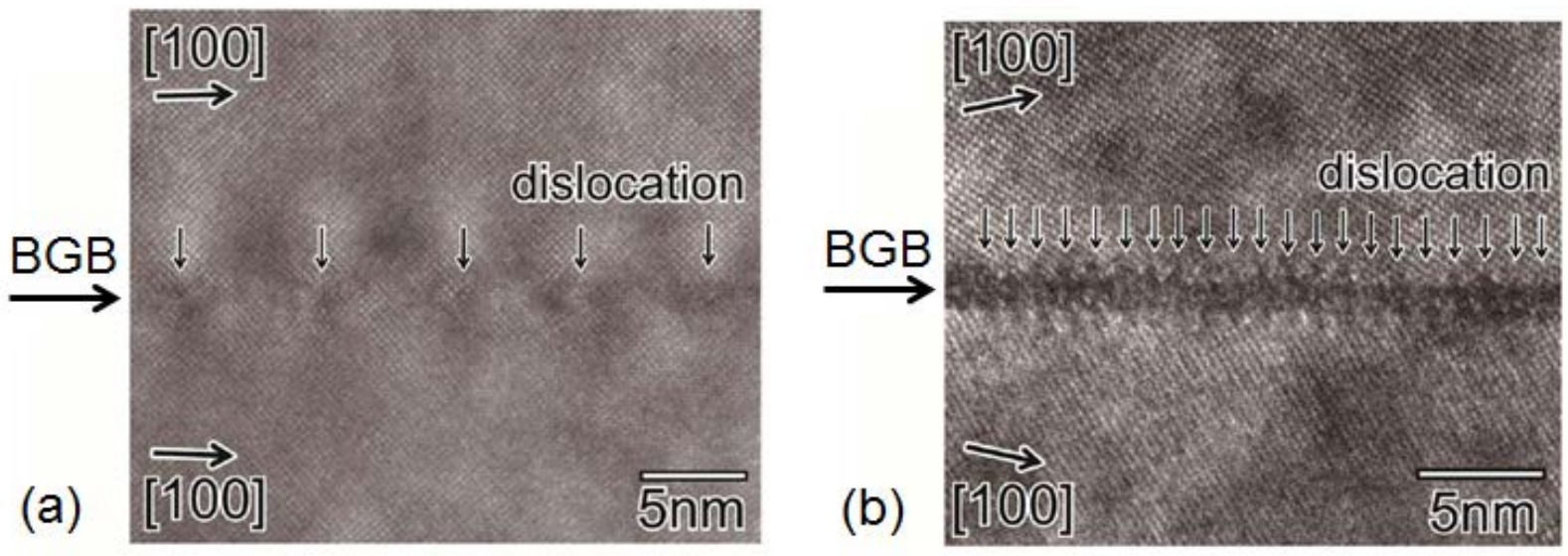

**Figure 94.** [001] plan-view HR-TEM images of the Ba122:Co BGB junctions on MgO bicrystal substrates with $\theta_{GB}$ = (a) 4$^o$, and (b) 24$^o$. Misfit dislocations are marked by the down-pointing arrows [434].

The critical angle of 9° observed in Ba122:Co BGBs corresponds to the spacing between the misfit dislocations of approximately 2.8 nm, which is comparable to or slightly larger than the coherence length $\xi_{ab}$(T) of 2.6 nm at 4 K [319]. This is consistent with notion that strong supercurrent channels still remain between the dislocations for the case of $\theta_{GB} < \theta_c$, while coherent supercurrent cannot pass through the BGBs at $\theta_{GB} > \theta_c$, giving rise to the weak-link behavior. For the case of YBCO, the critical angle of 5° corresponds to $D$ = 4.5 nm, which is much larger than $\xi_{ab}$(T) of 1.6 nm at 4 K. It was previously pointed out that strain near the dislocation cores induces a local transition to an antiferromagnetic phase and forms insulating regions near dislocation cores in cuprate superconductors [446]. The smaller critical angle for YBCO could be explained by the formation of such insulating regions.

For higher-angle BGBs, the distance between the dislocations becomes smaller and eventually dislocations overlap each other. For the case of YBCO, the GB regions become carrier-depleted and thus insulating, though several origins for this, such as shift of chemical potential due to excess ion charge [446], band bending [431], and local structural distortion [449], were proposed. Since carrier-depleted IBSc materials are antiferromagnetic metals (or semi-metals) not insulators, a different nature of high-angle GBs is expected. Actually, as indicated in Fig. 88, the Ba122:Co high-angle BGB junctions exhibit normal-metal-like behavior of the junction barrier. The exponential decay of $J_{c\ BGB}$ in Ba122:Co BGB junctions more gradual than that for YBCO BGB junctions in Fig. 93 can be attributed to this metallic nature of the GB region. The metallic nature of the GB region could also explain the critical angle $\theta_c$ which is substantially larger than that of YBCO and consistent with the dislocation distance $D$.

The observed larger critical angle appears to afford a great advantage for application to superconducting tapes since less strict control of the in-plane alignment for buffer layers would be required to obtain high-$J_c$ IBSc films on flexible metal substrates. Moreover, this would be also advantageous even to fabrication of superconducting wires, for example, based on powder-in-tube (PIT) method where grains are not highly oriented.

*4.2.5. Approach to electrostatic field induced device*

The modulation of electronic properties by applying electric fields is a commonly used technique in semiconducting materials. The electric double-layer transistor (EDLT) type device composed by an atomically flat film of insulator/semiconductor and an ionic liquid (or a polymer electrolyte) as a gate electrode (shown in Fig. 95 as an example) is one of the efficient electric-field devices because such device can accumulate extremely high curries density up to $10^{15}$ cm$^{-2}$ and insulator-to-metal transitions have been demonstrated for various materials such as organic polymers, $InO_x$ polycrystalline films, and ZnO single-crystal films [450-453].

In 2008, Ueno et al. [454] reported that the superconductivity of pristine $SrTiO_3$ single-crystal surface emerged using an EDLT structure. The critical temperature ($T_c$ =0.4 K) was comparable to the maximum value for a chemically doped bulk crystal [455]. This success proposed a new route to evolve superconductivity. Inspired by this result, insulator-superconductor transitions using EDLT devices have been reported in $KTaO_3$ ($T_c$ =0.047 K) [456], ZrNCl ($T_c$ =15.2 K) [457], and $MoS_2$ ($T_c$ =9-11 K) [458, 459]. The similar device has been applied on cuprate superconductors. In the case of epitaxial film of $La_{2-x}Sr_xCuO_4$, its underdoped and insulating film changed into a superconductor ($T_c$ =30 K) by applying -4.5 V to the gate electrode [460]. In the case of $YBa_2Cu_3O_{7-x}$ (YBCO), the enhancement of $T_c$ (from 83 to 134 K by applying

-3 V (hole accumulation) of gate voltage ($V_G$)) was observed but its decrease in resistivity from onset $T_c$ to lower temperature (*i.e.*, superconducting transition) was very broad [461]. When the positive $V_G$ was applied (*i.e.*, electron accumulation), $T_c$ decreased with increasing $V_G$ [461-463]. Whereas the resistivity changes seriously by keeping the device under the constant $V_G$. From this phenomenon, it is considered that the modulation of $T_c$ in the YBCO devices is due to not only the carrier accumulation but also the changing in the defect concentration (possibly oxygen) by the electro-chemical effect.

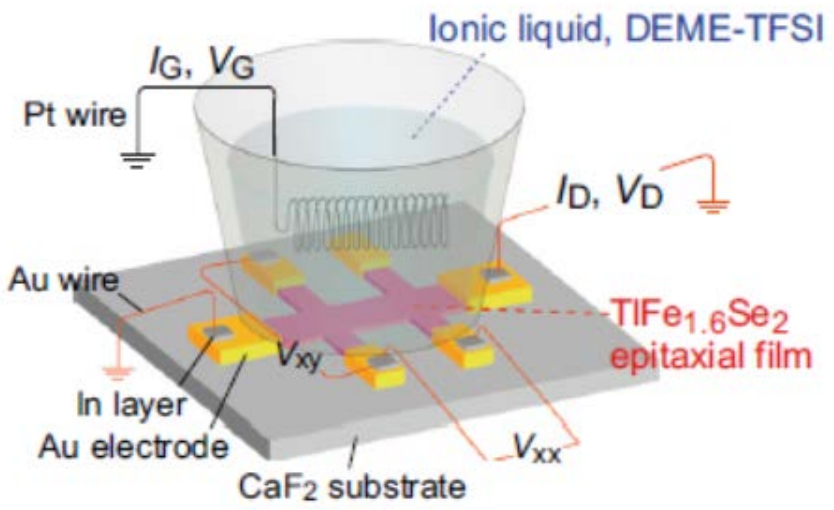

**Figure 95.** Schematic of the EDLT using $TlFe_{1.6}Se_2$ epitaxial film with a six-terminal Hall bar structure on a $CaF_2$ substrate. VG was applied via a Pt counter electrode through the ionic liquid, DEME-TFSI, contained in a silica glass cup. Electrical contacts were formed using Au wires and In/Au pads [464].

Though the parent materials of IBScs are generally antiferromagnetic (AFM) metals, only $A_{0.8}Fe_{1.6}Se_2$ ($A$: K, Rb, Cs, Tl) is an AFM Mott insulator, where iron vacancy ($V_{Fe}$) order forms a $\sqrt{5} \times \sqrt{5} \times 1$ supercell [155]. This material is called as the 245 type of IBSc of which crystal structure of fundamental cell (tetragonal $ThCr_2Si_2$-type) is the same as that of the 122 type (see Section 2.1). By decreasing $V_{Fe}$, it changes to metal and subsequently reveals superconductivity (maximum $T_c$ = 32 K) [154].

The FIRST PJ attempted to tune the transport property of $TlFe_{1.6}Se_2$ by carrier accumulation using EDLT structure [464]. The reason to select $TlFe_{1.6}Se_2$ is its chemical stability comparing other alkaline metal-based (K, Rb and Cs) 245-type compounds. Employing $CaF_2$ substrate, $TlFe_{1.6}Se_2$ film was deposited by PLD method with epitaxial relationships of [001] $TlFe_{1.6}Se_2$//[001] $CaF_2$ (out of plane) and [310] $TlFe_{1.6}Se_2$//[100] $CaF_2$ (in plane). The ordering of $V_{Fe}$, which indicates the insulating phase, was clearly observed in the HAADF-STEM image. The EDLT type device was fabricated using the 20 nm thick $TlFe_{1.6}Se_2$ film. The ionic liquid, N,N-diethyl-N-methyl N-(2-methoxyethyl)-ammonium bis-(trifluoromethylsulfonyl) imide, covered the EDLT device and a Pt coil electrode was inserted into the ionic liquid to act as a gate electrode. Figure 96 shows the R-T curves under applying $V_G$ at 0, 2 and 4 V and the activation energy in the high temperature region. Unfortunately superconductivity transition did not emerge but the electrostatic carrier doping has been controlled successfully by the EDLT structure. Moreover the EDLT structure induced the phase transition (resistance humps marked by arrow in Fig. 96(a)) assignable to a magnetic phase transition or the formation of an orbital-sensitive Mott phase, which has been commonly observed as a precursory phenomenon of transition to superconducting state. This demonstration of carrier doping of the Mott insulator by the electrostatic method offers a way to extend the exploration of high $T_c$ superconductors even to insulating materials.

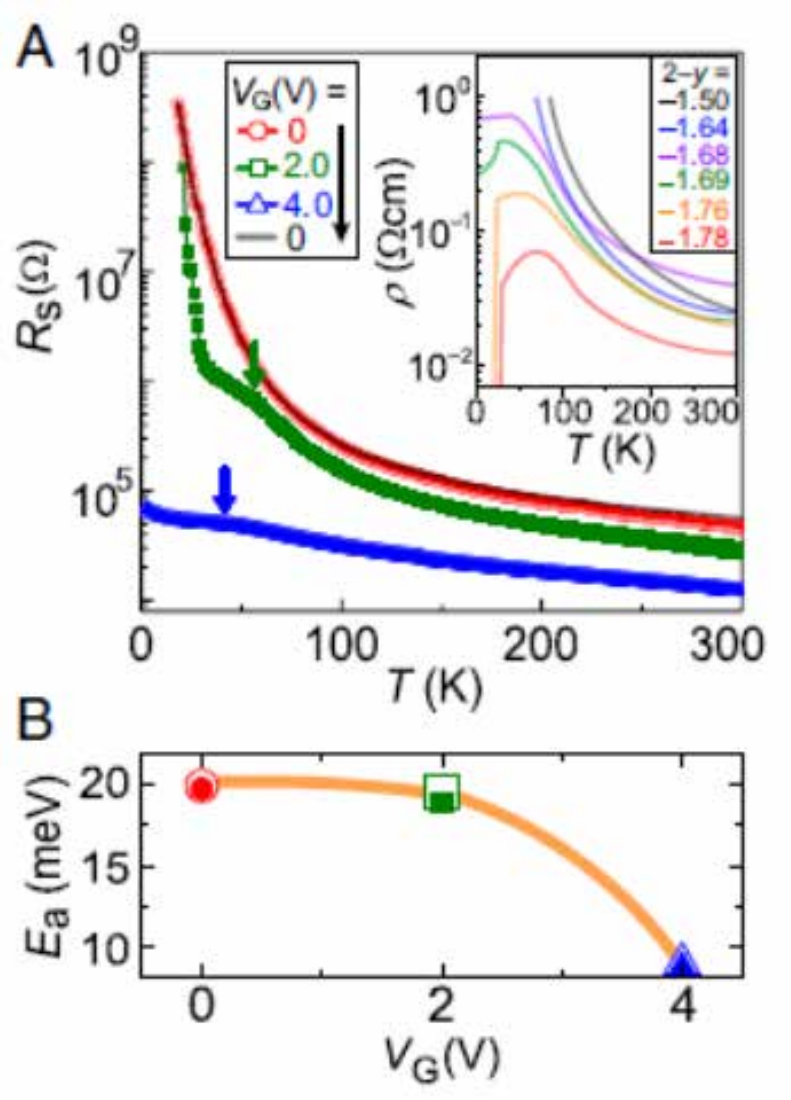

**Figure 96.** (A) $T$ dependences of $R_s$ for the $TlFe_{1.6}Se_2$ EDLT measured with increasing T (open symbols) and decreasing T (filled symbols) at $V_G$ = 0→+2.0→+4.0→0 V. The arrows indicate the positions of resistance humps. The reported $\rho$–$T$ curves of (Tl,K) $Fe_{2-y}Se_2$ bulk materials [156] are shown for comparison (Inset). A resistance hump appears at $2 - y \geq 1.68$, and superconductivity emerges at $2 - y \geq 1.76$. (B) The $E_a$ estimated from A in the high $T$ region as a function of $V_G$. Reprinted from [464].

### *4.3 Fabrication of superconducting tapes by deposition*

#### *4.3.1. Fabrication of short-length $BaFe_2(As, P)_2$ tape*

Superconducting tapes or coated conductors which are fabricated by depositing superconducting films on flexible metal tapes with proper buffer layers have been demonstrated to be practical conductors with high critical current for the case of REBCO cuprate superconductors [465]. In order to realize a high-$J_c$ REBCO film layer on a biaxially-textured oxide buffer layer, the ion-beam-assisted deposition (IBAD) technique [447] or rolling-assisted biaxially-textured substrate (RABiTS) technique [466] has been employed. For the case of $MgB_2$ with no significant GB problem, the PIT technique enables production of round superconducting wires which are more favorable for magnet application [467], though they can be only used at low temperatures below 20 K. Since IBScs also exhibit a weak-link behavior at GBs with the misorientation angle larger than the critical angle of approximately 9°, as described in the previous section [434], the coated conductor technique would be one promising candidate for production of practical conductors.

The first trial fabrication of iron-based 122 compound films on flexible metal substrates with biaxially textured buffer layers was reported by Iida *et al.* [468]. By employing the Fe buffer architecture, they realized the biaxially textured growth of Ba122:Co thin films on IBAD-MgO-buffered Hastelloy substrates, which are typically used for the fabrication of REBCO coated conductors. The films exhibited in-plane misorientation $\Delta\phi$ of about 5°, which was slightly smaller than that of the homoepitaxial MgO/IBAD-MgO layer. They also showed a broader transition width and a substantially lower self-field $J_c$ than those for films on MgO single-crystal substrates, although substantial improvement in $J_c$ has recently been reported [425].

Katase *et al.* succeeded in preparing biaxially textured Ba122:Co thin films directly on IBAD-MgO-buffered Hastelloy substrates [406]. Figure 97 (a) schematically shows the film on the structure.

They used 10 x 10 mm$^2$ substrates with moderate in-plane alignment for the epitaxial-MgO layer on IBAD MgO $\Delta\phi_{\mathrm{MgO}}$ of 5.5 – 7.3°. X-ray diffraction revealed that the films had a substantially smaller $\Delta\phi$ of approximately 3°, irrespective of the $\Delta\phi_{\mathrm{MgO}}$ value. As shown in Fig. 97 (b), the Ba122:Co films exhibited a resistive transition as sharp as that for films on MgO single crystal substrates and high self-field $J_c$ values of 1.2–3.6 MA/cm$^2$ at 2 K. Figure 98 shows the magnetic field dependence of $J_c$ ($J_c$-$H$) at 4-18 K for the film on the substrate with $\Delta\phi_{\mathrm{MgO}}$ = 6.1°. At 4 K, $J_c(H//ab)$ is larger than $J_c(H//c)$ in almost the whole field range, while crossovers are observed at 5.0 T at 12 K and 2.9 T at 16 K, respectively. At a higher $T$ of 18 K, $J_c(H//c)$ is greater than $J_c(H//ab)$ in the whole $H$ region. These results suggest the existence of naturally-formed relatively strong vortex pinning centers along the $c$-axis in the Ba122 film on the metal substrate, as observed in the Ba122:Co epitaxial films on LSAT substrates [400]. The pinning force density $F_p$ was found to show the largest value of ~8 GN/m$^3$ at 4 K which is smaller than those of Ba122:Co epitaxial films on LSAT substrates. This suggests that much higher in-field $J_c$ would be realized also on IBAD substrates by further optimizing or introducing the $c$-axis correlated pinning centers. These results demonstrated a possibility of fabricating high-$J_c$ coated conductors with Ba122 compounds by a rather simple low-cost process using less textured templates with a large $\Delta\phi$.

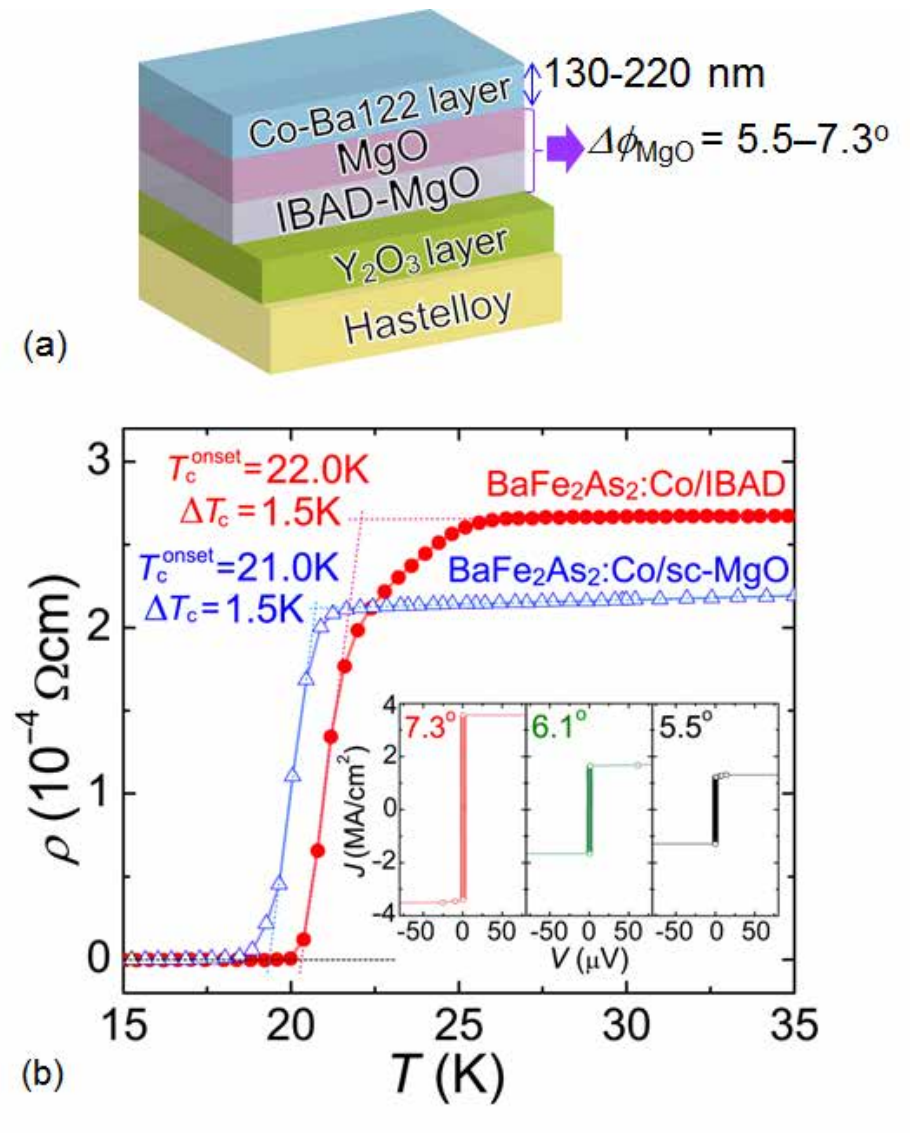

**Figure 97.** (a) Schematic cross-section and (b) $\rho$-$T$ curves for Ba122:Co thin films on IBAD-MgO substrate (circles) and single-crystal MgO (triangles). The inset shows the $J$-$V$ characteristics at 2 K of the films on IBAD-MgO with (left) $\Delta\phi_{\mathrm{MgO}}$ = 7.3°, (middle) 6.1°, and (right) 5.5°[406].

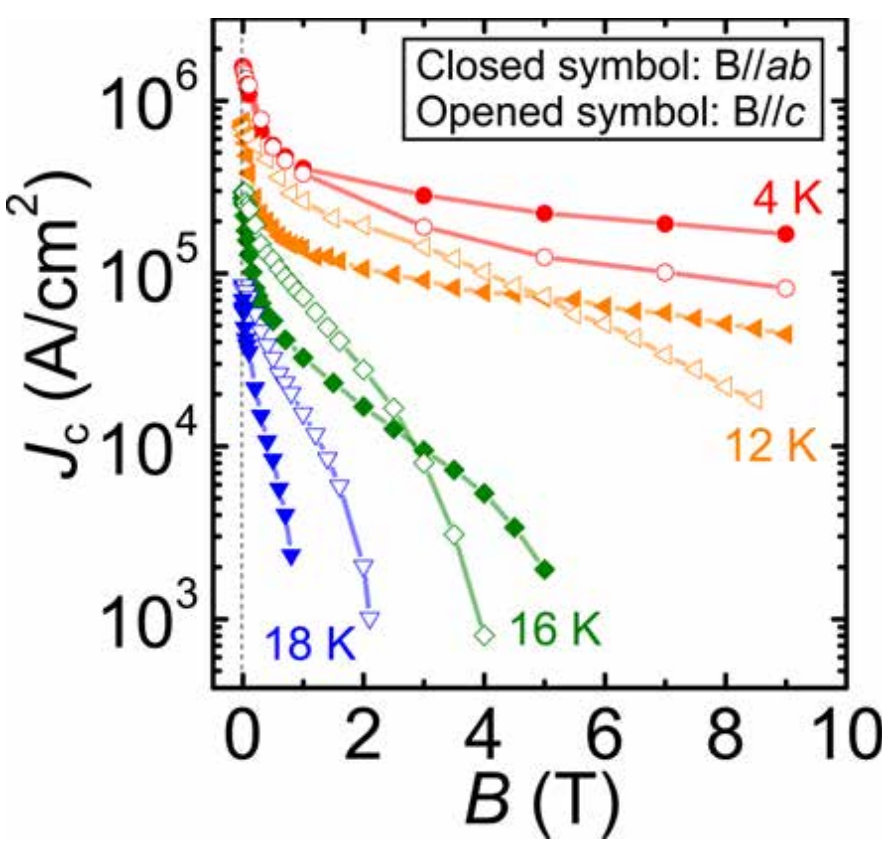

**Figure 98.** Magnetic field dependence of $J_c$ at 4, 12, 16, and 18 K for a Ba122:Co thin films on IBAD-MgO substrate. The closed and open symbols represent $J_c$ for fields parallel to the *a*, *b*-axes and *c*-axis, respectively [406].

*4.3.2. Fabrication of long-length $BaFe_2(As, P)_2$ tape*

As described in the sections 4.1.2 and 4.1.3, Ba122:P epitaxial films exhibited a higher $T_c$ than Ba122:Co films and a substantially higher in-field $J_c$ by introducing artificial pinning centers consisting of oxide nanoparticles or tuning naturally formed pinning centers along the *c*-axis [415, 422]. Thus Miyata *et al.* [469] tried fabricating long-length coated conductors using Ba122:P films on flexible metal tapes using a PLD system shown in Fig. 99. The system is equipped with a reel-to-reel tape feeding mechanism that enables deposition on a tape longer than 1 m. Two second-harmonic Nd:YAG lasers can simultaneously generate two laser plumes on a target. The target-tape distance is variable between 32 and 57 mm. The energy density of one plume on the target was 2 – 3 J/cm$^2$ and the repetition rate was 2.5 – 20 Hz. The target could be changed to a new one during a short interval of deposition without breaking vacuum.

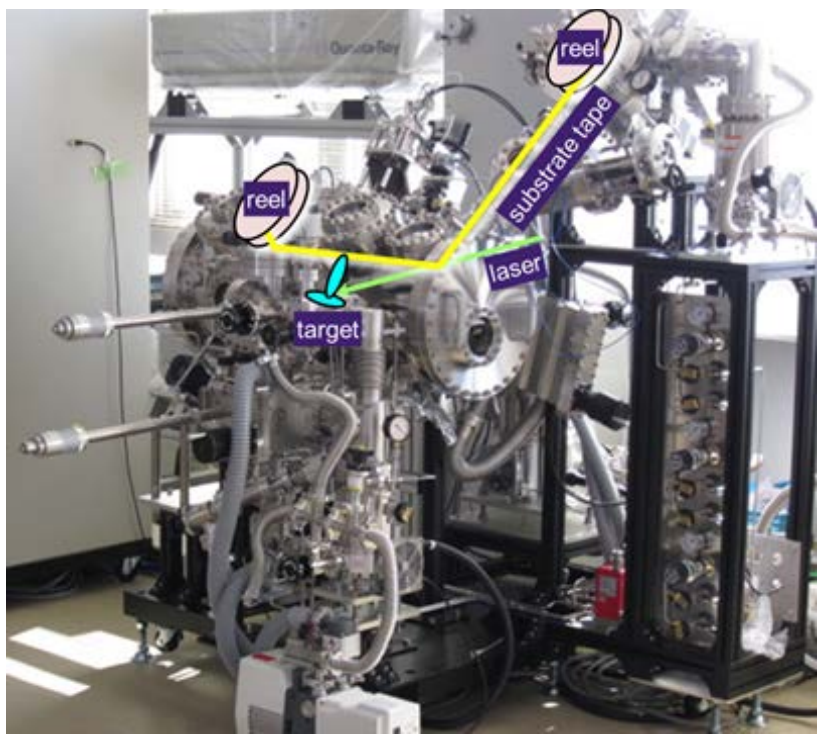

**Figure 99.** Photograph of a reel-to-reel PLD system designed for fabrication of long-length Ba122:P coated conductors.

Two types of IBAD-MgO-buffered Hastelloy tapes, commercial and home-made ones, were used. The commercial tape employed homoepitaxial-MgO (epi-MgO)/IBAD-MgO/ solution-deposited (SDP) $Y_2O_3$/Hastelloy architecture, while the latter employed epi-MgO/IBAD-MgO/sputtered $Gd_2Zr_2O_7$ (GZO)/Hastelloy architecture. In either case, 50-100 nm thick epi-MgO layers exhibited $\Delta\phi$ of ~5°. As the first step, deposition was performed using a Ba122:P target with the nominal P content of 0.40 on a fixed tape without travelling at a substrate temperature of approximately 850 °C. The deposited film exhibited biaxial texture and a resistive transition with a Tc zero of 17.6 K, which was substantially lower that for the film on single-crystal (sc) MgO (~24 K) [404]. In Fig. 100 (a), x-ray diffraction patterns for the film on IBAD-MgO and sc-MgO are compared. A clear shift of the (008) diffraction peak to a lower angle side for the former film is observed. This peak position is close to that for Ba122:P with $x$ = 0.2, suggesting a loss of P in the film. Cross-sectional elemental map analysis suggested existence of a phase consisting of Ba-P-O which grew along the c-axis. It was also found that the (004) as well as (008) peak position was sensitive to the background pressure before deposition as shown in Fig. 100 (b). Since a more than 1 m-long metal tape was loaded between two reels even for deposition of short-length sample, careful degasing of whole the tape was required to obtain low residual gas pressure.

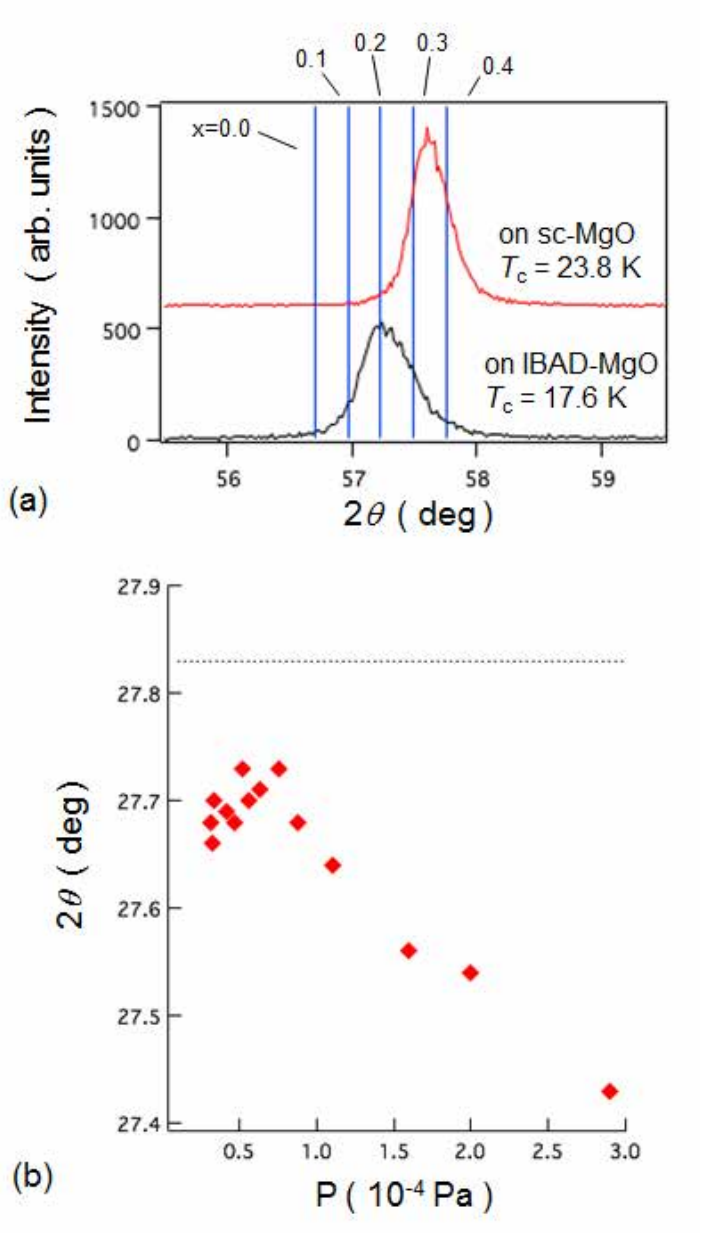

**Figure 100.** (a) Comparison of x-ray diffraction patterns for typical Ba122:P films prepared on IBAD-MgO and single crystal (sc-) MgO substrates using a PLD target with the nominal P content $x$ of 0.33. (b) Correlation between the position of (004) diffraction peak with the residual pressure before deposition for Ba122:P films on IBAD-MgO substrates.

Figure 101 (a) shows the temperature dependence of resistance for films deposited on fixed tapes under low residual pressure below $10^{-4}$ Pa at heater setting temperatures of 1150-1250 °C. The films deposited at a temperature of 1200 °C exhibited the highest $T_c$ ($R$ 10%) of 20.7 K, while the film $I_c$ tends to decrease with increasing the temperature, as shown in Fig. 101 (b), suggesting a change in film morphology. Figures 102 (a) and 102 (b) show the $R$-$T$ curve and the $I$-$V$ curve at 4.2 K, respectively, for a 2 mm wide rectangular piece cut from a 5 cm long coated conductor which was fabricated at a tape travelling speed of 6 mm/min using another Ba122:P ($x$ = 0.40) target. Although its $T_c$ ($R$ 10%) of 17.8 K is lower than that for the film deposited on a fixed tape, the observed self-field $I_c$ of 0.55 mA corresponds to a $J_c$ of 1.1 x $10^5$ A/cm$^2$. Figure 103 (a) shows a picture of a 15 cm long coated conductor fabricated at a tape travelling speed of 6 mm/min. The coated conductor showed a $T_c$ ($R$ 10%) of 18.7 K and an overall self-field $I_c$ at 4.2 K for 1cm width and 10 cm length of 0.47 mA, as shown in Fig. 103 (b). This $I_c$ corresponds to a $J_c$ of 4.7 x $10^4$ A/cm$^2$. This overall $J_c$ lower than that for the shorter sample indicates inhomogeneity of the film properties along the length and/or possibly across the width.

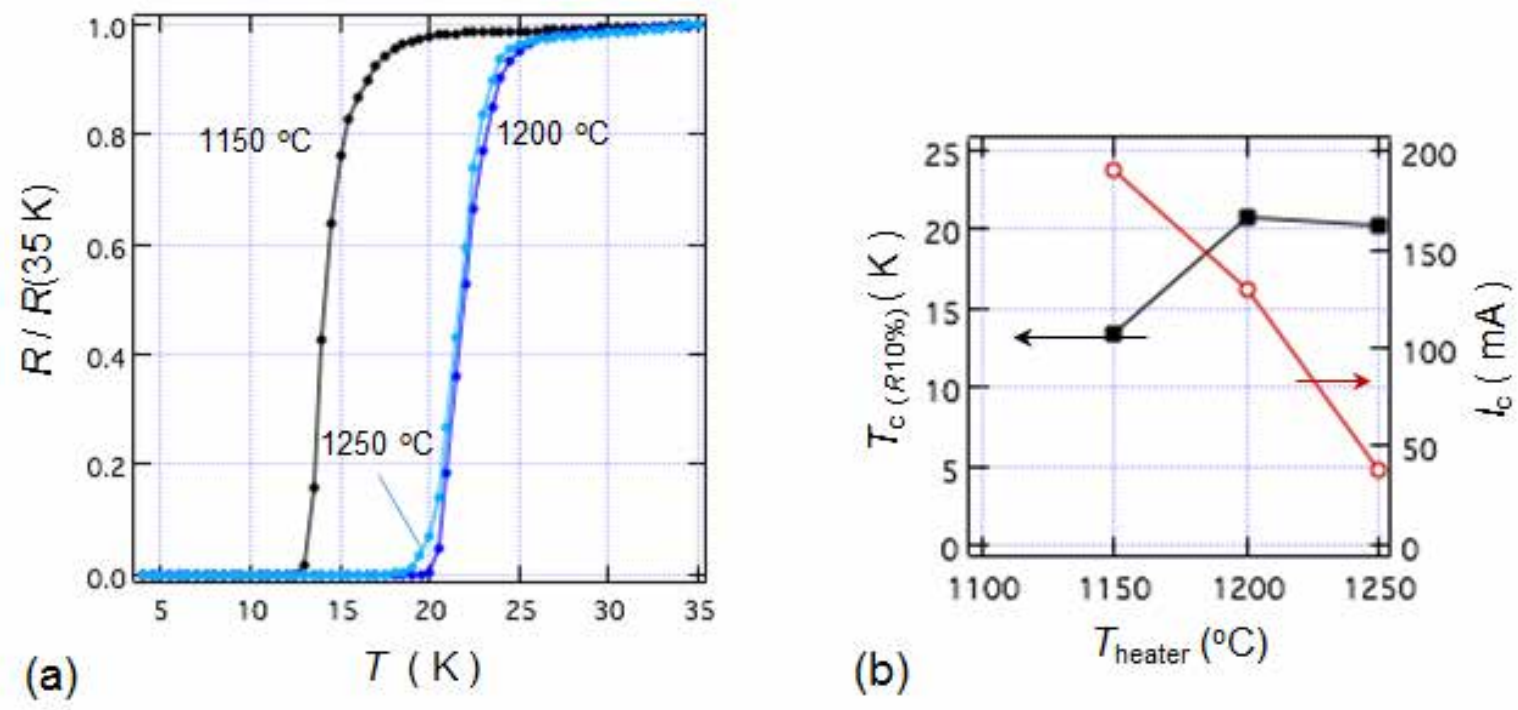

**Figure 101**. (a) Temperature dependence of resistance for Ba122:P films deposited on fixed IBAD-MgO-buffered metal tapes under low residual pressure below $10^{-4}$ Pa at heater setting temperatures

of 1150-1250 ° C. (b) Dependence of $T_c$ ($R$ 10%) and $I_c$ on the heater setting temperature for 2 mm wide Ba122:P films on IBAD-MgO-buffered metal tapes.

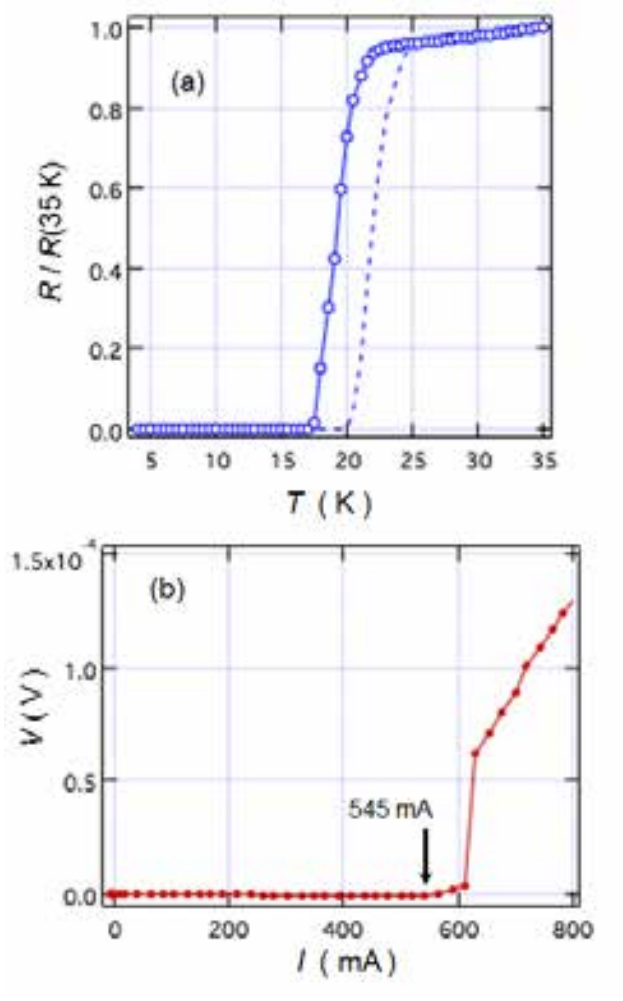

**Figure 102.** (a) $R$-$T$ curve (open circle) and (b) $I$-$V$ curve at 4.2 K for a 2 mm wide rectangular piece cut from a 5 cm long Ba122:P coated conductor which was fabricated at a tape travelling speed of 6 mm/min using a Ba122:P ($x$ = 0.40) target.

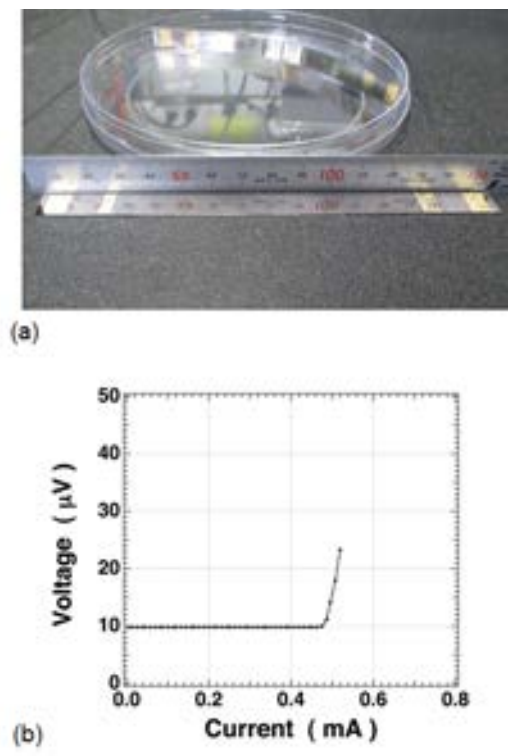

**Figure 103.** (a) Photograph and (b) $I$-$V$ curve measured at 4.2 K for a 15 cm long Ba122:P coated conductor fabricated at a tape travelling speed of 6 mm/min. The width and the distance between voltage taps are 1 cm and 10 cm, respectively.

Although the fabrication conditions have not been optimized yet, the superconducting properties of the trial-fabricated coated conductors are inferior to those for the Ba122:P films on sc-MgO. One reason for the inferior properties could be a difference in the film morphology. SEM observation of the surface of a coated conductor fabricated on a moving tape revealed a rather rough surface with bumps. This is clearly different from the surface for the high-$J_c$ films on sc-MgO with average surface roughness $R_a$ smaller than 4 nm. The lower $T_c$ values for the coated conductors also suggest that the actual P content is still lower than the optimal P content in the target. In order to examine the influence of the P content, Ishimaru *et al.* [470] fabricated multilayer films on the fixed IBAD-MgO-buffered tapes, as schematically shown in Fig. 104, using Ba122:P ($x$ = 0.33) and $Fe_3P$ targets. Figure 105 shows the $R$-$T$ curve and the $I$-$V$ curve at 4.2 K for a 280 nm thick $Fe_3P$/Ba122:P bilayer film. The film exhibits a $T_c$ (R 10%) of approximately 24.0 K, which is closer to the Tc for the film on sc-MgO. The self-field $I_c$ corresponds to a $J_c$ of 1.75 x $10^5$ A/cm$^2$, which is substantially improved as compared with the single-layer film. The four-layer film showed an even higher $T_c$ (R 10%) of 29.5 K, though its self-field $J_c$ was decreased (0.88 x $10^5$ A/cm$^2$). These results confirm that the films on IBAD-MgO-buffered tapes are actually P-deficient. Since the EDS results for the films fabricated under higher residual gas pressure indicated the formation of a Ba-P-O phase, the surface oxygen on the epi-MgO layer might react with the Ba122:P films to some extent, leading to the P-deficient composition. The superconducting properties of Ba122:P coated conductors could be further improved by employing more stable thin buffer layer, for example, BZO, in addition to optimization of deposition parameters.

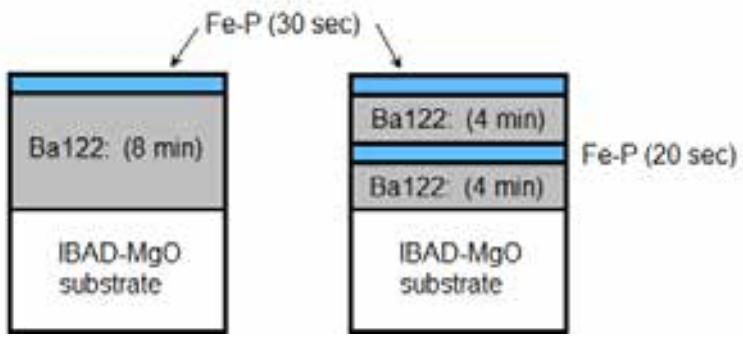

**Figure 104.** Schematic cross-section of Ba122:P/Fe-P multilayer films on IBAD-MgO substrates.

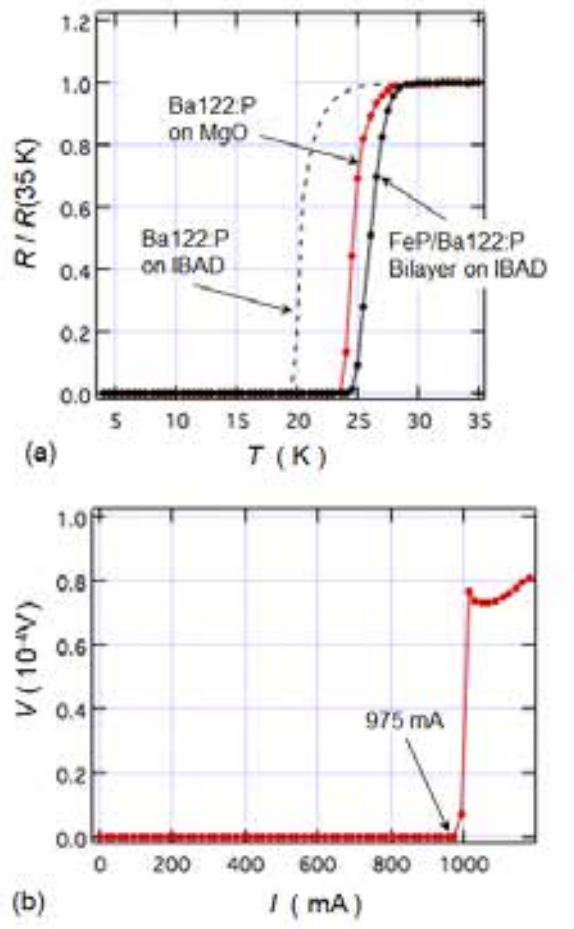

**Figure 105.** (a) *R-T* curve and (b) *I-V* curve at 4.2 K for a 280 nm thick $Fe_3P$/Ba122:P bilayer film.

*4.4. Fabrication of tape and wire by powder-in-tube method*

Among IBScs, K-doped (Ba,K)$Fe_2As_2$(Ba-122) and (Sr,K)$Fe_2As_2$(Sr-122) are most potentially useful for high field applications due to their high critical temperature ($T_c$) value of ~39 K and high upper critical field ($H_{c2}$) of over 50 T [319]. Relatively smaller anisotropy of Ba-122 compounds than those of cuprate superconductors [194] is also attractive for magnet applications because it is expected to bring about a higher irreversibility field, $H_{irr}$. Furthermore, the critical angle ($\theta_c$) of the transition from a strong link to a weak link for Ba-122 is substantially larger than that for YBCO-based conductors [434]. For the high magnetic field magnet applications, we have to develop high performance superconducting tape or wire conductor. In order to evaluate the potential for tape and wire applications, the development of tape and wire processing technique is essential.

For Ba-122 and Sr-122, powder-in-tube (PIT) and coated conductor processes have been developed for tape and wire fabrications. Iron-based coated conductors have been grown by several groups utilizing existing YBCO coated conductor technology and have been found to have a self-field $J_c$ of over 1 MA/cm$^2$ [425, 471, 472]. Although, at an early stage of development, the transport $J_c$ in IBScs reported was disappointingly low due to the weak link grain boundary problem [448, 473-485], astonishing progress has been made for Ba(Sr)-122 tapes in the past several years. $J_c$ for Ba(Sr)-122 tapes and wires approaches $10^4$ A/cm$^2$ at 4.2 K and 10 T through metal addition plus rolling induced texture process, hot isostatic press method, cold press method, hot press method and so on [486-495]. These results demonstrated that mechanical deformation is critical for producing high quality superconducting tapes and wires, which plays an important role in densifying the conductor core and aligning the grains of the superconducting phase. An understanding of the influence of mechanical deformation on the microstructure and superconducting properties will accelerate the development of the appropriate process and will further improve the transport $J_c$ of Ba(Sr)-122 tapes and wires. In this section, we report our recent development of Ba(Sr)-122 tape conductors by applying a conventional powder-in-tube method. Emphasis was placed on the relation between the microstructure and critical current properties. Influence of superconducting core density, grain alignment, microstructure on $J_c$ in the tapes were systematically investigated.

*4.4.1 Fabrication of tape conductors*

Ba-122 and Sr-122 tapes were fabricated by applying standard *ex situ* PIT process [496]. First we prepared the precursor powders of Ba-122 and Sr-122 as shown in Fig. 106 [495]. Ba or Sr filings, K plates, Fe powder and As pieces were mixed to the nominal composition of $Ba_{0.6}K_{0.4}Fe_2As_{2.1}$ and $Sr_{0.6}K_{0.4}Fe_2As_{2.1}$ in an Ar gas atmosphere using ball-milling machine and the materials were put into a Nb tube of 6mm outer diameter and 5mm inner diameter. The Nb tube was put into a stainless steel tube, both ends of which were pressed and sealed by arc welding in an Ar-gas atmosphere. In order to compensate for loss of elements, the starting mixture contained 10-20 % excess K. The stainless steel tube was heat treated at 900 ºC for 10 h and then cooled to a room temperature in a box furnace. After the heat treatment, the precursor was removed from the Nb tube and ground into powder with an agate mortar in a glove box filled with high purity argon gas. Figure 107 shows X-ray diffraction (XRD) pattern of the precursor powder [495]. This analysis together with the magnetization measurement indicate that the precursor powder obtained by this method have fairly good quality.

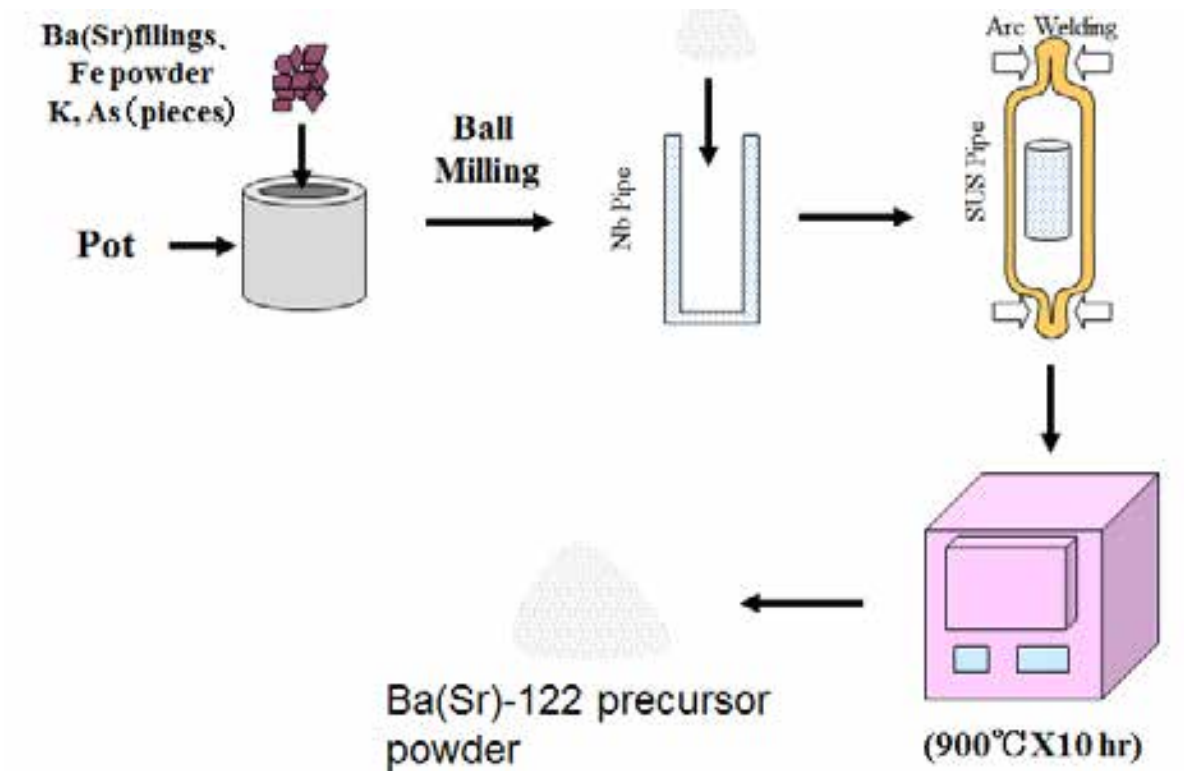

**Figure 106.** Fabrication of Ba(Sr)-122 precursor powder [495].

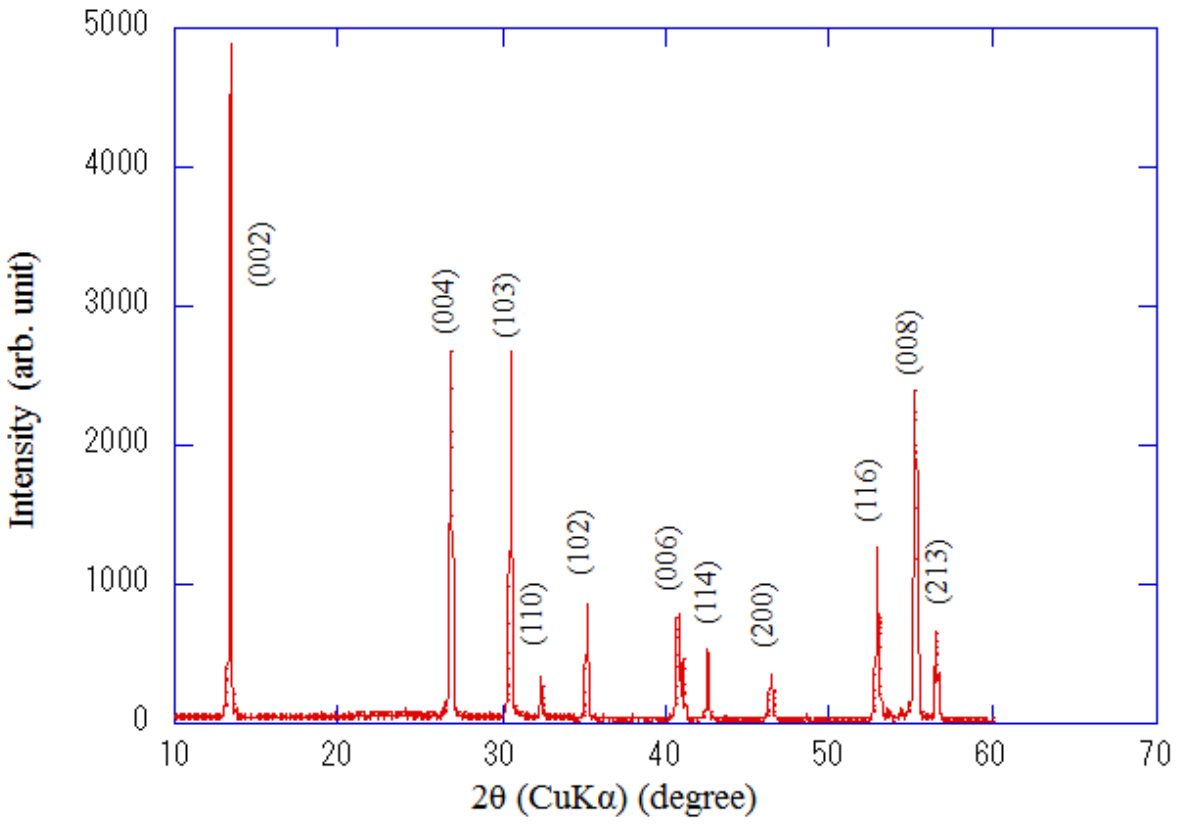

**Figure 107.** X-ray diffraction of Ba-122 precursor powder [495].

The precursor powder was packed into an Ag tube of outer diameter: 6-8 mm, inner diameter: 3.5-4 mm and length: 50 mm, and the Ag tube was groove rolled into a wire with a rectangular cross section of ~2 mm × ~2 mm. The wire was then cold rolled into a tape using a flat rolling machine, initially into 0.8 mm in thickness followed by intermediate annealing at 800 ºC for 2 h and then into 0.40 ~ 0.20 mm in thickness. For some of the tapes we applied uniaxial pressing. For this uniaxial pressing, the flat rolled tape was cut into samples of 35 mm in length and the sample was sandwiched between two hardened steel dies and pressed. The uniaxial pressure was widely changed from ~0.1 to ~4 GPa. The flat rolled and uniaxially pressed tapes were subjected to a final sintering heat treatment at 850 ºC for 2~10 h to obtain Ba-122/Ag and Sr-122/Ag superconducting tapes. All the sintering heat treatments were carried out by putting the tapes into a stainless steel tube, both ends of which were pressed and sealed by arc welding in an Ar atmosphere. We also fabricated seven-filamentary tapes. As cold worked mono-filamentary wire with 1.3 mm diameter was cut into seven short wires with length of 40 mm. The seven wires were bundled together and put into another Ag tube, and the assemblage was cold worked into tape with intermediate annealing, uniaxially pressed, and finally heat treated under the same condition as mono-filamentary tapes.

*4.4.2. Critical current and microstructure*

The transport critical current $I_c$ at 4.2 K in magnetic fields were measured by the standard four-probe resistive method in liquid helium(4.2K) using a 12T superconducting magnet. The criterion of $I_c$ determination was 1 μV/cm. The transport critical current density, $J_c$ was estimated by dividing $I_c$ by the cross sectional area of the Ba-122 or Sr-122 superconducting core. Magnetic fields up to 12 T were applied parallel to the tape surface and perpendicular to the tape axis.

Figure 108 shows $J_c$ versus magnetic field curves at 4.2K of two uniaxially pressed Ba-122/Ag tapes with final tape thicknesses of 0.40 and 0.47 mm [495] and Sr-122/Ag tape with 0.40 mm thickness [497]. For comparison, the $J_c$–$B$ curves of a groove rolled Ba-122/Ag wire with a rectangular cross-section (~2 mm×~2 mm) and a flat rolled tape (0.4 mm thickness) with no uniaxial pressing are also shown in the figure. All the wire and tapes were subjected to the final heat treatment of 850 °C for 10 h. The figure clearly indicates that the $J_c$ increases with the process from groove rolling to uniaxial pressing. The $J_c$ of the grooved rolled wire is as low as ~$10^3$ A/cm$^2$ in applied magnetic fields. Similar low $J_c$ values were reported for Ba-122 and Sr-122 round wires fabricated by the conventional PIT process [486, 487, 498]. A substantial $J_c$ increase was obtained by the application of the cycles of flat rolling and subsequent heat treatment [494]. However, the most remarkable result in this figure is that another large $J_c$ enhancement was obtained by the application of uniaxial pressing. All the pressed tapes show $J_c$ over $10^4$ A/cm$^2$ at 10 T, indicating that a high $J_c$ is obtained with good reproducibility. The data in Fig. 108 suggests that the application of higher uniaxial pressure will result in higher $J_c$ values. This can be realized as will be discussed later.

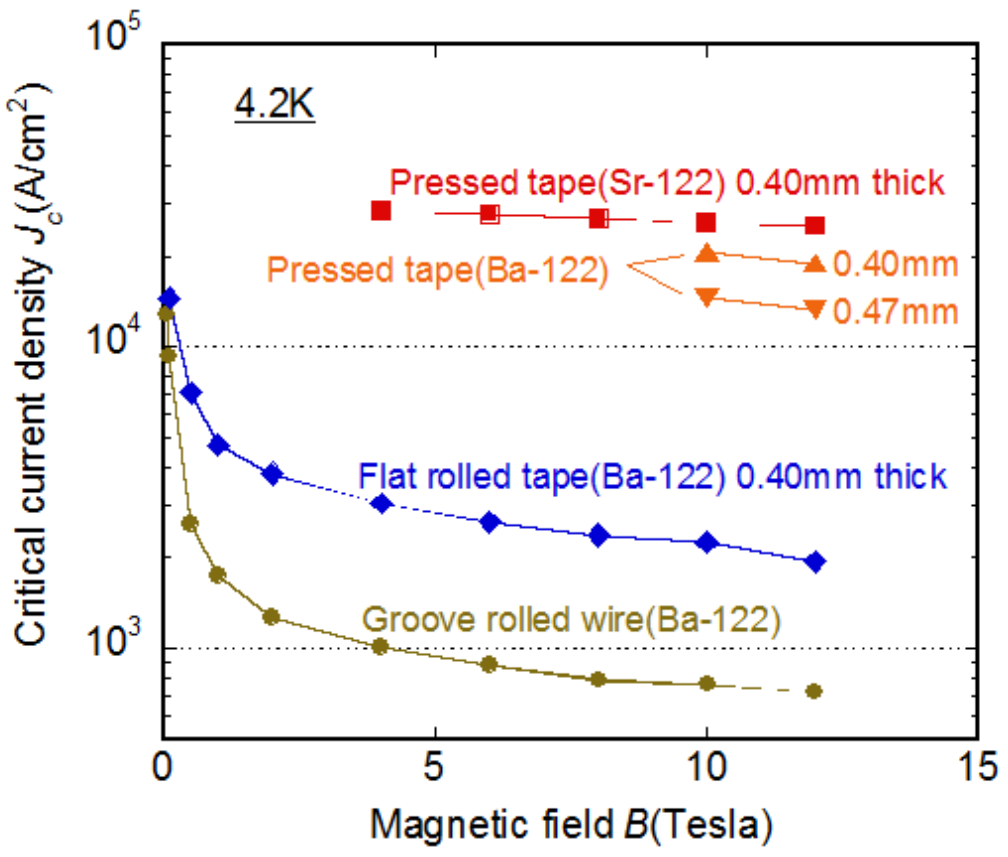

**Figure 108.** $J_c$-$B$ curves of uniaxially pressed Ba-122/Ag and Sr-122/Ag tapes [495, 497]. The data of groove rolled wire and flat rolled tape are also shown for comparison.

In order to investigate the mechanism of $J_c$ enhancement, we observed the microstructure change during the process. Fractured cross-sections of the wires and tapes were observed by scanning electron microscopy (SEM) using a SU-70 (Hitachi Co., Ltd.). Figures 109 (a) - (c) show the grain structures of the Ba-122 core for groove rolled wire, the flat rolled tape and uniaxially pressed tape, respectively [495]. All of the wire and tapes were finally heat treated at 850 °C for 10 h. The microstructure of the groove rolled wire shows non-uniformity in the grain size distributed widely from ~1 to ~10 μm. The flat rolled and pressed tapes show more uniform grain structure whose average size is a few micrometers. We consider that the cycles of deformation and subsequent heat treatment break up larger grains into smaller grains, resulting in the more uniform grain structure of the Ba-122 core. The difference of grain structure between flat rolled tape and pressed tape is not significant although large $J_c$ difference is obtained for rolled and pressed tapes. This large $J_c$ difference can be explained by the difference of Ba-122 core density as will be discussed later. Figure 110 (a) and (b) show XRD patterns of the Ba-122 core surface of the flat rolled and pressed tapes, respectively [495]. For the XRD observation the Ag sheath was peeled off. It should be noted that the relative intensities of the (00*l*) peaks of both tapes are not so high as those observed for pressed precursor powder in Fig. 106 and show more like random orientation. This is different from PIT processed Bi-2223/Ag tapes, in which the cycles of flat rolling and subsequent heat treatment produce stronger c-axis grain alignment due to the larger anisotropic morphology of the Bi-2223 crystal [499]. Our results for Ba-122/Ag tapes are also in contrast to the Fe-sheathed PIT Sr-122 tape [484, 489, 491], in which much stronger grain orientation was observed, similar to the Bi-2223 tape. This difference seems to be caused by the different sheath material and different processing parameters such as reduction ratio. There observed almost no difference of relative (00*l*) peaks between rolled and pressed tapes, indicating that the grain orientation of pressed tape is as low as the flat rolled tape. This is consistent with the grain structure shown in Fig. 109 (b) and (c). From almost the same grain morphology and XRD pattern between the rolled and pressed tapes, the grain orientation can be ruled out as a possible origin of the large enhancement of $J_c$ by the uniaxial pressing.

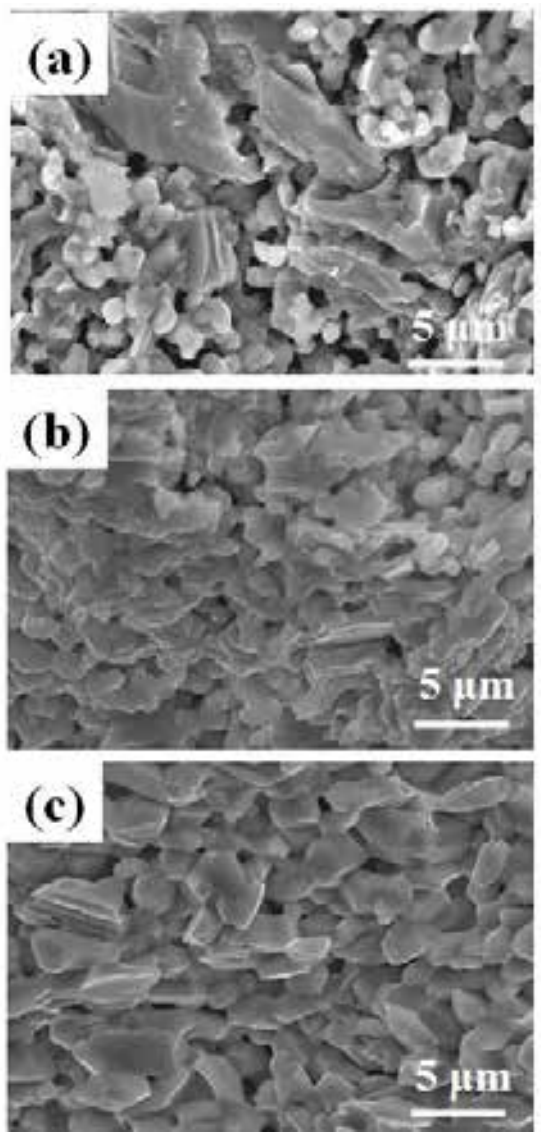

**Figure 109.** Grain structures of the Ba-122 core for (a) groove rolled wire, (b) flat rolled tape and (c) uniaxially pressed tape [495].

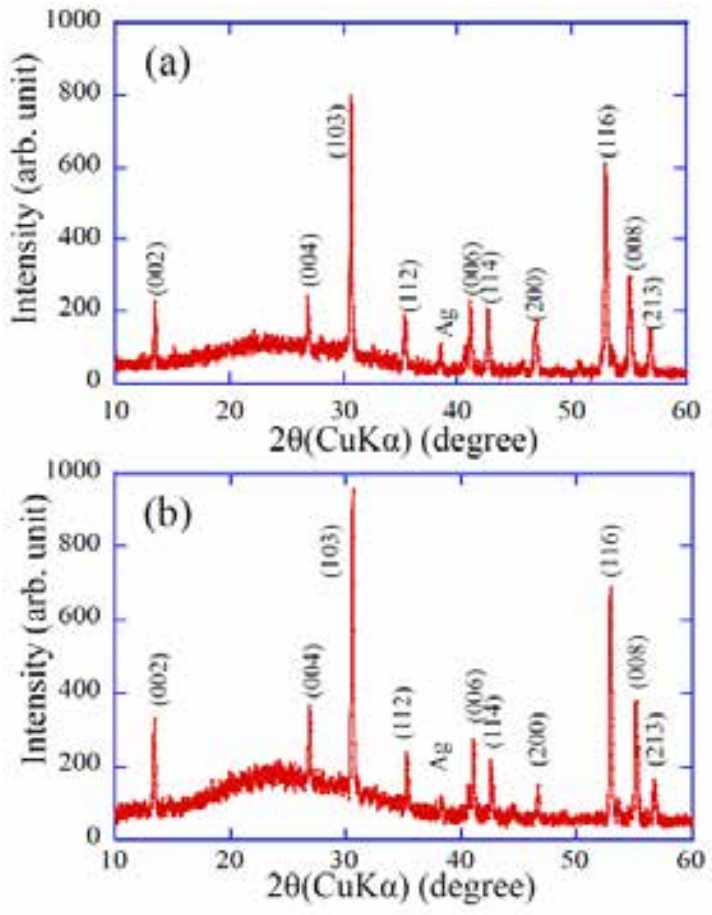

**Figure 110.** X ray diffraction patterns of (a) flat rolled and (b) uniaxially pressed tapes [495].

For the PIT processed superconducting wires and tapes, the density of the superconducting core seems to be one of the major factors that govern the $J_c$ values of tapes and wires. Because it is difficult to directly measure the density of the thin superconducting core, researchers sometimes use Vickers hardness as an indicator of the density of the superconducting core [500, 501]. In this study we performed the Vickers micro-hardness (*Hv*) measurements of the Ba-122 core in order to investigate the influence of the core density on the $J_c$ of our wire and tapes. Figure 111 shows *Hv* of groove rolled wire, flat rolled tape and uniaxially pressed tape [495]. All the wire and tapes were finally heat treated at 850 $^{o}$C for 10 h. The *Hv* measurements were made on polished transverse cross-section of each sample with 0.05 kg load and 10 s duration in a row at the center of the cross section. For the rolled tape, *Hv* at two different cross-sections were measured, because *Hv* varies more widely in the rolled tape depending on the position. Although the scattering of *Hv* is large, the figure clearly shows that the average *Hv* increases with the progression of deformation process from groove rolling to uniaxial pressing. The average *Hv* of the groove rolled wire, flat rolled tape and uniaxially pressed tape are 87.1, 94.0 and 117, respectively. This increase of *Hv* should be attributed to the increase of the Ba-122 superconducting core density. Thus, the increase of $J_c$ with the deformation process in Fig. 108 can be explained by the increase of Ba-122 core density in the wires and tapes.

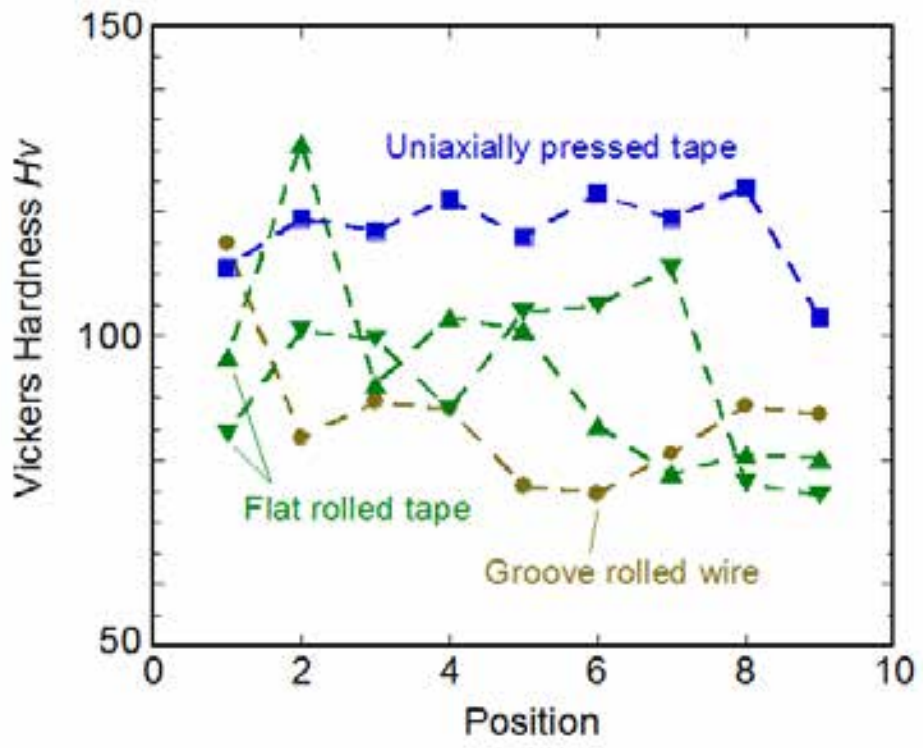

**Figure 111.** Vickers micro-hardness($Hv$) of groove rolled wire, flat rolled tape and uniaxially pressed tape [495].

In the case of PIT processed Bi-2223/Ag tapes, it is reported that cold uniaxial pressing is more effective in enhancing transport $J_c$ than flat rolling [502]. This can be attributed to a change in the micro-crack structure and more uniform deformation achieved by pressing rather than rolling. The uniaxial pressing of Bi-2223/Ag tape introduces micro-cracks along the longitudinal direction of the tape, while in the flat rolled tape cracks appear along the transverse direction of the tape. During rolling, pressure varies along the arc of contact between the two rolls and the tape. The stress induced by the inhomogeneous pressure in the tape promotes the alignment of cracks transverse to the length direction [495]. However, in the case of uniaxial pressing, inhomogeneous deformation of superconducting core occurs along the width of the tape, resulting in cracks along the direction of the tape length. In addition, the forces applied by uniaxial pressing are uniformly distributed perpendicular to the surface of the tape, thus resulting in higher and homogeneous compression. The higher uniform pressure reduces voids, improves texture formation, and thus further improves $J_c$. It is supposed that the longitudinal cracks have no influence on the transport superconducting currents along the tape, while the transverse cracks should be barriers of the superconducting currents. This influence of cracks on the superconducting current flow along the tape was directly evidenced by magneto-optical imaging.

It is expected that cracks are also introduced in our Ba-122 tape and that the different direction of residual cracks is another important reason for the difference of $J_c$ in the rolled and pressed tape in addition to the difference of densification of superconducting cores. In this study, an apparent difference in crack structures between the rolled and pressed tapes was also observed. Figure 112 (a) and (b) shows SEM images of the cracks in the flat rolled and pressed tape, respectively. The observation was carried out on the tape plane of the as-rolled or as-pressed tapes without subsequent heat treatment. The photographs were taken of the core surface after the Ag sheath was peeled off. As shown in Fig. 112 (a), all cracks observed in the as-rolled tape run transverse to the tape length, while they run parallel to the tape length in the as-pressed tape as shown in Fig. 112 (b). We consider that the observed difference in crack direction provides another key to elucidate the mechanism of the positive influence of applying the uniaxial pressing. It is speculated that the stress situation induced by the pressing together with the heat treatment after pressing acts so as to heal the transverse cracks produced by the previous rolling process, which run transverse to the tape length and reduce the effective cross-sectional area for superconducting current flow along the tape length.

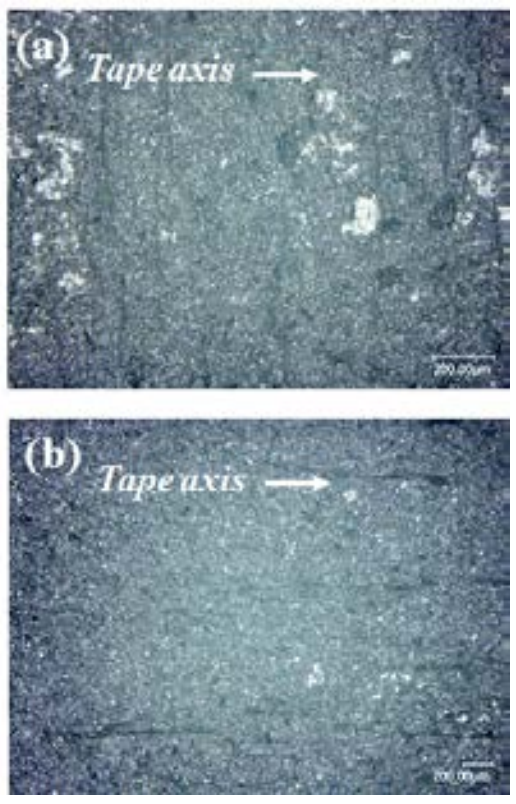

**Figure 112.** Optical microscope images of (a) flat rolled and (b) uniaxialy pressed tapes.

*4.4.3. Effects of high pressure uniaxial pressing and high degree of flat rolling*

As mentioned before, $J_c$ significantly enhanced by the uniaxial pressing under 0.4 GPa. This suggests that much higher $J_c$ values can be obtained by increasing the uniaxial pressure. Thus, we increased the uniaxial pressure up to 4 GPa [503]. Figure 113 shows $J_c$-$B$ curves of the single and seven filamentary Ba-122/Ag tapes uniaxially pressed under 2-4 GPa before final heat treatment at 850 °C for 2-4 h [503]. The $J_c$-$B$ curves of the Ba-122 tape pressed under 0.4 GPa, commercial Nb-Ti and $Nb_3Sn$ wires are also shown for comparison. Much higher $J_c$ values than the tape pressed under 0.4 GPa are obtained by increasing the uniaxial pressure. $J_c$ values of all the pressed Ba-122 tapes show very small field dependence as observed for the tapes shown in Fig. 108, and $J_c$ values well above $5.0\times10^4$ A/cm$^2$ in 10 T are obtained, indicating that a high $J_c$ values are obtained with good reproducibility. It should be noted that a high $J_c$ exceeding the practical level of $10^5$ A/cm$^2$ at 4.2 K is obtained in magnetic fields up to 6 T for mono-filamentary tape and $J_c$ maintains a high value of $8.6\times10^4$ A/cm$^2$ in 10 T. The seven-filamentary tape also sustains $J_c$ as high as $5.3\times10^4$ A/cm$^2$ at 10 T. These high $J_c$ values highlight the importance of uniaxial pressing for enhancing the $J_c$ of Ba-122 tape conductors.

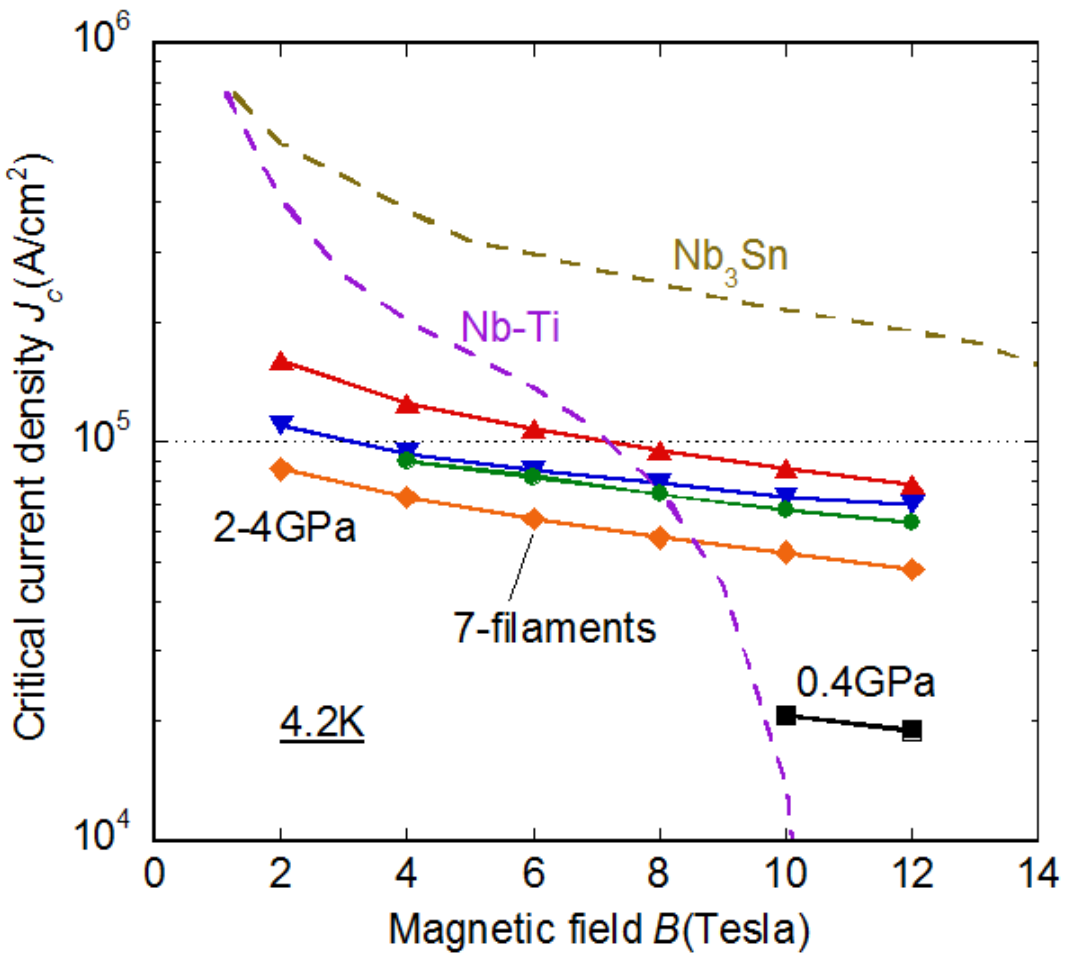

**Figure 113.** $J_c$-$B$ curves at 4.2K of uniaxially pressed Ba-122/Ag tapes. Pressure was changed from 2-4GPa. For comparison the data of Ba-122 tape in Fig. 108 and Nb-Ti and $Nb_3Sn$ commercial wires are also included in the figure [503].

The $J_c$ values of the Ba-122/Ag tapes exceed the values of the Nb-Ti conductor in magnetic fields higher than ~8 T. Furthermore, in comparison with the $Nb_3Sn$, the $J_c$ of Ba-122 tapes show very small magnetic field dependence, thus Ba-122 has great potential to surpass the $J_c$ performance of $Nb_3Sn$ in high magnetic fields in the near future. Our previous study [486, 487] indicates that the magnetic field dependence of $J_c$ at 4.2K of flat rolled Ba-122 tape is very small, comparable to that of melt-textured Bi-2212/Ag tape [504]. These results suggest that the Ba-122 superconducting wires will be competitive with well-established Nb-based superconductors and Bi-based oxide for high magnetic field applications in the near future.

The large reduction of tape thickness only by applying flat rolling is also found to be effective in enhancing $J_c$ values [503]. Figure 114 shows the $J_c$–$B$ curves of the flat rolled Ba-122 tapes. These $J_c$ values are much higher than that of the flat rolled tape shown in Fig. 108. Figure 114 clearly indicates that the $J_c$ significantly increases when the tape thickness is reduced by the flat rolling. $J_c$ reached a maximum value of $4.5\times10^4$ A/cm$^2$ at 10 T when the thickness of the tape reduced to 0.26 mm. But when the thickness is reduced smaller, degradation in $J_c$ was observed. However, further improvement in $J_c$ values could be achieved for thinner tapes when we applied uniaxial pressing instead of flat rolling as shown in Fig. 113. SEM observation of these thin tapes indicates that the transverse micro-cracks observed in Fig. 112 appear to decrease in number with decreasing the tape thickness. With decreasing the tape thickness, the stress and strain distribution in the rolling process seems to become similar to those in uniaxial pressing when the rolling diameter is constant, leading to the disappearance of transverse micro-cracks. This is one of the reasons why $J_c$ increases with decreasing the tape thickness in flat rolling process.

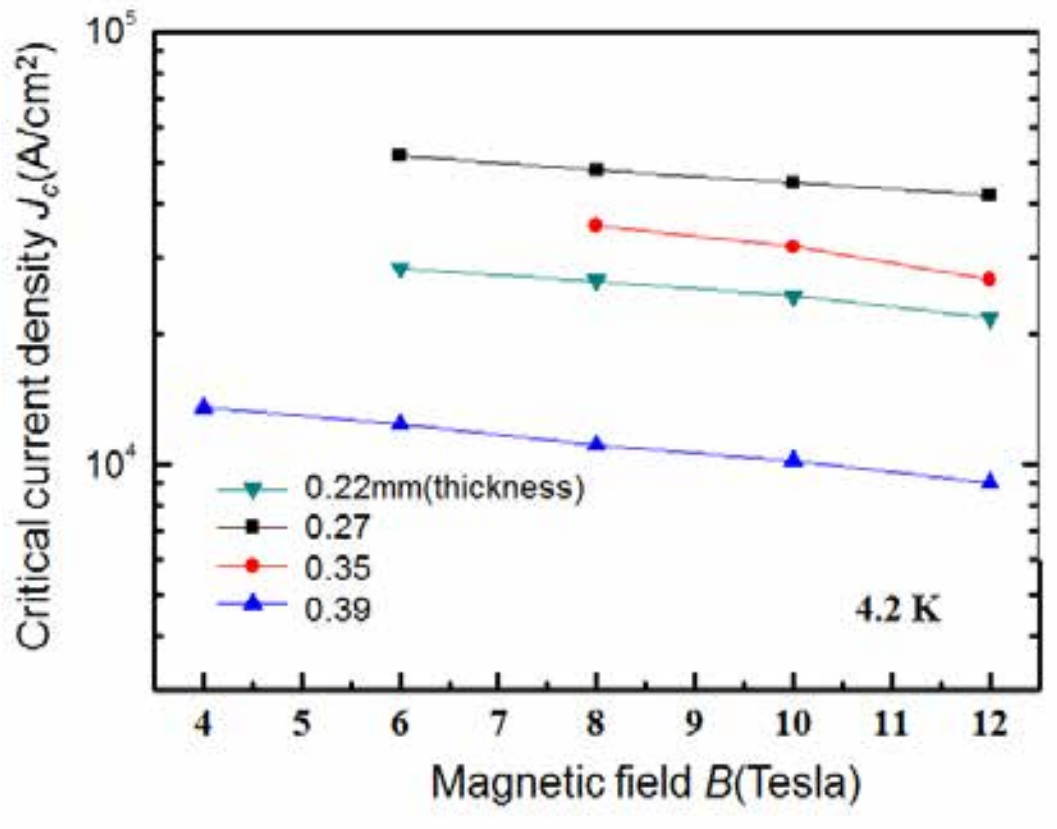

**Figure 114.** $J_c$-$B$ curves at 4.2K of flat rolled Ba-122/Ag tapes [503].

Figure 115 shows XRD patterns of the flat rolled and uniaxially pressed Ba-122/Ag tapes shown in Figs. 113 and 114 [503]. The data of randomly orientated Ba-122 precursor powder is also included in the figure for comparison. All the tapes consist of a main phase, $Ba_{1-x}K_xFe_2As_2$, however, Ag peaks from the sheath material are also detected. The relative intensities of the (*00l*) peaks with respect to that of the (*103*) peak in all tapes are strongly enhanced, when compared to the randomly oriented powder, indicating that a well-defined *c*-axis grain orientation is obtained by flat rolling and pressing. The relative intensities of the (*00l*) peak are higher than those of the flat rolled tape in Fig. 110 (a), indicating that large degree of flat rolling is effective in enhancing c-axis grain orientation. However, it remains at almost the same level when decreasing the tape thickness below 0.39 mm as shown in Fig. 115, suggesting that the grain orientation is hardly further improved by further flat rolling process. Therefore, grain orientation can be ruled out as a possible mechanism of the $J_c$ enhancement with decreasing the tape thickness in Fig. 114. In contrast, higher relative intensity of (00*l*) peaks was observed in the pressed tape. This relative intensity in Fig. 115 is much higher than that of the uniaxially pressed tape in Fig. 110 (b). Thus, the high pressure uniaxial pressing is effective in enhancing c-axis grain orientation. These results indicate that high pressure uniaxial cold pressing is more effective in improving c-axis grain orientation than the large degree of flat rolling. However, it should be noted that the degree of c-axis grain orientation in our pressed tapes is still lower than that in the Fe sheathed PIT processed Sr-122 tapes [489, 491]. This suggests that the grain orientation could be further enhanced by optimizing processing parameters of the tape conductors or by using harder sheath materials.

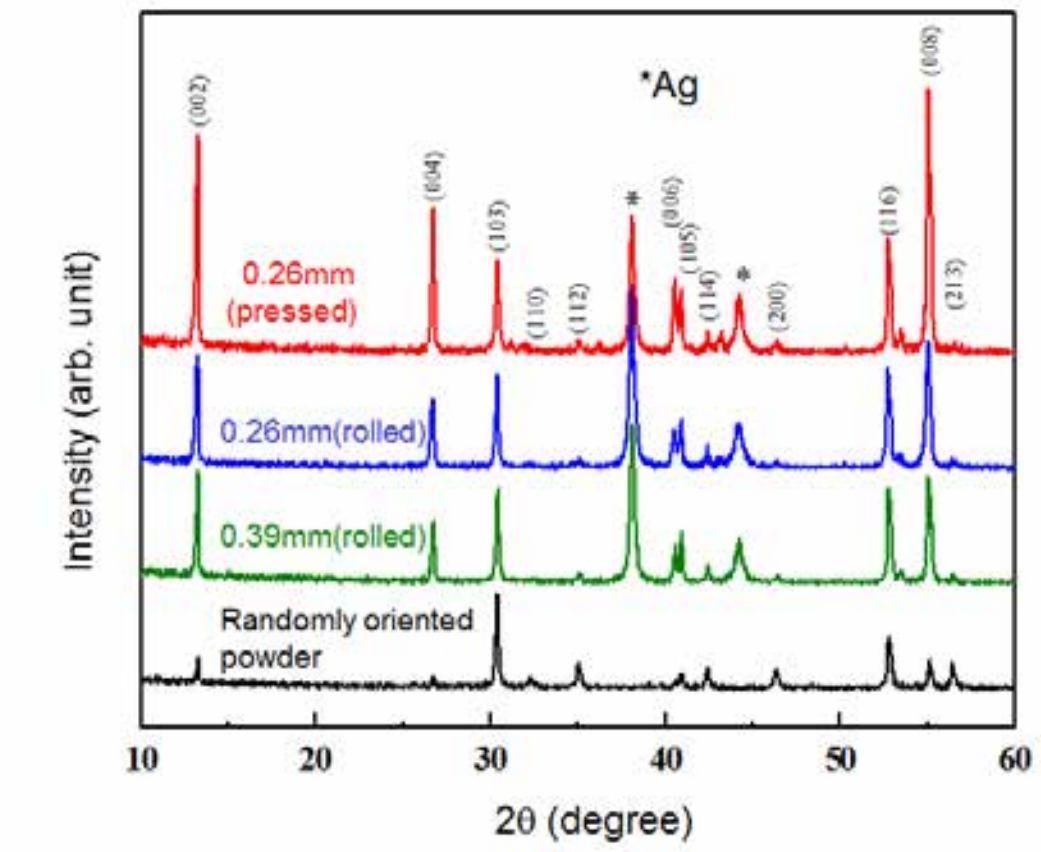

**Figure 115.** X-ray diffraction patterns of rolled and pressed Ba-122/Ag tapes. The data of Ba-122 powder with random grain orientation are also shown for comparison [503].

We evaluated the effect of flat rolling on *Hv* and $J_c$. Both the *Hv* and $J_c$ clearly increase with decreasing the tape thickness. This indicates that in addition to the disappearance of transverse cracks, the increase of Ba-122 density might be another reason for the $J_c$ enhancement by the hard flat rolling. Figure 116 shows $J_c$ (10 T, 4.2 K) as a function of *Hv* for both rolled and pressed tapes [503]. $J_c$ increases with increasing *Hv* and, hence, with increasing Ba-122 core density. There observed a strong correlation between the hardness of the tapes and $J_c$. With increasing the hardness, the $J_c$ of the Ba-122 core increased, however the hardness and $J_c$ of flat rolled tapes does not surpass the hardness and $J_c$ of the uniaxially pressed tapes respectively. This suggests that uniaxial pressing of tapes yield much better $J_c$-*H* characteristics than the flat rolling [495]. However, it seems that there is no discontinuity in $J_c$-*Hv* curve between flat rolling and uniaxial pressing. This suggests that there is no essential difference in the superconducting current limiting mechanism between flat rolling and uniaxial pressing. Thus, we can expect high $J_c$ values comparable to or higher than the pressed tapes can be obtained only by applying a high degree of flat rolling. As is well known, flat rolling is more convenient and useful than uniaxial pressing for the fabrication of practical level long tape conductors.

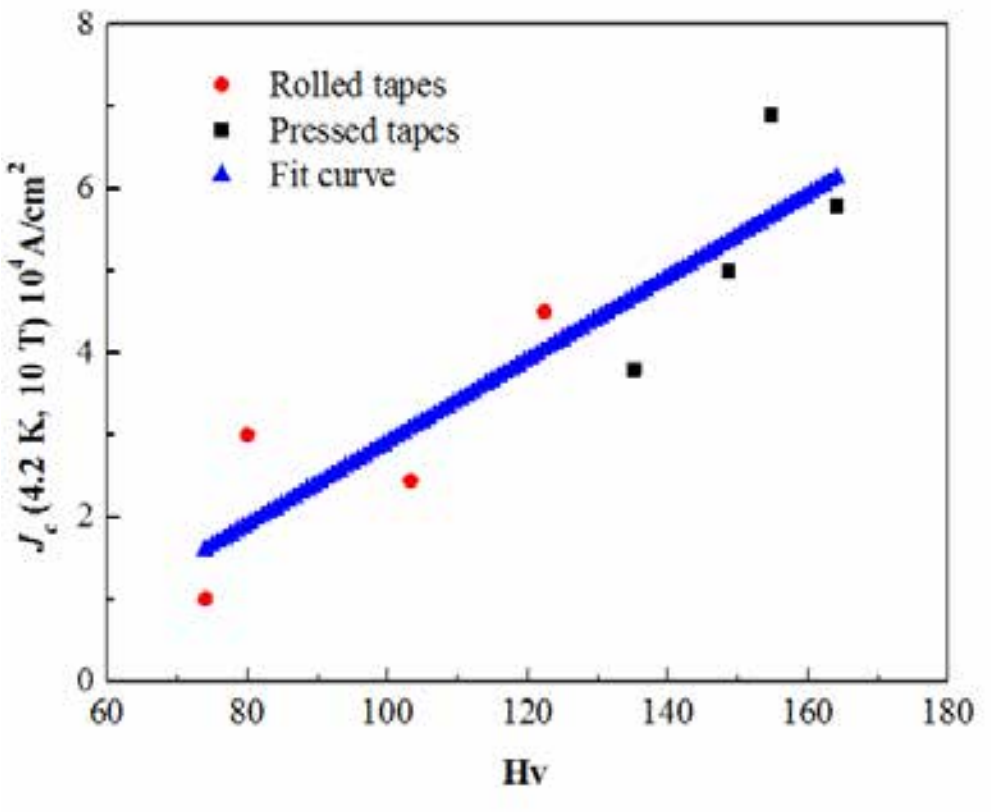

**Figure 116.** $J_c$ at 4.2K and 10Tesla as a function of Vickers hardness $Hv$ for the flat rolled and uniaxially pressed Ba-122/Ag tapes [503].

Figure 117 (a) and (b) shows typical SEM images of the polished surface for the flat rolled and uniaxially pressed Ba-122/Ag tapes. These observations were carried out on the tape plane of the tapes. Although rolling can reduce voids and improve the density of the Ba-122 core, the microstructures are still porous and quite inhomogeneous. In contrast, the pressed tapes with higher hardness and $J_c$ apparently have a denser and uniform microstructure than the rolled tapes with lower hardness and $J_c$. This result is consistent with the $Hv$ analysis in Fig. 116. Generally speaking, the pressure introduced by the uniaxial pressing is higher and more uniform than that introduced by the rolling. Thus, the microstructure in the pressed sample is denser and more uniform than that in the rolled one.

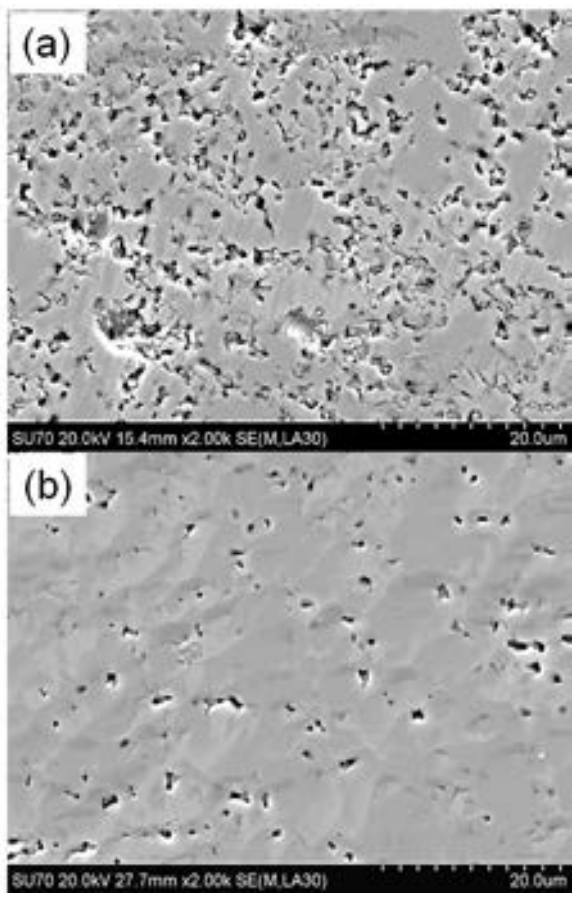

**Figure 117.** Scanning electron microscope images of (a) flat rolled and (b) uniaxially pressed Ba-122/Ag tapes.

In the previous sections, we used Ag tube as a metal sheath of Ba-122 tapes. The Ag sheath was completely annealed by the heat treatment and became very soft after the heat treatment. Thus, the Ag sheathed Ba-122 tape is not practical from the aspect of mechanical strength. In order to solve this problem we fabricated Ba-122 tape conductors applying a new sheath structure of double sheaths, stainless steel (SS) as an outer sheath and Ag as an inner sheath [505]. The inner Ag sheath was used to avoid the reaction between SS and Ba-122. Figure 118 shows $J_c$-$B$ curves at 4.2K of the double sheathed Ba-122 tapes fabricated with flat rolling and uniaxial pressing. The inset shows transverse cross sections of flat rolled (upper) and pressed (lower) tapes. For comparison data of commercial superconductors, Nb-Ti and Nb3Sn, are also shown in the figure. We found that the rolled tapes show $J_c$ values of 7.7x10$^4$A/cm$^2$ at 4.2K and 10T with high homogeneity. These $J_c$ are the highest values reported so far for IBSc tapes and wires fabricated by scalable rolling process. It should be noted that the use of hard SS as outer sheath increases not only mechanical strength of the tape but also the density of Ba-122 core. The application of uniaxial pressing to the double sheathed tape further increased $J_c$ at 4.2K and 10T up to 9.0x10$^4$A/cm$^2$. The transport $J_c$-$B$ curves for both rolled and pressed tapes show extremely small magnetic field dependence and the $J_c$ values exceed 3×10$^4$A/cm$^2$ in 28T, that are much higher than those of low-temperature commercial superconductors. The microstructure investigations indicate that such high $J_c$ values were achieved by higher density of the core, more uniform deformation of Ba-122 core and high degree of grain orientation [505]. These results indicate that the combination of the double sheath and rolling is very promising for fabricating long Ba-122 tape conductors for a high magnetic field magnet which can generates fields higher than 20T.

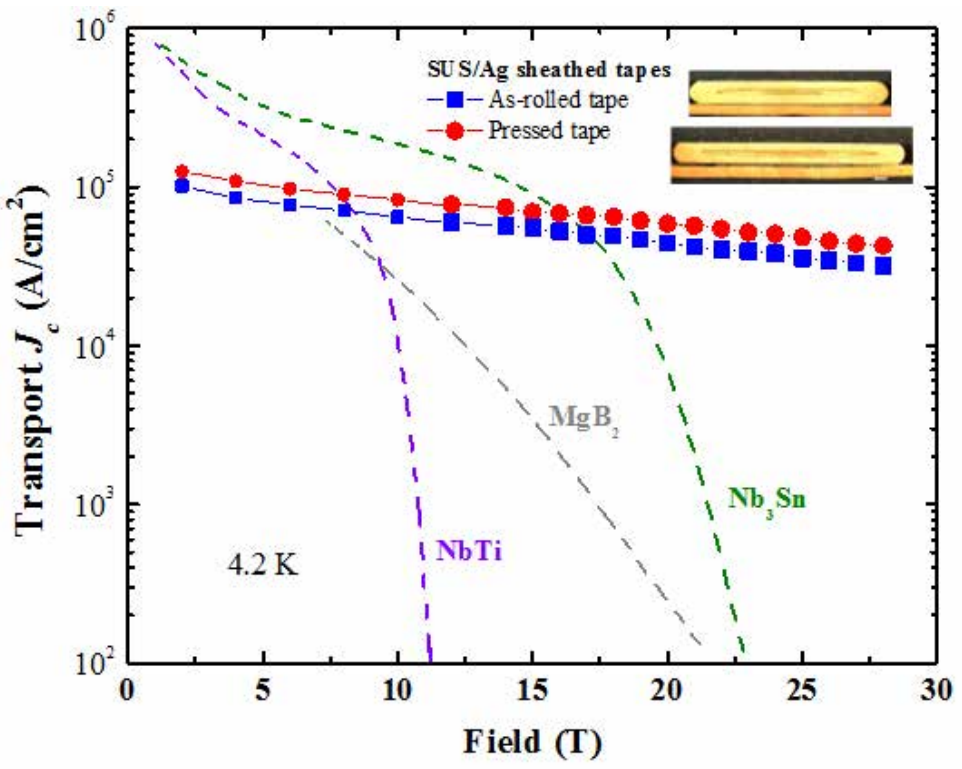

**Figure 118.** $J_c$ vs. magnetic field curves at 4.2 K of double sheathed Ba-122 tapes fabricated with rolling and pressing. For comparison $MgB_2$ wires are shown in the figure [505].

*4.4.4. Temperature dependence of critical current density*

The temperature dependence of the transport critical current was measured using a 10 T split type superconducting magnet with a variable temperature insert cryostat that can control the temperature between 10 and 40 K [497]. We also investigated the magnetic field orientation dependence of $J_c$ for Sr-122/Ag tapes to evaluate the $J_c$ anisotropy of the tape conductors.

Figure 119 shows $J_c$–$B$ curves of a uniaxally pressed Sr-122/Ag wire at four different temperatures of 10, 20, 25 and 30 K in external magnetic fields applied parallel and perpendicular to the tape surface [497]. $J_c$ was estimated by dividing the $I_c$ by the cross sectional area of the Sr-122 superconducting core of the tape conductor. The uniaxial pressure was 0.4 GPa, which brought a thickness reduction ratio of 20%. At 20 K, $J_c$ value was almost 10$^4$ A /cm$^2$ at 0 T. This $J_c$ decreases gradually with increasing magnetic field applied parallel to the tape surface. $J_c$ values in perpendicular fields are lower than those in parallel fields, however, field dependence is still small. The $J_c$–$B$ curves at temperatures below 25 K show small slopes, nearly equal to those at 10 K, suggesting that the *ex situ* Sr-122/Ag tape has a high potential in high magnetic fields at temperatures below 25 K. Furthermore, even at 30K, which is close to the $T_c$ of this superconducting material, $J_c$ was still observed in fields up to around 7 T. These results are due to the high $B_{c2}$ values of Sr-122. The $J_c$ value in a magnetic field applied perpendicular to the tape surface was about half the value in the magnetic field applied parallel to the tape surface at temperatures below 20 K. This $J_c$ anisotropy of Sr-122/Ag tape is smaller than those of Bi-2212/Ag and Bi-2223/Ag tape conductors. This result also suggests that the Sr-122/Ag tape shows lower anisotropy in $J_c$ and useful for practical magnet applications.

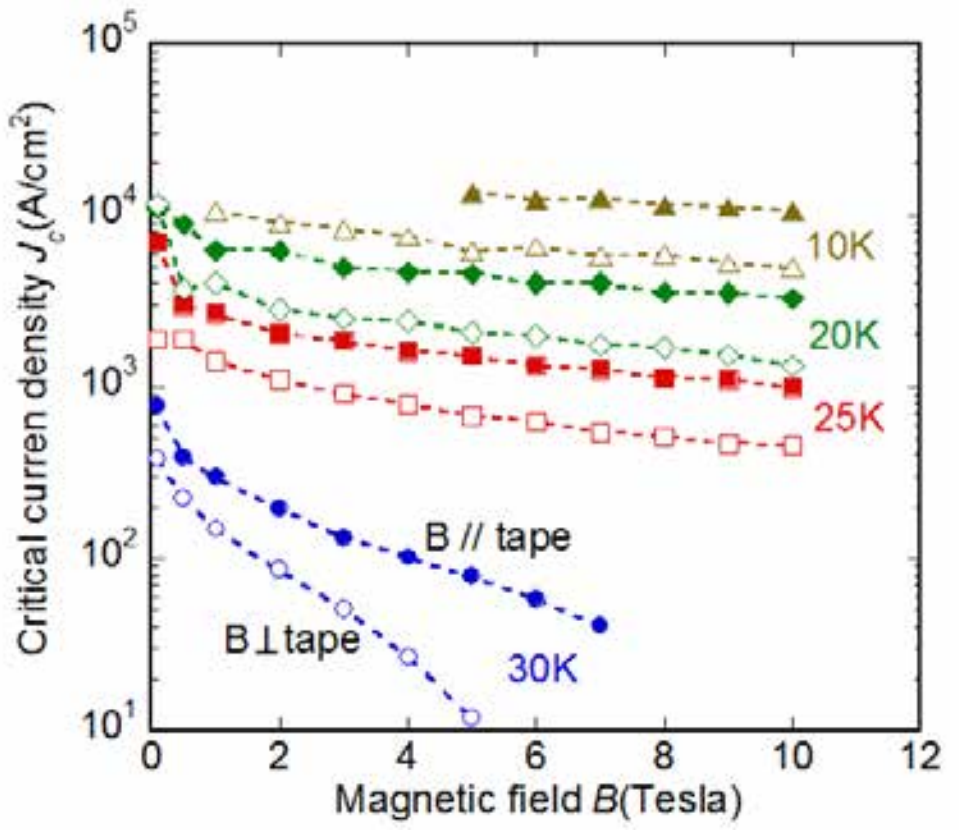

**Figure 119.** Temperature dependence of $J_c$-$B$ curves of Sr-122/Ag tape. Magnetic field was applied parallel (perpendicular to the tape axis) and perpendicular to the tape surface [497].

Figures 120 (a) and (b) show $J_c$ of a pressed Sr-122/Ag tape as a function of external magnetic field directions to the tape surface at 20 and 30 K, respectively, and in several magnetic fields [497]. Zero and 90$^o$ in the figure corresponds to the magnetic field directions perpendicular and parallel to the tape surface, respectively. Maximum and minimum $J_c$ values were obtained at around 0$^o$ and 90$^o$ in every magnetic field. This angular dependence is similar to the anisotropy of high-$T_c$ oxide superconducting tapes, Y-123(coated conductor), 2212/Ag and B-2223/Ag tapes. The ratio between maximum and minimum $J_c$ values ($J_c$ anisotropy) is almost two at every magnetic field up to 3 T. However, the $J_c$ anisotropy increased with increasing magnetic field as shown in Fig. 118.

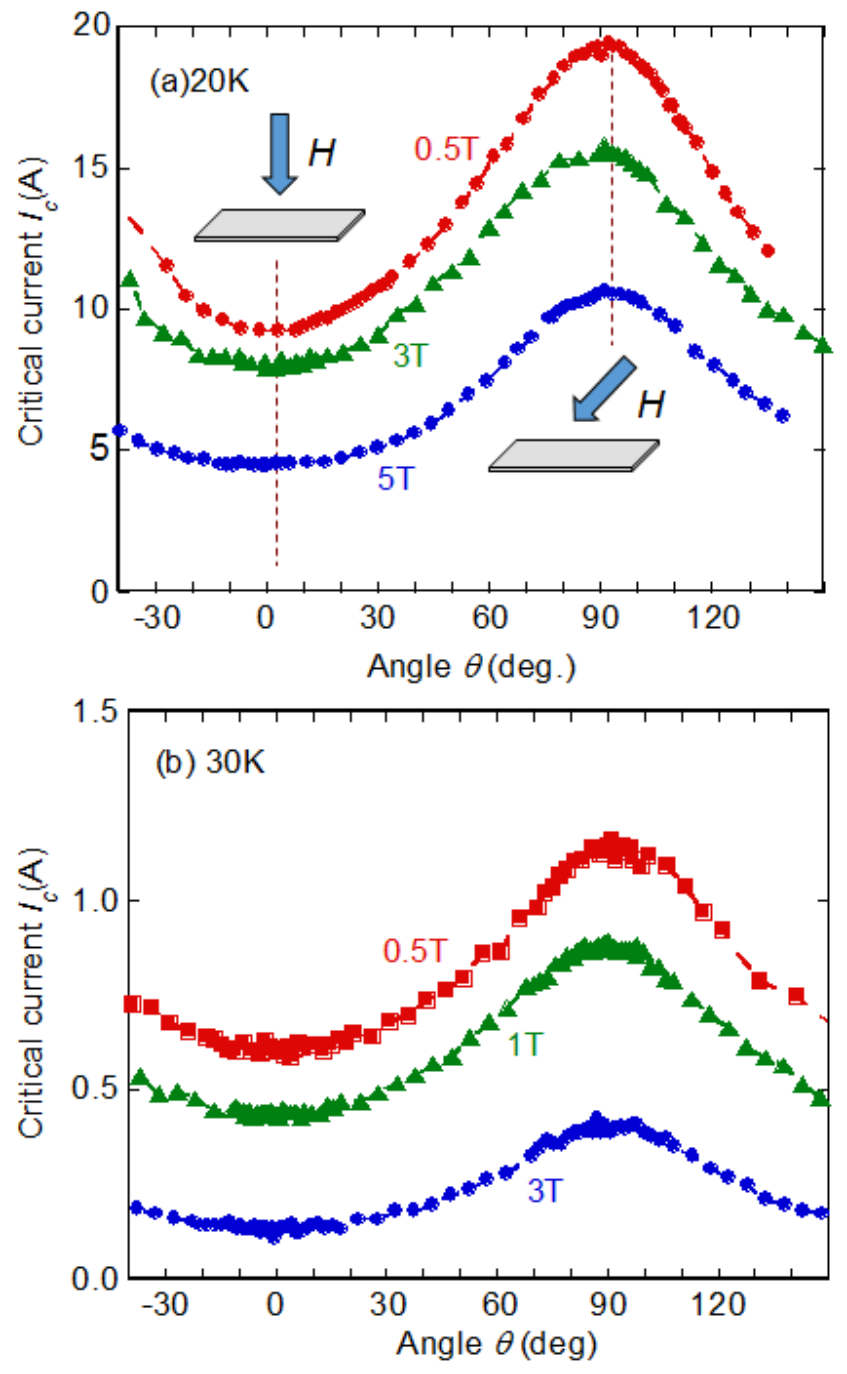

**Figure 120.** $I_c$ at 20K and 30K in several magnetic fields as a function of external magnetic field directions [497].

*4.4.5. Summary and future prospects*

Among the many fabrication stages of superconducting tapes, flat rolling has been commonly used to densify and to realize the grain orientation of superconducting core [494, 496, 505]. Our results showed a large increase in $J_c$ due to the improvement of the Ba(Sr)-122 core density and preferred orientation in the initial step by the rolling process. Upon further rolling to smaller tape thicknesses, degradation of critical current density was observed due to the introduction of transverse micro-cracks. However, larger reduction of tape thickness by the flat rolling tends to decrease the transverse micro-cracks and to enhance $J_c$ values. This suggests that higher $J_c$ can be realized by only applying flat rolling.

Furthermore, when the tape was uniaxially pressed, $J_c$ values were significantly increased by another improvement of the core density and grain orientation. Excellent transport $J_c$ values of ~$10^5$ A/cm$^2$ under magnetic fields up to 6 T were obtained in uniaxially pressed Ba-122/Ag tapes. A higher core density and more grain orientation are responsible for this high $J_c$ performance in the pressed samples. The use of hard metal as an outer sheath of double sheathed tape is also effective in increasing the core density and $J_c$ values. However, it should be emphasized that practical applications of uniaxial pressing for the manufacture of long length tapes require specialized machines for continuous pressing of the tape. Fortunately, there have been successful attempts at producing long Bi-2223/Ag wires by periodic pressing [506] and eccentric rolling [507], which might be also applied for the production of long length Ba-122 wires with high transport $J_c$.

Ba(Sr)-122/Ag tape conductors show very small field dependence of $J_c$ with small $J_c$ anisotropy. With further improvements in the critical current density and wire fabrication technology, Ba(Sr)-122 superconductors will have a very promising future for high-field applications.

## 5. New functional materials and devices found by this Project

Exploration of novel superconductors needs a non-conventional approach in various aspects such as the material system and synthetic processes as high pressure and electric field effect. As a result, it should have much higher probability than other materials research to find new functionalities or to encounter new phenomena during the research [508]. A well-known example is the finding of high performance thermoelectric material, $NaCo_2O_4$ in the course of comparative study high $T_c$ cuprates with layered cobaltites [509]. This feature comes from the fertility richness of materials we are engaging in.

In this project, this team proceeded exploration of new superconductors along with seeking new functional materials. Some of representative achievements are briefly described in this section.

### *5.1. Catalytic activity for ambient pressure $NH_3$ synthesis*

Ammonia is the most commercially produced chemical, reaching 160 million tons per year. Most ammonia is consumed as ammonium sulfate, which is used as an essential fertilizer in crop production. While industrial ammonia synthesis from $N_2$ and $H_2$ is conducted using the Haber-Bosch process using iron-based catalysts [510] at 400-600 °C & 20-40 MPa, such high reaction temperatures are disadvantageous with respect to the equilibrium and exothermic reaction (46.1 kJ mol$^{-1}$) of ammonia synthesis. The rate-determining step of ammonia synthesis is cleavage of the N≡N bond, because the bond energy is extremely large (945 kJ mol$^{-1}$). Transition metals such as Fe or Ru are indispensable for the promotion of N≡N bond cleavage, also in addition to electron donors that provide electrons to the transition metals. An $N_2$ molecule is fixed to form a bond with a transition metal by donating electrons from its bonding orbitals and accepting electrons to the antibonding π orbitals (back-donation) [511]. This back-donation is effectively enhanced by electron donors, which further weakens the N≡N bond and results in the cleavage of $N_2$. We assume electron injection is essentially indispensable to enhancing the efficiency of ammonia synthesis using Fe or Ru-catalysts. However, the situation is not so simple, i.e., it is extremely difficult to realize a low work function and chemical and thermal stability, both are generally incompatible. Although the catalytic activity of Ru is drastically enhanced by adding alkali or alkaline earth metals with small work functions [512], these metals are practically inapplicable for ammonia synthesis because these metals are so chemically active that the reaction with produced ammonia and/or $N_2$ forms metal nitrides and amides.

We examined the catalytic activity of a stable electride, C12A7:$e^-$ as an efficient promoter with high electron donating power and chemical stability for a Ru catalyst utilizing unique properties of this material, i.e., low work function (2.4eV) comparable metal potassium but chemical inertness [513]. Electride is an ionic crystal in which electrons serve as anions. $12CaO•7Al_2O_3$ (C12A7), which is a constituent of commercial alumina cement, works as complexant to electron and the resulting material became the first electride that is stable at temperatures above room temperature and an ambient atmosphere [514, 515]. As shown Fig. 121, the unit cell of C12A7 has a positively charged framework structure composed of 12 sub-nanometer-sized cages which are connected each other by sharing a mono-oxide layer to embrace $4O^{2-}$ in the 4 cages as the counter anions to compensate the electro-neutrality. Chemical reduction processes are used to inject four electrons into 4 of the 12 cages by extracting two $O^{2-}$ ions accommodated in the cavities as counter anions to compensate the positive charge on the cage wall. The resultant chemical formula is represented by $[Ca_{24}Al_{28}O_{64}]^{4+}(e-)_4$. The injected electrons occupy a unique conduction band called "the cage conduction band (CCB)", which is derived from the 3-dimensionally connected cages by sharing one oxide monolayer, and can migrate through the thin cage wall by tunneling, which leads to metallic conduction (ca. 1,500 S $cm^{-1}$ at RT). This electron-trapped cage structure of the bulk is retained up to the top surface if the sample is appropriately heated. Such electrons encapsulated in the cages of C12A7:$e^-$ can be readily replaced with hydride ion ($H^-$) by heating in $H_2$ gas. The incorporated $H^-$ ions desorb as $H_2$ molecules at ca. 400 °C, leaving electrons in the positively charged framework of C12A7: the incorporation and release of $H^-$ ions on C12A7:$e^-$ are entirely reversible. Please note that the electride formation and reversible storage ability of $H^-$ totally originate from the very unique crystal structure of C12A7 described above. No such formation was impossible for other oxides bearing $Al_2O_3$ and CaO.

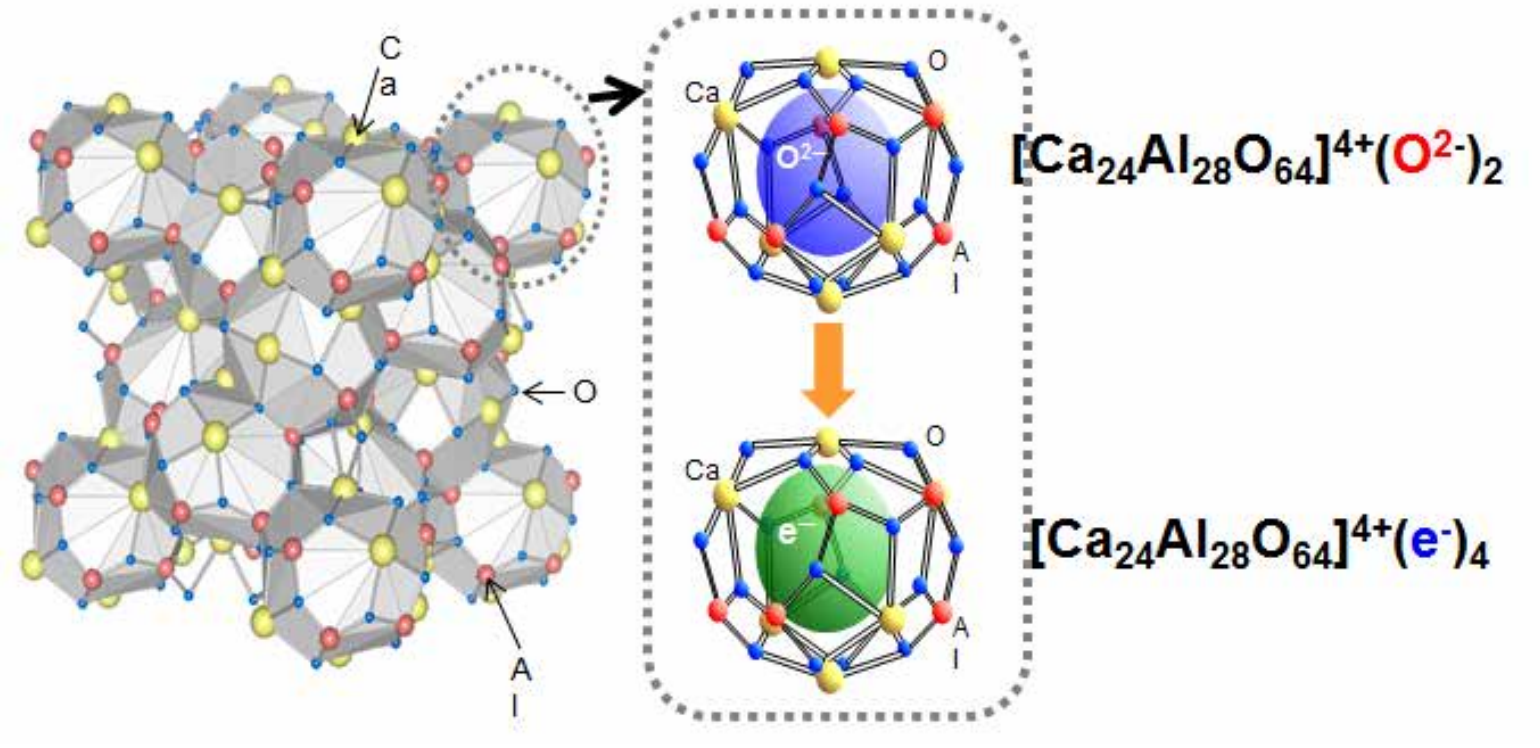

**Figure 121.** Crystal structure of $12CaO·7Al_2O_3$(C12A7). C12A7 has a cubic crystal and 2 formula units with the chemical composition of $Ca_{24}Al_{28}O_{64}$ and is a band insulator having a bad gap of ~7 eV.

Stoichiometric C12A7 (C12A7:$O^{2-}$) accommodates two $O^{2-}$ ions in two cages out of 12 cages constituting a unit cell to preserve electro-neutrality of the positively charged cage walls and its chemical formula may be expressed as $[Ca_{24}Al_{28}O_{63}]^{4+}(O^{2-})_2$. All of these in-cage oxygen ions may be replaced with electrons keeping the original case structure by an appropriate chemical treatment. The resulting material with a formula $[Ca_{24}Al_{28}O_{63}]^{4+}(4e^-)$ may be regarded as an electride. C12A7 electride, abbreviated as C12A7:e, is a metallic conductor and exhibits metal-superconductor transition at 0.2-0.4 K.

We found C12A7:e- exhibits superconductor transition at 0.2-0.4 K depending on the carrier concentration under an ambient atmosphere [516]. Although $T_c$ is very low, this is a first s-metal superconductor under an ambient pressure. Alkali and alkaline-earth metals do not show superconductivity under an ambient pressure but some of high pressure phases exhibit $T_c$. We found the crystal structure of C12A7 is similar to that of superconducting phases and considered that admixture of Ca d-orbitals to the s-orbital of the cage electrons is the origin of the superconductivity of C12A7:$e^-$ [517].

We deposited nano-sized Ru particles to C12A7:$e^-$ powders to capture $N_2$ on their surfaces. Figure 122 shows comparison in catalytic activity for $NH_3$ synthesis. The activation energy over Ru-loaded C12A7:$e^-$ is reduced to almost half of that over other Ru-catalysts and the turn-over-frequency (TOF, the measure of catalytic activity per active site) is larger by an order of magnitude than that of the latter including the best Ru-based catalyst (Ru/Cs-loaded MgO) [108] The excellent activity of C12A7:$e^-$ demonstrates the electron-donating effect plays an essential role in $NH_3$ synthesis. Very recently, it has been clarified by examining kinetics of isotope exchange reaction of $N_2$ with DFT calculations that the rate determining step of $NH_3$ synthesis over Ru-loaded C12A7:$e^-$ shifts from dissociation of N≡N bond to N-H bond formation process [518].

Another surprising finding is that Ru-loaded C12A7:$e^-$ does not exhibit hydrogen-poisoning [108] as shown in Fig. 123. Obstacle for industrial ammonia synthesis with Ru-loaded catalyst is hydrogen-poisoning in high $H_2$ pressure. Because $NH_3$ synthesis over Ru catalysts in general degrades by hydrogen ad atoms formed on Ru surfaces, the reaction order for $H_2$ on Ru catalysts often approaches -1 (i.e., the reaction rate *decreases* with the partial pressure of $H_2$). Such $H_2$ poisoning on Ru catalysts is a serious obstacle for industrial ammonia production which requires high pressure conditions to collect resulting $NH_3$ in the form of liquid ($NH_3$ easily becomes liquid above 0.85MPa at RT). The chemical industry is therefore currently searching for a supported Ru catalyst that promotes $N_2$ dissociation but suppresses $H_2$ poisoning. Figure 124 shows a tentative mechanism for ammonia synthesis over Ru-loaded C12A7:$e^-$. The robustness of Ru-loaded C12A7:$e^-$ to $H_2$-poisoning comes from its ability of reversible H-storage and release. C12A7:$e^-$ reacts with $H_2$ to form C12A7:$H^-$ in which H is incorporated into the cage as $H^-$. Although this reaction over Ru-free C12A7:$e^-$ is irreversible in temperature range below ~450℃、the H-release temperate is drastically reduced to ~300 ºC [519]. Desorption of $H^-$ to react with activated nitrogen species on Ru surfaces to form N-H bonds, leaving an electron in the cage. We consider that hydrogen-poisoning on Ru-surfaces is suppressed by preferential entrapping of H ad atom in the cage to form $H^-$ and the incorporated $H^-$ in the cage is activated by strong driving force for stable N-H bond formation.

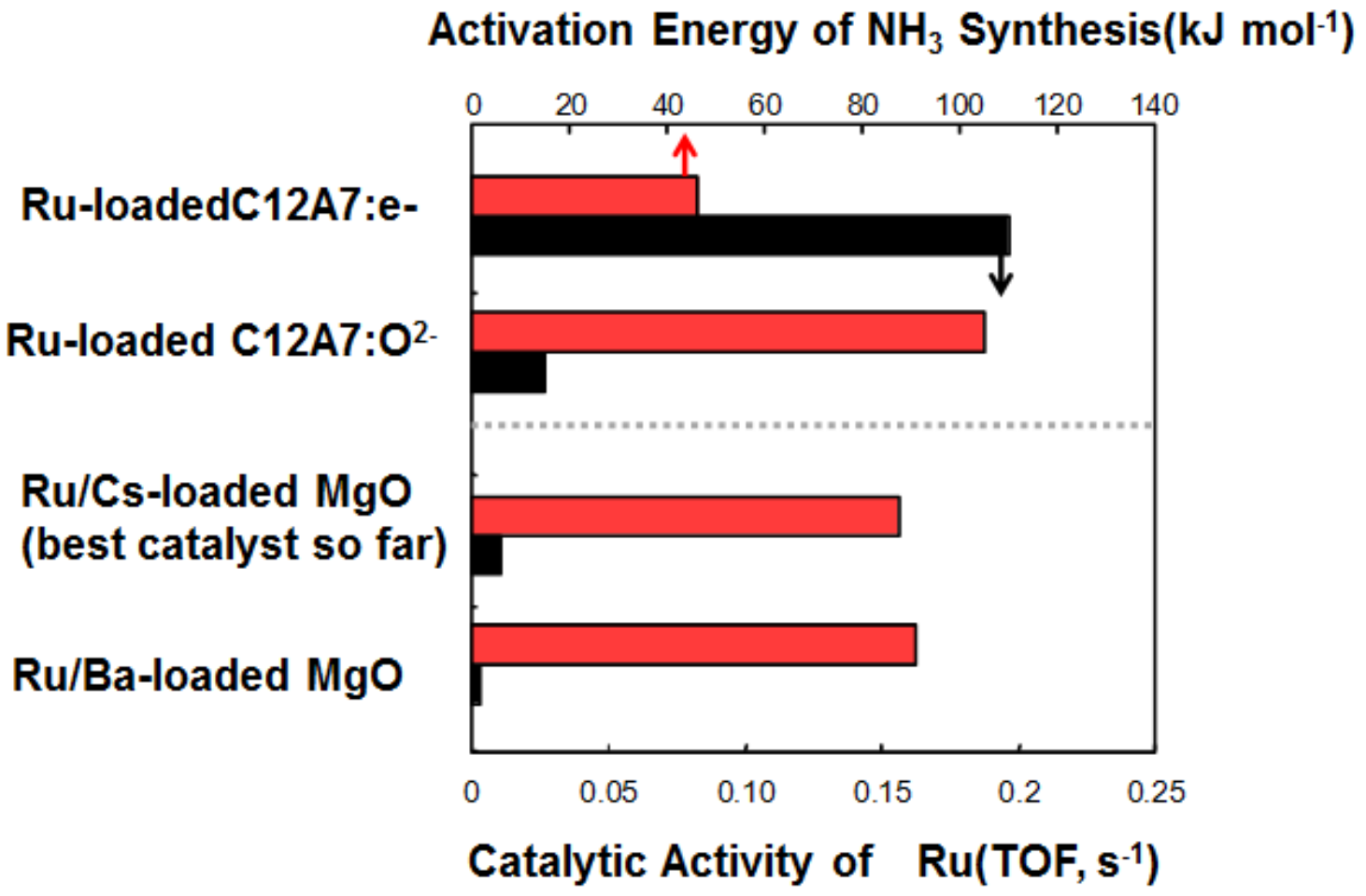

**Figure 122.** Comparison in performance between various Ru-loaded catalysts for ammonia synthesis. The reaction conditions; catalyst 0.3 g, flow rate of $H_2/N_2$ (3:1) gas 60 ml/min, temperature 400 °C, and pressure 0.1 MPa [108].

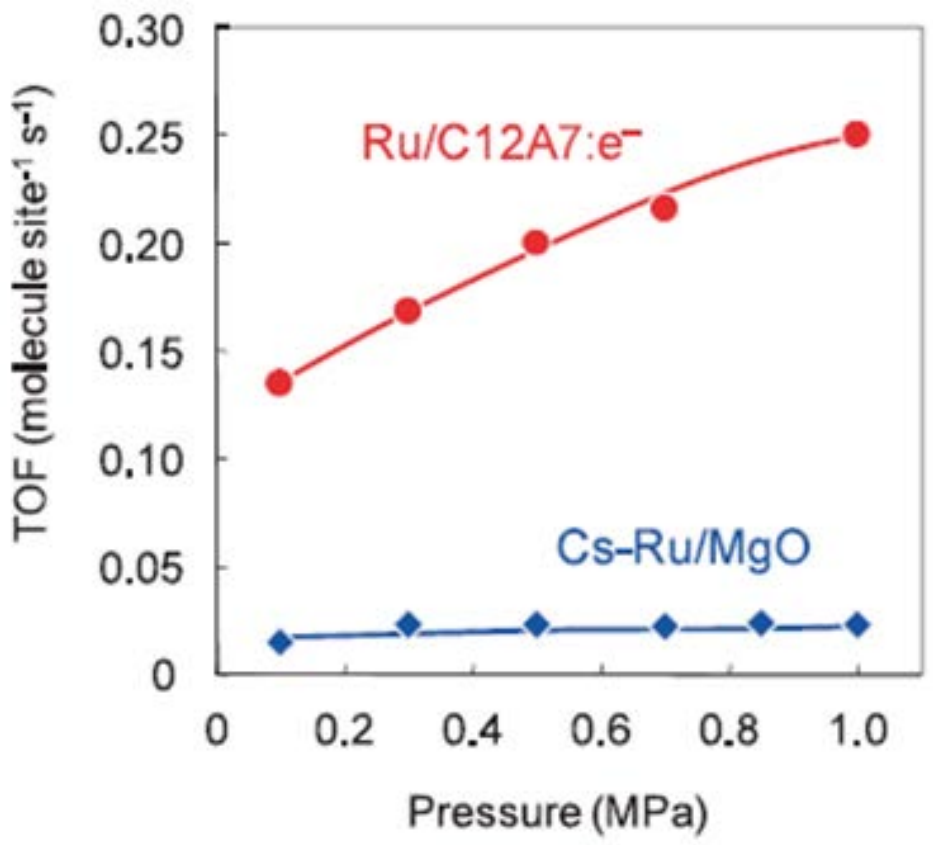

**Figure 123.** Catalytic activity of Ru-loaded C12A7:$e^-$ as a function of partial hydrogen pressure. Data of a representative Ru-catalyst are shown for comparison. Note that the activity is proportional to partial $H_2$ pressure. No serious $H_2$-poisoning which is a common drawback of Ru-catalysts [108].

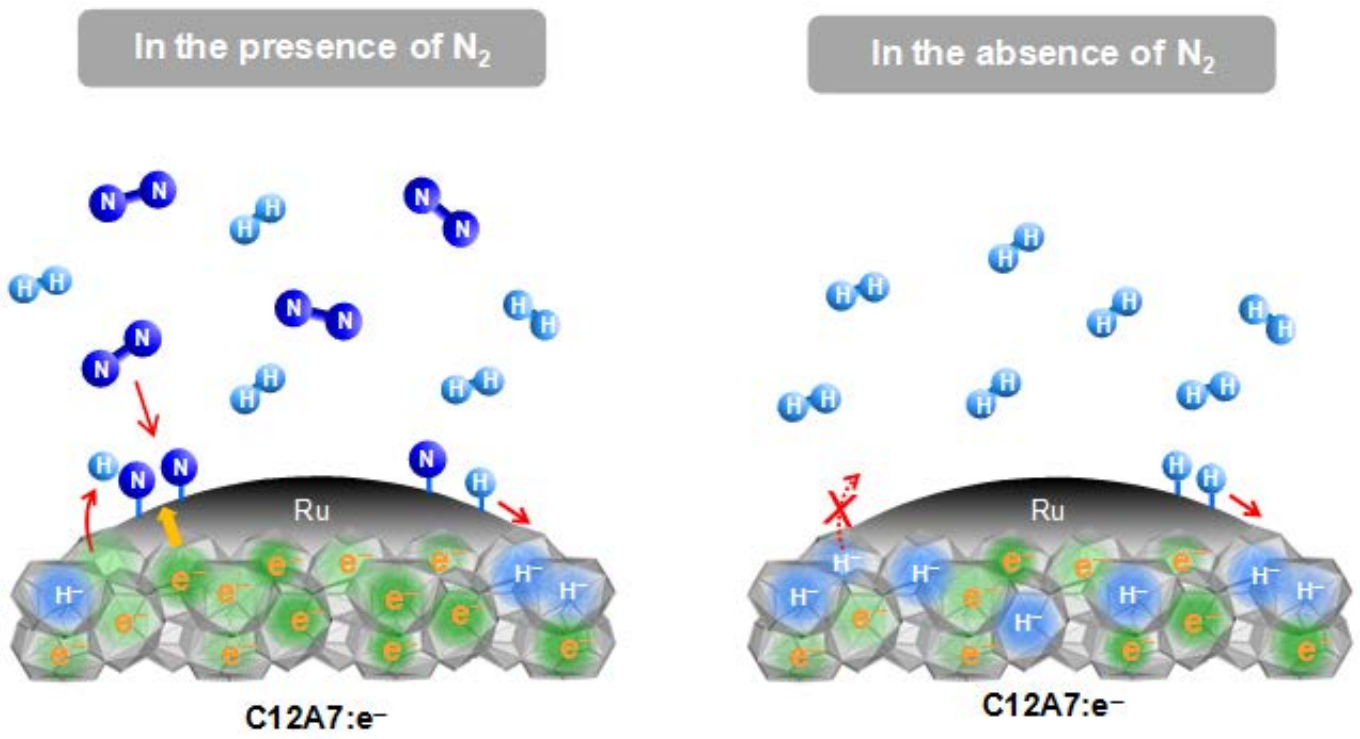

**Figure 124.** Tentative reaction mechanism to explain suppression of $H_2$-poisoning in Ru-loaded C12A7:$e^-$ [108].

### *5.2. Bipolar oxide thin film transistor*

Thin film transistor (TFT) is a fundamental building block of modern integrated circuits. Oxide semiconductor based TFT is emerging technology to take over the current Si TFT technology because they have superior properties such as larger electron mobility, lower defect density, low temperature fabrication and high optical transparency [520]. In particular, progress in TFTs using transparent amorphous oxide semiconductors-such as InGaZnOx (IGZO) [521, 522] is so remarkable that the application of IGZO-TFTs has started to drive energy-saving and high resolution displays for mobile phones, table PCs, PC monitors and large-sized OLED-TVs. A next challenge in oxide TFTs is application to logic circuits. So far, Oxide TFTs work only as unipolar devices mostly n-type, or recently realized p-type [523], but do not exhibit inversion/ambipolar operation; i.e., complementary circuits cannot be made from TFTs of the same oxide material. Therefore, the next challenge is to realize oxide-based ambipolar TFTs and complementary circuits using only a single oxide semiconductor channel.

In this project, we succeeded in fabricating an ambipolar oxide TFT using an SnO channel, and demonstrated operation of a complementary-like inverter [111]. It is of quite interest to note that SnO has the same crystal structure as FeSe, which is a parent compound of IBSc, and exhibits insulator-metal-superconductor transition under high pressure [524]. This is the first success in oxide TFTs. We achieved clear ambipolar operation with saturation mobilities of ~0.81 for the p-channel and ~$10^{-3}$ $cm^2$ $(Vs)^{-1}$ for n-channel modes in the SnO TFTs. The maximum voltage gain of ~2.5 was obtained in the complementary-like inverter circuit as shown in Fig. 125. This is the first demonstration of a complementary-like circuit using a single oxide semiconductor channel and would provide an important step toward practical oxide electronics. Further, the low temperature process (250 ºC at maximum) is compatible with emerging flexible electronics technology.

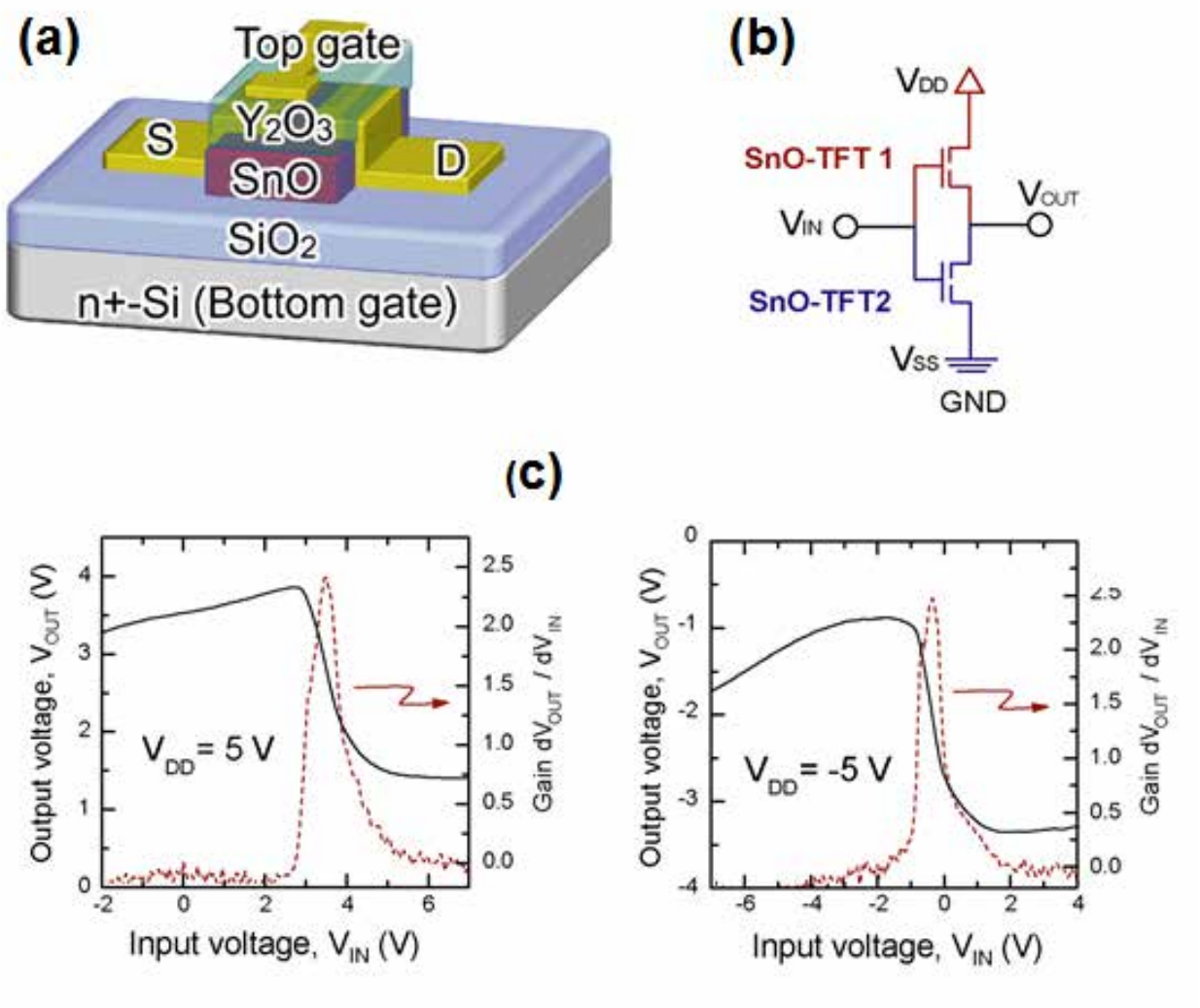

**Figure 125.** Ambipolar oxide TFTs using SnO. (a) TFT structure, (b) equivalent circuit of SnO C-MOS inverter, and (c) their characteristics [111].

*5.3. Metallic ferroelectritic materials: Slater insulators*

In 1995, Anderson and Blount [525] predicted the possibility of a ferroelectric metal, in which a ferroelectric-like structural transition occurs in the metallic state. They were motivated by the BCS theory and considered a possibility of such a material which would improve $T_c$ utilizing strong electron-phonon coupling. It is almost a consensus that metals do not exhibit ferroelectricity because static internal electric fields are screened by conduction electrons. In fact, no clear example of such a material had been reported up to 2013. In this project, we could identify this type of materials. One is $LiOsO_3$ [106] and another is $LaFeAsO_{0.5}H_{0.5}$ [211]. Since the latter was described in chapter 3.1, we focus on the former hereafter.

We found [106] that the high-pressure-synthesized material $LiOsO_3$ exhibits a structural transition at a temperature ($T_s$) of 140 K as shown in Fig. 126. A complementary usage of XRD and neutron diffraction revealed that structural transition at 140K is due to centrosymmetric ($R\bar{3}c$) to non-centrosymmetric (R3c) phase in metallic $LiOsO_3$. This transition is structurally equivalent to the ferroelectric transition of $LiNbO_3$ [526], involving a continuous shift in the mean position of $Li^+$ ions on cooling below 140 K. Its discovery realizes the scenario suggested by the Anderson and Blount, i.e., the existence of ferroelectric-like soft phonons could stabilize non-centrosymmetric superconductivity at enhanced temperatures.

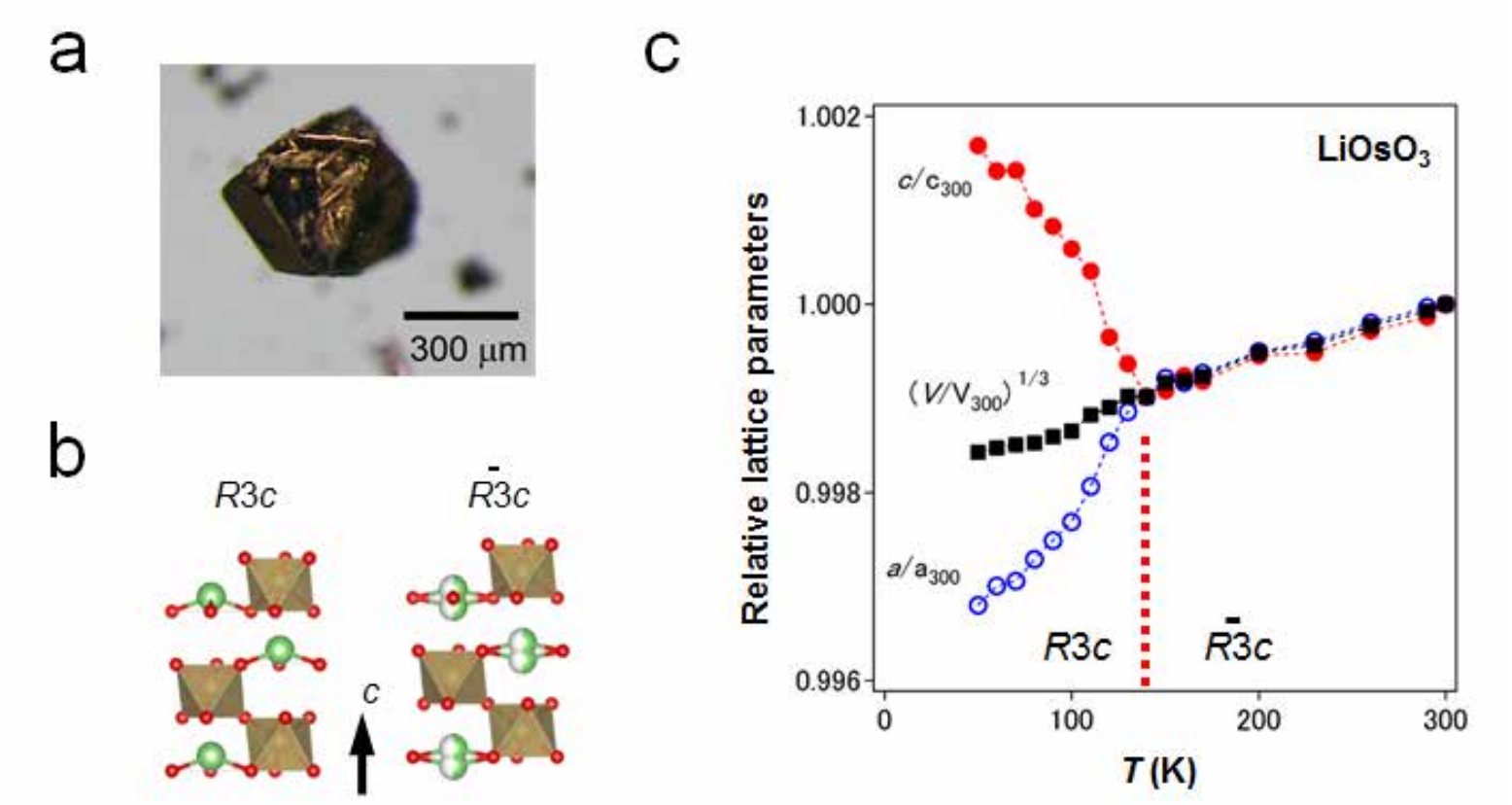

**Figure 126.** Slater insulator $LiOsO_3$. (a) photo of a flex-grown single crystal, (b) schematic views of centrosymmetric [right] and non-centrosymmetric [left] $LiNbO_3$-type structures, (c) thermal evolution of the rhombohedral lattice parameters of $LiOsO_3$ measured by neutron diffraction. The parameters are normalized by the 300 K values of $a$ = 5.0699(1) Å and $c$ = 13.2237(2) Å [106].

## 6. Perspective on superconducting tapes and wires

Among recently discovered new superconductors, the IBSCs exhibit the properties most attractive for tape or wire application. Although their $T_c$ of 30-56 K is below liquid nitrogen temperature, their upper critical fields ($H_{c2}$) of 60-100 T are comparable to those for cuprate superconductors with $T_c$ values over 90 K [319]. In particular, the 122 compounds such as Ba122 doped with K, P, or Co have very low anisotropy comparable to or lower than that of $MgB_2$ [319, 395, 407], implying that high irreversibility fields close to $H_{c2}$ or high in-field $J_c$ are expected. Moreover, the 122 compounds were found to have the critical angle for the transition from strong-link to weak-link behavior approximately twice of the value for REBCO, suggesting less sensitivity to the grain misorientation at GBs [434].

Actually the in-field $J_c$ of Ba(Sr)122:K wires having superconducting cores without biaxial textures which are fabricated by an *ex situ* powder-in-tube (PIT) method has recently been improved very much, as shown in Fig. 127. $J_c$ at 4.2 K and 10 T for Ba(Sr)122:K wires has reached $10^4$ A/cm$^2$ through metal addition plus rolling induced texture process such as hot isostatic or cold press method [490, 491, 493, 495]. Such mechanical deformation was found to be effective for densifying the conductor core and aligning the grains of the superconducting phase, resulting in high quality superconducting wires. By applying uniaxial pressure of 2-4 GPa, $J_c$ has been further improved and approached $10^5$ A/cm$^2$, which is a practical level as a superconducting wire [503]. In section 4.6, We show the magnetic field dependence of $J_c$ at 4.2 K for Ba122:K wires with a Ag sheath (Fig. 113) which were fabricated by NIMS group using a uniaxial pressure process and a more practical flat rolling process. The wires fabricated by employing the former process exhibit a $J_c$ at 10 T close to $10^5$ A/cm$^2$ for a single core wire and a high $J_c$ of 5 x $10^4$ A/cm$^2$ even for a 7-filaments wire. Though $J_c$ of the wires through the latter process is still lower, 4 x $10^5$ A/cm$^2$ at 10 T, these $J_c$ values are well over that of the Nb-Ti practical conductor. Moreover, in comparison with the $Nb_3Sn$ conductor, the $J_c$ of Ba122:K tapes through both the processes show very small magnetic field dependence, indicating that Ba122 has great potential to surpass the $J_c$ performance of $Nb_3Sn$ in high magnetic fields.

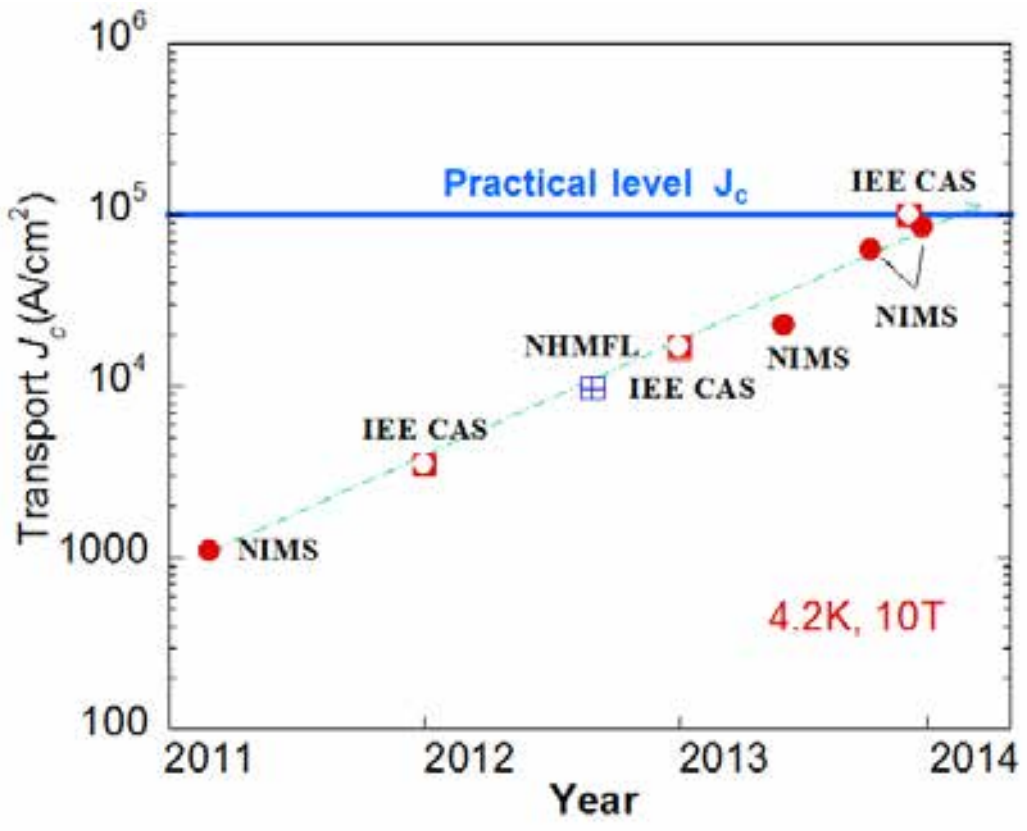

**Figure 127.** Recent evolution of transport $J_c$ at 4.2 K, 10 T for Ba(Sr)122:K PIT wires or tapes reported by groups of NIMS, IEE CAS and NHMFL.

The superior $J_c$ properties of Ba122 at higher fields have been demonstrated more clearly in thin film works. In Fig. 128, $J_c$ - $H$ properties at 4.2 K for various Ba122 films on single-crystal and IBAD-MgO-buffered metal substrates are compared with those for Nb-Ti and $Nb_3Sn$ conductors. $Nb_3Sn$ exhibits a steep decrease of $J_c$ at fields near 20 T, which is close to its $H_{c2}$. The Ba122:P film on MgO fabricated by an MBE method shows $J_c$ (H//c) over $10^5$ and $10^4$ A/cm$^2$ at 20 and 35 T, respectively [527]. The Ba122:P films with dense *c*-axis-correlated pinning centers (line dislocations in the the mother phase) [422] or $BaZrO_3$ nanoparticles [415] by a PLD method exhibit even higher $J_c$ values at fields below 9 T and a rather slow decay, suggesting that their $J_c$ values at higher fields would be higher than that for the MBE film. Another significance of these results is that the in-field performance of IBScs, in particular Ba122, can be remarkably improved by introduced nanometer-size vortex pinning centers, as already demonstrated in REBCO. This would be true not only for films or coated conductors but also for PIT wires. Although the in-field $J_c$ performance of Ba122 coated conductors, in particular longer-length ones, is still substantially lower than that for high-quality films on single crystal substrates, this difference could be mainly attributed to technical problems and should be overcome in the near future. In fact, for the case of 11 compound Fe(Se,Te) of which films can be grown at lower substrate temperatures, coated conductors, not long-length, with $J_c$ over $10^5$ A/cm$^2$ at 30 T have been demonstrated [528].

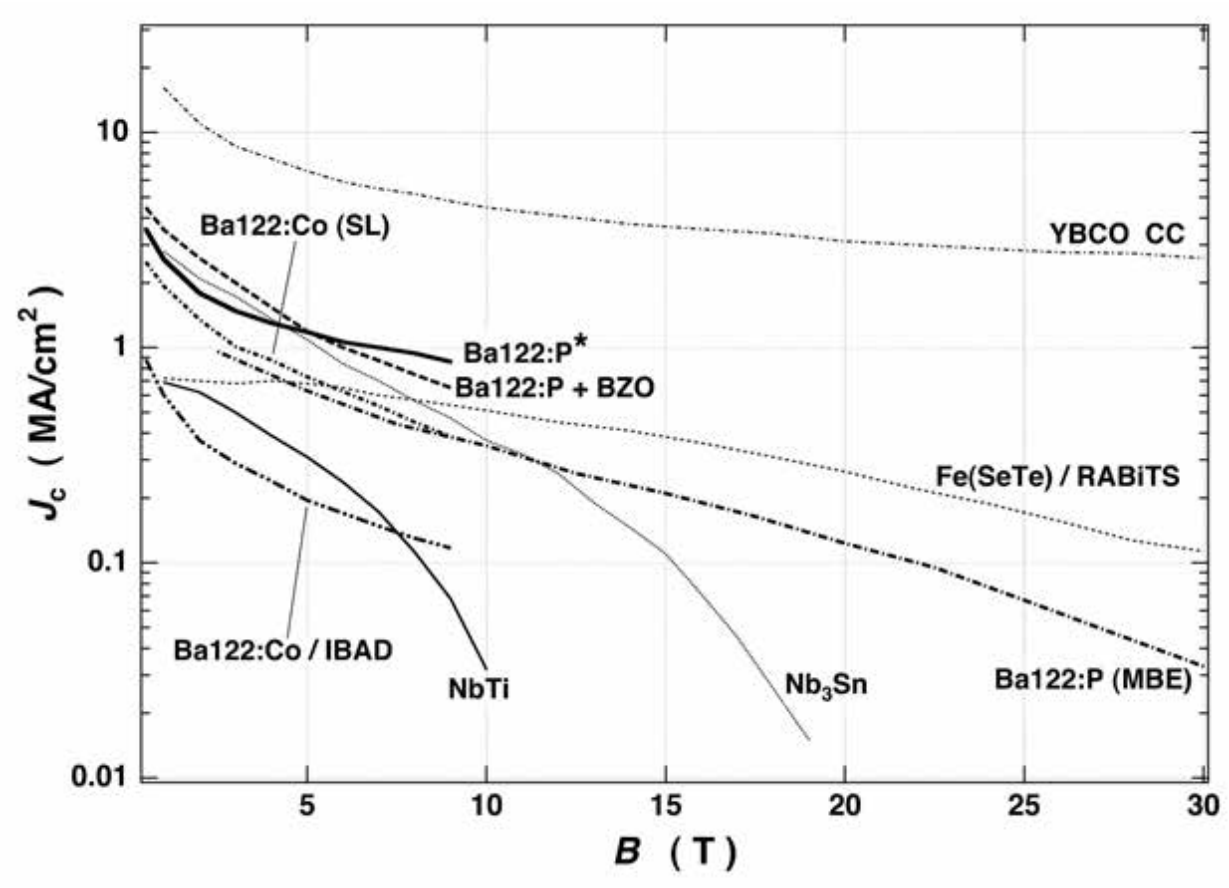

**Figure 128.** $J_c$-$B(//c)$ curves at 4.2K reported for Ba122 and Fe(Se,Te) films on single-crystal substrates and some technical substrates such as IBAD and RABiTS. The data of Nb-Ti and $Nb_3Sn$ commercial wires and YBCO coated conductors are also shown for comparison.
Ba122:P*; TIT [422], Ba122:P + BZO; ISTEC [415], Ba122:Co(SL); Wisconsin Univ. and NHMFL [417], Ba122:P(MBE); Nagoya Univ. and IFW Dresden [527], Ba122:Co/IBAD; IFW Dresden [425], Fe(Se,Te)/RABiTS; BNL and NHMFL [528].

The superior $J_c$ performance of IBScs such as Ba122 mentioned above indicates that their most promising application would be wires or tapes for magnets used at low temperatures and high fields well over 20 T. The competitive candidate for such application is REBCO coated conductor which exhibits even higher in-field $J_c$ at low temperatures. However, one of the current critical issues of REBCO coated conductors for this application is suppression of shielding current which is required to generate stable and highly-precise fields [529]. In order to overcome this difficulty, the fabrication technique for narrow multi-filamentary structures is being developed [530]. Another problem is the mechanical strength against delamination which originates from its multilayer structure [531]. If the $J_c$ performance of iron-based multifilamentary PIT wires is further improved by refining the fabrication process and reaches a practical level even at fields of 20-30 T, high $J_c$ in higher magnetic fields of IBScs could be a concrete advantage over REBCO coated conductors, though there are some more issues such as the sheath material with higher mechanical strength and lower cost. Concerning the iron-based coated conductors, more technical challenges seem to still remain to demonstrate an advantage over practical conductors, especially REBCO coated conductors. These include high in-field $J_c$ on less-textured buffer layers, narrow and thicker conductors with a lower aspect ratio, higher mechanical strength, and so on. Although coated conductors have been fabricated using the same oxide buffer-layer technology as for REBCO, special buffer-layer materials more compatible with IBScs would be necessary to demonstrate their own potential.

Another challenge common to PIT wires and coated conductors is development of IBScs with a less amount of toxic elements, which would further stimulate motivation toward their application. Considering the history that it needed more than 20 years to develop commercial REBCO coated conductors, further continuing research and development is definitely required to realize practical wires or tapes based on IBScs discovered only seven years ago. Fortunately, material's variety of IBScs is the largest among all the superconductor families and discovery of new types of superconducting materials have been continuing to date. The intrinsic nature of this materials system [508, 532] is to have Fermi surface composed of multi-pockets arising from multi-orbitals of Fe 3d. This nature provides a wide opportunity in which various degrees of freedom can contribute to emergence of superconductivity. Recent research has clarified the contribution of orbital [533], charge [534] and phonons [535] in addition to spin [159, 160] which was initially proposed. The current status may be regarded as “actors are ready!”. These actors may work cooperatively to raise $T_c$, which has been experimentally suggested recently. We may expect discovery of new IBScs which exhibit higher $J_c$ keeping less anisotropy from the background. *Iron is still hot!*

**Acknowledgements**

This study was executed under JSPS FIRST (Funding Program for World-Leading Innovative R&D on Science and Technology) program initiated by CSTP. The results described herein were obtained by many collaborators and graduate students participated. One of the authors (HH) appreciates staffs of administrative of the project at TIT and a funding from MEXT Element Strategy Initiative to form a core research center for harvesting achievements.

**References**

[1] Rogalla H and Kes P H (*Ed.*) 2012 *100 Years of Superconductivity* (London: CRC Press)

[2] A calculation method of Tc using DFT was developed by Gross *et al.*

Lüders M, Marques M A L, Lathiotakis N N, Floris A, Profeta G, Fast L, Continenza A, Massidda S and Gross E K U 2005 *Phys. Rev. B* **72** 024545

Marques M A L, Lüders M, Lathiotakis N N, Profeta G, Floris A, Fast L, Continenza A, Gross E K and Massidda S 2005 *Phys. Rev. B* **72** 024546

[3] Kamihara Y, Hiramatsu H, Hirano H, Kawamura R, Yanagi H, Kamiya T, and Hosono H 2006 *J. Am. Chem. Soc.* **128** 10012

[4] Kamihara Y, Watanabe T, Hirano M, and Hosono H 2008 *J. Am. Chem. Soc.* **130** 3296

[5] Watanabe T, Yanagi H, Kamiya T, Kamihara Y, Hiramatsu H, Hirano M and Hosono H 2007 *Inorg Chem.* **46** 7719

[6] Hosono H 2007 *Thin Solid Films* **515** 5745

[7] Hosono H 2009 *Physica C* **469** 314

[8] Hosono H 2013 *Jpn. J. Appl. Phys.***52** 090001

[9] Hanna H, Muraba Y, Matsuishi S, Igawa N, Kodama K, Shamoto S and Hosono H 2011 *Phys. Rev. B.* **84** 024521

[10] Matsuishi S, Hanna T, Muraba Y, Kim S W, Kim J E, Takata M, Shamoto S, Smith R I and Hosono H 2012 *Phys. Rev. B* **85** 014514

[11] Iimura S, Matsuishi S, Sato H, Hanna T, Muraba Y, Kim S W, Kim J E, Takata M and Hosono H 2012 *Nat. Commun.* **3** 943

[12] Matsuishi S, Maruyama T, Iimura S and Hosono H 2014 *Phys. Rev. B* **89** 094510

[13] Ishida J, Iimura S, Matsuishi S and Hosono H 2014 *J.Phys.: Condens Matter* **26** 435702

[14] Guo Y F, Wang X, Li J, Sun Y, Tsujimoto Y, Belik A A, Matsushita Y, Yamaura K and Takayama-Muromachi E 2012 *Phys. Rev. B* **86** 054523

[15] Guo Y F *et al* 2010 *Phys. Rev. B* **82** 054506

[16] Muraba Y, Matsuishi S and Hosono H 2014 *Phys. Rev. B* **89** 094501

[17] Muraba Y, Matsuishi S and Hosono H 2014 *J. Phys. Soc. Jpn.* **83** 033705

[18] Hanna T, Muraba Y, Matsuishi S and Hosono H 2013 *Appl. Phys. Lett.* **103** 142601

[19] Muraba Y, Matsuishi S, Kim S.-W, Atou T, Fukunaga O and Hosono H 2010 *Phys. Rev. B* **82** 180512(R)

[20] Katase T, Iimura S, Hiramatsu H, Kamiya T and Hosono H 2012 *Phys. Rev. B* **85** 140516(R)

[21] Katase T, Hiramatsu H, Kamiya T and Hosono H 2013 *New Journal of Physics* **15** 073019

[22] Hiramatsu H, Katase T, Kamiya T and Hosono H 2013 *IEEE Transactions on Applied Superconductivity* **23** 7300405

[23] Kudo K, Iba K, Takasuga M, Kitahama Y, Matsumura J. Danura M, Nogami Y and Nohara M 2013 *Sci. Rep.* **3** 1478

[24] Guo Y, Wang X, Li J, Zhang S, Yamaura K and Takayama-Muromachi E 2012 *J. Phys. Soc. Jpn.* **81** 064704

[25] Katayama N, Kudo K, Onari S, Mizukami T, Sugawara K, Sugiyama Y, Kitahama Y, Iba K, Fujimura K, Nishimoto N, Nohara M and Sawa H 2013 *J. Phys. Soc. Jpn.* **82** 123702

[26] Kudo K, Mizukami T, Kitahama Y, Mitsuoka D, Iba K, Fujimura K, Nishimoto N, Hiraoka Y and Nohara M 2014 *J. Phys. Soc. Jpn.* **83** 025001

[27] Kudo K, Kitahama Y, Fujimura K, Mizukami T, Ota H and Nohara M 2014 *J. Phys. Soc. Jpn.* **83** 093705

[28] Kudo K, Mitsuoka D, Takasuga M, Sugiyama Y, Sugawara K, Katayama N, Sawa H, Kubo H S, Takamori K, Ichioka M, Fujii T, Mizokawa T and Nohara M 2013 *Sci. Rep.* **3** 3101

[29] Guo J G, Lei C H, Hayashi F and Hosono H 2014 *Nat. Comm.* **5**, 4756

[30] Ryu G, Kim S W, Mizoguchi H, Matsuishi S and Hosono H 2012 *Europhys. Lett.* **99** 27002

[31] Yajima T, Nakano K, Takeiri F, Ono T, Hosokoshi Y, Matsushita Y, Hester J, Kobayashi Y, Kageyama H 2012 *J. Phys. Soc. Jpn.* **81** 103706

[32] Yajima T, Nakano K, Takeiri F, Nozaki Y, Kobayashi Y, Kageyama H 2013 *J. Phys. Soc. Jpn.* **82** 033705

[33] Nakano K, Yajima T, Takeiri F, Green M A, Hester J, Kobayashi Y, Kageyama H 2013 *J. Phys. Soc. Jpn.* **82** 074707

[34] Mizoguchi H, Kuroda T, Kamiya T and Hosono H 2011 *Phys. Rev. Lett.* **106**, 237001

[35] Kudo K, Takasuga M, Okamoto Y, Hiroi Z and Nohara M 2012 *Phys. Rev. Lett.* **109** 097002

[36] Matsuishi S, Nakamura A, Muraba Y and Hosono H 2012 *Supercond. Sci. Technol.* **25** 084017

[37] Mizoguchi H, Matsuishi S, Hirano M, Tachibana M, Takayama-Muromachi E, Kawaji H and Hosono H 2011 *Phys. Rev. Lett.* **106** 057002

[38] Imamura N, Mizoguchi H and Hosono H 2012 *J. Am. Chem. Soc.* **134** 2516

[39] Ganesanpotti S, Yajima T, Tohyama T, Li Z, Nakano K, Nozaki Y, Tassel C, Kobayashi Y and Kageyama H 2014 *J. Alloy. Compd.* **583** 151

[40] Ganesanpotti S, Yajima T, Nakano K, Nozaki Y, Yamamoto T, Tassel C, Kobayashi Y and Kageyama H 2014 *J. Alloy. Compd.* **613** 370

[41] Imai M, Emura S, Nishio M, Matsushita Y, Ibuka S, Eguchi N, Ishikawa F, Yamada Y, Muranaka T and

Akimitsu J 2013 *Supercond. Sci. Tecnol.* **26** 075001
[42] Imai M, Ibuka S, Kikugawa N, Terashima T, Uji S, Kageyama H, Yajima T and Hase I 2015 *Phys. Rev. B* **91** 014513
[43] Mizoguchi H and Hosono H 2011 *Chem. Commun.* **47** 3778
[44] Lei H and Hosono H 2013 *Europhys. Lett.* **104** 17003
[45] Tanaka M, Zhang S, Tanaka Y, Inumaru K and Yamanaka S 2013 *J. Solid State Chem*. **198** 445
[46] Ji S, Imai M, Zhu H and Yamanaka S 2013 *Inorg. Chem.* **52** 3953
[47] Ibuka S, Imai M, Naka T and Nishio M 2014 *Supercond. Sci. Technol.* **27** 025012
[48] Kudo K, Fujimura K, Ota H and Nohara M 2015 *Phys. Rev. B* submitted
[49] Isobe M, Yoshida H, Kimoto K, Arai M and Takayama-Muromachi E 2014 *Chem. Mater.* **26** 2155
[50] Pyon S *et al.* 2014 *J. Phys. Soc. Jpn.* **83** 093706
[51] Qi Y, Guo J, Lei H, Xiao Z, Kamiya T and Hosono H 2014 *Phys. Rev*. B **89** 024517
[52] Mizugichi Y *et al.* 2012 *Phys. Rev. B* **86** 220510
[53] Demura S *et al.* 2013 *J. Phys. Soc. Jpn.* **82** 033708
[54] Yajima T, Takeiri F, Nozaki Y, Li Z, Tohyama T, Green M A, Kobayashi Y, Kageyama H 2014 *J. Phys. Soc. Jpn*. **83** 073705
[55] Yamanaka S, Umemoto K, Zheng Z, Suzuki Y, Matsui H, Toyota N and Inumaru K 2012 *J. Mater. Chem.* **22** 10752
[56] Zhang S, Tanaka M and Yamanaka S. 2012 *Phys. Rev. B* **86** 024516
[57] Zhang S, Tanaka M, Watanabe E, Zhu H, Inumaru K and Yamanaka S 2013 *Supercond. Sci. Technol.* **26** 122001
[58] Zheng Z and Yamanaka S 2011 *Chem Mater*. **23** 1558
[59] Zhang S, Tanaka M, Zhu H and Yamanaka S 2013 *Supercond. Sci. Technol.* **26** 085015
[60] Zhang S, Tanaka M, Onimaru T, Takabatake T, Isikawa Y, Yamanaka S 2013 *Supercond. Sci. Technol.* **26** 045017
[61] Kitagawa S, Kotegawa H, Tou H, Ishii H, Kudo K, Nohara M and Harima H 2013 *J. Phys. Soc. Jpn.* **82** 113704
[62] Kudo K, Ishii H, Takasuga M, Iba K, Nakano S, Kim J, Fujiwara A and Nohara M 2013 *J. Phys. Soc. Jpn.* **82** 063704
[63] Pyon S, Kudo K and Nohara M 2012 *J. Phys. Soc. Jpn.* **81** 053701
[64] Kudo K, Kobayashi M, Pyon S and Nohara M 2013 *J. Phys. Soc. Jpn.* **82** 085001
[65] Qi Y, Matsuishi S, Guo J, Mizoguchi H and Hosono H 2012 *Phys. Rev. Lett.* **109** 217002
[66] Guo J, Qi Y, Matsuishi S and Hosono H 2012 *J. Am. Chem. Soc.* **134** 20001
[67] Guo J, Qi Y and Hosono H 2013 *Phys. Rev. B* **87** 224504
[68] Sathish C I *et al* 2012 *J. Solid State Chem.* **196** 579
[69] Sathish C I, Shirako Y, Tsujimoto Y, Feng H L, Sun Y, Akaogi M and Yamaura K 2014 *Solid State Commun.* **177** 33
[70] Ji S, Tanaka M, Zhang S and Yamanaka S 2012 *Inorg. Chem*. **51** 10300
[71] Tanaka M, Zhang S, Inumaru K and Yamanaka S 2013 *Inorg. Chem*. **52** 6039
[72] Ryu G, Kim S W, Matsuishi S, Kawaji H, Hosono H 2011 *Phys. Rev. B* **84** 224518

[73] Guo J, Yamaura J, Lei H, Matsuishi S, Qi Y and Hosono H 2013 *Phys. Rev. B* **88**, 140507(R)

[74] Jeong S, Matsuishi S, Lee K, Toda Y, Kim S W and Hosono H 2014 *Supercond. Sci. Technol.* **27** 055005

[75] Kudo K, Kobayashi M, Kakiya S, Danura M and Nohara M 2012 *J. Phys. Soc. Jpn.* **81** 035002

[76] Yajima T *et al.* 2013 *J. Phys. Soc. Jpn.* **82** 013703

[77] Park S W, Mizoguchi H, Kodama K, Shamoto S, Otomo T, Matsuishi S, Kamiya T and Hosono H 2013 *Inorg. Chem.* **52** 13363

[78] Liu X, Matsuishi S, Fujitsu S and Hosono H 2012 *Phys. Rev. B* **85** 104403

[79] Liu X, Matsuishi S, Fujitsu S and Hosono H 2012 *Phys. Rev. B* **84** 214439

[80] Liu X, Matsuishi S, Fujitsu S, Ishigaki T, KamiyamaT and Hosono H 2012 *J. Am. Chem. Soc.* **134** 11687

[81] Mizoguchi H and Hosono H 2011 *J. Am. Chem. Soc.* **133** 2394

[82] Lei H, Yamaura J, Guo J, Qi Y, Toda Y and Hosono H 2014 *Inorg. Chem.* **53** 5684

[83] Lee K, Kim S-W, Toda Y, Matsuishi S and Hosono H 2013 *Nature* **494** 336

[84] Anzai A, Fuchigami M, Yamanaka S, Inumaru K, 2012 *Mater. Res. Bull.* **47** 2062

[85] Shirako Y, Shi Y G, Aimi A, Mori D, Kojitani H, Yamaura K, Inaguma Y and Akaogi M 2012 *J. Solid State Chem.* **191** 167

[86] Shirako Y *et al.* 2011 *Phys. Rev. B* **83** 174411

[87] Wang X X, Guo Y F, Shirako Y, Yamaura K and Takayama-Muromachi E 2011 *Physica C* **471** 763

[88] Shirako Y, Kojitani H, Oganov A R, Fujino K, Miura H, Mori D, Inaguma Y, Yamaura K,and Akaogi M 2012 *American Mineralogist* **97** 159

[89] Zhang S, Yoshikawa M, Inumaru K and Yamanaka S 2013 *Inorg. Chem.* **52** 10571

[90] Yajima T, submitted.

[91] Sun Y S, Guo Y F, Wang X X, Tsujimoto Y, Matsushita Y, Shi Y G, Wang C, Belik A A and Yamaura K 2012 *Appl. Phys. Lett.* **100** 161907

[92] Feng H L, Guo Y, Sathish C I, Wang X, Yuan Y-H and Yamaura K 2014 *JPS Conf. Proc.* **1** 012002

[93] Feng H L, Tsujimoto Y, Guo Y, Sun Y, Sathish C I and Yamaura K 2013 *High Pressure Research* **33** 221

[94] Mizoguchi H, Kamiya T, Matsuishi S and Hosono H 2011 *Nat. Commun.* **2** 470

[95] Yajima T, Kitada A, Kobayashi Y, Sakaguchi T, Bouilly G, Kasahara S, Terashima T, Takano M and Kageyama H 2012 *J. Am. Chem. Soc.* **134** 8782

[96] Tassel C, Goto Y, Kuno Y, Hester J, Green M, Kobayashi Y, Kageyama H 2013 *Angew. Chem. Int. Ed.* **53** 10377

[97] Tanaka M, Zhang S, Onimaru T, Takabatake T, Inumaru K and Yamanaka S 2014 *Carbon* **73** 125

[98] Yamanaka S, Komatsu M, Tanaka M, Sawa H and Inumaru K 2014 *J. Amer. Chem. Soc.* **136** 7717

[99] Nishikubo Y, Nakano S, Kudo K and Nohara M 2012 *Appl. Phys. Lett.* **100** 252104

[100] Kitada A, Kasahara S, Terashima T, Kobayashi Y, Yoshimura K and Kageyama H 2011 *Appl. Phys.Exp.* **4** 035801

[101] Sakurai H, Kolodiazhnyi T, Michiue Y, Takayama-Muromachi E, Tanabe Y and Kikuchi H 2012 *Angew. Chem. Int. Ed.* **51** 6653

[102] Wang X X *et al.* 2011 *Phys. Rev. B* **83** 100410

[103] Wang X X, Guo Y G, Shi Y G, Li J J, Zhang S B and Yamaura K 2012 *J. Phys. Conf. Ser* **400** 032109

[104] Lei H, Yin W-G, Zhong Z and Hosono H 2014 *Phys. Rev. B* **89** 020409

[105] Wang X *et al.* 2012 *Inorg. Chem.* **51** 6868

[106] Shi Y *et al.* 2013 *Nat. Mat.* **12** 1024

[107] Shi Y, Guo Y, Yu S, Arai M, Sato A, Belik A A, Yamaura K and Takayama-Muromachi E 2010 *J. Am. Chem. Soc.* **132** 8474

[108] Kitano M *et al.* 2012 *Nat. Chem.* **4** 934

[109] Toda Y, Hirayama H, Kuganathan N, Torrisi A, Sushko P V and Hosono H 2013 *Nat. Commun.* **4** 2378

[110] Kobayashi Y *et al.* 2012 *Nat. Mater.* **11** 507

[111] Nomura K, Kamiya T, and Hosono H 2011 *Adv. Mater.* **23** 3431

[112] Takahashi H, Igawa K, Arii K, Kamihara Y, Hirano M, and Hosono H 2008 *Nature* **453** 376

[113] Liu R H *et al* 2008 *Phys. Rev. Lett.* **101** 087001

[114] Chen X H, Wu T, Wu G, Liu R H, Chen H and Fang D F 2008 *Nature* **453** 761

[115] Z-A. Ren *et al* 2008 *Chin. Phys. Lett.* **25** 2215

[116] M. Fujioka *et al* 2013 *Supercond. Sci. Technol.* **26** 085023

[117] Z-A. Ren *et al* 2008 *Europhys. Lett.* **83** 17002

[118] Wang C *et al* 2008 *Europhys. Lett.* **83** 67006

[119] Rotter M, Tegel M and Johrendt D 2008 *Phys. Rev. Lett.* **101** 107006

[120] Hosono H, Matsuishi S, Nomura N and Hiramatsu H 2009 *Butsuri* **64** 807 (Review in Japanese).

[121] Aswathy P M, Anooja J B, Sarum P M and Syamaprasad U 2010 *Supercond. Sci. Technol.* **23** 073001 (Review)

[122] Johnston D C 2010 *Advances in Physics* **59** 803 (Review)

[123] Peglione J and Greene R L 2010 *Nat. Phys.* **6** 645 (Review)

[124] Johrendt D 2011 *J. Mat. Chem.* **21** 13726 (Review)

[125] Fujitsu S, Matsuishi S and Hosono H 2012 *Int. Mater. Rev.* **57** 311 (Review)

[126] Wang N L, Hosono H and Dai P C *"Iron-based Superconductors - Materials, Properties and Mechanism"* 2013 Pan Stanford Publishing (Singapore)

[127] Johnson P D, Xu G and Y W G ed, "*Iron-Based Superconductivity*" 2015 Springer (New York)

[128] Pottgen R, and Johrendt D, 2008 *Z. Naturforsch* **63b** 1135

[129] Johnson V, Jeitschko W 1974 *J. Solid State Chem.* **11** 161

[130] Zimmer B I, Jeitschko W, Albering J H, Glaum R, Reehuis M 1995 *J. Alloys Compd.* **229** 238

[131] Yanagi H, Watanabe T, Kodama K, Iikubo S, Shamoto S, Kamiya T, Hirano M and Hosono H 2009 *J. Appl. Phys.* **105** 093936

[132] Yanagi H, Kawamura R, Kamiya T, Kamihara Y, Hirano M, Nakamura T, Osawa H and Hosono H 2008 *Phys. Rev. B* **77** 224431

[133] Watanabe T, Yanagi H, Kamihara Y, Kamiya T, Hirano M and Hosono H 2008 *J. Sol. State Chem.* **181** 2117

[134] Kayanuma K, Kawamura R, Hiramatsu H, Yanagi H, Hirano M, Kamiya T and Hosono H 2008 *Thin Solid Films* **516** 5800

[135] Kayanuma K, Hiramatsu H, Hirano M, Kawamura R, Yanagi H, Kamiya T and Hosono H 2007 *Phys Rev. B* **76** 195325

[136] Quebe P, Terbüchte L J and Jeitschko W 2000 *J. Alloys Compd.* **302** 70

[137] Matsuishi S, Inoue Y, Nomura T, Yanagi H, Hirano M and Hosono H, 2008 *J. Am. Chem. Soc.* **130** 14428

[138] Just G, and Paufler P 1996 *J. Alloys Compd.* **232** 1

[139] Wang X C, Liu Q Q, Lv Y X, Gao W B, Yang L X, Yu R C, Li F Y and Jin C Q 2008 *Solid State Commun.* **148** 538

[140] Parker D R, Pitcher M J, Baker P J, Franke I, Lancaster T, Blundell S J, and Clarke S J 2009 *Chem. Commun.* 2189

[141] Ogino H, Matsumura Y, Katsura Y, Ushiyama K, Horii S, Kishio K and Shimoyama J 2009 *Supercond. Sci. Technol.* **22** 075008

[142] Ogino H, Katsura Y, Horii S, Kishio K and Shimoyama J 2009 *Supercond. Sci. Technol.* **22** 085001

[143] OginoH, Sato S, Kishio K, Shimoyama J, Tohei T, and Ikuhara Y 2010 *Appl. Phys. Lett.* **97** 072506

[144] Chen G F, Xia T-L, Yang H X, Li J Q, Zheng P, Luo J L, and Wang N L 2009 *Supercond. Sci. Technol.* **22** 07001

[145] Zhu X, Han F, Mu G, Zeng B, Cheng P, Shen B, and Wen H-H 2009 *Phys. Rev. B* **79** 024516

[146] Shirage P M, Kihou K, Lee C H, Kito H, Eisaki H and Iyo A 2011 *J. Am. Chem. Soc.* **133** 9360

[147] Hsu F-C *et al* 2008 *Proc. Natl. Acad. Sci. USA* **105** 14262

[148] Medvedev S, McQueen T M, Troyan I A, Palasyuk T, Eremets M I, Cava R J, Naghavi S, Casper F, Ksenofontov V, Wortmann G and Felser C 2009 *Nat. Mat.* **8** 630

[149] Wang Q-Y *et al* 2012 *Chin. Phys. Lett.* **29** 037402

[150] He S L *et al* 2013 *Nat. Mater.* **12** 605

[151] Tan S Y *et al* 2013 *Nat. Mater.* **12** 634

[152] Liu D *et al* 2012 *Nat. Commun.* **3** 931

[153] Ge J-F, Liu Z-L, Liu C, Gao C-L, Qian D, Xue Q-K, Liu Y, Jia J-F 2014 *Nat. Mat.* **14** 285

[154] Guo J, Jin S, Wang G, Wang S, Zhu K, Zhou T, He M and Chen X 2010 *Phys. Rev. B* **82** 180520

[155] Zavalij P *et al* 2011 *Phys. Rev. B* **83** 132509

[156] Fang M-H, Wang H-D, Dong C-H, Li Z-J, FengC-M, Chen J and Yuan H Q 2011 *Europhys. Lett.* **94** 27009

[157] Ivanovskii A L 2011 *Physica C* **471** 409

[158] Malaeb W *et al* 2008 *J. Phys. Soc. Jpn.* **77** 093714

[159] Mazin I I, Singh D J, Johannes M D and Du M H 2008 *Phys. Rev. Lett.* **101** 057003

[160] Kuroki K, Onari S, Arita R, Usui H, Tanaka Y, Kontani H and Aoki H 2008 *Phys. Rev. Lett.* **101** 087004

[161] Chubukov A V, Efremov D V and Eremin I, 2008 *Phys. Rev. B* **78** 134512

[162] Graser S, Maier T A, Hirschfeld P J and Scalapino D J 2009 *New J. Phys.* **11** 025016

[163] Ikeda H, Arita R and Kuneš J 2010 *Phys. Rev. B* **81** 054502

[164] Daghofer M, Moreo A, Riera J A, Arrigoni E, Scalapino D J and Dagotto E 2008 *Phys. Rev. Lett.* **101** 237004

[165] Thomale R, Platt C, Hanke W and Bernevig B A 2011 *Phys. Rev. Lett.* **106** 187003

[166] Wang F, Zhai H, Ran Y, Vishwanath A and Lee D-H 2009 *Phys. Rev. Lett.* **102** 047005

[167] Qian T *et al* 2011 *Phys. Rev. Lett.* **106** 187001

[168] Kontani H. and Onari S., 2010 *Phys. Rev. Lett.* **104** 157001

[169] Suzuki K, Usui H, Iimura S, Sato Y, Matsuishi S, Hosono H and Kuroki K 2014 *Phys. Rev. Lett.* **113** 027002

[170] Nomura T, Kim S W, Kamihara Y, Hirano M, Sushko P V, Kato K, Takata M, Shluger A L and Hosono H 2008 *Supercond. Sci. Technol.* **21** 125028

[171] de la Cruz C *et al* 2008 *Nature* **453** 899

[172] Malavasi L, Artioli G A, Ritter C, Mozzati M C, Maroni B, Pahari B and Canesch A 2010 *J. Am. Chem. Soc.* **132** 2417

[173] Hess C, Kondrat A, Narduzzo A, Hamann-Borrero J E, Klingeler R, Werner J, Behr G and Büchner B 2009 *Europhys. Lett.* **87** 17005

[174] Kamihara Y *et al* 2010 *New J. Phys.* **12** 033005

[175] Margadonna S, Takabayashi Y, McDonald M T, Brunelli M, Wu G, Liu R H, Chen X H and Prassides K 2009 *Phys. Rev. B* **79** 014503

[176] Drew A J *et al* 2009 *Nat. Mater.* **8** 310

[177] Kito H, Eisaki H and Iyo A 2008 *J. Phys. Soc. Jpn* **77** 063707

[178] Kodama K, Ishikado M, Esaka F, Iyo A, Eisaki H and Shamoto S 2011 *J. Phys. Soc. Jpn* **80** 034601

[179] Zhigadlo N D, Katrych S, Weyeneth S, Puzniak R, Moll P J W, Bukowski Z, Karpinski J, Keller H and Batlogg B 2010 Phys. Rev. B **82** 064517

[180] Sefat A S, Huq A, McGuire M A, Jin R, Sales B C, Mandrus D, Cranswick L M D, Stephens P W and Stone K H 2008 *Phy. Rev. B* **78** 104505

[181] Dong X L *et al* 2010 *Phys. Rev. B* **82** 212506

[182] Cao G *et al* 2009 *Phys. Rev. B* **79** 174505

[183] Maroni B, Malavasi L, Mozzati M C, Grandi M S, Hill A H, Chermisi D, Dore P and Postorino P 2010 *Phys. Rev. B* **82** 104503

[184] Canfield P C and Bud'ko S L 2010 *Annu. Rev. Condens. Matter Phys.* **1** 27

[185] Wen H-H, Mu G, Fang L, Yang H and Zhu X 2008 *Europhys. Lett.* **82** 17009

[186] Sefat A S, Jin R, McGuire M A, Sales B C, Singh D J and Mandrus D 2008 *Phys. Rev. Lett.* **101** 117004

[187] Canfield P C, Bud'ko S L, Ni N, Yan J Q and Kracher A 2009 *Phys. Rev. B* **80** 060501

[188] Ni N, Thaler A, Kracher A, Yan J Q, Bud'ko S L and Canfield P C 2009 *Phys. Rev. B* **80** 024511

[189] Jiang S, Xing H, Xuan G, Wang C, Ren Z, Feng C, Dai J, Xu Z and Cao G 2009 *J. Phys.: Condens. Matter* **21** 382203

[190] Sharma S, Bharathi A, Chandra S, Reddy V R, Paulraj S, Satya A T, Sastry V S, Gupta A and Sundar C S 2010 *Phys. Rev. B* **81** 174512

[191] Tarascon J M, Greene L H, Barboux P, McKinnon W R, Hull G W, Orlando T P, Delin K A, Foner

S and McNiA E J, Jr 1987 *Phys. Rev. B* **36** 8393

[192] Sato M, Kobayashi Y, Lee S C, Takahashi H, Satomi E and Miura Y 2010 *J. Phys. Soc. Jpn.* **11** 014710

[193] Ni N, Tillman M E, Yan J-Q, Kracher A, Hannahs S T, Bud'ko S L and Canfield P C 2009 *Phys. Rev. B* **78** 214515

[194] Tanatar M A *et al* 2009 *Phys. Rev. B* **79** 094507

[195] Moll P J W, Puzniak R, Balakirev F, Rogacki K, Karpinski J, Zhigadlo N D and Batlogg B 2010 *Nat. Mater.* **9** 628

[196] Nagamatsu J, Nakagawa N, Muranaka T, Zenitani Y and Akimitsu J 2001 *Nature* **410** 63

[197] Karpinski J, Zhigadlo N D, Katrych S, Puzniak R, Rogacki K and Gonnelli R 2007 *Phys. C* **456** 3

[198] Eltsev Yu, Nakao K, Lee S, Masui T, Chikumoto N, Tajima S, Koshizuka N and Murakami M 2003 *J. Low Temp. Phys.* **131** 1069

[199] Wu M K, Ashburn J R, Torng C J, Hor P H, Meng R L, Gao L, Huang Z J, Wang Y Q and Chu C W 1987 *Phys. Rev. Lett.* **58** 908

[200] Maeda H, Tanaka Y, Fukutomi M and Asano T 1988 *Jpn J. Appl. Phys.* **27** L209

[201] Schilling A, Cantoni M, Guo J D and Ott H R 1993 *Nature* **363** 56

[202] Asada Y 1991 *Bull. Jpn. Inst. Metals* **30** 832 (in Japanese).

[203] Tozer S W, Kleinsasser A W, Penney T, Kaiser D and Holtzberg F 1987 *Phys. Rev. Lett.* **59** 1768

[204] Iye Y, Tamegai T, Sakakibara T, Goto T, Miura N, Takeya H and Takei H 1988 *Physica C* **153-155** 26

[205] Hagen S J, Jing T W, Wang Z Z, Horvath J and Ong N P 1988 *Phys. Rev. B* **37** 7928

[206] Farrell D E, Bonham S, Foster J, Chang Y C, Jiang P Z, Vandervoort K G, Lam D J and Kogan V G 1989 *Phys. Rev. Lett.* **63** 782

[207] Köhler A and Behr G 2009 *J. Supercond. Novel Magn.* **22** 565

[208] Artioli G A, Malavasi L, Mozzati M C and Fernandez Y D 2009 *J. Am. Chem. Soc.* **131** 12044

[209] Lee C-H, Iyo A, Eisaki H, Kito H, Fernandez-Diazi M T, Ito T, Kohou K, Matsuhata H, Brenden M and Yamada K 2008 *J. Phys. Soc. Jpn.* **77** 083704

[210] Iimura S *et al* 2013 *Phys. Rev. B* **88** 060501

[211] M. Hiraishi *et al* 2014 *Nat. Phys.* **10** 300.

[212] Sunagawa M *et al.* 2014 *Sci. Rep.* **4** 4381

[213] Harnagea L *et al.* 2011 *Phys. Rev. B* **83** 094523

[214] Kumar N, Chi S, ChenY, Rana K G, Bigam A K, Thamizhavel A, William Ratcli. II, Dhar S K and Lynn J W 2009 *Phys. Rev. B* **80** 144524

[215] Leithe-Jasper A, Schnelle W and Rosner H 2008 *Phys. Rev. Lett.* **101** 207004

[216] Saha S R, Butch N P, Kirshenbaum K and Paglione J 2009 *Phys. Rev. B* **79** 224519

[217] Li L J *et al.* 2009 *New J. Phys.* **11** 025008

[218] Danura M, Kudo K, Oshiro Y, Araki S, Kobayashi T C and Nohara M 2011 *J. Phys. Soc. Jpn.* **80** 103701

[219] Kudo K, Matsumura J, Danura M and Nohara M *unpublished data*

[220] Schnelle W, Leithe-Jasper A, Gumeniuk R, Burkhardt U, Kasinathan D and Rosner H 2009 *Phys.*

*Rev. B* **79** 214516

[221] Han F *et al.* 2009 *Phys. Rev. B* **80** 024506

[222] Qi Y, Gao Z, Wang L, Zhang X, Wang D, Yao C, Wang C, Wang C and Ma Y 2011 *Europhys. Lett.* **96** 47005

[223] Nishikubo Y, Kakiya S, Danura M, Kudo K and Nohara M 2010 *J. Phys. Soc. Jpn.* **79** 095002

[224] Wang X L, Shi H Y, Yan X W, Yuan Y C, Lu Z -Y, Wang X Q and Zhao T -S 2010 *Appl. Phys. Lett.* **96** 012507

[225] Saha S R, Drye T, Kirshenbaum K, Butch N P, Zavalij P Y and Paglione J 2010 *J. Phys.: Condens. Matter* **22** 072204

[226] Shirage P M, Miyazawa K, Kito H, Eisaki H and Iyo A 2008 *Appl. Phys. Express* **1** 081702

[227] Cortes-Gil R and Clarke S J 2011 *Chem. Mat.* **23** 1009

[228] Sasmal K, Lv B, Lorenz B, Guloy A M, Chen F, Xue Y -Y and Chu C -W 2008 *Phys. Rev. Lett.* **101** 107007

[229] Lv B, Gooch M, Lorenz B, Chen F, Guloy A M and Chu C W 2009 *New J. Phys.* **11** 025013

[230] Aswartham S *et al.* 2012 *Phys. Rev. B* **85** 224520

[231] Bukowski Z, Weyeneth S, Puzniak R, Moll P, Katrych S, Zhigadlo N D, Karpinski J, Keller H and Batlogg B 2009 *Phys. Rev. B* **79** 104521

[232] Wang C *et al.* 2009 *Phys. Rev. B* 79 054521

[233] Wu G *et al.* 2008 *Europhys. Lett.* **84** 27010

[234] Saha S R, Butch N P, Drye T, Magill J, Ziemak S, Kirshenbaum K, Zavalij P Y, Lynn J W and Paglione J 2012 *Phys. Rev. B* **85** 024525

[235] Gao Z, Qi Y, Wang L, Wang D, Zhang X, Yao C, Wang C and Ma Y 2011 *Europhys. Lett.* **95** 67002

[236] Lv B, Deng L, Gooch M, Wei F, Sun Y, Meen J K, Xu Y -Y, Lorenz B and Chu C -W 2011 *Proc. Natl Acad. Sci. USA* **108** 15705

[237] Qi Y, Gao Z, Wang L, Wang D, Zhang X, Yao C, Wang C, Wang C and Ma Y 2012 *Supercond. Sci. Technol.* **25** 045007

[238] Katase T, Hiramatsu H, Kamiya T and Hosono H 2012 *Supercond. Sci. Technol.* **25** 084015

[239] Yakita H *et al* 2014 *J. Am. Chem. Soc.* **136** 846

[240] Sala A *et al* 2014 *Appl. Phys. Express* **7** 073102

[241] Cheng P, Shen B, Mu G, Zhu X, Han F, Zeng B and Wen H -H 2009 *Europhys. Lett.* **85** 67003

[242] Kakiya S, Kudo K, Nishikubo Y, Oku K, Nishibori E, Sawa H, Yamamoto T, Nozaka T and Nohara M 2011 *J. Phys. Soc. Jpn.* **80** 093704

[243] Nohara M, Kakiya S, Kudo K, Oshiro Y, Araki S, Kobayashi T C, Oku K, Nishibori E and Sawa H 2012 *Solid State Cummun.* **152** 635

[244] Zhou W, Zhuang J, Yuan F, Li X, Xing X, Sun Y and Shi Z 2014 *Appl. Phys. Express* **7** 063102

[245] Rutzinger D, Bartsch C, Doerr M, Rosner H, Neu V, Doert T and Ruck M 2010 *J. Solid State Chem.* **183** 510

[246] Kasahara S, Shibauchi T, Hashimoto K, Nakai Y, Ikeda H, Terashima T and Matsuda Y 2011 *Phys. Rev. B* **83** 060505(R)

[247] Löhnert C, Stürzer T, Tegel M, Frankovsky R, Friederichs G and Johrendt D 2011 *Angew. Chem.*

*Int. Ed.* **50** 9195
[248] Ni N, Allred J M, Chan B C and Cava R J 2011 *Proc. Natl. Acad. Sci. USA* **108** E1019
[249] Stürzer T, Derondeau G and Johrendt D 2012 *Phys. Rev. B* **86** 060516(R)
[250] Hieke C, Lippmann J, Stürzer T, Friederichs G, Nitsche F, Winter F, Pöttgen R and Johrendt D 2013 *Philos. Mag.* **93** 3680
[251] Katayama N, Sugawara K, Sugiyama Y, Higuchi T, Kudo K, Mitsuoka D, Mizokawa T, Nohara M and Sawa H 2014 *J. Phys. Soc. Jpn.* **83** 113707
[252] Sawada K *et al* 2014 *Phys. Rev. B* **89** 220508(R)
[253] Li J *et al.* 2012 *J. Am. Chem. Soc.* **134** 4068
[254] Krzton-Maziopa A, Shermadini Z, Pomjakushina E, Pomjakushin V, Bendele M, Amato A, Khasanov R, Luetkens H and Conder K 2011 *J. Phys. Condens. Matter* **23** 052203
[255] Wang, A. F. et al. *Phys. Rev. B* **83** 060512
[256] Park J T *et al* 2011 *Phys. Rev. Lett.* **107** 177005
[257] Buffinger D R, Ziebarth R P, Stenger V A, Recchia C and Pennington C H, 1993 *J. Am. Chem. Soc.* **115** 9267
[258] Ying T P, Chen X L, Wang G, Jin S F, Zhou T T, Lai X F, Zhang H and Wang W Y 2012 *Sci. Rep.* **2** 426
[259] Ying T P, Chen X L, Wang G, Jin S F, Lai X F, Zhou T T, Zhang H, Shen S J and Wang W Y 2013 *J. Am. Chem. Soc.* **135** 2951
[260] Burrard-Lucas M *et al* 2013 *Nat. Mater.* **12** 15
[261] Sedlmaier S J *et al* 2014 *J. Am. Chem. Soc.* **136** 630
[262] Zhang A M, Xia T L, Liu K, Tong W, Yang Z R and Zhang C M 2013 *Sci. Rep.* **3** 1216
[263] Bednorz J G and Müller K A 1986 *Z. Phys. B* **64** 189
[264] Hebard A F, Rosseinsky M J, Haddon R C, Murphy D W, Glarum S H, Palstra T T M, Ramirez A P, Kortan A R 1991 *Nature* **350** 600
[265] Deslandes F, Nazzal A I, Torrance J B 1991 *Physica C* **179** 85
[266] Arita R, Yamasaki A, Held K, Matsuno J, Kuroki K 2007 *Phys. Rev. B* **75** 174521
[267] Axtell III E A, Ozawa T, Kauzlarich S M, Singh R R P 1997 *J. Solid State Chem.* **134** 423
[268] Adam A, Schuster H U 1990 Z. *Anorg. Allg. Chem.* **584**, 150
[269] Wang X F, Yan Y J, Ying J J, Li Q J, Zhang M, Xu N, Chen X H 2010 *J. Phys.: Condens. Matter* **22** 075702
[270] Sun Y L, Jiang H, Zhai H F, Bao J K, Jiao W H, Tao Q, Shen C Y, Zeng Y W, Xu Z A, Cao G H 2012 *J. Am. Chem. Soc*. **134** 12893
[271] Ozawa T C, Kauzlarich S M 2008 *Sci. Technol. Adv. Mater*. **9** 033003
[272] Ozawa T C, Kauzlarich S M 2001 *Chem, Mater.* **13** 1804
[273] Liu R H, Tan D, Song Y A, Li Q J, Yan Y J, Ying J J, Xie Y L, Wang X F, Chen X H 2009 *Phys. Rev. B* **80** 144516
[274] Doan P, Gooch M, Tang Z, Lorenz B, Moller A, Tapp J, Chu P C W, Guloy A M 2012 *J. Am. Chem. Soc.* **134** 16520
[275] Kitagawa S, Ishida K, Nakano K, Yajima T, Kageyama H 2013 *Phys. Rev. B*. **87** 060510(R)

[276] Von Rohr F, Schilling A, Nesper R, Baines C, Bendele M 2013 *Phys. Rev. B* **88** 140501(R)

[277] Nozaki Y *et al.* 2013 *Phys. Rev. B* **88** 214506

[278] Gooch M, Doan P, Tang Z, Lorenz B, Guloy A M, Chu P C W 2013 P*hys. Rev. B* **88** 064510

[279] Pachmayr U and Johrendt D 2014 *Solid State Sci.* **28** 31

[280] Von Rohr F, Nesper R, Schilling A 2014 *Phys. Rev. B* **89** 094505

[281] Shi Y G, Wang H P, Zhang X, Wang W D, Huang Y, Wang N L 2013 *Phys. Rev. B* **88** 144513

[282] Huang Y, Wang H P, Wang W D, Shi Y G, Wang N L 2013 *Phys. Rev. B* **87** 100507

[283] Xu H C *et al* 2014 *Phys. Rev. B* **89** 155108

[284] Fan G, Zhang X, Shi Y, Luo J 2013 *Science China Physics, Mechanics and Astronomy* **56** 2399

[285] Huang Y, Wang H P, Chen R Y, Zhang X, Zheng P, Shi Y G, Wang N L 2014 *Phys. Rev. B* **89** 155120

[286] Gooch M, Doan P, Lorenz B, Tan Z J, Guloy A M, Chu C W 2013 *Supercond. Sci. Technol.* **26** 125011

[287] Yu W, Dong X-L, Ma M-W, Yang H-X, Zhang C, Zhou F, Zhou X-J, Zhao Z-X 2014 *Chinese Phys. Lett.* **31** 077401

[288] Litvinchuk A P, Doan P, Tang Z, Guloy A M 2013 *Phys. Rev. B* **87** 064505

[289] Zhai H F *et al.* 2013 *Phys. Rev. B* **87** 100502(R)

[290] Singh D J 2012 *New J. Phys.* **14** 123003

[291] Yan X W, Lu Z Y 2013 *J. Phys. Condens. Matter* **25** 365501

[292] Subedi A 2013 *Phys. Rev. B* **87** 054506

[293] Frandsen B A *et al* 2014 *Nat. Commun.* **5** 5761

[294] Wang G, Zhang H, Zhang L, Liu C 2013 *J. Appl. Phys.* **113** 243904

[295] Kodenkandath T A, Lalena J N, Zhou W L, Carpenter E E, Sangregorio C, Falster AU, Simmons W B, O'Connor C J, Wiley J B 1999 *J. Am. Chem. Soc.* **121** 10743

[296] Tassel C, Kang J, Lee C, Hernandez O, Qiu Y, Paulus W, Collet E, Lake B, Guidi T,. Whangbo M H, Ritter C, Kageyama H, Lee S H 2010 *Phys. Rev. Lett.* **105** 167205

[297] Fleming R M *et al.* 1991 *Nature* **352** 701

[298] Yamanaka S 2000 *Annual Review of Materials Science* **30** 53

[299] Yamanaka S 2010 *J. Mater. Chem.* **20** 2922

[300] Yamanaka S, Hotehama K and Kawaji H 1998 *Nature* **392** 580

[301] Yamanaka S, Kawaji H, Hotehama K and Ohashi M 1996 *Adv. Mater.* **8** 771

[302] Yamanaka S and Tou H 2001 *Current Opinion in Solid State & Materials Science* **5** 545

[303] Kasahara Y, Kuroki K, Yamanaka S, Taguchi Y. 2015 *Physica C* DOI:10.1016/j.physc.2015.02.022

[304] Tou H, Maniwa Y, Koiwasaki T and Yamanaka S 2001 *Phys. Rev. Lett.* **86** 5775

[305] Taguchi Y, Hisakabe M and Iwasa Y 2005 *Phys. Rev. Lett.* **94** 217002

[306] Sugimoto A, Shohara K, Ekino T, Zheng Z and Yamanaka S 2012 *Phys. Rev. B* **85** 144517

[307] Tou H, Maniwa Y and Yamanaka S 2003 *Phys. Rev. B* **67** 100509

[308] Kuroki K 2010 *Phys. Rev. B* **81** 104502

[309] Bill A, Morawitz H and Kresin V Z 2002 *Phys. Rev. B* **66** 100501

[310] Bill A, Morawitz H and Kresin V Z 2003 *Phys. Rev. B* **68** 144519

[311] Chen X, Zhu L P, Yamanaka S 2002 *J. Solid State Chem.* **169** 149

[312] Shamoto S, Takeuchi K, Yamanaka S, Kajitani T, 2004 *Physica C* **402**, 283

[313] Tou H, Maniwa Y, Koiwasaki T and Yamanaka S 2001 *Phys. Rev. B* **63** 020508

[314] Tou H *et al.* 2005 *Phys. Rev. B* **72** 020501

[315] Hotehama K, Koiwasaki T, Umemoto K, Yamanaka S and Tou H 2010 *J. Phys. Soc. Jpn.* **79** 014707

[316] Yamanaka S, Okumura H and Zhu L-P 2004 *J. Phys. Chem. Solids* **65** 565

[317] Yamanaka S, Yasunaga T, Yamaguchi K and Tagawa M 2009 *J. Mater. Chem.* **19** 2573

[318] Marouchkine A 2004 *Room-Temperature Superconductivity* (Cambridge: Cambridge International Science Publishing)

[319] Putti M et al 2010 *Supercond. Sci. Technol.* **23** 034003

[320] Welp U, Xie R, Koshelev A E, Kwok W K, Cheng P, Fang L and Wen H H 2008 *Phys. Rev. B* **78** 140510

[321] Jaroszynski J et al 2008 Phys. Rev. B **78** 174523

[322] Welp U, Chaparro C, Koshelev A E, Kwok W K, Rydh A, Zhigadlo N D, Karpinski J and Weyeneth S 2011 *Phys. Rev. B* **83** 100513

[323] Hunte F, Jaroszynski J, Gurevich A, Larbalestier D C, Jin R, Sefat A S, McGuire M A, Sales B C, Christen D K and Mandrus D 2008 *Nature* **453** 903

[324] Chen G F, Li Z, Dong J, Li G, Hu W Z, Zhang X D, Song X H, Zheng P, Wang N L and Luo J L 2008 *Phys. Rev. B* **78** 224512

[325] Altarawneh M M, Collar K, Mielke C H, Ni N, Bud'ko S L and Canfield P C 2008 *Phys. Rev. B* **78** 220505

[326] Yuan H Q, Singleton J, Balakirev F F, Baily S A, Chen G F, Luo J L and Wang N L T 2009 *Nature* **457** 565

[327] Yamamoto A et al 2009 *Appl. Phys. Lett.* **94** 062511

[328] Yin Q A, Ylvisaker E R, Pickett W E 2011 *Phys. Rev. B* **83** 014509

[329] Kanamaru F, Shimada M, Koizumi M, Takano M and Takada T 1973 *J. Solid State Chem.* **7** 297

[330] Yamanaka S, Enishi E, Fukuoka H, Yasukawa M 2000 *Inorg. Chem*. **39** 56

[331] Yamanaka S, Izumi S, Maekawa S and Umemoto K 2009 *J. Solid State Chem*. **182** 1991

[332] Yamanaka S and Maekawa S 2006 *Z. Naturforsch. B* **61** 1493

[333] Fukuoka H, Ueno K and Yamanaka S 2000 *J. Organomet. Chem.* **611** 543

[334] Kurakevych O O, Strobel T A, Kim D Y, Muramatsu T, Struzhkin V V 2013 *Cryst Growth Des*. **13** 303

[335] Andersen S J, Marioara C D, Froseth A, Vissers R and Zandbergen H W. 2005 *Mat. Sci. Eng*. **390** 127

[336] Yamanaka S 2010 *Dalton Trans*. **39** 1901

[337] Yamanaka S 2014 *The Physics and Chemistry of Inorganic Clathrates* G S Nolas *ed.* Springer Dordrecht, Chap. 7

[338] Ekimov E A *et al* 2004 *Nature* **428** 542

[339] Connetable D, Timoshevskii V, Masenelli B, Beille J, Marcus J, Barbara B, et al. 2003 *Phys. Rev.*

*Lett.* **91** 247001
[340] Yamanaka S, Kubo A, Inumaru K, Komaguchi K, Kini N S, Inoue T, et al. 2006 *Phys. Rev. Lett.* **96** 076602
[341] Yamanaka S, Kini N S, Kubo A, Jida S and Kuramoto H 2008 *J. Am. Chem. Soc.* **130** 4303
[342] Ronning F, Bauer E D, Park T, Baek S-H, Sakai H and Thompson J D 2009 *Phys. Rev. B* **79** 134507
[343] Mine T, Yanagi H, Kamiya T, Kamihara K, Hirano M and Hosono H 2008 *Solid State Commun.* **147** 111
[344] Hirai D, Takayama T, Higashinaka R, Aruga-Katori H and Takagi H 2009 *J. Phys. Soc. Jpn.* **78** 023706
[345] Jeitschko W, Glaum R and Boonk L 1987 *J. Solid State Chem.* **69** 93
[346] Han J-T, Zhou J-S, Cheng J-G and Goodenough J B 2010 *J. Am. Chem. Soc.* **132** 908
[347] Bauer E D, Ronning F, Scott B L and Thompson J D 2008 *Phys. Rev. B* **78** 172504
[348] Ronning F, Kurita N, Bauer E D, Scott B L, Park T, Klimczuk T, Movshovich R and Thompson J D 2008 *J. Phys.: Condens. Matter* **20** 342203
[349] Parthe E, Chabot B, Braun H F, Engel N 1983 *Acta. Cryst.* B **39** 588
[350] Kudo K, Nishikubo Y and Nohara M 2010 *J. Phys. Soc. Jpn* **79** 123710
[351] Imre A, Hellmann A, Wenski G, Graf J, Johrendt D and Mewis A 2007 *Z. Anorg. Allg. Chem.* **633** 2037
[352] Shein I R and Ivanovskii A L 2011 *Phys. Rev. B* **83** 10450
[353] Hoffman W K and Jeistchko W 1985 *Monatsh. Chemi.* **116** 569
[354] Uhoya W O, Montgomery J M, Tsoi G M, Vohra Y K, McGuire M A, Sefat A S, Sales B C and Weir S T 2011 *J. Phys.: Condens. Matter* **23** 122201
[355] Saha S R, Butch N P, Drye T, Magill J, Ziemak S, Kirshenbaum K, Zavalij P Y, Lynn J W and Paglione J 2012 *Phys. Rev. B* **85** 24525
[356] Yamazaki T, Takeshita N, Kobayashi R, Fukazawa H, Kohori Y, Kihou K, Lee C, Kito H, Iyo A and Eisaki H 2010 *Phys. Rev. B* **81** 224511
[357] Uhoya W O, Tsoi G M, Vohra Y K, McGuire M A and Sefat A S 2011 *J. Phys.: Condens. Matter* **23** 365703
[358] Anand V K, Kim H, Tanatar M A, Prozorov R and Johnston D C 2013 P*hys. Rev. B* **87** 224510
[359] Tsutsumi K, Takayanagi S, Ishikawa M and Hirano T 1995 *J. Phys. Soc. Jpn.* **64** 2237
[360] Takada K, Sakurai H, Takayama-Muromachi E, Izumi F, Dilanian R A and Sasaki T 2003 *Nature* **422** 53
[361] Niihara K, Shishido T and Yajima S 1973 *Bull. Chem. Soc. Jpn.* **46** 1137
[362] Sefat A S, McGuire M A, Jin R, Sales B C, Mandrus D, Ronning F, Bauer E D and Mozharivskyj Y 2009 *Phys. Rev. B* **79** 094508
[363] Kurita N, Ronning F, Tokiwa Y, Bauer E D, Subedi A, Singh D J, Thompson J D and Movshovich R 2009 *Phys. Rev. Lett.* **102** 147004
[364] Subedi A and Singh D J 2008 *Phys. Rev. B* **78** 132511
[365] Shein I R and Ivanovskii A L 2009 *Phys. Rev. B* **79** 054510
[366] Kim J S, Boeri L, Kremer R K and Razavi F S 2006 *Phys. Rev. B* **74** 214513

[367] Gauzzi A, Takashima S, Takeshita N, Terakura C, Takagi H, Emery N, He´rold C, Lagrange P and Loupias G 2007 *Phys. Rev. Lett.* **98** 067002

[368] Gauzzi A *et al.* 2008 *Phys. Rev. B* **78** 064506

[369] Mauri F, Zakharov O, de Gironcoli S, Louie S G and Cohen M L 1996 *Phys. Rev. Lett.* **77** 1151

[370] Matsumoto N, Taniguchi K, Endoh R, Takano H and Nagata S 1999 *J. Low Temp. Phys.* **117** 1129

[371] Yang J J, Choi Y J, Oh Y S, Hogan A, Horibe Y, Kim K, Min B I and Cheong S-W 2012 *Phys. Rev. Lett.* **108** 116402

[372] Ootsuki D *et al.* 2012 *Phys. Rev. B* **86** 014519

[373] Ootsuki D *et al.* 2013 *J. Phys. Soc. Jpn.* **82** 093704

[374] Toriyama T *et al.* 2014 *J. Phys. Soc. Jpn.* **83** 033701

[375] Pyon S, Kudo K and Nohara M 2011 presented at *Int. Workshop Novel Superconductors and Super Materials 2011* (*NS*$^2$ *2011*)

[376] Pyon S, Kudo K and Nohara M 2011 presented at *Int. Conf. Novel Superconductivity in Taiwan 2011 (ICNSCT 2011).*

[377] Pyon S, Kudo K, Nohara M 2013 *Physica C* **494** 80

[378] Kamitani M, Bahramy M S, Arita R, Seki S, Arima T, Tokura Y and Ishiwata S 2013 *Phys. Rev. B* **87** 180501(R)

[379] Tunell G and Pauling L 1952 *Acta Crystallogr.* **5** 375

[380] Schutte W J and de Boer J L 1988 *Acta Crystallogr. B* **44** 486

[381] Janner A and Dam B 1989 *Acta Crystallogr. A* **45** 115

[382] Jobic S, Evain M, Brec R, Deniard P, Jouanneaux A and Rouxel J 1991 *J. Solid State Chem.* **95** 319

[383] Leger J M, Pereira A S, Haines J, Jobic S and Brec R 2000 *J. Phys. Chem. Solids* **61** 27

[384] Gor'kov L P and Rashba E I 2001 *Phys. Rev. Lett.* **87** 037004

[385] Frigeri P A, Agterberg D F, Koga A and Sigrist M 2004 *Phys. Rev. Lett.* **92** 097001

[386] Frigeri P A, Agterberg D F and Sigrist M 2004 *New J. Phys.* **6** 115

[387] Bauer E, Hilscher G, Michor H, Paul C, Scheidt E W, Gribanov A, Seropegin Y, Noël H, Sigrist M and Rogl P 2004 *Phys. Rev. Lett.* **92** 027003

[388] Settai R, Sugitani I, Okuda Y, Thamizhavel A, Nakashima M, Ōnuki Y and Harima H. 2007 *J. Magn. Magn. Mater.* **310** 844

[389] Settai R, Miyauchi Y, Takeuchi T, Lévy F, Sheikin I and Ōnuki Y 2008 *J. Phys. Soc. Jpn.* **77** 073705

[390] Kimura N, Ito K, Aoki H, Uji S and Terashima T 2007 *Phys. Rev. Lett.* **98** 197001

[391] McMillan W L 1968 *Phys. Rev.* **167** 331

[392] White J G and Hockings E F 1971 *Inorg. Chem.* **10** 1934

[393] Sologub O L, Salamakha P S, Sasakawa T, Chen X, Yamanaka S and Takabatake T 2002 *J. Alloys Compd.* **34** 6

[394] Loehken A, Reiss G J, Johrendt D and Mewis A 2005 Z. *Anorg. Allg. Chem.* **631** 1144

[395] Tanabe K and Hosono H 2012 *Jpn. J. Appl. Phys.* **51** 010005

[396] Hiramatsu H, Katase T, Kamiya T, Hirano M and Hosono H 2008 *Appl. Phys. Express* **1** 101702

[397] Katase T, Hiramatsu H, Yanagi H, Kamiya T, Hirano M and Hosono H 2009 *Solid State Commun.*

**149** 2121

[398] Katase T, Hiramatsu H, Kamiya T, and Hosono H 2010 *Appl. Phys. Express* **3** 063101

[399] Iida K, Hänisch J, Hühne R, Kurth F, Kidszun M, Haindl S, Werner J, Schultz L and Holzapfel B 2009 *Appl. Phys. Lett.* **95** 192501

[400] Maiorov B, Katase T, Baily S A, Hiramatsu H, Holesinger T G, Hosono H and Civale L 2011 *Supercond. Sci. Technol.* **24** 055007

[401] Iida K, Hänisch J, Thersleff T, Kurth F, Kidszun M, Haindl S, Hühne H, Schultz S and Holzapfel B 2010 *Phys. Rev. B* **81** 100507

[402] Takeda S, Ueda S, Yamagishi T, Agatsuma S, Takano S, Mitsuda A and Naito M 2010 *Appl. Phys. Express* **3** 093101

[403] Kasahara K *et al* 2010 *Phys. Rev. B* **81** 184519

[404] Adachi S, Shimode T, Miura M, Chikumoto N, Takemori A, Nakao K, Oshikubo Y and Tanabe K 2012 *Supercond. Sci. Technol.* **25** 105015

[405] Kawaguchi T, Sakagami A, Mori Y, Tabuchi M, Takeda Y and Ikuta H 2014 *Supercond. Sci. Technol.* **27** 065005

[406] Katase T, Hiramatsu H, Matias V, Sheehan C, Ishimaru Y, Kamiya T, Tanabe K and Hosono H 2011 *Appl. Phys. Lett.* **98** 242510

[407] Miura M, Adachi S, Shimode T, Wada K, Takemori A, Chikumoto N, Nakao K and Tanabe K 2013 *Appl. Phys. Express* **6** 093101

[408] Shimoyama J, Kitazawa K, Shimizu K, Ueda S, Horii S, Chikumoto N and Kishio K 2003 *J. Low Temp. Phys.* **131** 1043

[409] Fang L *et al* 2012 *Appl. Phys. Lett.* **101** 012601

[410] Lee S *et al* 2010 *Nat. Mater.* **9** 397

[411] Tarantini C *et al* 2010 *Appl. Phys. Lett.* **96** 142510

[412] Haugan T, Barnes P N, Wheeler R, Meisenkothen F and Sumption M 2004 *Nature* **430** 867

[413] Gutiérrez J *et al* 2007 *Nat. Mater.* **6** 367

[414] Miura M, Maiorov B, Baily S A, Haberkorn N, Willis J O, Marken K, Izumi T, Shiohara Y and Civale L 2011 *Phys. Rev. B* **83** 184519

[415] Miura M, Maiorov B, Kato T, Shimode T, Wada K, Adachi S and Tanabe K 2013 *Nat. Commun.* **4** 2499

[416] Maiorov B, Baily S A, Zhou H, Ugurlu O, Kennison J A, Dowden P C, Holesinger T G, Foltyn S R and Civale L 2009 *Nat. Mater.* **8** 398

[417] Lee S *et al* 2013 *Nat. Mater.* **12** 392

[418] Cooly L D, Lee P J and Larbalestier D C 1996 *Phys. Rev. B* **53** 6638

[419] Zhuang C G, Meng S, Yang H, Jia Y, Wen H H, Xi X X, Feng Q R and Gan Z Z 2008 *Supercond. Sci. Technol.* **21** 082002

[420] Godeke A 2006 *Supercond. Sci. Technol.* **19** R68

[421] Miura M, Maiorov B, Willis J O, Kato T, Sato M, Izumi T, Shiohara Y and Civale L 2013 *Supercond. Sci. Technol.* **26** 035008

[422] Sato H, Hiramatsu H, Kamiya T and Hosono H 2014 *Appl. Phys. Lett.* **104** 182603

[423] Sakagami A, Kawaguchi T, Tabuchi M, Ujihara T, Takeda Y and Ikuta H 2013 *Physica C* **494** 181

[424] Iida K et al 2010 *Appl. Phys. Lett.* **97** 172507

[425] Trommler S, Hänisch J, Matias V, Hühne R, Reich E, Iida K, Haindl S, Schultz L and Holzapfel B 2012 *Supercond. Sci. Technol.* **25** 084019

[426] Hiramatsu H, Katase T, Ishimaru Y, Tsukamoto A, Kamiya T, Tanabe K and Hosono H 2012 *Mater. Sci. Eng. B* **177** 515

[427] Iida K *et al* 2013 *Sci. Rep.* **3** 2139

[428] Braccini V *et al* 2013 *Appl. Phys. Lett.* **103** 172601

[429] Zhang Y *et al* 2011 *Appl. Phys. Lett.* **98** 042509

[430] Gurvitch M, Washington M A and Huggins H A 1983 *Appl. Phys. Lett.* **42** 472

[431] Hilgenkamp H and Mannhart J 2002 *Rev. Mod. Phys.* **74** 485

[432] Clarke J and Braginski A I (*ed.*) 2004 *The SQUID Handbook* (Weinheim: WILEY-VCH)

[433] Katase T, Ishimaru Y, Tsukamoto A, Hiramatsu H, Kamiya T, Tanabe K and Hosono H 2010 *Appl. Phys. Lett.* **96** 142507

[434] Katase T, Ishimaru Y, Tsukamoto A, Hiramatsu H, Kamiya T, Tanabe K and Hosono H 2011 *Nat. Commun.* **2** 409

[435] Ambegaokar V and Halperin B I 1969 *Phys. Rev. Lett.* **22** 1364

[436] Saitoh K, Ishimaru Y, Fuke H and Enomoto Y 1997 *Jpn. J. Appl. Phys.* **36** L272

[437] De Gennes P G 1964 *Rev. Mod. Phys.* **36** 225

[438] Delin K A and Kleinsasser A W 1996 *Supercond. Sci. Technol.* **9**, 227

[439] Katase T, Ishimaru Y, Tsukamoto A, Hiramatsu H, Kamiya T, Tanabe K and Hosono H 2010 *Supercond. Sci. Technol.* **23** 082001

[440] Lee L P, Longo J, Vinetskiy V and Cantor R 1995 *Appl. Phys. Lett.* **66** 1539

[441] Wakana H, Adachi S, Hata K, Hato T, Tarutani Y and Tanabe K 2009 *IEEE Trans. Appl. Supercond.* **19** 782

[442] Ishimaru Y, Murai Y, Adachi S and Tanabe K 2013 *Extended Abstracts of International Workshop on Novel Superconductors and Super Materials 2013* (Tokyo) pp. 104

[443] Poppe U, Divin Y Y, Faley M I, Wu J S, Jia C L, Shadrin P and Urban K 2001 *IEEE Trans. Appl. Supercond.* **11** 3768

[444] Sarnelli E, Testa G, Crimaldi D, Monaco A and Navacerrada M A 2005 *Supercond. Sci. Technol.* **18** L35

[445] Larbalestier D, Gurevich A, Feldmann D M, Polyanskii A 2001 *Nature* **414** 363

[446] Gurevich A and Pashitskii E A 1998 *Phys. Rev. B* **57** 13878

[447] Iijima Y, Tanabe N, Kohno O and Ikeno Y 1992 *Appl. Phys. Lett.* **60** 769

[448] Lee *et al* 2009 *Appl. Phys. Lett.* **95** 212505

[449] Browning N D, Buban J P, Nellist P D, Norton D P, Chisholm M F and Pennycook S J 1998 *Physica C* **294** 183

[450] Dhoot A S, Yuen J D, Heeney M, McCulloch I, Moses D, J. Heeger A 2006 *Proc. Natl. Acad. Sci. USA* **103** 11834

[451] Panzer M J, Frisbie C D 2006 *Adv. Funct. Mater.* **16** 1051

[452] Misra R, McCarthy M, Hebard A F 2007 *Appl. Phys. Lett.* **90** 052905
[453] Shimotani H, Asanuma H, Tsukazaki A, Ohtomo A, Kawasaki M, Iwasa Y 2007 *Appl. Phys. Lett.* **91** 082106
[454] Ueno K, Nakamura S, Shimotani H, Ohtomo A, Kimura N, Nojima T, Aoki H, Iwasa Y, Kawasaki M 2008 *Nat. Mater.* **7** 855
[455] Schooley J F, Hosier W R, Ambler E, Becker J H 1965 *Phys. Rev. Lett.* **14** 305
[456] Ueno K, Nakamura S, Shimotani H, Yuan H T, Kimura N, Nojima T, Aoki H, Iwasa Y, Kawasaki M 2011 *Nat. Nonotechnol.* **6** 408
[457] Ye J T, Inoue S, Kobayashi K, Kasahara Y, Yuan H T, Shimotani H, Iwasa Y 2010 *Nat. Mater.* **9** 125
[458] Taniguchi K, Matsumoto A, Shimotani H, Takagi H 2012 *Appl. Phys. Lett.* **101** 042603
[459] Ye J T, Zhang Y J, Akashi R, Bahramy M S, Arita R, Iwasa Y 2012 *Science* **338** 1193
[460] Bollinger A T, Dubuis G, Yoon J, Pavuna D, Misewich J, Božović I 2011 *Nature* **472** 458
[461] Dhoot A S, Wimbush S C, Benseman T, MacManus-Driscoll J L, Cooper J R, Friend R H, 2010 *Adv. Mater.* **22** 2529
[462] Leng X, Garcia-Barriocanal J, Bose S, Lee Y, Goldman A M 2011 *Phys. Rev. Lett.* **107** 027001
[463] Nojima T, Tada H, Nakamura S, Kobayashi N, Shimotani H, Iwasa Y 2011 *Phys. Rev. B* **84** 020502
[464] Katase T, Hiramatsu H, Kamiya, T, Hosono H 2014 *Proc. Natl Acad. Sci. USA* **111** 3979
[465] Selvamanickam V *et al* 2009 *IEEE Trans. Appl. Supercond.* **19** 3225
[466] Goyal A *et al* 1996 *Appl. Supercond.* **4** 403
[467] Tomsic M, Rindfleisch M, Yue J, McFadden K, Phillips J, Sumption M D, Bhatia M, Bohnenstiehl S and Collings E W 20027 *Int. J. Appl. Ceramic Technol.* **4** 250
[468] Iida K et al 2011 *Appl. Phys. Express* **4** 013103
[469] Miyata S, Ishimaru Y, Adachi S, Shimode T, Murai Y, Chikumoto N, Nakao K and Tanabe K 2013 *Extended Abstracts of International Workshop on Novel Superconductors and Super Materials 2013* (Tokyo) pp. 101
[470] Ishimaru Y, Miyata S, Adachi S, Shimode T, Murai Y, Chikumoto N, Nakao K and Tanabe K *in preparation*
[471] Tarantini C *et al* 2010 *Appl. Phys. Lett.* **96** 142519
[472] Si W, Han S J, Shi X, Ehrlich S N, Jaroszynski J, Amit Goyal A and Li Q 2013 *Nature Commun.* **4** 2337
[473] Yamamoto A et al 2008 *Appl. Phys. Lett.* **92** 252501
[474] Moore J D *et al* 2008 *Supercond. Sci. Technol.* **21** 092004
[475] Kametani F *et al* 2009 *Supercond. Sci. Technol.* **22** 015010
[476] Gao Z, Wang L, Qi Y, Wang D, Zhang X, Ma Y, Yang H and Wen H 2008 *Supercond. Sci. Technol.* **21** 112001
[477] Wang L, Gao Z, Qi Y, Zhang X, Wang D and Ma Y 2009 *Supercond. Sci. Technol.* **22** 015019
[478] Qi Y, Zhang X, Gao Z, Zhang Z, Wang L, Wang D and Ma Y 2009 *Physica C* **469** 717
[479] Mizuguchi Y, Deguchi K, Tsuda S, Yamaguchi T, Takeya H, Kumakura H and Takano Y 2009 *Appl. Phys. Express* **2** 083004
[480] Ma Y, Gao Z, Qi Y, Zhang X, Wang L, Zhang Z and Wang D 2009 *Physica C* **469** 651

[481] Wang L, Qi Y, Wang D, Zhang X, Gao Z, Zhang Z, Ma Y W, Awaji S, Nishijima G and Watanabe K 2010 *Physica C* **470** 183

[482] Wang L, Qi Y, Zhang Z, Wang D, Zhang X, Gao Z, Yao C and Ma Y W 2010 *Supercond. Sci. Technol.* **23** 054010

[483] Durrell J H, C-B Eom, Gurevich A, Hellstrom E E, Tarantini C, Yamamoto A and Larbalestier D C 2011 *Rep. Prog. Phys.* **74** 124511

[484] Wang L, Qi Y P, Zhang X P, Wang D L, Gao Z S, Wang C L, Yao C and Ma Y W 2011 *Pysica C* **471** 1689

[485] Fujioka M, Kota T, Matoba M, Ozaki T. Takano Y, Kumakura H and Kamihara Y 2011 *Appl. Phys. Express* **4** 063102

[486] Togano K, Matsumoto A and Kumakura H 2011 *Appl. Phys. Express* **4** 043101

[487] Matsumoto A, Togano K and Kumakura H 2012 *Supercond. Sc. Technol.* **25** 125010

[488] Togano K, Matsumoto A and Kumakura H 2012 *Solid State Communications* **152** 740

[489] Gao Z S, Wang L, Chao Y, Qi Y P, Wang C L, Zhang X P, Wang D L, Wang C D and Ma Y W 2011 *Appl. Phys. Lett.* **99** 242506

[490] Weiss J D, Tarantini C, Jiang J, Kametani F, Polyanskii A A, Larbalestier D C and Hellstrom EE 2012 *Nature Materials* **11** 682

[491] Gao Z S, Ma Y W, Yao C, Zhang X P, Wang C L, Wang D L, Awaji S and Watanabe K 2012 *Sci. Rep.* **2** 998

[492] Ma Y 2012 *Supercond. Sci. Technol.* **25** 113001

[493] Yao C, Lin H, Zhang X, Wang D, Zhang Q, Ma Y, Awaji S and Watanabe K 2013 *Supercond. Sci. Technol.* **26** 075003

[494] Togano K Gao Z S, Taira H, Ishida S, Iyo A, Kihou K, Eisaki H, Matsumoto A and Kumakura H 2013 *Supercond. Sci. Technol.* **26** 065003

[495] Togano K, Gao Z S, Matsumoto A and Kumakura H 2013 *Supercond. Sci. Technol.* **26** 115007

[496] Kumakura H, Matsumoto A, Fujii H and Togano K 2001 *Appl. Phys. Lett.* **79** 2435

[497] Matsumoto A, Gao Z, Togano K and Kumakura H 2014 *Supercond. Sci. Technol.* **27** 025011

[498] Ma Y W, Wang L, Qi Y P, Gao Z S, Wang D L and Zhang X P 2011 *IEEE Trans. Appl. Supercond.* 21 2878

[499] Flukiger R, Graf T, Decroux M, Groth C and Yamada Y 1991 *IEEE Trans. Magn.* **2** 1258

[500] Satou M, Yamada Y, Murase S, Kitamura T and Kamisada Y 1994 *Appl. Phys. Lett.* **64** 640

[501] Parrell J A, Dorris S E and Larbalestier D C 1994 *Physica C* **231** 137

[502] Parrell J A, Polyanskii A A, Pashitski A E and Larbalestier D C 1996 *Supercond. Sci. Technol.* **9** 393

[503] Gao Z S, Togano K, Matsumoto A and Kumakuta H 2014 *Scientific Reports* **4** 4065

[504] Kumakura H, Togano K, Maeda H, Kase J and Morimoto T 1991 *Appl. Phys. Lett.* **58** 2830

[505] Gao Z, Togano K, Matsumoto A and Kumakura H 2015 *Supercond. Sici. Technol.* **28** 012001

[506] Marti F, Grasso G, Huang Y B and Flükiger R 1998 *Supercond. Sci. Technol.* **11** 1251

[507] Kopera L, Kovác P and Husek I 1998 *Supercond. Sci. Technol.* **11** 433

[508] Hosono H and Kuroki K 2015 Physica C, in press (Review)

[509] Terasaki I, Sasago Y and Uchinokura K 1997 *Phys. Rev. B* **56** R12685

[510] Mittasch A, 1950, *Adv. Catal.* **2** 81

[511] Rao C.N.R and Rao G.R 1991 *Suf. Sci. Rep.* **13** 221-263

[512] Urabe K, Aika K and Ozaki A 1975 *J. Catal.* **38** 430

[513] TodaY, Yanagi H, Ikenaga E, Kim J-J, Kobata M, Ueda S, Kamiya T, Hirano M, Kobayashi K and Hosono H 2007 *Adv. Mat.* **19** 3564

[514] Matsuishi S, Toda Y, Miyakawa M, Hayashi K, Kamiya T, Hirano M, Tanaka I and Hosono H 2003 *Science* **301** 626

[515] Kim S-W and Hosono H 2012 *Philo. Mag.* **92** 2596

[516] Miyakawa M, Kim S-W, Hirano M, Kohama Y, Kawaji H, Atake T, Ikegami H, Kono K and Hosono H 2007 *J. Am. Chem. Soc.* **129** 7270

[517] Hosono H, Kim S-W, Matsuishi S, Tanaka, S Miyake A, Kagayama T and Shimizu:K 2015 *Phil. Trans. R. Soc. A* **373** 20140450

[518] Kitano M, Kanbara S, Inoue Y, Kuganathan N, Sushko P.V, Yokoyama T, Hara M and Hosono H 2015 *Nat. Commun.* **6** 7731

[519] Hayashi F, Toda Y, Kanie Y, Kitano M, Inoue Y, Yokoyama T, Hara M and Hosono H 2013 *Chem. Sci.* **4** 3124

[520] Nomura K, Ohta H, Ueda K, Kamiya T, Hirano M and Hosono H 2003 *Science* **300** 1269

[521] Nomura K, Ohta H, Takagi A, Kamiya T, Hirano M and Hosono H 2004 *Nature* **432** 488

[522] Hosono H 2006 *J. Non-Cryst. Sol.* **352** 851

[523] Ogo Y, Hiramatsu H, Nomura K, Yanagi H, Kamiya T, Hirano M, and Hosono H 2008 *Appl. Phys. Lett.* **93** 032113

[524] Forthaus M.K, Sengupta K, Heyer O, Christensen N.E, Svane A, Syassen K, Khomskii D.I, Lorenz T, and Abd-Elmeguid M.M 2010 *Phys. Rev. Lett.* **105** 157001

[525] Anderson, P. W. and Blount, E. I. 1995 *Phys. Rev. Lett.* **14** 217

[526] Boysen H and Altorfer, F. A 1994 *Acta Crystallogr.* **B50** 405

[527] Kurth F et al 2014 *presented in ASC2014* (Charlotte)

[528] Si W, Han S J, Shi X, Ehrlich S N, Jaroszynski J, Goyal A and Li Q 2013 *Nat. Commun.* **4** 1347

[529] Yanagisawa Y, Nakagome H, Uglietti D, Kiyoshi T, Hu R, Takematsu T, Takao T, Takahashi M and Maeda H 2010 *IEEE Trans. Appl. Supercond.* **20** 744

[530] Machi T, Nakao K, Kato T, Hirayama T and Tanabe K 2013 *Supercond. Sci. Technol.* **26** 105016

[531] Takematsu T, Hua R, Takao T, Yanagisawa Y, Nakagome H, Uglietti D, Kiyoshi T, Takahashi M, Maeda H 2010 *Physica C* **470** 674

[532] Aoki H and Hosono H 2015 *Physics World* **2015** Feb-31 (Review)

[533] Onari S and Kontani H 2013 *Phys. Rev. Lett.* **109** 137001

[534] Misawa T and Imada M 2014 *Nat. Commun.* **5** 5738

[535] Lee J J *et al* 2014 *Nature* **515** 245

**Table 1** Organization of FIRST Hosono Project

| | Team Leader | Mission |
|---|---|---|
| Exploration of Superconductors | Hideo Hosono (Core-Researcher) (Tokyo Institute of Technology) | Exploration of superconductors of transition metal compounds with layered structure collaborated with Minoru Nohara (Okayama Univ.)<br>Development of superconducting device by thin film technique<br>Development of novel functional materials (especially C12A7 electride) collaborated with Michikaze Hara (Tokyo Tech.), Tohru Setoyama (MCC), Alex Shluger (UCL) and Sung Wng Kim (SKKU) |
| | Eiji Takayama Muromachi (NIMS) | Exploration of superconductors using high pressure synthesis technique and their characterization |
| | Hiroshi Kageyama (Kyoto University) | Exploration of superconductors using low temperature synthesis technique and their characterization |
| | Shoji Yamanaka (Hiroshima University) | Exploration of superconductors with layered or clathrate structure and their characterization |
| Development of superconducting wire | Keiichi Tanabe (Sub-Core-Researcher) (ISTEC) | Development and evaluation of superconducting wire and superconducting device prepared by thin film technique |
| | Hiroshi Kumakura (NIMS) | Development and evaluation of superconducting wire prepared by PIT method collaborated with Yoichi Kamihara (Keio Univ.) |

NIMS: National Institute of materials Science, ISTEC: International Superconductivity Technology Center, MCC: Mitsubishi Chemical Corporation, UCL: University College London, SKKU: Sungkyunkwan University

**Table 2** Exploration of new superconductor in this project

(a) Materials which exibited superconductivity

[shaded]: noted in this review, (HP): synthesized using high pressure technique

| | Material | $T_c$ (K) | Comment |
|---|---|---|---|
| **IBSc** | | | |
| 1 | $SmFeAsO_{1-x}H_x$ [9] | 56 | large solubility limit of dopant $H^-$ (HP) |
| 2 | $NdFeAsO_{1-x}H_x$ [9] | 54 | large solubility limit of dopant $H^-$ (HP) |
| 3 | $CeFeAsO_{1-x}H_x$ [10] | 48 | large solubility limit of dopant $H^-$ (HP) |
| 4 | $LaFeAsO_{1-x}H_x$ [11] | 36 | two $T_c$ domes (HP) |
| 5 | $SmFeAs_{1-y}P_yO_{1-x}H_x$ [12] | <56 | $T_c$ decreased by replacing with P (HP) |
| 6 | $LaFe_{1-y}Zn_yAsO_{1-x}H_x$ [13] | <36 | $T_c$ decreased by replacing with Zn (HP) |
| 7 | $LaFeAsO_{0.85}H_x$ ($x$ =0-0.85) [14] | 35 | indirect electron doping by $V_O$ and $H^-$ (HP) |
| 8 | $LaFeAsO_{1-x}$ [15] | 22 | indirect electron doping by $V_O$ (HP) |
| 9 | $CaFe_{1-x}Co_xAsH$ [16] | 23 | direct electron doping into CaFeAsH (HP) |
| 10 | $Ca_{1-x}La_xFeAsH$ [17] | 47 | indirect electron doping into CaFeAsH (HP) |
| 11 | $Ca_{1-x}Sm_xFeAsH$ [17] | 46 | indirect electron doping into CaFeAsH (HP) |
| 12 | $CaFeAsF_{1-x}$ [18] | 27 | indirect electron doping (HP) |
| 13 | $Sr_{1-x}La_xFe_2As_2$ [19] | 23 | indirect electron doping (HP) |
| 14 | $(Ba,La)Fe_2As_2$ [20] | 22.4 | indirect electron doping (thin film) |
| 15 | $(Ba,Ce)Fe_2As_2$ [21] | 13.5 | indirect electron doping (thin film) |
| 16 | $(Ba,Pr)Fe_2As_2$ [21] | 6.1 | indirect electron doping (thin film) |
| 17 | $(Ba,Nd)Fe_2As_2$ [21] | 5.8 | indirect electron doping (thin film) |
| 18 | $(Sr,La)Fe_2As_2$ [22] | 20 | indirect electron doping (thin film) |
| 19 | $(Ca,La)Fe_2(As,P)_2$ [23] | 45 | bulk SC achieved by co-doping of La and P |
| 20 | $Ba(Fe,Pt)_2As_2$ [24] | 24 | dome-like relation between $T_c$ and dopant Pt |
| 21 | $(Ca,La)FeAs_2$ [25] | 34 | new type IBSc: 112 |
| 22 | $(Ca,La)Fe(As,Sb)_2$ [26] | 43 | increase in $T_c$ by co-doping |
| 23 | $(Ca,RE)Fe(As,Sb)_2$ [27] | 47 | highest $T_c$ in 112 type ($RE$=La, Ce, Pr, Nd) |
| 24 | $Ca_{10}(Ir_4As_8)(Fe_2As_2)_5$ [28] | 16 | analogous to $Ca_{10}(Pt_4As_8)(Fe_{2-x}Pt_xAs_2)_5$ |
| 25 | $Na_{0.65}Fe_{1.93}Se_2$ [29] | 37 | intercalating Na into FeSe |
| 26 | $(Na,NH_3)Fe_2Se_2$ [29] | 42 | co-intercalating Na and $NH_3$ into FeSe |
| 27 | $LaFeAs(O,C_2)$ | 27 | electron doping by $C_2$ (HP) |
| **Analogous structure to IBSc** | | | |
| | Material | $T_c$ (K) | Comment |
| 28 | $SrAl_2Si_2$ | 4.6 | IBSc 122 type structure (HP) |
| 29 | NbSiAs [30] | 8.2 | square net of Si |
| 30 | $BaTi_2Sb_2O$ [31] | 3 | square net of Ti (measured under HP) |
| 31 | $BaTi_2(Sb_{1-x}Bi_x)_2O$ [32] | 1.2 | square net of Ti, SC by doping of isovalent ion. |
| 32 | $BaTi_2(Sb_{1-x}Sn_x)_2O$ [33] | 2.5 | square net of Ti, SC by doping of aliovalent ion |
| 33 | $La(Co_{1-x}Fe_x)_2B_2$ [34] | 4 | square net of Co |
| 34 | $(La_{1-x}Y_x)Co_2B_2$ [34] | 4 | square net of Co |
| 35 | $BaNi_2(As,P)_2$ [35] | 3.3 | analogous to 122 type of IBSc |
| 36 | $Ba(Ni,Cu)_2As_2$ | 3.3 | analogous to 122 type of IBSC |
| 37 | $LnNiAsO_{1-x}H_x$ [36] | 3.7 | H doping to LnNiAsO ($Ln$=La-Nd) (HP) |
| 38 | $NdNi_{0.64}Bi_2$ | 4 | square net of Ni and Bi |
| 39 | $YNi_xBi_2$ [37] | 4 | square net of Ni and Bi |
| 40 | $LaNi_xBi_2$ [37] | 4 | square net of Ni and Bi |
| 41 | $CeNi_xBi_2$ [37] | 4 | square net of Ni and Bi |
| 42 | LaNiBN [38] | 4.1 | square net of Ni (HP) |
| 43 | CaNiBN [38] | 2.2 | square net of Ni (HP) |
| 44 | LaPtBN [38] | 6.7 | square net of Pt (HP) |
| 45 | $La_3Ni_2B_2N_3$ [38] | 15 | square net of Ni (HP) |
| 46 | $LaPd_2As_2$ [39] | 1 | collapsed 122 structure |
| 47 | $LaPd_2Sb_2$ [40] | 1.4 | $CaBe_2Ge_2$ structure |
| 48 | $SrPt_2Sb_2$ [41] | 2.1 | $CaBe_2Ge_2$ structure |
| 49 | $BaPt_2Sb_2$ [42] | 1.9 | $CaBe_2Ge_2$ structure |
| 50 | $SrPd_2Bi_2$ | 2.2 | $CaBe_2Ge_2$ structure |
| 51 | $CaPd_2Bi_2$ | 2.6 | $CaBe_2Ge_2$ structure |
| 52 | $SrPt_2Bi_2$ | 2.6 | $CaBe_2Ge_2$ structure |
| 53 | $La_2Sb$ [43] | 5 | square net of La |
| **Other layered structure** | | | |
| | Material | $T_c$ (K) | Comment |
| 54 | $La_3Ru_8B_6$ [44] | 3.2 | homologous structure |
| 55 | CrNbN | 11 | Cr layer (multiphase) |
| 56 | $Ca_2Al_3Si_4$ [45] | 6.4 | high pres. phase of Ca-Al-Si (HP) |
| 57 | $Mg_4AlSi_3$ [46] | 5.2 | new layered structure (HP) |
| 58 | $Sr(Si_{1-x}Ni_x)_2$ | 3 | Si layer |
| 59 | $Ba(Si_{1-x}Cu_x)_2$ | 3 | Si layer |
| 60 | $LaSi_2H_{0.03}$ | 3 | H doping to $ThSi_2$ structure ($T_c$ increased) |
| 61 | $BaGe_2H_{0.27}$ | 4 | H doping to $ThSi_2$ structure ($T_c$ decreased) |
| 62 | $LaGe_{1.7}$ | 2 | $ThSi_2$ structure with cage formed by $sp^2$ orbital |
| 63 | $Zr_2Ru_3Si_4$ [47] | 5.7 | analogous structure to ZrRuP superconductor |
| 64 | MgPtSi [48] | 2.5 | honeycomb lattice |
| 65 | $SrAuSi_3$ [49] | 1.54 | noncentrosymmetric structure |
| 66 | $Li_2IrSi_3$ [50] | 3.8 | noncentrosymmetric structure |
| 67 | $LaIr\mathit{Pn}$ [51] | 5.3 | noncentrosymmetric structure ($\mathit{Pn}$ = P and As) |
| 68 | LaRhP [51] | 2.5 | noncentrosymmetric structure |
| 69 | $Bi_4O_4S_3$ [52] | 4.5 | double BiS plane |
| 70 | $\mathit{Ln}O_{1-x}F_xBiS_2$ ($\mathit{Ln}$=La,Ce,Nd) [53] | 3 (10) | triple BiS plane. (measured under HP) |
| 71 | $KMo_6Se_8$ | 9 | new Chevrel phase compound |
| 72 | $Nb_5Ir_3O$ | 10.5 | O doped $Mn_5Si_3$ structure |
| 73 | $Ba_2(Bi_{1-x}Sb_x)_3$ [54] | 4.3 | square/honeycomb lattice |
| 74 | $Y_2CrC_3$ | 1.5 | Cr/C layer |
| 75 | $SrNiSn_3$ | 5 | noncentrosymmetric structure |
| 76 | β-ZrNF | 15 | ion exchange of β-ZrNCl by F |
| 77 | $Zr_5(N_4O)F_6$ | 2-5 | low SVF |
| Intercalation to layered structure | | | |
| | Material | $T_c$ (K) | Comment |
| 78 | TiNCl← amines [55] | 17 | intercalation of long alkylene diamine →high $T_c$ |
| 79 | TiNCl← alkali metal (Li-Rb) [56] | 18.0 | highest $T_c$ by Na intercalation |
| 80 | TiNCl← alkali metal + solvent [56] | 10.5 | highest $T_c$ by Na+THF co-intercalation |
| 81 | TiNBr← alkali metal [57] | 17.2 | highest $T_c$ by K intercalation |
| 82 | ZrNCl← K + THF, PC [58] | 16 | highest $T_c$ by K+THF co-intercalation |
| 83 | α-ZrNCl← alkali metal | 10 | superconductivity of α-type structure |
| 84 | HfNCl← *AE*(Ca-Ba), solvent [59] | 26 | highest $T_c$ by Ca+THF co-intercalation |
| 85 | *M*NCl←Yb, Eu (*M*: Zr, Hf) [60] | 24.3 | highest $T_c$ by Eu+$NH_3$ co-intercalation into HfNCl |
| 86 | $Hf_2N_2O$←Li | 4.6 | $La_2O_3$ type structure |
| Chalcogenides | | | |
| | Material | $T_c$ (K) | Comment |
| 87 | 1T-$TaS_2$ | 3 | suppressing CDW (HP) |
| 88 | $AuTe_2$ [61] | 2.3 | SC by dissociation of $Te_2$ dimer ($CdI_2$ structure) |
| 89 | $(Au,Pt)Te_2$ [62] | 4 | SC by dissociation of $Te_2$ dimer ($CdI_2$ structure) |
| 90 | $(Au,Pd)Te_2$ | 3 | SC by dissociation of $Te_2$ dimer ($CdI_2$ structure) |
| 91 | $(Ir,Pt)Te_2$ [63] | 3.1 | SC by dissolving of orbital ordering ($CdI_2$ structure) |
| 92 | $Cu_xIrTe_2$ | 3 | SC by dissolving of orbital ordering ($CdI_2$ structure) |
| 93 | $(Ir,Rh)Te_2$ [64] | 2.6 | SC by dissolving of orbital ordering ($CdI_2$ structure) |
| 94 | $Ir_xSe_2$ [65] | 8 | pyrite structure (HP) |
| 95 | $Ir_{1-x}Rh_xSe_2$ [66] | 10 | pyrite structure (Rh doping) (HP) |
| 96 | $Ir_{0.95-x}Rh_xTe_2$ [67] | 4.6 | pyrite structure (small effect of Rh doping) (HP) |
| Others | | | |
| | Material | $T_c$ (K) | Comment |
| 97 | $NbBe_2$ | 2.6 | Laves phase |
| 98 | $MoC_{1-x}$ [68] | 12 | formed $MoC_{0.681}$ under 6 GPa |
| 99 | $MoC_{0.75}$ [69] | 13 | formed $MoC_{0.75}$ under 17 GPa |
| 100 | $YFe_2SiC$ | 3.5 | having $YFe_2Si$ framework |
| 101 | $ScC_xO_y$ | 8 | having C-C bonding (low SVF) |
| 102 | $Ti_{4-x}N_3$ | <5 | decomposition of TiNCl in $NH_3$ |
| 103 | $Zr_{4-x}N_3$ | <10 | decomposition of ZrNCl in $NH_3$ |
| 104 | $Hf_{4-x}N_3$ | <6 | decomposition of HfNCl in $NH_3$ |
| 105 | $Mg(Mg_{1-x}Al_x)Si$ [70] | 6.2 | anticotunnite or TiNiSi structure (HP) |
| 106 | $Ca(Al, Si)_2$ [71] | 2.6 | Laves phase (HP) |
| 107 | $Nb_4MSi$ (M=Ni, Co, Fe) [72] | 7.7 | 2 dimensional network of Nb |
| 108 | $Ba_3Ir_4Ge_{16}$ [73] | 6.1 | aving cage structure |
| 109 | $Ba_4Ir_8Ge_{28}$ [73] | 3.2 | having cage structure |
| 110 | $Ca_2InN$ [74] | 0.6 | doping of $In^{-1}$ into $Ca_2N$ |
| 111 | SnAs | 2 | hole in As4*p* orbital |
| 112 | $CuZr_2$ | 1 | |
| 113 | $HfZr_2$ | 1 | |

(b) Materials which exhibited no superconductivity

: possible to obtain the target phase, : impossible to obtain the target phase

| | | | | |
|---|---|---|---|---|
| IBSc (Square net of Fe) | | | | |
| MgFeAsH | SrFeAsH | BaFeAsH | EuFeAsH | MgFePH |
| CaFePH | $CaFeAsF_{1-x}$ | $CaFeAsF_{1-x}O_x$ | $CaFeAsF_{1-x}H_x$ [9] | $Ca_{1-x}La_xFeAsF$ |
| $Sr_{1-x}La_xFeAsF$ | $Ca_{1-x}Na_xFeAsH$ | LaFeAsO:N | $CaFeAsH_{1-x}O_x$ | LaFeOSb |
| CaFeOSe | SrFeOSe | KFeSeF | LaFeSiF | $Ba(Fe,Re)_2As_2$ |
| $Ca(Fe,Au)_2As_2$ | $Ca(Fe,Mg)_2As_2$ | $Ca(Fe,Al)_2As_2$ | $Ca(Fe,Pt)_2As_2$ [75] | $Ca(Fe,Ru)_2As_2$ |
| $(Ca,La)Fe_2As_2$ | $YFe_{2-x}Co_xGe_2$ | $BaFe_{2-x}Zn_xAs_2$ | $BaFe_2Sb_2$ | $BaFe_2Se_2$ |
| $(Fe,Mn)_2AlB_2$ | $(Fe,Co)_2AlB_2$ | $(Fe,Ni)_2AlB_2$ | $(Fe,Cr)_2AlB_2$ | $ZrFe_2Si_2$ |
| $ZrFe_2B_2$ | $LuFe_{2-x}Si_2$ | $YFe_{2-x}Si_2$ | $TlFe_{2-x}Co_xSe_2$ | $TlFe_{2-x}Ni_xSe_2$ |
| $Tl_{1-x}Fe_2Se_2$ | $MgFe_2As_2$ | $CdFe_2As_2$ | $YFe_2B_2$ | $Fe_{2-x}Cu_xAs$ |
| (Fe,Pt)Se | Fe(Te,As) | Fe(Se,As) | $FeP_{0.5}X_{0.5}0.5$ ($X$ = Cl, Br) | $FeSe_{1-x}X_x$ ($X$ = P, Cl) |
| $FeSb_{1-x}$ | La- $Ca_{10}(Ir_4As_8)(Fe_2As_2)_5$ | Ge- $Ca_{10}(Ir_4As_8)(Fe_2As_2)_5$ | P- $Ca_{10}(Ir_4As_8)(Fe_2As_2)_5$ | Re- $Ca_{10}(Ir_4As_8)(Fe_2As_2)_5$ |
| K- $Ca_{10}(Ir_4As_8)(Fe_2As_2)_5$ | Sr- $Ca_{10}(Ir_4As_8)(Fe_2As_2)_5$ | Ba- $Ca_{10}(Ir_4As_8)(Fe_2As_2)_5$ | $Ca_{10}(M_4As_8)(Fe_2As_2)_5$ ($M$=Pd,Rh) | $Ba_{10}(Pt_4As_8)(Fe_2As_2)_5$ |
| $Sr_{10}(Pt_4As_8)(Fe_2As_2)_5$ | $Ca_{10}Pt_{5.8}Fe_{8.2}Sb_{18}$ | $Ba_2InO_3FeAs$ | $Sr_2InO_3FeAs$ | $Sr_2FeO_3FeAs$ |
| $Ba_2FeO_3FeAs$ | $Sr_4V_2O_5Fe_2As_2$ | $Bi_2SrFe_2O_4Se_2$ | $La_3O_2Fe_4As_4$ | $Ca_3F_2Fe_4As_4$ |
| $La_2O_3Fe_2(S_{1-x}F_x)_2$ | $La_2O_3Fe_2(Se_{1-x}F_x)_2$ | $La_2O_3Fe_2(Se_{1-x}As_x)_2$ | $La_2O_3Fe_2(Se_{1-x}Ge_x)_2$ | $La_2O_3(Fe_{1-x}Co_x)_2Se$ |
| $La_2O_3(Fe_{1-x}Mn_x)_2Se$ | $NdFe_xSb_2$ | $Fe_{1-x}Cu_xAs_{0.5}$ | $EuFeAsO_{1-x}$ | $Fe_{1-x}Si_2$ |
| $(Fe_{1-x}Mn_x)_{1-y}Si_2$ | $(Fe_{1-x}Co_x)_{1-y}Si_2$ | $(Fe_{1-x}Mo_x)_{1-y}Si_2$ | AgFeAs | $La(Fe_{1-x}Co_x)_2Si_2$ |
| $Fe_{0.82}(Si_{1-x}Ge_x)_2$ | $RFeC_2$ | $Sc(Fe,Co)C_2$ | $Sc(Fe,Ni)C_2$ | $La(Fe,Co)Si_2$ |
| $La(Fe,Mn)Si_2$ | $LaFe_{0.5}Sb_2$ | $CaFe_4As_3$ | $CaFe_3AgAs_3$ | $CaFe_3CuAs_3$ |
| $CaFe_3LiAs_3$ | $La_3O_4Fe_4As_3$ | $Fe(Se,As)_2$ | $(Fe,Mn)As_2$ | $(Fe,Ru)As_2$ |
| $(Fe,Rh)As_2$ | $(Fe,Co)As_2$ | $(Fe,Ni)As_2$ | $(Fe,Pd)As_2$ | $La_{3-x}RE_xCu_2O_{6-y}Fe_2As_2$ |
| $Sr_3Fe_2Cu_2Se_2O_5$ | $Sr_{2.6}K_{0.4}Fe_2Cu_2Se_2O_5$ | $La_3O_4Fe_4As_4$ | $Ca_3Fe_4As_4F_3$ | $Ca_2FeOsO_6$ |
| $Ba_{1-x}K_xFe_2X$ ($X$ = S, Se) | $NaFe_2O_4$ | $BaFe_{2-x}Co_xSe_3$ | $Tl_{1-x}Ba_xFeSe_2$ | $La_2Fe_2Se_2O_{3-x}H_x$ |
| $La_2Fe_2S_2O_{3-x}H_x$ | CaBaFe2Pn2O (Pn = As Sb, Bi), | $SrFe_2Pn_2O$ ($Pn$ = As, Sb, Bi) | $BaFe_2Pn_2O$ ($Pn$ = As, Sb, Bi) | $BaFe_2Se_2O$ |
| $t$-FeS | | | | |
| Analogous structure to IBSc material | | | | |
| Square net of transition metal elements | | $BaSc_2Pn_2O$ ($Pn$=As, Sb, Bi) | $SrTi_2As_2O$ | $BaTi_2(As,Sb)_2O$ [31] |
| $EuTi_2As_2O$ | $BaTi_2Sb_2O$ | $(SrF)_2Ti_2Bi_2O$ [76] | $(SrF)_2Ti_2(Sb_{1-x}Bi_x)_2O$ | $Zr_2Ti_2As_2H$ |
| $Zr_2Ti_2Pn_2H$ ($Pn$ = Sb, Bi) | $ZrTiPn$ ($Pn$ = Sb, Bi) | ZrTiAs | $BaV_2Ge_2O$ | $BaV_2Sn_2O$ |
| $BaV_2Pn_2O$ ($Pn$ = As, Sb, Bi) | $LnCrAsO$ ($Ln$=La-Nd, Sm) [77] | SrCrAsF | EuCrAsF | $La_{1-x}CaMnAsO_{1-y}F_y$ |
| $Ce_{1-x}CaMnAsO_{1-y}F_y$ | $Sm_{1-x}CaMnAsO_{1-y}F_y$ | SrMnAsF | BaMnAsF | EuMnAsF |
| $LaMnAsO_{1-x}H_x$ | $CeMnAsO_{1-x}H_x$ | $DyMnAsO_{1-x}H_x$ | $LaZn_{1-x}Mn_xAsO_{0.75}$ | $CaMn_2As_2$ |
| $CaMn_2Sb_2$ | $CaMnBi_2$ | $(Ca,La)Mn_2Sb_2$ | $BaMn_2As_2$ | $(Ba,K)Mn_2As_2$ |
| $BaMnRuAs_2$ | $BaMnFeAs_2$ | $BaMnCoAs_2$ | $BaMnRhAs_2$ | $BaMnIrAs_2$ |
| $BaMnBi_2$ | $LaMn_2Si_2$ | $Sr(Mn,Fe)_2P_2$ | $Sr(Mn,Cr)_2P_2$ | $Sr(Mn,Al)_2P_2$ |
| $Sr(Mn,Zn)_2P_2$ | $(Ca,Eu)Mn_2P_2$ | $La_2O_2Mn_2Te_2O$ | $CaMn_2Pn_2O$ ($Pn$ = As, Sb, Bi) | SrMn2$Pn$2O ($Pn$ = As, Sb, Bi) |
| $BaMn_2Pn_2O$ ($Pn$ = As, Sb, Bi) | $BaMn_2Se_2O$ | $Ca_{10}(Pt_4As_8)(Mn_{2-x}Pt_xAs_2)_5$ | $LaCoAsO_{1-x}H_x$ | $(Ca,La)CoAs_2$ |
| $(Sr_{1-x}Ca_x)Co_2Ge_2$ | $BaCo_2Ge_2$ | $YCo_2B_2$ | $YCo_2Ge_2$ | $LaCo_{0.6}Sb_2$ |
| $LaCo_2Si_2$ | $La(Co_{1-x}Ni_x)_2Si_2$ | $LaCo_2(B_{1-x}Si_x)_2$ [37] | $LaCo_2Ge_2$ | $NdCo_xSb_2$ |
| $GdCo_2B_2$ | $LaCo_2B_2C_x$ | $ZrCo_2Si_2$ | $HfCo_2Si_2$ | $HfCo_2Ge_2$ |
| $CaCo_2Pn_2O$ ($Pn$ = As, Sb, Bi) | $SrCo_2Pn_2O$ ($Pn$ = As, Sb, Bi) | $BaCo_2Pn_2O$ ($Pn$ = As, Sb, Bi) | $Ca_{10}(Pt_4As_8)(Co_{2-x}Pt_xAs_2)_5$ | $(Ca,La)NiAs_2$ |
| CaNiAsH | $Ba_{1-x}La_xNi_2As_2$ | MgNiGe [78] | $CaNi_{1-x}Mn_xGe$ [79] | $CaNi_{1-x}Mn_xGeH$ [80] |
| $Y_3NiSi_3$ | LaNiAs | $La_3NiSi_3$ | $CeNiBiO_{1-x}$ | $ZrNi_{0.75}P$ |
| $ZrNi_{0.75}As$ | $KNi_2Se_2$ | $ZrNi_2Si_2$ | RhNiP | $GdNi_xBi_2$ |
| $Ni_3(Te_{1-x}Se_x)_2$ | $CaNi_2Pn_2O$ ($Pn$ = As, Sb, Bi) | $SrNi_2Pn_2O$ ($Pn$ = As, Sb, Bi) | $BaNi_2Pn_2O$ ($Pn$ = As, Sb, Bi) | $Ca_{10}(Pt_4As_8)(Ni_{2-x}Pt_xAs_2)_5$ |
| $Cu_2As$ | $Cu_2Sb$ | MgCuAs | $Ca_2Cu_6P_5$ | $BaCu_{1.8}As_2$ |
| $BaCu_2As_2$ | $Ba_2Cu_3P_4$ | $BaCu_6As_2$ | $YCuSb_2$ | $YCuAs_2$ |
| CeCuPO | $GdCu_{1.25}P_{1.75}$ | $ZrCu_2P_2$ | $ZrCu_2As_2$ | ZrCuSiAs |
| $HfCu_2P_2$ | $Hf_{0.5}Cu_{1-x}Mn_xP$ | $HfCuSi_2$ | $HfCuGe_2$ | BiOCuS |
| $Ca_2Cu_2Pn_2O$ ($Pn$ = As, Sb, Bi) | $Sr_2Cu_2Pn_2O$ ($Pn$ = As, Sb, Bi) | $Ba_2Cu_2Pn_2O$ ($Pn$ = As, Sb, Bi) | $La_3(O,F)_2Cu_4As_4$ | $Bi_2LnO_4Cu_{2-x}Fe_xSe_2$ ($Ln$=lamtanide elements) |
| $BaZnBi_2$ | $K_2NbO_{3-x}F_{1+x}$ | $(Ca,La)RuAs_2$ | $La_3Ru_2B_2N_3$ [38] | LaRuBN [38] |
| LaRhBN [38] | (La, Sr)$Pd_2As_2$ | LaIrBN [38] | (Sr,K)$Pt_2As_2$ | (Ca,La)$PtAs_2$ |
| $[Sr_2MO_2][Ti_2As_2O]$, $[Sr_3M_2O_5][Ti_2As_2O]$ ($M$ = Cr, Mn, Fe, Co, Ni, Cu) | | | | |
| Square net of traditional elements | | TiSiAs | $SrGa_2$ | $BaGa_2$ |
| NbGeAs | TaGeAs | $BaNiSn_3$ | $LaMn_xSb_2$ | $La_2O_2Sb$ |
| $Ce_2O_2Sb$ | $CeAgSb_2$ | $Nd((Fe_{1-x}Co_x)_ySb_2$ | $La_2O_2Te$ | $Y_2O_2Bi$ [81] |
| $La_2O_2Bi$ [81] | $LaAgBi_2$ | $Ce_2O_2Bi$ [81] | $CeMn_xBi_2$ | $CeZn_xBi_2$ |
| $CeCu_xBi_2$ | $CeAgBi_2$ | $Pr_2O_2Bi$ [81] | $Nd_2O_2Bi$ [81] | $Sm_2O_2Bi$ [81] |
| $Eu_2O_2Bi$ | $Gd_2O_2Bi$ | $Er_2O_2Bi$ [81] | $Yb_2O_2Bi$ | $Y_5Si_3$ |
| $La_5Si_3$ | | | | |
| Other layered structure | | | | |
| Honeycomb lattice | | $Mg_2PtSi$ | SrPdAs | (Sr,Y)PtAs |
| SrPtSb | BaPtSb | SrPdSb | YPtAs | LaPtSb |
| MgAgAs | SrCoP | SrNiP | SrCuP | SrPdP |
| SrAgP | SrIrP | SrAuSn | SrPtSn | SrAuSb |
| CaAuP | SrAuP | SrAuSi | | |
| $Mn_5Si_3$ structure | | $V_5P_3$ | $V_5P_3N_x$ | $V_5As_3N$ |
| $Nb_5Si_3$ | $Ta_5Si_3$ | $V_5Ge_3$ | $Nb_5Ge_3N_x$ | $Ta_5Ge_3N_x$ |
| $Nb_5Ge_3C_x$ | $Zr_5Pb_3$ | $Ta_5Ir_3O$ | | |
| $Mg_4AlSi_3$ structure | | $LiMg_4Si_3$ | $NaMg_4Si_3$ | $SrMg_4Si_3$ |
| $BaMg_4Si_3$ | $YMg_4Si_3$ | $BaGe_3Mg_4$ | | |
| B, Si layer | | $CrB_2$ | $MnB_2$ | $FeB_2$ |
| $ZnB_2$ | $SrNiSi_3$ | $Ba(Si_{1-x}Zn_x)_2$ | $Ba(Si_{1-x}Ni_x)_2$ | $SiTe_2$ |
| $SmGa_{2-x}Si_x$ | CaCuSi | $YbGa_{0.9}Si_{1.1}$ | $LaGa_{2-x}Si_x$ | $Lu_2AlSi_2$ |
| $CaSrSi_4$ | $Ba_3Ga_{0.7}Si_{4.3}$ | $Sr_{1-x}Na_xAl_2Si_2$ | $Ba_{1-x}K_xAl_2Si_2$ | $Ca_{1-x}Na_xAl_2Si_2$ [45] |
| $SrAl_{2-x}Si_{2+x}$ | $BaSi_2$ (HP phase) | | | |
| Skutterdite structure | | $Sr_xRhGe_{1.5}Se_{1.5}$ | $Ba_xIrGe_{1.5}S_{1.5}$ | $Ba_xIrGe_{1.5}Se_{1.5}$ |
| $BaRh_2Ge_4S_6$ [82] | $BaRh_2Ge_4Se_6$ [82] | $BaIr_2Ge_4S_6$ [82] | $BaIr_2Ge_4Se_6$ [82] | $La_xIrGe_{1.5}S_{1.5}$ |
| $La_xRhGe_{1.5}S_{1.5}$ | $LaFe_4Bi_{12}$ | $LaCo_4Bi_{12}$ | $LaNi_4Bi_{12}$ | $LaRu_4Bi_{12}$ |
| $LaRh_4Bi_{12}$ | $LaPd_4Bi_{12}$ | $LaIr_4Bi_{12}$ | $LaPt_4Bi_{12}$ | |
| Others | | $Y(Co,Fe)_3B_2$ | $Y(Co,Ni)_3B_2$ | $Ca_2N$ [83] |
| $Ca_xBN$ (film) [84] | $Ba(Ge_{1-x}Cu_x)_2$ | $Cr_2N$ | CuNCN | NbCrN |
| TiNF | $Ti_2SbP$ | $Zr_2SbP$ | $Cr_2Ti$ | $Cr_2Zr$ |
| $Cr_2Hf$ | CrTaN | $NiTe_2$ | $Ni(Te_{1-x}Se_x)_2$ | $NaNiF_3$ [85] |
| SmNiAs | ZrAs | $Ni_xBi_3$ | $CaRuO_3$ [86] | $CaRhO_3$ [87] |
| $CaOsO_3$ | $SrOsO_3$ | $BaOsO_3$ | $CaRhO_3$ (HP phase) [88] | $MAN$ phase ($M$=Ti,V,Zr,Nb,Hf,Ta) ($A$=Si,P,S,Ga,In,Sn,Pb) |
| $La_3Ni_2O_7$ | $Sr_2SnO_{4-x}$ | $Ba_2SnO_{4-x}$ | SnO: F | $Cu(Al,Mg)O_2$ |
| $SrCr_2O_4$ | $LaSrMnO_3H$ | $(CuCl)LaNb_2O_{7-x}F_x$ | $Bi_2WO_6$ | $Ho_{0.25}Sr_{0.75}FeO_{2+d}$ |
| $Na_2La_2Ti_3O_{10-x}$ | $Sr_2TiO_3F_2$ | $CaCu_3Cr_4O_{12}$ | $SrCuO_2$: F | |
| Intercalation to layered structure | | | | |
| FeOCl | VOCl | TiOCl | CrOCl | TiNCl←mono-amine [55] |
| $Zr_2N_2S$ [89] | $Hf_2N_2S$ [89] | $Ca_xHf_2N_2S$ | α-HfNBr | $Ti_2PTe_2$ ← Ag, Cu, Zn [90] |
| $Bi_2PdO_4F_x$ | $Bi_{1.67}Pb_{0.33}PtO_4F_x$ | $Pb_3O_4$ ← F | $Zr_5(N_4O)F_{6-x}$ | $Ti_2N_2S$ |
| $Ti_2N_2O$ | $Hf_5(N,O)_5F_{6-x}$ | $Zr_2N_2O$ | Graphite← KH | graphite← B |
| $CsLaTe_2O_5Cl$ | | | | |
| Perovskite and its related structure | | | | |
| anti-Perovskite structure | | $Mn_3AlC$ | $Mn_3SnC$ | $Mn_3AlN$ |
| $Mn_2CrN$ | $Mn_3ZnN$ [91] | $AlNi_3C$ | $Mn_3ZnN_{1-x}C_x$ | $Mn_3InN_{1-x}C_x$ |
| $Mn_3Sb_{1-x}Sn_xN$ | $Mn_3Ag_{1-x}Sn_xN$ | $Ca_3SiO$ | $Ca_3SnO$ | $Ca_2PbO$ |
| $SrAg_3P$ | $SrIr_3P$ | $SrAu_3P$ | $Mg_2CNi_4$ | $Ca_4OSb_2$ |
| $RhNCr_3$ | | | | |
| double Perovskite structure | | $Ca_3OsO_6$ | $Ca_2MnOsO_6$ | $Ca_2CoOsO_6$ |
| $Ca_2ZnOsO_6$ | $Sr_3OsO_6$ | $Sr_2LiOsO_6$ [92] | $Sr_2MnOsO_6$ | $Sr_2FeOsO_6$ [93] |
| $Sr_2CoOsO_6$ [93] | $Sr_2NiOsO_6$ | $Sr_2YOsO_6$ | $Sr_2PbOsO_6$ | $Sr_2DyOsO_6$ |
| $Ba_2(Sr_2,Ca_2)FeOsO_6$ | $Ba_2YScOsO_6$ | $Sr_4FeMoO_8$ | | |
| normal Perovskite | | $SrGeO_3$ [94] | $(Ca_{1-x}Y_x)GeO_3$ | $(Sr_{1-x}La_x)GeO_3$ [94] |
| $CaTiO_{3-x}H_x$ | $SrTi(O,H)_3$ (thin film) | $EuTiO_{3-x}H_x$ | $CaVO_3$:H | $SrVO_3$:H |
| | [95] | | | |
| $SrCrO_2H$ [96] | $Sr_{1-x}Ba_xCrO_2H$ ($x$ = 0.1-0.4) | $LnCrO_3$ ($Ln$ = La, Sm, Nd) | $BaTiO_{3-x}$ | $BaZrO_3$:H |
| $SrSnO_{3-x}$ | $SrNbO_2N$ | $BaSnO_{3-x}$ | $NaIrO_3$ | $Na_{1-x}Ca_xIrO_3$ |
| $SrMoO_{2.5}N_{0.5}$ | $BaBiS_3$ | | | |
| other Perovskite | | $A_2Ru_2O_5$ | $LaBaMn_2O_6$ | $YBaMn_2O_6$ |
| $A_2Cr_2O_5$ | | | | |
| cuprate superconductor and its related structure | | | | |
| $(Sr,Eu)CuO_2$ | $(Sr,Pb)CuO_2$ | $Bi_2CuO_4$_$CaH_4$ | $Bi_2CuO_4$-CuF2 | BiOCuS |
| YBCO:H | $(Cu_2S_2)Sr_3Fe_2O_5$ | $Sr_3FeNiCu_2Se_2O_5$ | $Sr_3Ni_2Cu_2Se_2O_5$ | $Sr_3Cu_4Se_2O_5$ |
| $Bi_2YCu_2O_4Se_2$ | $Bi_2SrCu_2O_4Se_2$ | $SrLa_2Cu_4Se_2O_5$ | $A_2CuO_x(OH)_y$ | $LaSrTiO_4$ |
| $Sr_2CrO_4$ | $Ca_2CrO_4$ | $Sr_3Cr_2O_7$ | $Ca_3Cr_2O_7$ | $A_4Cr_3O_{10}$ ($A$ = Sr, Ca) |
| $Sr_{2-x}Na_xCrO_4$ | $YCaCrO_4$ | $Srn_{1-x}K_xMn_nO_{3n+1}$ ($n$ = 1, ∞) | | |
| clathrate structure | | | | |
| $Ca_3Ni_{12}B_6$ | $Sr_3Ni_{12}B_6$ | $KC_{24}$ | $Li_xC_{60}$ | $K_xC_{60}$ |
| $Ca_xC_{60}$ | $Ba_xC_{60}$ | $Mg_2C_{60}$ | $Ba_3C_{60}$ [97] | $Ca_xAlN$ |
| $Li_xSi_{136}$ | $Na_{32}Si_{136}$ [98] | $K_xSi_{136}$ | $Mg_xSi_{136}$ | $Ca_xSi_{136}$ |
| $Ba_xSi_{136}$ | | | | |
| Pyrite structure | | | | |
| $FeS_{2-x}$ | $PtSb_2$ | $(Pt,Pd)Sb_2$ | $(Pt,Ir)Sb_2$ [99] | $Pt(Sb,As)_2$ |
| $Pt(Sb,Te)_2$ | $Pt(Sb,Sn)_2$ | $(Pt,Au)Sb_2$ | $IrSb_2$ | |
| other materials | | | | |
| carbide | | $CaC_2$ | $BaC_2$ | $YC_2$ |
| nitride | | $Ti_3N_4$ | $Zr_3N_4$ | $Hf_3N_4$ |
| VCrN | $Mg_xAlN$ | MnN (thin film) | CoN (thin film) | |
| oxide | | $TiO_x$ (anatase) [100] | $Cr_3O$ | $Cr_nO_{2n-1}$ ($n$ = 3, 4) |
| $MnO_2$ (nanosheet) | $Co_3O_4$: H | NiO: H | $GeO_2$ (HP phase) | h-$WO_3$: H |
| $BaScO_2H$ | $Ti_3O_{4.9}N_{0.1}$ | $Ti_4O_{6.99}N_{0.01}$ | $(Ir,Ru)O_2$ | $CuAl(O,N)_2$ |
| $LaSiO_3H$ | $NaTi_2O_4$ | $Bi_4Ti_3O_{12}$: H | $Sc_4Ti_3O_{12}$ | $Cs_{0.7}Ti_{1.825}O_4$: F |
| $La_2Ti_2O_{7-x}$ | Ba-Ti-O glass + e | $Y_2Ti_2O_{2-x}$ | $Sr_2TiO_3F_2$ | $Cr_{1-x}Ti_xO_2$ |
| $Cr_{1-x}V_xO_2$ | $n$SrO・$V_2O_3$ ($n$= 4, 2, 1) | $Bi_xV_8O_{16}$ | $CaV_2O_5$ | $LiV_2O_4$ |
| $LaVO_3$: H | $LaVO_4$: H | $NaCr_2O_4$ [101] | $Ca_{1-x}Na_xCr_2O_4$ | $CaCr_2O_4$ |
| $BaCr_2O_4$ | $SrCu_3Cr_4O_{12}$ | $BaCu_3Cr_4O_{12}$ | $YCu_3Cr_4O_{12}$ | $LaCu_3Cr_4O_{12}$ |
| $CeCu_3Cr_4O_{12}$ | $LuCu_3Cr_4O_{12}$ | $EuCu_3Cr_4O_{12}$ | $BiCu_3Cr_4O_{12}$ | $Cr_{1-x}Mn_xO_2$ |
| $Ag_2MnO_2$ | $Pb_2MnO_4$: F | $LiMn_2O_4$ | $Sr_3Co_2O_6$ [102] | $Ba_3Co_2O_6(CO_3)_x$ |
| $Sr_{3-x}Y_xCo_2O_6$ [103] | $Sr_{3-x}Ca_xCo_2O_6$ | $Sr_3Co_{2-x}Zn_xO_6$ | $Sr_3Co_{2-x}Fe_xO_6$ | Y-$ZrO_{2-x}$ |
| $SrNb_2O_6$: F | $Ba_5Nb_4O_{15-x}$ | $Ba_2Nb_5O_9$: F | $PbMoO_4$: N | $Ca_2Nb_2O_{6-x}F_{1+x}$ |
| $Sr_2NbO_3N$ | $Li_2Ir_{1-x}Ru_xO_3$ [104] | $La_4Ru_2O_{10}$ | $Bi_3Ru_3O_{11}$ | $Sr_3Ru_2O_7C_{12}$ |
| $Sr_3Ru_2O_7F_2$ | $LaCu_3Ru_4O_{12}$ | $CdRh_2O_4$ [105] | $Cd_{1-x}Na_xRh_2O_4$ | $La_2Pd_2O_5$ |
| $BaSbO_3$ | $CsW_2O_6$ | $SrWO_2N$ | $LiOsO_3$ [106] | $Na_{1-x}Ca_xOsO_3$ |
| $KOsO_3$ | $Ca_3LiOsO_6$ [107] | $Ca_2Os_2O_7$ | $Ca_5Os_2O_{12}$ | $Ba_2OsO_5$ |
| $Bi_2OsO_6$ | $Pb_2Os_2O_7$ | $Pb_2Ir_2O_7$ | $Ag_2BiO_3$ | $Ba_{1-x}K_xBiO_3$ (junction) |
| chalcogenide | | $Cu_xS$ | Cu(S,As) | $KMo_3S_3$ |
| $SrPt_3S$ | $BaPt_3S$ | $InV_6Se_8$ | $Pt_xBi_2Se_3$ | $KMo_3Se_3$ |
| $Cu_xBi_2Se_3$ | $Cu_xBi_2Te_3$ | $Pd_xBi_2Te_3$ | $Li_xZrTe_3$ | SnTe |
| pnictide | | CoP | NiP | (Co,Ni)P |
| $La_xIr_4P_{12}$ | $Mg_xAlP$ | $BaNi_2P_4$ | $BaPd_2P_4$ | $(Y_{1-x}Eu_x)P$ |
| SrPt3Sb | $BaPt_3Sb$ | $La_3TiSb_5$ | $La_3VSb_5$ | $La_3CrSb_5$ |
| $La_3MnSb_5$ | $La_3TiBi_5$ | $La_3VBi_5$ | $La_3CrBi_5$ | $La_3ZtBi_5$ |
| $La_3HfBi_5$ | | | | |
| silicide, germanide, stannide, plumbide | | (Ca, Mg)$Si_2$ | $Fe_3Al_2Si_4$ | $MgAl_2Si_2$ |
| $Mg_4AlSi_6$ | $Mg_{1.3}Ga_{0.7}Si$ | $Li_xPtSi_3$ | $BaMg_2Si_2$ | $Ti_2Ru_3Si_4$ |
| BaPdSi | SrPdSi | $BaMg_2Ge_2$ | $Rh_3Ge_7$ | $BaMg_2Sn_2$ |
| $BaMg_2Pb_2$ | | | | |
| halide | | $K_{1+x}TiF_4$ | $K_{1-x}TiF_4$ | $Zr_5(N_4Na)F_6$ |
| $Zr_5(N_4H-)F_6$ | $CsSnI_3$ | $La_2TeI_2$ | | |
| metal | | $ZrMn_2$ | $Zr_5Pb_4$ | |

**Table 3** Summary of electromagnetic properties of La$T_M$$Pn$O
($T_M$: 3d transition metal, and $Pn$=P or As)

| | | | | | | | | | | | | | | | | | |
|---|---|---|---|---|---|---|---|---|---|---|---|---|---|---|---|---|---|
| H(+) | | | | | | | | | | | | | | | | H(-) | He |
| Li | Be | | | | | | | | | | | B | C | N | O | F | Ne |
| Na | Mg | | | | | | | | | | | Al | Si | P | S | Cl | Ar |
| K | Ca | Sc | Ti | V | **Cr** | **Mn** | **Fe** | **Co** | **Ni** | **Cu** | **Zn** | Ga | Ge | As | Se | Br | Kr |

| $T_M^{2+}$ (electron configuration) | **Cr(3d[4])** | | **Mn(3d[5])** | | **Fe(3d[6])** | | **Co(3d[7])** | | **Ni(3d[8])** | | **Cu** | **Zn(3d[10])** | |
|---|---|---|---|---|---|---|---|---|---|---|---|---|---|
| *Pn* | P | As | P | As | P | As | P | As | P | As | P, As | P | As |
| Elect. Prop. | | Metal | Mott Insulator | | Superconductor | | Metal | | Superconductor | | | Semiconductor | |
| Magnetism | | AF(CB) | AF(CB) | | | | FM | | | | | Non-magnetic | |
| $E_g$ | | - | ~1 eV | | - | | - | | - | | | ~1.5 eV | |
| $T_c$(SC) | | | | | undoped 4 K | F-doped 26 K | | | undoped 3 K | undoped 2.4 K | | | |
| $T_{NC}$(Mag) | | > 300K | > 400K | | | | 43K | 66K | | | | | |
| Ref. | 77 | | 131 | | 3, 4 | | 132 | | 5, 133 | | | 134, 135 | |

: impossible to synthesize, CB: checker board type

**Table 4** Typical example of emergence of superconductivity by doping ($T_c$ and composition)

| | 1111 | | 122 | |
|---|---|---|---|---|
| doping type | indirect | direct | indirect | direct |
| electron | 55 K ($SmFeAsO_{0.9}F_{0.1}$) [115]<br>55 K ($SmFeAsO_{0.85}$) [117]<br>55 K ($SmFeAsO_{0.8}H_{0.2}$) [9]<br>56 K ($Gd_{0.8}Th_{0.2}FeAsO$) [118] | 14 K ($LaFe_{0.89}Co_{0.11}AsO$) [180]<br>22 K($CaFe_{0.9}Co_{0.1}AsF$) [137]<br>6 K ($LaFe_{0.96}Ni_{0.04}AsO$) [182]<br>18 K ($SmFe_{0.89}Ir_{0.11}AsO$) [183] | 45 K ($Ca_{0.83}La_{0.17}Fe_2(As,P)_2$ [23]<br>22 K ($Sr_{0.6}La_{0.4}Fe_2As_2$) [19]<br>22 K ($Ba_{0.93}La_{0.07}Fe_2As_2$) [20] | 22 K ($BaFe_{1.8}Co_{0.2}As_2$) [186]<br>19 K ($BaFe_{1.908}Ni_{0.092}As_2$) [187]<br>24 K ($BaFe_{1.886}Rh_{0.114}As_2$) [188]<br>18 K ($BaFe_{1.914}Pd_{0.086}As_2$) [188] |
| hole | 13 K(?) ($Nd_{0.8}Sr_{0.2}FeAsO$) [185] | no | 38 K ($Ba_{0.6}K_{0.4}Fe_2As_2$) [119] | no |
| isoelectronic | no | no | no | 30 K ($BaFe_2(As_{0.68}P_{0.32})_2$) [189]<br>22 K ($Ba(Fe_{0.56}Ru_{0.44})_2As_2$) [190] |

**Table 5** Properties of high $T_c$ superconductors

| | IBSc's | $MgB_2$ | Cuprates |
|---|---|---|---|
| parent material | antiferromagnetic metal ($T_N$~150 K) (excepting for the 245-type) | non-magnetic metal | antiferromagnetic insulator ($T_N$~400 K) |
| Orbitals composing Fermi level | five Fe3*d* orbitals | two B2*p* orbitals | single Cu3*d* orbital |
| $T_c$ | 56 K ($Gd_{0.8}Th_{0.2}SmFeAsO$) [118]<br>38 K ($Ba_{0.6}K_{0.4}Fe_2As_2$) [119] | 39 K (pure $MgB_2$) [196] | 92 K ($YBa_2Cu_3O_{7-\delta,}$) [199]<br>105 K ($Bi_2Sr_2Ca_2Cu_3O_{10}$) [200]<br>134 K ($HgBa_2Ca_2Cu_3O_8$) [201] |
| $H_{c2}(0)^{//ab*}$ | ~90 T (($Ba_{0.55}K_)Fe_2As_2$) [193]<br>~150 T ($SmFeAsO_{0.7}F_{0.25}$) [193] | ~40 T ($MgB_{1.87}C_{0.13}$) [197]<br>~18 T (pure $MgB_2$) [197] | 210 T ($YBa_2Cu_3O_{7-\delta,}$) [202]<br>400 T ($Bi_2Sr_2CaCu_2O_8$) [202] |
| $\gamma_\rho=\rho(c)/\rho(ab)$ | 3-5 ($Ba(Fe_{0.936}Co_{0.074})_2As_2$) [194]<br>8-10 ($SmFeAsO_{0.7}F_{0.25}$) [94] | ~3.5 (pure $MgB_2$) [198] | 30-250 ($YBa_2Cu_3O_{7-\delta,}$) [203-205]<br>>1000 ($Bi_2Sr_2CaCu_2O_8$) [206] |
| $\gamma_H=H_{c2}(0)^{//ab}/H_{c2}(0)^{//c}$ | 1.1-1.2 (($Ba_{0.55}K_)Fe_2As_2$) [193]<br>~1.5 ($SmFeAsO_{0.7}F_{0.25}$) [195] | ~4 ($MgB_{1.87}C_{0.13}$) [197]<br>~6 (pure $MgB_2$) [197] | 6 ($YBa_2Cu_3O_{7-\delta,}$) [202]<br>21 ($Bi_2Sr_2CaCu_2O_8$) [202] |

*$H_{c2}(0)^{//ab}$: estimated by extrapolating the $H_{c2}$-T curve, where $H_{c2}$ is measured by applying a magnetic field along the *ab* plane.

**Table 6** Characteristics of two domes in $LaFeAsO_{1-x}H_x$ [11]

| Dome | First | Second |
|---|---|---|
| $x$ | $0.05 \leq x \leq 0.2$ | $0.2 \leq x \leq 0.5$ |
| Exponent, $n$ | $0.0 \leq n \leq 2.3$ | $0.7 \leq n \leq 2.0$ |
| $T_c^{max}$ | 29 K | 36 K |
| $T_c$ sensitivity to $x$ | High | Low |
| Under high pressure | Unified | Unified |
| FS nesting between hole and electron pockets | Strong | Weak |
| DOS($E_F$) | No shoulder | Shoulder |

The exponent ($n$) was estimated from the curve fitting of $\rho$-$T$ plots to $\rho=AT^n+\rho_0$ near $T_c$.

**Table 7** The maximum superconducting transition temperature $T_c$ (K) of transition metal (TM)-doped $AE(Fe_{1-x}TM_x)_2As_2$, where AE = Ca, Sr and Ba; and TM = Co, Ni, Ru, Rh, Pd, Ir and Pt. For instance, $T_c$ = 20 K for $Ca(Fe_{1-x}Co_x)_2As_2$. Pt-doped $Ca(Fe_{1-x}Pt_x)_2As_2$ does not exhibit superconductivity [75].

| | Fe | Co | Ni |
|---|---|---|---|
| Ca | | 20 [213] | 15 [214] |
| Sr | | 19.2 [215] | 9.8 [216] |
| Ba | | 23 [186, 193] | 20.5 [217] |
| | Ru | Rh | Pd |
| Ca | | 14 [218] | 10 [219] |
| Sr | 19.3 [220] | 21.9 [221] | 8.7 [221] |
| Ba | 22 [190] | 23.2 [188] | 19 [188] |
| | | Ir | Pt |
| Ca | | 22 [222] | no SC [75] |
| Sr | | 24.2 [221] | 16.5 [223] |
| Ba | | 28 [224] | 23 [225] |

**Table 8** The maxim superconducting transition temperature $T_c$ (K) of the alkali metal (A)-doped $(AE_{1-x} A_x)Fe_2As_2$, where $AE$ = Ca, Sr and Ba; and $A$ = Na, K, Rb and Cs.

| | Na | K | Rb | Cs |
|---|---|---|---|---|
| Ca | 26 [226] | | | |
| Sr | 26 [227] | 36.5 [228, 229] | | 37.2 [228] |
| Ba | 34 [230] | 38 [119] | 23 [231] | |

**Table 9** The maximum superconducting transition temperature $T_c$ (K) of rare-earth ($RE$) doped $(AE_{1-x} RE_x)Fe_2As_2$, where $AE$ = Ca, Sr and Ba; and $RE$ = La, Ce, Pr and Nd.

| | La | Ce | Pr | Nd |
|---|---|---|---|---|
| Ca | 45 [23] | 37 [234] | 49 [236] | |
| Sr | 22 [19] | | | |
| Ba | 22.4 [20] | 13.4 [21] | 6.2 [21] | 5.8 [21] |

**Table 10** The superconducting transition temperature $T_c$ (K) of RE-doped $(Ca_{1-x}RE_x)FeAs_2$ and $RE$- and Sb-doped $(Ca_{1-x}RE_x)Fe(As_{1-y}Sb_y)_2$ as determined by magnetic measurements.

| $RE$ | $(Ca_{1-x}RE_x)FeAs_2$ | $(Ca_{1-x}RE_x)Fe(As_{1-y}Sb_y)_2$ |
|---|---|---|
| La | 34 [25] | 47 [27] |
| Ce | | 43 [27] |
| Pr | 20 [239] | 43 [27] |
| Nd | 11.9 [240] | 43 [27] |
| Sm | 11.6 [240] | |
| Eu | 9.3 [240] | |
| Gd | 12.6 [240] | |

**Table 11** Intercalated FeSe Superconducting phases

| | NH3-rich | | NH3-poor | | | | NH3-free | |
|---|---|---|---|---|---|---|---|---|
| | | | Metal-poor | | Metal-rich | | | |
| | *d* | $T_c$ | *d* | $T_c$ | *d* | $T_c$ | *d* | $T_c$ |
| $Li^+$ | ~9.0 | 39 | - | - | ~8.3 | 44 | - | - |
| $Na^+$ | ~11.1 | 42 | - | - | ~8.7 | 45 | ~6.8 | 37 |
| $K^+$ | ~10.2 | ? | ~7.8 | 44 | ~7.4 | 30 | 7.14 | ~44 |
| $AE^{2+}$ | - | - | 8.0-8.4 | 35-40 | ~10.3 | 38-39 | - | - |
| $RE^{2+}$ | - | - | ~8.1 | 42 | ~10.2 | 40-42 | - | - |

"-": none, "?": unknown

The separation of nearest Fe layers d (Å) and $T_c$ (K) of intercalated $(A/AE/RE)_x(NH_3)_y(NH_2)_zFe_2Se_2$ (*A*: alkali metals, *AE*: alkali earth metals, *RE*: rare earth metals) superconductors synthesized by the ammonothermal method shown in literatures [29, 258-261].

**Table 12** Characteristic superconducting parameters of the α- and the β-structured layered nitride superconductors [57]

| Compound | $T_c$ (K) | *d* (Å) | $\xi_{ab}$ (Å ) | $\xi_c$ (Å ) | $\lambda_{ab}$ (Å) | γ | Reference |
|---|---|---|---|---|---|---|---|
| α-$K_{0.21}$TiNBr | 17.2 | 9.5 | 53 | 41 | 3045 | 1.3 | [57] |
| α-$Na_{0.16}$TiNCl | 18.1 | 8.4 | 33 | 28 | 4746 | 1.2 | [56] |
| α-$Na_{0.16}(THF)_y$TiNCl | 10.2 | 13.1 | 55 | 35 | - | 1.5 | [56] |
| β-$ZrNCl_{0.7}$ | 13 | 9.8 | 71 | 16 | - | 4.5 | [314] |
| β-$Li_{0.48}(THF)_y$HfNCl | 25.5 | 18.7 | 60 | 16 | 4630 | 3.7 | [313] |
| β-$Eu_{0.08}(NH_3)_y$HfNCl | 24.3 | 11.9 | 61 | 15 | - | 4.1 | [60] |
| β-$Ca_{0.11}(THF)_y$HfNCl | 26.0 | 15.0 | 47 | 12 | - | 4.1 | [59] |

**Table 13** Superconducting and lattice parameters for the LaMX compounds [51].

| Compound | LaIrP | LaIrAs | LaRhP |
|---|---|---|---|
| Superconducting parameters | | | |
| $T_C$(K) | 5.3 | 3.1 | 2.5 |
| $H_{C2}$(0) (WHH) (kOe) | 13.8 | 5.5 | 2.1 |
| $H_{C2}$(0) (GL) (kOe) | 16.4 | 6.4 | 2.7 |
| $\xi_{GL}$(0) (Å) | 14.2 | 22.7 | 34.9 |
| $\Delta C(T_C)/\gamma T_C$ | 0.65 | 0.84 | 0.73 |
| $\lambda_{ep}$ | 0.67 | 0.58 | 0.52 |
| Lattice parameters (tetragonal $I4_1md$) | | | |
| a (Å) | 4.2065 | 4.1505 | 4.1846 |
| b (Å) | 14.9379 | 14.3277 | 14.9358 |

**Table 14.** Superconducting properties of representative iron-based superconductors.

| Material | $T_c$ (K) | $\mu_0 H_{c2}^{//c}(0)^*$ ( T ) | $\gamma_H$ |
|---|---|---|---|
| *R*FeAs(O,F) (*R* = Nd,Sm) | 47-55 | 80-100 | 5-10 |
| $(Ba,K)Fe_2As_2$ | 37-38 | 70-135 | 1.5-2 |
| $Ba(Fe,Co)_2As_2$ | 22-25 | 47-50 | 1.5-1.9 |
| $BaFe_2(As,P)_2$ | 30-31 | ~ 60 | 1.5-1.9 |
| Fe(Se,Te) | 14-16 | ~ 50 | 1.1-1.9 |